\documentclass[11pt, oneside]{book}
\usepackage[a4paper, margin=2.5cm]{geometry}
\usepackage{fancyhdr}
\usepackage{titlesec}
\usepackage{setspace}

\fancypagestyle{plain}{%
  \fancyhf{}
  \fancyfoot[C]{--\thepage--}

}

\titleformat{\chapter}[display]
  {\normalfont\huge\bfseries}
  {Chapter~\thechapter}{1ex}
  {\titlerule\vspace{1ex}}
  [\vspace{1ex}\titlerule]

\usepackage{fontspec}      
\usepackage{polyglossia}   
\setdefaultlanguage{english}
\setotherlanguage{polish}
\usepackage[stable]{footmisc}
\newcommand{\breakslash}{\slash\hspace{0pt}}

\usepackage[normalem]{ulem}
\usepackage{enumitem}
\usepackage{graphicx}
\usepackage{xcolor}
\usepackage[hidelinks]{hyperref}

\usepackage{physics}
\usepackage{amsmath,amssymb}
\usepackage{tikz-feynman}
\tikzfeynmanset{compat=1.1.0}
\tikzfeynmanset{warn luatex=false} 
\tikzset{
    cross/.style={path picture={\draw[black]
        (path picture bounding box.south east) -- (path picture bounding box.north west)
        (path picture bounding box.south west) -- (path picture bounding box.north east);}}
}

\usepackage{listings}
\newcommand{\sumint}{\mathop{\mathchoice
  {\ooalign{$\displaystyle\sum$\cr\hidewidth$\displaystyle\int$\hidewidth\cr}}
  {\ooalign{$\textstyle\sum$\cr\hidewidth$\textstyle\int$\hidewidth\cr}}
  {\ooalign{$\scriptstyle\sum$\cr\hidewidth$\scriptstyle\int$\hidewidth\cr}}
  {\ooalign{$\scriptscriptstyle\sum$\cr\hidewidth$\scriptscriptstyle\int$\hidewidth\cr}}
}\nolimits}

\makeatletter
\newsavebox\myboxA
\newsavebox\myboxB
\newlength\mylenA

\newcommand*\xoverline[2][0.75]{%
    \sbox{\myboxA}{$\m@th#2$}%
    \setbox\myboxB\null
    \ht\myboxB=\ht\myboxA%
    \dp\myboxB=\dp\myboxA%
    \wd\myboxB=#1\wd\myboxA
    \sbox\myboxB{$\m@th\overline{\copy\myboxB}$}
    \setlength\mylenA{\the\wd\myboxA}
    \addtolength\mylenA{-\the\wd\myboxB}%
    \ifdim\wd\myboxB<\wd\myboxA%
       \rlap{\hskip 0.5\mylenA\usebox\myboxB}{\usebox\myboxA}%
    \else
        \hskip -0.5\mylenA\rlap{\usebox\myboxA}{\hskip 0.5\mylenA\usebox\myboxB}%
    \fi}
\makeatother
\newcommand{\nn}{\nonumber \\}

\newcommand{\pprime}{\prime\prime}

\newcommand{\veva}{v^a}
\newcommand{\vJtilde}{\tilde{v} _\rmii{$\tilde{J}$} }
\newcommand{\be}{\begin{equation}}
\newcommand{\ee}{\end{equation}}
\newcommand{\Seff}{S_{\text{eff}}}
\newcommand{\pathD}[1]{\mathcal{D} #1 }
\newcommand{\intx}{\int_\textbf{x}}
\newcommand{\intr}{\int_\textbf{r}}
\newcommand{\SLO}{S^{\rmii{LO}}}
\newcommand{\SNLO}{S^{\rmii{NLO}}}
\newcommand{\field}{\varphi}
\newcommand{\fpisq}{(4\pi)^2}

\newcommand{\Tint}[1]{{\hbox{$\sum$}\!\!\!\!\!\!\!\int\,}_{\!\!\!\!\raise-0.9ex\hbox{$\scriptstyle{#1}$}}}
\newcommand{\Tinti}[1]{{{\Sigma}\!\!\!\!\raise0.3ex\hbox{$\int$}_\rmii{${#1}$}}}
\newcommand{\Tintip}[1]{{{\Sigma'}\!\!\!\!\!\raise0.3ex\hbox{$\int$}_\rmii{${#1}$}}}

\newcommand{\phithree}{\phi_\rmii{3}}
\newcommand{\lamthree}{\lambda_\rmii{3}}
\newcommand{\Lambdam}{\Lambda_\rmii{m}}

\newcommand{\SthreeSoft}{S_3 ^{(\rmii{soft})}}
\newcommand{\Xzero}{X_\rmi{0}}
\newcommand{\Xspat}{X_\rmi{$i$}}
\newcommand{\muthree}{\mu_\rmi{3}}

\newcommand{\gammaE}{\gamma_E}
\newcommand{\vthree}{v_\rmii{3}}
\newcommand{\vbreak}{v_\rmii{break}}
\newcommand{\vbarrier}{v_\rmii{barrier}}

\newcommand{\gxthree}{g_\rmii{$X$,3}}
\newcommand{\hthree}{h_\rmii{3}}
\newcommand{\kappathree}{\kappa_\rmii{3}}

\newcommand{\mXthree}{m_{\rmii{$X$,3}}}
\newcommand{\mXtemporal}{m_{\rmii{$X_0$,3}}}
\newcommand{\ZX}{Z_{\rmii{$X$,3}}}
\newcommand{\ZXtemporal}{Z_{\rmii{$X_0,3$}}}

\newcommand{\VLOvectors}{V^\rmii{LO} _\rmii{nucl,$V$}}
\newcommand{\VLOspatial}{V^\rmii{LO} _\rmii{nucl,$X$} }
\newcommand{\VLOtemporal}{V^\rmii{LO} _{\rmii{nucl,$X_0$ } }}

\newcommand{\CX}{C_\rmii{$X$} ^\rmii{LO}}
\newcommand{\CXtemporal}{C_\rmii{$X_0$} ^\rmii{LO} }
\newcommand{\Cvec}{C_{\rm{v}} ^\rmii{LO}}

\newcommand{\VLO}{V^{\rmii{LO}}_\rmii{nucl} }
\newcommand{\VDRNLO}{V^{\rmii{NLO}}_\rmii{nucl} }
\newcommand{\SDRLO}{S^{\rmii{LO}}_\rmii{nucl} }
\newcommand{\SDRNLO}{S^{\rmii{NLO}}_\rmii{nucl} }

\newcommand{\Snucl}{S_\rmii{nucl}}
\newcommand{\GammaT}{\Gamma_\rmii{$T$}}
\newcommand{\Astat}{A_\rmii{stat}}
\newcommand{\Adyn}{A_\rmii{dyn}}
\newcommand{\phib}{\phi_\rmii{b}}
\newcommand{\varphibLO}{\varphi_\rmii{b} ^{\rmii{LO}}}
\newcommand{\varphib}{\varphi_\rmii{b} } 
\newcommand{\rtail}{r_{\rmii{t}}}
\newcommand{\varphiF}{\varphi_{\rmii{F}}}
\newcommand{\Ainf}{A_\rmii{$\infty$} }

\newcommand{\rmdet}{\mathrm{det}}
\newcommand{\detOS}{\det\mathcal{O}_{\rmii{$\phi$}} }
\newcommand{\detOSp}{\mathrm{det}^{\prime}\mathcal{O}_{\rmii{$\phi$}}  }
\newcommand{\detOX}{\det\mathcal{O}_\rmii{$X$} }
\newcommand{\detOXzero}{\det\mathcal{O}_{\rmii{$X_0$}} }
\newcommand{\detS}{\mathrm{det}_{\rmii{$S$}} }
\newcommand{\detX}{\mathrm{det}_{\rmii{$X$}} }
\newcommand{\detXzero}{\mathrm{det}_{\rmii{$X_0$}} }
\newcommand{\detV}{\mathrm{det}_{\rm{v}} }
\newcommand{\aS}{a_{\rmii{$S$}}}
\newcommand{\aL}{a_{\rmii{$L$}}}
\newcommand{\aT}{a_{\rmii{$T$}}}
\newcommand{\aG}{a_{\rmii{$G$}}}
\newcommand{\aGfv}{a_{\rmii{$G$;F}}}
\newcommand{\aSfv}{a_{\rmii{$S$;F}}}
\newcommand{\aLfv}{a_{\rmii{$L$;F}}}
\newcommand{\TS}{T_{\rmii{$S$}}}
\newcommand{\TL}{T_{\rmii{$L$}}}
\newcommand{\TG}{T_{\rmii{$G$}}}

\newcommand{\yS}{y_{\rmii{$S$}}}
\newcommand{\yL}{y_{\rmii{$L$}}}
\newcommand{\yG}{y_{\rmii{$G$}}}

\newcommand{\OpX}{\mathcal{O}_\rmii{$X$} }
\newcommand{\OpXT}{\mathcal{O}_{\rmii{$X_T$}} }
\newcommand{\Opg}{\mathcal{O}_{\rm{g}} }
\newcommand{\OpXG}{\mathcal{O}_{\rmii{$XG$}}}
\newcommand{\OpXzero}{\mathcal{O}_{\rmii{$X_0$}} }
\newcommand{\OpS}{\mathcal{O}_{\rmii{$\phi$}}}

\newcommand{\WX}{W_\rmii{$X$}}
\newcommand{\WG}{W_\rmii{$G$}}

\newcommand{\hrefNLOdet}{\hyperref[it:NLO-det]{[NLO~det]}}
\newcommand{\hrefdaisy}{\hyperref[it:daisy]{[daisy]}}
\newcommand{\hrefNLOgrad}{\hyperref[it:NLO-grad]{[NLO~$\nabla$]}}

\newcommand{\mX}{m_\rmii{$X$}}
\newcommand{\mx}{m_\rmii{$X$}}
\newcommand{\gxbar}{\overline{g}_\rmii{$X$} }
\newcommand{\mxbar}{\overline{m}_\rmii{$X$} }
\newcommand{\lamh}{\lambda_h}
\newcommand{\lamphi}{\lambda_\varphi}
\newcommand{\lamphih}{\lambda_{\varphi h}}
\newcommand{\mufour}{\mu_\rmii{$4$}}

\newcommand{\mF}{m_\rmii{F}}
\newcommand{\mD}{m_\rmii{D}}

\newcommand{\MH}{M_\rmii{$H$}}

\newcommand{\MZ}{M_\rmii{$Z$}}

\newcommand{\mthree}{m_\rmii{$3$}}

\newcommand{\mxzero}{m_\rmii{$X_0$}}
\newcommand{\MX}{M_\rmii{$X$}}

\newcommand{\gx}{g_\rmii{$X$}}
\newcommand{\gX}{g_\rmii{$X$}}

\newcommand{\Tc}{T_\text{c}}
\newcommand{\Tv}{T_\rmii{$V$}}
\newcommand{\Tst}{T_*}
\newcommand{\Tn}{T_\text{n}}
\newcommand{\Tp}{T_\text{p}}
\newcommand{\Treh}{T_\text{reh}}
\newcommand{\Tshrink}{T_\rmii{shrink}}
\newcommand{\RH}{R_*H_*}

\newcommand{\PLO}{P_{1 \rightarrow 1}}
\newcommand{\PLL}{P_{1\rightarrow N}}
\newcommand{\kcol}{\kappa_\rmii{col}}

\newcommand{\Tqcd}{T_\rmii{QCD}}
\newcommand{\SUTwoX}{\mathrm{SU}(2)_\rmii{X}}

\newcommand{\lb}{\lambda_2}
\newcommand{\lc}{\lambda_3}
\newcommand{\g}{\,\mathrm{GeV}}

\newcommand{\f}{\varphi}

\newcommand{\mDX}{m_{\rmii{D},\rmii{$X$}}}

\newcommand{\rmi}[1]{{\mbox{\scriptsize #1}}}
\newcommand{\rmii}[1]{{\mbox{\tiny\rm{#1}}}}

\usepackage[font=it]{caption}

\usepackage[
    backend=biber,
    style=phys,
    maxbibnames=5,
    articletitle=true,
    biblabel=brackets,
    eprint=true,
    sorting=none,
    defernumbers=true
]{biblatex}
\usepackage{xurl}
\begin{document}

\frontmatter

\begin{titlepage}
    \begin{center}
        
        {\Huge
            Theoretical consistency and phenomenology of supercooled cosmological phase transitions
        }
            
        \vspace{1.cm}
        
        {\LARGE Maciej Kierkla}
        
        \vspace{1.5cm}
        
        \includegraphics[width=0.6\linewidth]{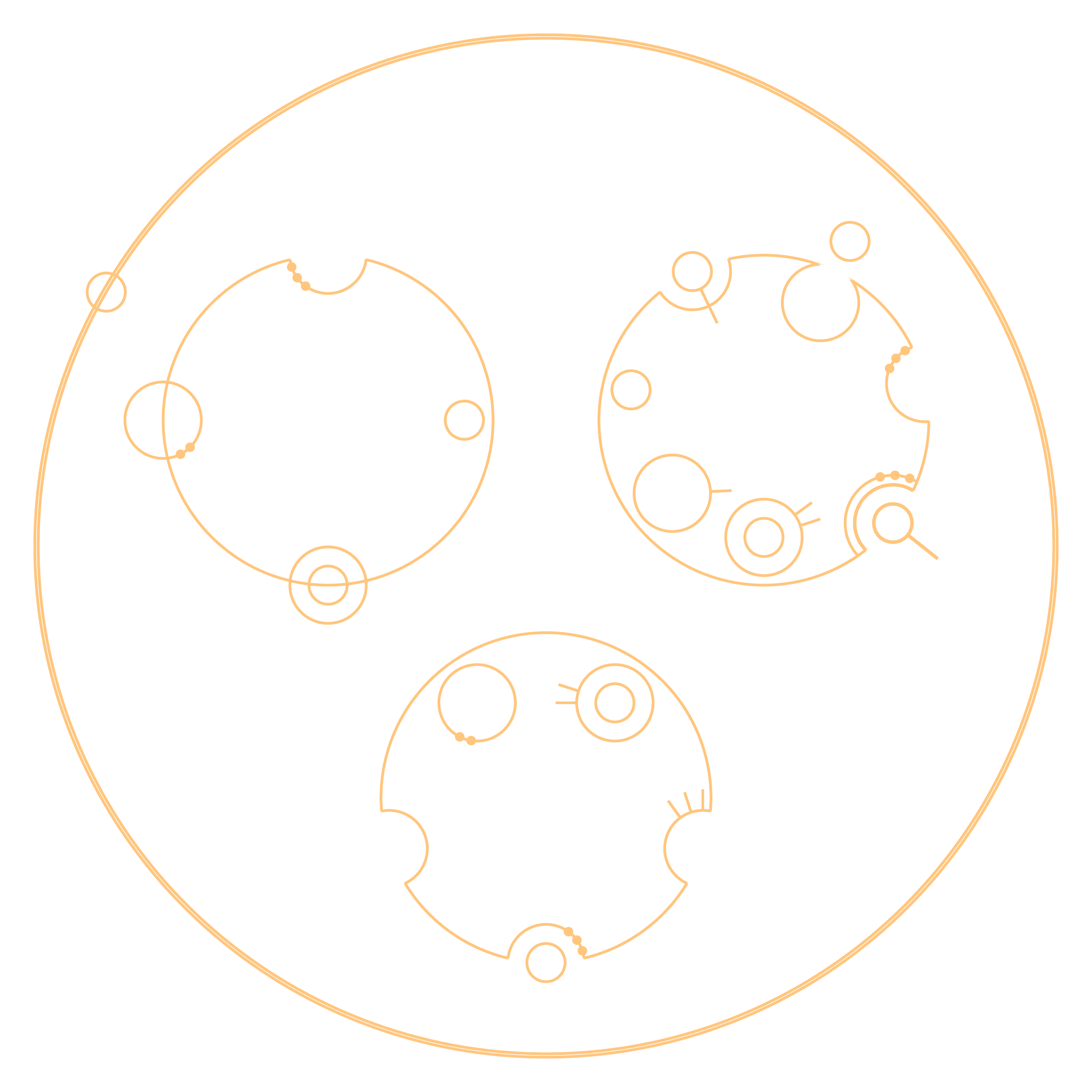}
        \vspace{2cm}
        
        Doctoral dissertation prepared under the supervision of \\
        \textbf{ prof. dr hab. Zygmunt Lalak, dr Bogumiła Świeżewska}
            
        \vspace{0.8cm}
     

        \vspace{0.8cm}
        
        Institute of Theoretical Physics, Faculty of Physics\\
        University of Warsaw

        \vspace{1.5cm}
        November 2025

    
   \end{center}
\end{titlepage}

\chapter*{Acknowledgements}

First and foremost, I want to thank \textit{Zygmunt Lalak }and \textit{Bogumiła Świeżewska} for their supervision and collaboration. I am very grateful for the opportunity to pursue my PhD studies in the topic of cosmological phase transitions. I would also like to thank you for the support and understanding throughout my studies at the University of Warsaw and during the preparation of this thesis.

In particular, I would like to express my sincere gratitude to \textit{Bogumiła Świeżewska}. You were the best supervisor I could have imagined, and I feel incredibly fortunate to have had the opportunity to be your PhD student.
All the things I have managed to achieve during that time would not have been possible without your patience, guidance, and support. I am grateful that I could freely ask you the most basic and random physics questions, career advice, and opinions on numerous presentation slides.
Your kindness and diligence will always inspire me.

I also want to thank \textit{Marek Lewicki} for keeping the doors to your office always open. I am grateful for all the discussions about the physics of bubbles, as I have learned a lot from you as well. I also thank you for all the coffee you have shared with me.

I would also like to acknowledge all the current and past members of our ``Cosmo'' group at the University. Thank you for all the networking and discussions during meetings and social events.  

One of the best aspects of my PhD studies was the opportunity to travel and meet new people. I am grateful for all the new friendships I have formed during my time at various schools, conferences and research visits. The complete list would be too long, but I wanted to especially thank \textit{
Giulio Barni, 
Nathan Borak, 
Matthias Carosi,
Robin F. Diedrichs, 
Clara Fernandez, 
Silvia Gasparotto,
Edoardo Giangrandi, 
Ángel Gil Muyor, 
Mehmet Asım Gümüş, 
Maya Hager, 
Antonio J.~Iovino, 
Ioanna Kourkoulou, 
Rémi Faure, 
Rémy Larue, 
Giorgio Laverda, 
David Maibach, 
Marco Matteini,
Nadine Nussbaumer, 
George A.~Parker, 
Álvaro Pastor Gutiérrez, 
Daniele Perri, 
Matteo Piani, 
Flavio Riccardi, 
Nicklas Ramberg, 
Francesco Sciotti, 
Francesco Serra, 
Daniel Schmitt, 
Mia West
} for such great memories together.
EFT stands for ``effective friendship theory''. 

To all my collaborators, I thank you for all the exciting research we have done together. In particular, I feel deeply grateful to \textit{Tuomas V.I.~Tenkanen, Jorinde van de Vis,} and \textit{Philipp Schicho} for allowing me to learn so much about bubbles and effective theories from you. I feel glad that, after these projects, I can consider you my friends. 
I also want to acknowledge all the other senior colleagues from the ``dimensional reduction gang'': \textit{Andreas Ekstedt, Oliver Gould} and \textit{Joonas Hirvonen}. I learned a great deal from discussing with you and from reading your papers. 

I am grateful to have had the opportunity to spend my years at the University of Warsaw alongside all my friends from the RL study group. I will always cherish the memories of studying together and all the karaoke nights we had. 
I also want to thank \textit{Paulina Michalak, Michał Ryczkowski, Michał Łukawski, Paweł Pajewski}, and \textit{Arkadiusz Szpyra} for sharing the excitement and struggles of learning theoretical particle physics. 
I also feel obliged to express my deep gratitude to \textit{Jan Klamka} for causing two major events in my life. First, I thank you for sending me an announcement of Bogumila’s PhD position, and second, for inviting one of your high school friends to a certain house party in November 2021.  

To my beloved \textit{Agacia}, thank you for your love and support, and for always being here for me. 
Quantum field theory has not yet explained what our souls are made of, but yours and mine are the same. Thank you for coming to that house party. 

I also want to thank all the friends outside my physics bubble whom I met here in Warsaw. This list would definitely be too long.
However, here I would like to thank all the participants in my ``29th birthday quiz’’, which provided a great relief during the writing of this thesis. The main prize of the quiz was being acknowledged here. Hereby, on this page, I honour the winning team -- \textit{``Labubu squad''}. 
 
Chciałbym również podziękować pani \textit{Ewie Białous} za efektywne popchnięcie mnie w stronę fizyki na lekcjach w gimnazjum. 

Wreszczie, dziękuję swoim kochanym rodzicom za wsparcie podczas tych wszystkich lat. Zawdzięczam wam wszystko i bez waszej pomocy nigdy nie byłbym w stanie podążać tą ścieżką i spełniać swoich marzeń.  

This thesis is the culmination of a ten-year journey at the University of Warsaw. A non-negligible factor for my decision to enrol on the physics program at the University was watching a lot of Doctor Who in high school. 
In particular, I remember watching a scene where the 11th Doctor is reading a book titled ``Advanced Quantum Mechanics''. I thought it was the most ``supercool'' thing, and I felt encouraged to learn theoretical physics and become a scientist. Thus, I feel compelled to also thank \textit{Matthew Robert Smith} for playing the best Doctor.
In connection to this story, I have used the translator from Gallifreyan to English\footnote{
The official BBC Gallifreyan Translator is publicly available on the website \url{https://www.doctorwho.tv/gallifreyan-translator}
} to generate a cover illustration for this dissertation. Meaning of the message encoded by the circular glyphs on the cover can be found as the last sentence at the end of section \ref{sec:dr_phi4}.  

This work was partially supported by the Polish National Science Centre (NCN) grants 
2018\breakslash31\breakslash D\breakslash ST2\breakslash 03302, 
2023\breakslash49\breakslash B\breakslash ST2\breakslash 02782, 
2023\breakslash50\breakslash E\breakslash ST2\breakslash 00177, 
the Polish National Agency for Academic Exchange (NAWA) grants PPN\breakslash PPO\breakslash 2020\breakslash 1\breakslash 00013\breakslash U\breakslash 00001, 
BPN\breakslash BEK\breakslash 2023\breakslash 1\breakslash 00311\breakslash U\breakslash 00001, 
BPI\breakslash STE\breakslash 2021\breakslash 1\breakslash 00020\breakslash U\breakslash 00001, 
and the Polish ``Excellence Initiative – Research University'' (IDUB) grants 
501-D111-20-0023110, 
PSP:501-D111-20-0004410 .

\newpage

\chapter*{Abstract}
%
In this dissertation, we investigate a supercooled phase transition (PT) in the early Universe. Using high-temperature dimensional reduction, we calculate the next-to-leading order (NLO) nucleation rate, paying particular attention to the proper inclusion of higher-order corrections and renormalisation scale (RG-scale) dependence. In our approach, we explicitly evaluate fluctuation determinants, providing a state-of-the-art description of thermal bubble nucleation at NLO.  
We study a concrete model called SU(2)cSM, which extends the conformal Standard Model by an additional SU(2)$_X$ gauge sector. The new sector also contains an additional scalar field that acquires a non-zero vacuum expectation value via radiative symmetry breaking. 
This breaking of the SU(2)$_X$ proceeds through a supercooled first-order phase transition. 

The first part of this thesis provides the theoretical background. We begin by introducing effective actions and discussing RG-scale dependence of effective potentials. 
Then, we turn to quantum field theory at finite temperature. We describe the relation between free energy and the effective actions. We review perturbation theory at finite temperature and perturbative expansions within the framework of effective field theory (EFT). Finally, we discuss the thermal bubble nucleation.

In the second part of this dissertation, we apply the theoretical framework we have introduced to describe the SU(2)cSM model and supercooled phase transition. 
We establish a power-counting scheme and construct the leading-order (LO) effective potential. Next, we discuss the symmetry breaking and show the available parameter space of the model. 
To address the issue of RG-scale dependence, we derive an RG-improved effective potential for the dark SU(2) sector.
Then, we include the thermal corrections and show how high-temperature effects can influence even a supercooled transition. Then, for the first time, we apply high-temperature dimensional reduction to a model with classical scale-invariance (and supercooled PT). We discuss the resulting 3d EFT and its validity.    
Finally, we derive an NLO thermal nucleation rate. We examine various approximations to this nucleation rate and their limitations. In particular, we numerically evaluate fluctuation determinants, which leads to a state-of-the-art description of bubble nucleation. We also present a direct comparison of different methods for a chosen benchmark.

In the last part, we collect the phenomenological results. We describe all phase transition parameters and perform parameter space scans for SU(2)cSM. Then, we discuss the gravitational wave (GW) signals resulting from the phase transition, and show quantitative predictions of the signal in SU(2)cSM.
We compute the signal-to-noise ratio for Laser Interferometer Space Antenna (LISA) and find that the supercooled phase transition in SU(2)cSM leads to a strong signal observable in LISA across the entire parameter space considered. This makes the model experimentally testable in future experiments. 
Finally, we discuss the impact of theoretical uncertainties on the GW predictions. We show that inclusion of higher-order corrections can substantially modify both the PT parameters and resulting GW spectra.

\newpage

\begin{otherlanguage*}{polish}
\chapter*{Streszczenie}

W niniejszej rozprawie badamy przechłodzoną przemianę fazową (PF) zachodzącą we wczesnym Wszechświecie. Wykorzystując wysokotemperaturową redukcję wymiarową (DR ang. dimensional reduction), obliczamy prawdopodobieństwo nukleacji do następnego rzędu po wiodącym (NLO, ang. next-to-leading order) w rachunku perturbacyjnym, zwracając uwagę na uwzględnienie poprawek wyższych rzędów oraz zależności od skali renormalizacji (skali RG, ang.~renormalisation scale). 
Obliczamy także wyznaczniki fluktuacyjne, co prowadzi do najbardziej precyzyjnego opisu nukleacji bąbli na poziomie NLO.
Przedmiotem naszych badań jest model zwany SU(2)cSM, który rozszerza konforemną wersję Modelu Standardowego o dodatkowy sektor z symetrią SU(2)$_X$. 
Nowy sektor zawiera również dodatkowe pole skalarne, które uzyskuje niezerową wartość oczekiwaną w próżni poprzez radiacyjne łamanie symetrii. Łamanie symetrii SU(2)$_X$ zachodzi w wyniku przechłodzonej PF pierwszego rzędu.

Pierwsza część pracy stanowi wprowadzenie teoretyczne. Opisujemy działanie efektywne oraz zależność potencjału efektywnego od skali RG. 
Następnie przechodzimy do teorii pola w skończonej temperaturze, omawiając związek między energią swobodną a potencjałem efektywnym. Przedstawiamy podstawy rachunku zaburzeń w skończonej temperaturze oraz rozwinięcia perturbacyjne w ramach efektywnej teorii pola. Na końcu omawiamy termiczną nukleację bąbli.

W drugiej części rozprawy stosujemy ten formalizm do modelu SU(2)cSM oraz zachodzącej w nim przechłodzonej PF. 
Uwzględniając wszystkie istotne wkłady, konstruujemy potencjał efektywny, po czym omawiamy łamanie symetrii oraz dostępny obszar przestrzeni parametrów modelu. 
Aby rozwiązać problem zależności od skali RG, wyprowadzamy ulepszony potencjał efektywny (ang.~RG-improved) dla ciemnego sektora SU(2).
Następnie dodajemy poprawki termiczne i pokazujemy, w jaki sposób efekty wysokotemperaturowe mogą wpływać nawet na przechłodzoną PF. Po raz pierwszy w literaturze stosujemy DR do modelu z klasyczną symetrią skalowania oraz z przechłodzoną PF. Następnie omawiamy otrzymaną efektywną teorię oraz zakres jej stosowalności.
W dalszej części wyprowadzamy prawdopodobieństwo nukleacji na poziomie NLO, omawiamy różne przybliżenia stosowane w literaturze oraz ich ograniczenia. 
W szczególności numerycznie obliczamy wyznaczniki fluktuacyjne, co prowadzi do najbardziej precyzyjnego opisu nukleacji bąbli. Pokazujemy także bezpośrednie porównanie różnych metod dla wybranego punktu w przestrzeni parametrów modelu SU(2)cSM.

W trzeciej części przedstawiamy wyniki fenomenologiczne. Opisujemy wszystkie parametry PF oraz prezentujemy wyniki skanów przestrzeni parametrów w modelu SU(2)cSM.~Następnie omawiamy sygnały fal grawitacyjnych generowane przez kosmologiczną PF i przedstawiamy przewidywania sygnału w modelu SU(2)cSM. 
Obliczamy stosunek sygnału do szumu dla detektora LISA (ang. Laser Interferometer Space Antenna) i pokazujemy, że przechłodzona PF w modelu SU(2)cSM prowadzi do silnego sygnału fal grawitacyjnech, który będzie obserwowalny przez eksperyment LISA w całym rozważanym zakresie przestrzeni parametrów modelu. 
Dzięki temu model SU(2)cSM może zostać empirycznie zweryfikowany w przyszłych eksperymentach.
Na końcu omawiamy wpływ niepewności teoretycznych na przewidywania dotyczące sygnału fal grawitacyjnych i pokazujemy, że uwzględnienie poprawek wyższych rzędów może znacząco zmienić zarówno parametry przemiany, jak i ostateczne widmo fal grawitacyjnych.

\end{otherlanguage*}

\chapter*{List of publications}
\addcontentsline{toc}{chapter}{List of publications}

\nocite{Kierkla:2022odc, Kierkla:2023von, Kierkla:2025qyz, Kierkla:2025vwp}
    The work presented in this dissertation is mostly based on the following papers, 
    \nocite{Kierkla:2022odc, Karam:2023jge, Kierkla:2023von, Kierkla:2025qyz, Kierkla:2025vwp}
    \printbibliography[heading=none, resetnumbers=false, keyword=mypaper]
\noindent
In addition, the following works were completed during the course of the author's doctoral studies; however, they are beyond the scope of this dissertation
    \nocite{Kierkla:2023uzo}
    \printbibliography[heading=none, resetnumbers=false, keyword=mypaper2]

\chapter*{Acronyms}
\addcontentsline{toc}{chapter}{Acronyms}

\begin{enumerate}[label={}, leftmargin=*]
    \item \textbf{1PI} one-particle irreducible
    \item \textbf{BSM} beyond the Standard Model
    \item \textbf{CW} Coleman--Weinberg
    \item \textbf{CSI} classical scale-invariance
    \item \textbf{DM} dark matter
    \item \textbf{DR} dimensional reduction
    \item \textbf{EFT} Effective Field Theory
    \item \textbf{EW} electroweak
    \item \textbf{GW} gravitational wave
    \item \textbf{LO} leading order
    \item \textbf{MS} minimal subtraction
    \item \textbf{NLO} next-to-leading order
    \item \textbf{PT} phase transitions
    \item \textbf{QFT} Quantum Field Theory
    \item \textbf{SM} Standard Model
    \item \textbf{SSB} spontaneous symmetry breaking
    \item \textbf{vev} vacuum expectation value
\end{enumerate}

\tableofcontents
\mainmatter


\chapter{Introduction}

The Standard Model (SM) of particle physics is often regarded as the best physical theory of all time. It provides an astonishing agreement with experiments and describes the most fundamental ingredients of matter and carriers of forces in nature. 
Despite its phenomenological success, it is widely known that it is not a final ``Theory of Everything'', as it lacks the description of (ultraviolet complete) quantum gravity (see e.g.~\cite{Buoninfante:2024yth} for a recent review ). 
Moreover, there are many open questions, e.g., regarding the nature of dark matter, sensitivity to ultraviolet (UV) corrections, the smallness of the cosmological constant, or the observed baryon asymmetry (see \cite{vandeVis:2025efm, Baumann:2022mni, Gouttenoire:2022gwi} for a pedagogical review). 
Most of these problems and puzzles can be connected to the Higgs sector of the Standard Model. Many proposed extensions, so-called beyond Standard Model (BSM) theories, propose solutions to these problems that involve the Higgs field, e.g., by coupling it to a new scalar field, or by using the Higgs as a portal to some yet unknown dark sector. Therefore, it is important to investigate the details of the Higgs potential. In the SM, its form allows for spontaneous breaking of electroweak symmetry due to the dimensionful parameter inserted by hand. 
Such a solution raises theoretical concerns about sensitivity to the ultraviolet physics often referred to as the hierarchy problem (see refs~\cite{Peskin:2025lsg, Garces:2025rgn} for a recent review). Moreover, there remains a question -- where does the electroweak scale come from? These questions further suggest that the Standard Model alone might be an effective description valid below some higher energy scale of \textit{new physics}. 

An interesting class of BSM theories are models with classical scale invariance.
In these models, the classical Lagrangian does not contain any dimensionful parameter, but rather the mass term of the scalar field is generated due to the quantum corrections from interactions with other fields \cite{PhysRevD.7.1888, Khoze:2014xha}. While this principle is not suitable for the Standard Model alone due to the large value of the top quark Yukawa coupling \cite{Hempfling_1996}, it is still a viable option for BSM models. 
In particular, a scale-invariant Standard Model with a dark sector could provide a very distinct feature -- a strong first-order electroweak phase transition ~\cite{Randall:2006, Konstandin:2010, Konstandin:2011, vonHarling:2017, Bruggisser:2018, Kubo:2016, Baldes:2021aph, Hambye:2013, Jaeckel:2016,  Jinno:2016, Marzola:2017, Baldes:2018, Prokopec:2018, Marzo:2018, Kang:2020jeg, Khoze:2014xha, Kierkla:2022odc, Ellis:2020}. 

Electroweak symmetry breaking in the context of cosmology is a physical process that occurred in the hot early Universe, during which the Higgs field transitioned to a new vacuum state \cite{Mukhanov:2005sc, Kolb:1990vq}. 
In the Standard Model alone, such a transition is a crossover; the Universe smoothly transitioned from being filled with massless radiation to containing massive particles \cite{Kajantie:1996mn}. However, many BSM theories predict that the nature of this process was instead a first-order transition. In such a case, when the Universe was cooling down, bubbles started to randomly appear everywhere, and inside them the Higgs (or other scalar) field was in its new vacuum state with non-zero expectation value \cite{Laine_2016}. 
Such bubbles would then expand, and particles would gain mass upon crossing the bubble wall. The bubbles would then further grow until they filled the whole Universe, thereby completing the transition. During the final moments of the transition, the bubbles would leave a particular imprint in the Universe -- a stochastic gravitational wave background \cite{Croon:2024mde, Caprini:2015, Caprini:2019pxz, LISACosWG:2022jok, Gowling:2021gcy, Boileau:2022ter, Gowling:2022pzb}. Thus, the detection of such a signal would be smoking-gun evidence of new physics.

With the first-ever detection of gravitational waves in 2015 \cite{Abbott:2016-2, Abbott:2016, Abbott:2017, Abbott:2017-2, LIGOScientific:2017ycc, LIGOScientific:2017vox}, we have thus opened a new window to look into the history of the Universe. 
With new experiments on the horizon \cite{AEDGE:2019nxb, Punturo:2010zz, Hild:2010id, Badurina:2019hst}, there are high hopes to look into that particular era of electroweak transition and learn both about the nature of the Higgs field and the early Universe. 
The most promising experiment, which has recently entered into the implementation phase, is the Laser Interferometer Space Antenna (LISA), which will start gathering data in the mid-2030s. With its 2.5 million kilometre arms length, this triangle-shaped interferometer will be sensitive to the gravitational wave signal exactly from the electroweak epoch \cite{Arun_2022, LISACosmologyWorkingGroup:2022jok}. 

This thesis is focused on the theoretical and phenomenological study of 
cosmological supercooled phase transitions. They appear naturally in models with classical scale invariance. In this thesis we will study in detail one concrete model which implements this symmetry.
We improve the theoretical description of supercooled transitions and perform a phenomenological study in a scenario where the GW signal is generically strong.
The use of tools based on effective field theory and phenomenological predictions of GW signals for LISA in this thesis presents the current state-of-the-art in this area of research.

With the prospect of LISA operating in the near future, a duty of the theoretical physics community is to provide a robust framework for studying first-order transitions in BSM theories and concrete predictions of possible gravitational wave signals. We hope that this thesis will substantially contribute to these goals.

The thesis is organised as follows. We start with reviewing the formalism of the effective action in chapter \ref{chapter:Seff_vacuum}. We show the basic definition based on the Legendre transform and then discuss the physical meaning of quantum, effective actions \cite{Schwartz:2014sze, Burgess_2020}. We then move to a specific example of a real scalar theory, where we explicitly show the calculation of the effective action and, in particular, the effective potential. Moreover, we discuss the validity of perturbation theory and the renormalisation scale dependence of the obtained objects. Finally, we show how radiative corrections can lead to spontaneous symmetry breaking in gauge theories, which leads to re-organisation of the perturbative expansion of the effective actions.

In chapter \ref{chapter:TFT} we review the basics of thermal field theory. We discuss the imaginary time formalism, which describes the equilibrium physics of quantum field theory at finite temperature \cite{Matsubara:1955ws, Laine_2016}. We show the relation of the effective potential to the free energy of a physical system. Then we commit to explaining the problems of perturbation theory in the limit of infrared energy scales. We then show how to reorganise the computations using an elegant framework based on effective field theory. We again use the model of a real scalar field to explicitly illustrate the matching procedure and general features. Finally, we discuss the computation of the thermal bubble nucleation rate, a key quantity that allows for the description of the nucleation process. 
We then show the relation of the nucleation rate to effective theories at high temperatures \cite{Gould:2021ccf, Ekstedt:2021kyx, Ekstedt:2022tqk}.

Chapters \ref{chapter:su2csm}-\ref{chapter:pheno} contain the main original results of this thesis. In chapter \ref{chapter:su2csm} we introduce the classically scale-invariant extension of the Standard Model. This model is called SU(2)cSM \cite{Hambye:2013, Hambye:2018qjv, Kierkla:2022odc, Prokopec:2018}, and it provides a benchmark for studying supercooled phase transitions in the early Universe. We discuss the general features of the model, the computation of the zero-temperature effective potential and the renormalisation group improvement.

In chapter \ref{chapter:supercoolEFT}, we discuss the thermal corrections for SU(2)cSM, and construct the effective theory capable of capturing thermal nucleation \cite{Kierkla:2023von}. We derive the thermal nucleation rate up to the next-to-leading order, discussing the accuracy of the so-called derivative expansion. We estimate the possible errors associated with theoretical uncertainties, and finally, we provide a direct comparison between different approaches for a chosen benchmark point. 

In chapter \ref{chapter:pheno}, we apply the theoretical apparatus we have provided in the previous section to obtain phenomenological predictions of the phase transition in SU(2)cSM \cite{Kierkla:2022odc, Kierkla:2023von, Kierkla:2025vwp, Kierkla:2025qyz}. We discuss in detail the parameters of phase transition and show how to calculate them for a model under consideration. We then show how they can be used to obtain predictions for the gravitational wave signal, and we assert the observability of the signal in the LISA interferometer by computing the signal-to-noise ratio. Next, we discuss the impact of theoretical uncertainties on predictions of the gravitational wave signal.  

Finally, chapter \ref{chapter:summary} contains a summary of the presented results and discussion on the open research questions. 

In the appendices, we have collected additional information and derivations. Appendix A contains a collection of relevant beta functions. Appendix B presents the details of dimensional reduction in the SU(2)cSM. Appendix C contains the details of calculations regarding fluctuation determinants.

\chapter{Effective actions}
\label{chapter:Seff_vacuum}
In this chapter, we will review the basics of effective actions in Quantum Field Theory. We will begin with a definition of the effective action as a Legendre transform of the generating functional. We will also shortly discuss the physical meaning of effective actions. Then, we will discuss another method based on a background field (see e.g. ref.~\cite{Abbott:1981ke}), which will then be used throughout the thesis. 
Finally, we will explicitly calculate the effective potential for a toy model, where we will also discuss renormalisation scale dependence. Finally, we will review radiative symmetry breaking \cite{PhysRevD.7.1888} in massless scalar electrodynamics. This chapter is based mostly on textbooks \cite{Schwartz:2014sze, Peskin:1995ev, Burgess_2020} and the seminal paper of S. Coleman and E. Weinberg \cite{PhysRevD.7.1888}.

\section{Effective action}

\subsection{Legendre transform}
The generating functional, $W[J]$, for a theory described by the action $S[\phi]$ is defined as
\begin{align}
    e^{iW[J]} \equiv
    \int\mathcal{D}\phi ~e^{iS[\phi] + i\int \dd[4]x J\phi},
\end{align}
where $J$ is the source current. 
Let us now consider a one-point function for a quantum field $\phi^a (x)$, i.e. vacuum expectation value (vev), in the presence of classical current $J_a(x)$ 
\begin{align}\label{eq:J_1point}
    \fdv{W[J]}{J_a(x)} = \frac{
    \matrixel{\Omega_{\rm out},J_a}{\phi^a (x)}{\Omega_{\rm in},J_a}
    }{
    \braket{\Omega_{\rm out},J_a}{\Omega_{\rm in},J_a}
    }
    = \expval{\phi^a (x)}_J  \equiv \veva,
\end{align}
where $\veva$ denotes the vev, and $a=1,2,3,\dots$ indexes the fields in a theory.
This expression gives a physical interpretation of the source current $J_a$ -- it corresponds to arbitrary interactions which try to change the vev of the $\phi^a$ field. Thus, for $J=0$, we would have obtained the actual vacuum of a given quantum theory.
Practically, one uses source currents to obtain the $n$-point correlator functions via functional derivatives of the generating functional, as we have just shown. However, it is possible to get rid of the currents altogether by performing the following Legendre transform \cite{Schwartz:2014sze, Peskin:1995ev, PhysRevD.7.1888}
\begin{align}
\label{eq:Gamma[v]_Legendre}
    \Gamma[v] \equiv W[J] - \int\dd[4]{x}~ \veva J_a,
\end{align}
where now $J_a(x)$ can be interpreted as a functional of, $\veva(x)$ with an implicit relation between them given by
\begin{align}
    \eval{\fdv{W[J]}{J(x)}}_{J=J_a} = \veva.
\end{align}
It is worth emphasising that we do not lose any information by this transformation. If we know the functional $\Gamma[v]$, we can easily obtain the source current
\begin{align}
    \fdv{\Gamma}{v^a (x)} &= 
    \int \dd[4]{y} \fdv{J_b (y)}{\veva (x)} \fdv{W}{J_b (y)} - J_a(x)
    - \int \dd[4]{y} v^b(y) \fdv{J_a(y)}{\veva(x)} 
    = -J_a(x).
\end{align}
The above equality also illustrates an important concept. The vev of the field $\phi$ in the presence of vanishing current $J =0$ is the stationary point of the functional $\Gamma[v]$.
Thus, the functional $\Gamma[v]$ seems to serve as a parallel to a classical action, $S[\phi_{\rm cl}]$, 
\begin{align*}
\begin{aligned}
        \mbox{Classical action } S[\phi]\\ 
        \mbox{Field configuration } \phi_{\rm cl} 
\end{aligned}
\quad
\longrightarrow 
\quad
\begin{aligned}
        \mbox{``Quantum'' action } \Gamma[v]\\ 
        \mbox{vacuum expectation value } v,
\end{aligned}
\end{align*}
for this reason, it is often called a \textit{quantum action} or \textit{effective action}, as it effectively includes quantum effects into a tree-level, action-like object. 
The classical action in general can be expressed as a difference between kinetic energy $K$ and potential energy $V$, i.e., $S=K-V$. A ground state of such action corresponds to a time-independent configuration with $K=0$, thus this configuration actually is the minimum of potential energy, $V=-S$. Similarly, if the ground state of quantum theory is well-behaved in the adiabatic approximation, a static configuration minimises $-\Gamma$, see e.g. refs.~\cite{Symanzik:1969ek, Coleman:1985rnk}. If we also consider additionally a homogenous configuration (so the total kinetic energy disappears), $\bar\phi$, then it will minimise the
\textit{quantum} or \textit{effective potential }
\begin{align}
\label{V_q_definition}
    V_{\rm{eff}}[\bar\phi] \equiv -\frac{1}{(\mathcal{VT})}\Gamma[\bar\phi],
\end{align} 
where $\mathcal{VT}$ denotes the volume of spacetime.
Moreover, it can be shown that for any configuration, $\bar\phi$, the quantity, $-\Gamma[\bar\phi]$, can be interpreted as the minimum of energy i.e. $\matrixel{\Psi}{H}{\Psi}$, where the states $\ket{\Psi}$ are such that $\matrixel{\Psi}{\phi^a(x)}{\Psi} = \bar\phi$, see e.g.~\cite{Burgess_2020, Coleman:1985rnk, Symanzik:1969ek}. Then, the value of $\bar\phi$ that is a global minimum of the effective potential, $V_{\rm{eff}}$, corresponds to the actual true vacuum of the quantised theory.

\subsection{Semi-classical expansion}

We have shown that the minimum of the effective action corresponds to the ground state of the full quantum theory in analogy to the classical action. 
This analogy can be extended, as the complete effective action contains terms which correspond to vertices containing the sums of $n$-particle irreducible diagrams of the quantum theory. 
In other words, it acts as an action which at its tree-level gives the same correlation functions that would result from calculating loop corrections of the quantised $S[\phi]$ action.

\begin{figure}
    \centering
    \begin{tikzpicture}
        \begin{feynman}
            \vertex[dot, label=left:$\tfrac{1}{\hbar}$]  (LT) at (-1, 1) {};
            \vertex[dot, label=above:$\tfrac{1}{\hbar}$] (L)  at (0, 0) {};
            \vertex[dot, label=above:$\tfrac{1}{\hbar}$] (R)  at (2, 0) {};
            \vertex[dot, label=right:$\tfrac{1}{\hbar}$] (RT) at (3, 1) {};
            \vertex[dot, label=left:$\tfrac{1}{\hbar}$]  (LB) at (-1,-1) {};
            \vertex[dot, label=right:$\tfrac{1}{\hbar}$] (RB) at (3,-1) {};
            \diagram* {
                (LT) -- [plain, edge label'=$\hbar$] (L),
                (L)  -- [plain, edge label=$\hbar$]  (R),
                (R)  -- [plain, edge label'=$\hbar$] (RT),
                (L)  -- [plain, edge label'=$\hbar$] (LB),
                (R)  -- [plain, edge label=$\hbar$]  (RB),
            };
        \end{feynman}
    \end{tikzpicture}
    \hspace{1.5cm}
    \begin{tikzpicture}
        \begin{feynman}
            \vertex[dot, label=left:$\tfrac{1}{\hbar}$]       (i1) at (0,0) {};
            \vertex[dot, label=above left:$\tfrac{1}{\hbar}$]  (a) at (1,0) {};
            \vertex[dot, label=above right:$\tfrac{1}{\hbar}$] (b) at (2,0) {};
            \vertex[dot, label=right:$\tfrac{1}{\hbar}$]      (f1) at (3,0) {};
            \diagram* {
                (i1) -- [plain, draw=black, edge label'=$\hbar$] (a),
                (a)  -- [plain, half left, edge label=\(\hbar\)] (b),
                (b)  -- [plain, half left, edge label=\(\hbar\)] (a),
                (b)  -- [plain, draw=black, edge label'=$\hbar$] (f1),
            };
        \end{feynman}
    \end{tikzpicture}
    \caption{Scheme of $\hbar$-scaling of example diagrams at tree- (left panel) and one-loop level (right panel).}
    \label{cw_vacuum_fig:hbar_loops}
\end{figure}
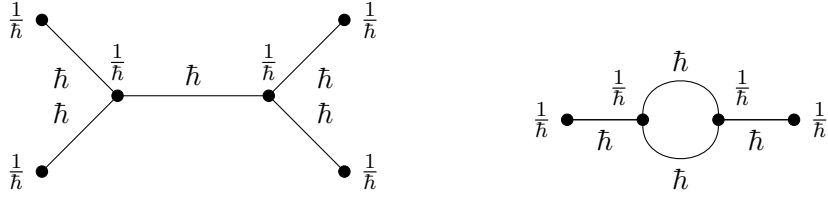

In the path integral formulation, there are exponents of $S[\phi]/\hbar$. 
By including the $1/\hbar$ factor inside the action, we can observe that each term inside will have a $1/\hbar$ factor. Thus, vertices and external states would come with $1/\hbar$, while propagators would come with a factor $\hbar$, see figure~\ref{cw_vacuum_fig:hbar_loops} for a schematic depiction. 
Tree-level terms correspond to ``classical'' contributions, collecting the $\hbar$-factors, they are of order $\order{\hbar^{-1}}$ i.e., they are the leading terms when $\hbar \rightarrow 0$, while loop corrections would be suppressed. In the literature, this is known as \textit{$\hbar$-expansion} or \textit{semi-classical expansion}, as it treats quantum corrections as a small correction to classical physics. 

In general, all connected diagrams can be computed using the generating functional $W[J]$, with the path integral containing the action $S[\phi]$ expanded to all orders in $\hbar$. 
Equivalently, such diagrams can be obtained from the generating functional $W[J]$ constructed with the effective action $\Gamma[v]$ in the $\hbar\rightarrow 0 $ limit
\begin{align}
    W[J] = \lim_{\hbar \to 0} (-i\hbar) \log{
    \qty[
        \int \mathcal{D}v 
        \exp{ 
            \frac{i}{\hbar} 
            \qty[
                \Gamma[v] + \int{\dd[4]{x} J_a \veva}
            ] 
        }
    ]
    }.
\end{align}
Notice that, here, taking the $\hbar\to 0$ limit isolates the tree-level diagrams computed with $\Gamma[v]$, but in the end, we obtain the functional $W[J]$ that can generate any correlation function. Hence, tree-level vertices of $\Gamma[v]$ indeed include the ``quantum'' corrections to $S[\phi]$.
Moreover, analogously to the generating functional, correlators can be obtained directly from the effective action by taking the $J$-derivatives. As an example, let us consider a propagator, given as the inverse of a two-point function, $G^{(2)}(x,y)$,
\begin{align}
    [G^{(2)}(x,y)]^{-1} = 
    \qty[
    -i \eval{
    \frac{\delta^2 W}{\delta J(x) \delta J(y)}
    }_{J=0} 
    ]^{-1}
    =\qty[\eval{-i \fdv{v(y)}{J(x)}}_{J=0}]^{-1},
\end{align}
where we used $\fdv{W[J]}{J(x)}=\phi(x)$.
Now, by differentiating the effective action twice, we can indeed obtain the same expression \cite{Schwartz:2014sze}
\begin{align}
    \eval{
    \frac{\delta^2 \Gamma[v]}{\delta v(x) \delta v(y)}
    }_{v=v_{0}}
    = - \eval{\fdv{J(y)}{v(x)}}_{J=0} = i \qty[G^{(2)}(x,y)]^{-1},
\end{align}
where $v_0$ is a vev with a vanishing current.
To summarise, performing the Legendre transform of the generating functional, we obtain a vev-dependent functional. This new functional acts as a generalisation of the classical action that accounts for the quantum effects. It can be used to generate correlators, its vertices correspond to one-particle irreducible (1PI) diagrams, and its global minimum corresponds to a true vacuum of the quantum theory. In the end, this ``effectiveness'' is the reason the functional $\Gamma[v]$ gets its well-deserved name.

\subsection{Background field method}
Let us now further investigate how to obtain an effective action in a given theory. Besides the Legendre transform, there are various other methods, e.g. matching of correlation functions (which is also very relevant for constructing effective field theories) or summing a certain class of $n$-particle irreducible diagrams. 
Here we will review a different one, which is based on so-called \textit{background fields}. In this approach, effective action can be obtained by considering an arbitrary background field configuration and integrating out all the quantum fluctuations on top of it. 

Let us now consider a field expanded on a background $\phi = \tilde{\phi} + \phib$, where $\phib(x)$ is a non-dynamical (it is not integrated over in the path integral) background field configuration. Note that it can still be inhomogeneous. Then, $\tilde{\phi}(x)$, is a quantum fluctuation.
Our initial action thus becomes $S[\tilde\phi+\phib]$.
We can now define the generating functional in the presence of a background field as 
\begin{align}
    e^{iW_b[J,\phib]} &\equiv 
    \int \mathcal{D}\tilde\phi~
    e^{
        iS[\tilde\phi+\phib] - \int \dd[4]{x} J \tilde\phi 
    }.
\end{align}
Moreover, the relation between source current and vev for $W_b[J,\phib]$ is
\begin{align}
    \frac{\delta W_b[J,\phib]}{\delta J} = \tilde{v}_\rmii{$J$},
\end{align}
where $\tilde{v}_\rmii{$J$} $ is the vev of $\tilde\phi$ in presence of a current $J$. Now, we can express the background-dependent functional $W_b[J,\phib]$ in terms of the original $W[J]$ if we make a shift in the path integral measure $\tilde \phi \rightarrow \tilde\phi - \phi_b $. Then we obtain
\begin{align}
    e^{iW_b[J,\phib]} &= 
    \int \mathcal{D}\tilde\phi~
    e^{
        iS[\tilde\phi] + i\int \dd[4]{x} J \tilde\phi 
    }
    e^{
        -i\int \dd[4]{x} J \phib 
    }\\
    &=
    e^{iW[J]} 
    e^{
        -i\int \dd[4]{x} J \phib 
    }\\
    \implies& 
    W_b[\phib, J] = \tilde{W}[J] - \int \dd[4]{x} J\phib
    ,
\end{align}
where $\tilde{W}[J]$ denotes a generating functional of the fluctuation only. 
Differentiating the above expression with respect to the source current, we get a relation between the vevs of the field, $\phi$, and the fluctuation $\tilde\phi$,
\begin{align}
    \label{eq:vev_phi_vev_phitild}
    \tilde{v}_\rmii{$J$} = v_\rmii{$J$} - \phib,
\end{align}
where $v_\rmii{$J$} $ is the vev of field $\phi$ in presence of current $J$. This is quite intuitive -- it means that the vev of fluctuation is just the vev of the total field shifted by the value of the background.

To obtain an effective action, we need to perform a Legendre transform (cf. eq.~\eqref{eq:Gamma[v]_Legendre}) of a generating functional. Thus, the effective action in the presence of the background field becomes
\begin{align}
    \Gamma_b[\phib, \vJtilde ] \equiv 
    W_b[\phib, \tilde{J}] - \int \dd[4]{x} \tilde{J} \vJtilde
    =
    W[\tilde{J}] 
    - \int \dd[4]{x} \tilde{J} \qty(\phib + \vJtilde )
    =\Gamma [\phib + \vJtilde].
\end{align}
Now, using the relation between the vevs, eq.~\eqref{eq:vev_phi_vev_phitild}, we obtain 
\begin{align}
    \Gamma_b[\phib, \vJtilde ]  
    = \Gamma [\phib+ \vJtilde] 
    = \Gamma[v_\rmii{$\tilde J$ }].
\end{align}
Since this holds for any value of the source current, we can rename $\tilde J = J$. Finally, if we now set the vev of the fluctuation to zero, $\vJtilde =0 $, then the vev of $\phi$ is just equal to the background field, and the effective action is 
\begin{align}
    \Gamma [\phib] = \Gamma_b[\phib, 0 ].  
\end{align}
This is the background field method of computing the effective action. It can also be understood in terms of diagrams. Calculating such 1PI effective action with the relation $\Gamma [\phib]=\Gamma_b[\phib, 0]$ means that its vertices contain diagrams where the static background $\phib$ is in the external legs, and the fluctuations $\tilde \phi$ are instead in the loops. 
Expanding in powers of $\tilde\phi$, we would see then that at zeroth order we would simply have $\Gamma[\phib] = S[\phib]$. Then the next contributions would come from vacuum bubbles with a constant background field on external legs, see figure~\ref{fig:Gamma_vac_diagrams}. 

Background-field method of computing the effective action,  $\Gamma[\varphi_b] = \Gamma_b[\varphi_b,0]$, can be written in a more explicit way using a path integral
\begin{align}
\label{eq:Gamma_background}
    e^{i\Gamma[\phib]} = 
    \int_{\rm restricted} \mathcal{D}\phi e^{iS[\phi+\phi_b]},
\end{align}
where ``restricted'' means here that we evaluate the path integral only on a restricted set of field configurations, such as the 1PI diagrams (as in figure~\ref{fig:Gamma_vac_diagrams}) in the case of perturbation theory \cite{Schwartz:2014sze}. This form also has a solid physical interpretation - we can obtain the effective action of a background field by integrating out all quantum fluctuations. 
%
	
	
	
	
	
	
	
	
	
	
	
	
\begin{figure}
\centering

\[
\Gamma_b[\varphib,0] =
\vcenter{\hbox{
\begin{tikzpicture}[scale=0.65]
\begin{feynman}
    \vertex (a) at (0,0) [dot] {};
\end{feynman}
\end{tikzpicture}
}}
\;+\;
\vcenter{\hbox{
\begin{tikzpicture}[scale=0.65]
\begin{feynman}
    \vertex (U) at (0,0.6);
    \vertex (D) at (0,-0.6);

    \diagram{
        (U) -- [plain, half left] (D)
           -- [plain, half left] (U)
    };
\end{feynman}
\end{tikzpicture}
}}
\;+\;
\vcenter{\hbox{
\begin{tikzpicture}[scale=0.65]
\begin{feynman}
    \vertex (i1) at (0, 1.2) [empty dot] {};
    \vertex (U)  at (0,0.6)  [dot] {};
    \vertex (D)  at (0,-0.6);
    \vertex (f1) at (0,-1.2)  {}; 

    \diagram{
        (i1) -- [plain] (U),
        (U) -- [plain, half left, looseness=1.4] (D)
           -- [plain, half left, looseness=1.4] (U)
    };
\end{feynman}
\end{tikzpicture}
}}
\;+\;
\vcenter{\hbox{
\begin{tikzpicture}[scale=0.65]
\begin{feynman}
    \vertex (i1) at (0, 1.2) [empty dot] {};
    \vertex (U)  at (0,0.6)  [dot] {};
    \vertex (D)  at (0,-0.6) [dot] {};
    \vertex (f1) at (0,-1.2) [empty dot] {};

    \diagram{
        (i1) -- [plain] (U),
        (U) -- [plain, half left] (D)
           -- [plain, half left] (U),
        (D) -- [plain] (f1)
    };
\end{feynman}
\end{tikzpicture}
}}
\;+\;
\vcenter{\hbox{
\begin{tikzpicture}[scale=0.65]
\begin{feynman}
    \vertex (i1) at (0,1.2) [empty dot] {};
    \vertex (U)  at (0,0.6)  [dot] {};
    \vertex (D)  at (0,-0.6) [dot] {};
    \vertex (f1) at (0,-1.2) [empty dot] {};

    \diagram{
        (i1) -- [plain] (U),
        (U) -- [plain, half left] (D)
           -- [plain, half left] (U),
        (D) -- [plain] (f1),
        (U) -- [plain] (D)
    };
\end{feynman}
\end{tikzpicture}
}}
\;+\;\dots
\]

	\caption{
		Diagrammatic representation of $\Gamma[\phib]$. The dot is the zeroth-order term $S[\phib]$. Solid lines are propagating $\tilde\phi$ fields, while external lines with empty dots correspond to the background field $\phib$.
	}
    \label{fig:Gamma_vac_diagrams}
\end{figure}

\section{Practical uses in real scalar theory}

\subsection{Effective potential}
Let us now demonstrate how to obtain the 1PI effective action in practice. Following \cite{Schwartz:2014sze}, we will consider a simple real scalar theory, defined by
\begin{align}
    S[\phi] &= \int\dd[4]{x} \mathcal{L}(\phi), \\
    \mathcal{L} &= \frac12(\partial_\mu \phi)^2 - \qty[
    \frac{1}{2}m^2 \phi^2 + \frac{1}{4!}\lambda \phi^4,
    ]
\end{align}
and the square brackets contain the potential $V(\phi)$. Here $m^2$ is positive; with that choice of the sign, the theory has a global minimum at $\phi=0$. One may wonder whether quantum corrections can change the position of the minimum.  This would be even more crucial for the case of $m^2=0$, as perhaps quantum fluctuation could destabilise the system. To answer such a question, we can use the formalism of effective action and then investigate whether its stationary point corresponds to the $\phi=0$ field configuration. 

We will use the background field method described in the previous section, and we will compute the 1PI effective action. First, we expand the field around a background $\phi \rightarrow \tilde\phi + \phi_b$, where $\tilde\phi$ denotes fluctuations and $\phi_b$ is a ``classical'' background. The Lagrangian now takes the form: 
\begin{align}
    \mathcal{L} =& 
    \frac{1}{2} (\partial_\mu \tilde\phi + \partial_\mu\phi_b)^2 -\frac{1}{2}m^2 (\tilde\phi + \phib)^2 + \frac{1}{4!}(\tilde\phi + \phib)^4\\
    =& 
    \frac12(\partial_\mu \phib) + \frac{1}{2}m^2 \phi_b^2 + \frac{1}{4!}\lambda \phib^4 \\
    &+\frac12(\partial_\mu \tilde\phi) + \frac{1}{2}m^2 (\tilde\phi)^2 + \frac{1}{4!}\lambda (\tilde\phi)^4
    + \frac14 (\tilde\phi)^2 \phib^2  
     + \dots
\end{align}
The dots denote ``mixed'' terms corresponding to the tadpole diagrams, and we can neglect them in our evaluation as they do not contribute to 1PI diagrams. Note that the total Lagrangian now contains a ``tree-level''  Lagrangian for the background field. In fact, this expression can be organised by considering the expansion in terms of derivatives over the potential,
\begin{align}
    \mathcal{L}[\phib, \tilde\phi] 
    = 
    \mathcal{L}[\phib]
    +\frac12(\partial_\mu \tilde\phi) 
    -\frac{1}{2} \tilde\phi^2 V^{\prime \prime}\left[\phi_b\right]
    -\frac{1}{3!} \tilde\phi^3 V^{\prime \prime \prime}\left[\phi_b\right]
    +\cdots,
\end{align}
where we have used a notation $V^\prime \equiv d{V}/d{\tilde\phi}$. Note that the first derivative corresponds to tadpole terms, and is thus contained among the terms denoted by dots.
Now, we can calculate the effective action in our theory by using the formula \eqref{eq:Gamma_background}, i.e., writing a path integral for the action expanded in fluctuations: 
\begin{align}
    e^{i \Gamma\left[\phi_b\right]} 
    &=
    e^{
        i \int d^4 x\left(
            \frac12(\partial_\mu \phib)^2
            -V\left[\phi_b\right]
        \right)
    } 
    \int_{\text {1PI}} \mathcal{D} \tilde\phi 
    e^{
        i \int d^4 x\left(
        \frac12(\partial_\mu \tilde\phi)^2
        -\frac{1}{2} \tilde\phi^2 V^{\prime \prime}\left[\phi_b\right]
        -\frac{1}{3!} \tilde\phi^3 V^{\prime \prime \prime}\left[\phi_b\right]
        +\cdots
        \right)
    }.
\end{align}
Note that the first term is just the tree-level action for the background field. Then, for a general potential, $V(\phi)$, such as the one under consideration in our toy theory, it is impossible to evaluate this path integral completely. Therefore, one can resort to perturbation theory, as then it is possible to evaluate the path integral up to finite order in the $\hbar$-expansion. Let us now then consider one-loop corrections, i.e., $\order{\hbar}$ order, we need only terms quadratic in fields, as only they contribute to one-loop topologies
\begin{align}
    e^{i \Gamma\left[\phi_b\right]}=
    e^{i S[\phib]} 
    \int \mathcal{D} \tilde\phi \exp 
    \left[
        i \int d^4 x\left(
        \frac12(\partial_\mu \tilde\phi) ^2
        -\frac{1}{2} \tilde\phi^2 V^{\prime \prime}\left(\phi_b\right)\right)
    \right]
\end{align} 
Notice that such a path integral is a Gaussian integral. Thus, we will also call quadratic fluctuations as \textit{Gaussian fluctuations}. Performing such an integral results in
\begin{align}
    e^{i \Gamma\left[\phi_b\right]}=
    \text {const.} \times 
    \qty[
        \operatorname{det}
        \left(
            \partial^2 +V^{\prime\prime} \left[\phi_b\right]
        \right)
    ]^{-\frac{1}{2}}
    e^{i S[\phib]}, 
\end{align}
where $\partial^2 = \partial_\mu \partial^\mu$. We see that the result consists of the exponential containing the tree-level action of the background field, and the prefactor with the so-called \textit{fluctuation determinant}. This determinant contains the effects of quantum fluctuations on the background field configuration. Consequently, the functional operator inside the determinant is called \textit{fluctuation operator}. Let us investigate it further now. After taking the logarithm and multiplying by the $(-i)$ factor, the effective action can be read off explicitly, as
\begin{align}
\label{eq:Seff_pathint_phi4_vacuum}
    \Gamma\left[\phi_b\right] &=
    S[\phib]
    + \frac{i}{2}
    \log{
    \qty[
        \operatorname{det}
        \left(
            \partial^2 +V^{\prime\prime} \left[\phi_b\right]
        \right)
    ]
    }+\text { const. }\\
    &= 
    S[\phib]+
    \underbrace{
    \qty[
        \frac{i}{2} \tr \log(\partial^2 + V^{\prime\prime}(\phi_b))+\text{const.}
    ]
    }_{\equiv \Delta\Gamma} 
\end{align}
We can now explicitly see how these terms constitute the $\hbar$-expansion. The first term here is of order $\order{\hbar^{-1}}$, i.e., it is the tree-level action. The second term, here denoted as $\Delta\Gamma$, is of order $\order{\hbar^{0}}$ and contains the first (one-loop) quantum corrections. Including higher orders is straightforward, as it would just require keeping the $V^{\prime\prime\prime}$ term in the path integral in eq.~\eqref{eq:Seff_pathint_phi4_vacuum}. 

Let us now discuss $\Delta\Gamma$ in more detail. To evaluate this term, we can express the trace as an integral in position space \cite{Schwartz:2014sze}:
\begin{align}
\label{eq:DeltaGamma_position_phi4}
    i\Delta\Gamma = -\frac{1}{2} \int \dd[4]{x} \ev{\log(1 + \frac{V^{\prime\prime}(\phi_b)}{\partial^2})}{x} + \text { const. }
\end{align}
The background field $\phi_b(x)$ is generally non-static and position-dependent, and one cannot find a general solution to the integral. 
However, for any such configuration, it is possible to use \textit{derivative expansion}\footnote{Other name one can encounter in literature is gradient expansion.} to expand the total effective action in the powers of momenta (derivatives) of the background. At zeroth order, i.e.,~zero momentum, the $\phi_b = \bar\phib$ is just a constant and thus $\Delta \Gamma[\bar\phib]$ will be a function (instead of a functional). It will contribute as the one-loop correction to the tree-level potential. Their sum is the effective potential $V_q=V_{\rm eff}$ that we have described before 
\begin{align}
    V_{\rm eff}[\bar\phib] = V_{\rm{tree}} (\bar\phib) + \Delta \Gamma (\bar\phib).
\end{align}
Therefore, the effective potential can also be understood as the effective action at zeroth order in derivative expansion 
\begin{align}
\label{eq:Gamma_momenta_expansion}
    \Gamma[\phib] = 
    \int \dd[4]{x} \qty[
        \underbrace{-V_{\rm eff}(\phib)}_{\order{\partial^0}\rm{~order} } 
        + 
        \underbrace{\frac{1}{2}(\partial_\mu \phib)^2 \delta Z(\phib)}_{\order{\partial^2}\rm{~order} }
        + \dots.
    ].
\end{align}
Let us now then consider the case of a function $\Delta \Gamma (\bar\phib)$ contribution. In such a scenario, the $V^{\prime\prime}(\bar\phib)$ is now just a function and not a functional anymore, and if we look at our toy theory, then we see that $V^{\prime\prime}(\phi_b)$ can be interpreted as an effective mass squared, $m^2_{\rm eff}$, which now is background-field dependent 
\begin{align}
    m^2_{\rm eff}(\bar\phib) 
    = V^{\prime\prime}(\bar\phib) 
    = m^2 + \frac{\lambda}{2}\phi_b^2 .
\end{align} 
Using this notation, and inserting the set of momentum states in eq.~\eqref{eq:DeltaGamma_position_phi4}, we can write the one-loop correction to the effective potential as 
\begin{align}\label{Delta_Gamma_k_int}
    i\Delta \Gamma(\bar\phib)  
    = -\frac{1}{2} \int \dd[4]{x} \int \frac{\dd[4]{k}}{\qty(2\pi)^4}
    \log(1 - \frac{m^2_{\rm eff}(\bar\phib)}{k^2}) + \text {const.}
\end{align}
The first integral is equal to the spacetime volume $\mathcal{VT}$ so we can now justify eq.~\eqref{V_q_definition} from previous section:
\begin{align}
    \Delta \Gamma[\bar\phib] = - (\mathcal{VT}) \cross \Delta V_{\rm eff}(\bar\phib),
\end{align}
where $\Delta V_{\rm eff}$ is the one-loop contribution to the total effective potential.
Now, the momentum integral in eq.~\eqref{Delta_Gamma_k_int} is divergent, so it must be regularised. It can be evaluated, e.g. in the dimensional regularisation. After the regularisation and renormalisation in the $\overline{\rm{MS}}$-scheme, one obtains the following result: 
\begin{align}
    \Delta V_{\rm eff}(\bar\phib) = 
    \frac{m^4_{\rm eff}(\bar\phib)}{64 \pi^2} \qty(\log{\frac{m^2_{\rm eff}(\bar\phib)}{\mu^2}} - \frac32),
\end{align}
where $\mu$ is the energy scale introduced in $\overline{\rm{MS}}$ renormalisation. We will refer to it as RG-scale. This contribution is the famous Coleman-Weinberg potential \cite{PhysRevD.7.1888}. Therefore, we will also denote one-loop corrections as
\begin{align}
    V_{\rm{CW}} (\bar\phib) \equiv \Delta V_{\rm eff}(\bar\phib).
\end{align}
Taking all the pieces together, the total effective potential at one-loop order for the $\lambda\phi^4$ theory is 
\begin{align}
\label{eq:Veff_phi4}
    V_{\rm eff}(\bar\phib) =
    \frac{1}{2}m^2 \bar\phib^2 + \frac{1}{4!}\lambda \bar\phib^4
    + \frac{(m^2 + \frac{\lambda}{2}\bar\phib^2)^2}{64 \pi^2} \qty(\log{\frac{m^2 + \frac{\lambda}{2}\bar\phib^2}{\mu^2}} - \frac32).
\end{align}
Now we can answer the question posed at the beginning of the section, whether the $\phi =0$ configuration is a true vacuum state of the theory. Let us consider a massless case for simplicity, $m^2 =0$. The stationary points of the effective potential are then given by 
\begin{align}
    \dv{V_{\rm eff}}{\bar\phib}  &= 0\\
    \implies
    \bar\phib = 0, 
    \quad& \text{or} \quad
    \lambda \log\frac{\frac12 \lambda \bar\phib^2}{\mu^2} = \lambda - \frac{32}{3}\pi^2.
\end{align}
Looking closer, we conclude that the global minimum exists for some non-zero $\bar\phib$. However, upon closer inspection, we see that such a field value would result in a large logarithm since on the RHS we have $32/3 \pi^2 \simeq 105$ factor. Thus, the perturbation theory breaks down, and we cannot trust the effective potential at such large field values. The root of this problem is associated with the RG scale. Let us now discuss the RG-scale dependence in more detail.

\subsection{Renormalisation scale dependence}

First, let us note that the effective potential computed to all orders, i.e., including all possible loop effects, should be completely independent of the renormalisation scale $\mu$. This statement can be formalised by the Callan-Symanzik equation \cite{Callan1970a, Symanzik:1970rt, Symanzik:1971vw}
\begin{align}
    \qty(
        \mu \pdv{}{\mu} 
        + \sum_i \beta_{g_i} \pdv{}{g_i} 
        - \gamma_\phi \phi \pdv{}{\phi}
    )
    V_{\rm{eff}} (\phi) = 0, 
\end{align}
where we implicitly assumed that $\phi$ written here is a background field. We will employ this notation here. In the equation, $g_i$ denotes $i$-th coupling in a theory under consideration. Then, $\beta_{g_i}$ are associated beta functions, which describe the running of coupling with the renormalisation scale
\begin{align}
    \beta_{g_i} \equiv \mu \pdv{g_i}{\mu}.
\end{align}
Then, $\gamma_\phi$ is the anomalous dimension.   
Let us now consider the case of our toy model. The Callan-Symanzik equation becomes
\begin{align}
    \qty(
        \mu \pdv{}{\mu} 
        + \beta_{\lambda} \pdv{}{\lambda}
        + \beta_{m^2} \pdv{}{m^2}
        - \gamma_\phi \phi \pdv{}{\phi}
    )
    V_{\rm{eff}} (\phi) = 0. 
\end{align}
Now we want to solve this equation perturbatively up to order $\order{\lambda^2}$ in the quartic coupling, which corresponds to one-loop corrections. Note that the anomalous dimension is associated with contributions to the kinetic term of the effective action, which at one-loop level is just zero, cf. eq.~\eqref{eq:Veff_phi4}. Then the Callan-Symanzik equation becomes
\begin{align}
    \qty(
        \mu \pdv{}{\mu} V_{\rm CW}
        + \beta_{\lambda} \pdv{}{\lambda}V_{\rm tree}
        + \beta_{m^2} \pdv{}{m^2} V_{\rm tree}
    )
     = 0, \\
     \implies 
    \qty(
        -\frac{\lambda m^2 }{2\fpisq}\phi^2
        +\frac14 \frac{\lambda^2}{\fpisq} \phi^4
        + \beta_{\lambda} \frac{1}{4!}\phi^4
        + \beta_{m^2} \frac12 \phi^2
    ) + \order{\lambda^4}
     = 0,
\end{align}
by collecting the powers of $\phi$ together, we see that the equation is satisfied when
\begin{align}
\label{eq:beta_funcs_phi4}
    \beta_{m^2} = \frac{\lambda m^2}{\fpisq}, \quad 
    \beta_{\lambda} = \frac{3\lambda^2}{\fpisq}. 
\end{align}
These are the one-loop beta functions in real scalar theory. Using them, we can learn about the behaviour of relevant coupling as we change the renormalisation scale. We will show it explicitly in the next section. 
In principle, beta functions could also be obtained by studying the pole structure of relevant diagrams (see e.g. \cite{Schwartz:2014sze, Peskin:1995ev}) without a single mention of the effective potential. However, arguably, it is much more convenient to derive them via the Callan-Symanzik equation for the effective potential. 
This equation also shows an important, general principle. The effective potential with running couplings is RG-scale independent. More specifically, running of couplings cancels the explicit scale-dependent logarithm up to a given order in perturbation theory. Thus, if we had computed two-loop corrections as well, then the two-loop running of couplings would cancel a two-loop logarithm with the RG-scale. This principle holds for all quantum theories. 
 
\subsection{RG-improved effective potential}
The large logarithm we have encountered before (as well as others that would appear at higher orders) can be resummed by running the couplings with the renormalisation group equations (RGE) \cite{Schwartz:2014sze}. In other words, after including running couplings in the effective potential, we will be able to investigate large field values, since there will be no more explicit large logarithms. The effect of logarithms will now be encoded in the running couplings of the theory. 

Let us now demonstrate this procedure. First, we need to solve the RGE for the real scalar theory. We will start with the quartic coupling
\begin{align}
     \mu \pdv{}{\mu} \lambda &= \beta_{\lambda} = \frac{3\lambda^2}{\fpisq} \\
     \implies
     \label{eq:lambda_1PIrunning_phi4}
     \lambda(\mu) &= \frac{\lambda_0}{1 - \frac{3 \lambda_0}{\fpisq}\log\frac{\mu}{\mu_0}}, 
\end{align}
where $\mu_0$ is the initial, reference scale, and $\lambda_0 = \lambda(\mu_0)$.
We see that there exists a scale at which the denominator vanishes and consequently, coupling blows up to infinity. This scale corresponds to \textit{the Landau pole} of the theory, and it suggests that the one-loop potential cannot be trusted for an arbitrary energy scale. However, it can be used for energy scales smaller than that of the Landau pole.
Now, similarly, for the mass parameter, we obtain 
\begin{align}
    \mu \pdv{}{\mu} m^2 &= \frac{\lambda m^2}{\fpisq} \\
     \implies
     m^2(\mu) &= m_0 ^2 \exp(\frac{1}{\fpisq} \int^t _{t_0} \dd{t^\prime} \lambda(t^\prime) ),
\end{align}
where we have used the notation $t \equiv \log\frac{\mu}{\mu_0}$. Furthermore, we can now use the explicit form of running quartic coupling to obtain
\begin{align}
    m^2(\mu) = m_0 ^2 \qty(
        \frac{\lambda(\mu)}{\lambda_0} 
    )^\frac13
    =
    m_0 ^2 \qty(
        1 - \frac{3\lambda_0}{\fpisq}\log\frac{\mu}{\mu_0}
    ).
\end{align}
This means that in scalar theory, mass parameter squared scales roughly like a $\lambda^\frac13$ at one-loop order. Taking all the ingredients together, we can now obtain the total effective potential with running couplings 
\begin{align}
\label{eq:Veff_running_phi4}
V_{\rm{eff}} (\phi, \mu) = 
\frac12 m^2(\mu) 
+\frac{1}{4!}\lambda(\mu)\phi^4 
+ \frac{m^4_{\rm eff}(\mu)}{4\fpisq}\qty[
    \log\frac{m^2_{\rm eff}}{\mu} -\frac32
].
\end{align}
One may wonder that since we are still setting the RG-scale both in the logarithm and the couplings to some arbitrary value, it is always possible to pick $\phi \gg \mu$, which will lead to a large logarithm. It will not happen, though, if the RG-scale is going to be chosen dynamically, i.e. $\mu \sim \phi$. Such a choice for the RG-scale, for the potential above, leads to \textit{renormalisation-group improved potential}, or RG-improved in short. Moreover, to further minimise the logarithm, we may choose 
\begin{align}
    \tilde \mu = m^2_{\rm eff} =  m^2 + \frac{1}{2}\lambda \phi^2,
\end{align}
which leads to 
\begin{align}
    V_{\rm RG}(\phi) = 
    \frac12 m^2(\tilde \mu) \phi^2 
    +\frac{1}{4!}\lambda(\tilde \mu)\phi^4 
    -\frac32 \frac{m^4_{\rm eff}(\tilde \mu)}{\fpisq}.
\end{align}
This potential effectively resums the logarithms to all orders, and allows for use for arbitrary background field values for which couplings are not singular. We can now address the question of stability of the $\phi=0$ state. Considering $m^2 = 0$ the RG-improved potential becomes
\begin{align}
    V_{\rm RG}(\phi) = 
     \frac{1}{4!} \lambda(\tilde\mu) \phi^4
     -\frac{3}{8} \frac{\lambda(\tilde\mu)^2}{4\fpisq}\phi^4.
\end{align}
Looking at the running of quartic coupling in eq.~\eqref{eq:lambda_1PIrunning_phi4}, we can see that in the perturbative regime, i.e., when $\frac{\lambda}{4\pi} \ll 1$, the potential is indeed stable, and the second minimum we have encountered before was a mere mirage caused by an improper use of perturbation theory.

\section{Radiative symmetry breaking}
In the previous section, we have seen that naively including one-loop corrections for the scalar theory led to a false minimum in the effective potential. That minimum was a mirage, because it required the one-loop correction to become as important as the tree level action, which signals breakdown of perturbation theory organised as $\hbar$-expansion (or loop expansion in case of scalar theory).  On the other hand, if we had another field present, then maybe there could be a coupling hierarchy such that one-loop terms coming from integrating out fluctuations of the new field could affect the tree-level scalar potential. Such a phenomenon was described in a seminal paper by S.Coleman and E.Weinberg \cite{Coleman:1977py}, and here we will provide a summary.

\subsection{Massless scalar electrodynamics}
Consider now a theory of massless scalar electrodynamics, i.e. Abelian Higgs model with $m^2=0$. This theory contain a complex scalar field $\Phi$ and a U(1) gauge boson $A_\mu$, and is described by a Lagrangian
\begin{align}
\mathcal{L} = 
-\frac14 (F_{\mu\nu})^2 + \frac12(D_\mu \Phi)^2 - \frac{1}{4!}\lambda(\Phi \Phi^*)^2,
\end{align}
where the gauge field strength tensor is defined as
\begin{align}
    F_{\mu\nu} \equiv \partial_\mu A_\nu - \partial_\nu A_\mu,
\end{align}
and the covariant derivative is 
\begin{align}
    D_\mu \Phi \equiv (\partial_\mu + igA_\mu)\Phi.
\end{align}
We can observe that expanding the complex field in terms of its real components, i.e., $\Phi = \phi_1 + i\phi_2$, results in canonical normalisation for both of them in the Lagrangian. Therefore, as the effective potential will depend only on the $\phi = \sqrt{\phi_1 ^2 + \phi_2 ^2}$, it will also have canonical normalisation for the real field $\phi$, (see also \cite{Schwartz:2014sze, PhysRevD.7.1888, Coleman:1985rnk}). 
The effective potential containing one-loop gauge correction, calculated in Landau gauge and the $\overline{\rm{MS}}$ renormalisation scheme, takes the following form (for derivation see refs.~\cite{PhysRevD.7.1888, Swiezewska:2016rrp})
\begin{align}
    V_{\rm eff} = 
    \frac{1}{4!}\lambda \phi^4
    + \frac{\frac14 \lambda^2 \phi^4}{4\fpisq}\qty[ 
        \log\frac{\frac12 \lambda \phi^2}{\mu^2} - \frac32
    ]
    + \frac{3 g^4 \phi^4}{4\fpisq}\qty[ 
        \log\frac{ g^2 \phi^2}{\mu^2} - \frac56
    ].
\end{align}
The effective potential of this form also has a minimum for some non-zero field value. However, now it can be an actual global minimum. If we consider a scenario where the couplings obey a scaling, 
\begin{align}
    \lambda \sim g^4 \ll 1,
\end{align}
Then we can organise the effective potential not in number of loops but in powers of the gauge coupling. In other words, in the $\hbar$-expansion, the leading order effective potential consisted of tree-level terms, i.e. $\mathcal{O}(\lambda)$ terms. However, now, noting the scaling $\lambda \sim g^4$, we would denote the LO effective potential (in a new expansion) as
\begin{align}
    V_{\rm eff} = 
    \underbrace{
    \frac{1}{4!}\lambda \phi^4
    + \frac{3 g^4 \phi^4}{4\fpisq}\qty[ 
        \log\frac{ g^2 \phi^2}{\mu^2} - \frac56
    ]
    }_{V_{\rm eff}^{\rm LO}}
    +\underbrace{
     \frac{\frac14 \lambda^2 \phi^4}{4\fpisq}\qty[ 
        \log\frac{\frac12 \lambda \phi^2}{\mu^2} - \frac32
    ]
    }_{ \order{g^8} }.
\end{align}
where we can observe that one-loop scalar corrections belong to higher orders.\footnote{
In fact, this contribution belongs to next-to-next-to-leading order (NNLO) corrections to the effective potential ${V_{\rm eff}^{\rm NNLO}}$, because the previous, next-to-leading order (NLO) corrections contain terms scaling as $\order{g^6}$. Such contributions are given as ``sunset'' diagrams containing internal lines with both gauge and scalar fields, see e.g.~\cite{Andreassen_2015}.
}
Thus, a quantum correction can substantially modify the ``classical'' theory as long as there is a hierarchy between couplings (which, as we will see, also implies hierarchy between masses of fields). 
Focusing on the LO potential, we can try to find the global minimum $\expval{\phi}$ as
\begin{align}
    \log\frac{ g^2 \phi^2}{\mu^2} &= 
    \frac13 - \frac{8\pi^2}{9} \frac{\lambda}{g^4}, \\
    \implies \log\frac{\expval{\phi}^2}{\mu^2} &\simeq
    \frac13 - \frac{8\pi^2}{9} \frac{\lambda}{g^4}.
\end{align}
We see that a non-zero global minimum was indeed generated by gauge radiative corrections. This phenomenon is called \textit{radiative symmetry breaking}.
Now, if we choose the renormalisation scale such that $\mu = \expval{\phi}$, then we arrive at a relation between couplings \cite{PhysRevD.7.1888}
\begin{align}
    \lambda  = \frac{3}{8\pi^2} g^4,
\end{align}
which is exactly the scaling we have postulated before. Using this relation, we can further re-express our LO effective potential as
\begin{align}
\label{eq:VLO_CWmodel_final}
    V_{\rm eff}^{\rm LO}  = 
    \frac{3 g^4 \phi^4}{4\fpisq}\qty[ 
        \log\frac{ g^2 \phi^2}{\expval{\phi} ^2} - \frac12
    ].
\end{align}
Note that we have started with the theory described by two free parameters: $\lambda, g$, but now we do not see the quartic coupling. In fact, it was not removed but rather replaced by $\expval{\phi}$, which is a dimensionful parameter. This phenomenon is called \textit{dimensional transmutation}~\cite{PhysRevD.7.1888}. This feature is inevitable in massless theories with radiative symmetry breaking. Physically, it can be interpreted as the change of value of the quartic coupling with different renormalisation scales, as it is closely related to the renormalisation group.

Finally, let us now consider the particle spectrum of the theory with RSB. Similar to the regular Abelian Higgs model, one of the real components of the complex scalar field becomes massive, while the other becomes the longitudinal component of the gauge field. We can obtain the masses of scalar and gauge particles as 
\begin{align}
    m_\phi ^2 &= 
    \eval{\qty(V_{\rm eff}^{\rm LO})^{\pprime} }_{ \expval{\phi}}
    =  \frac{3 g^4 }{8\pi^2} \expval{\phi} ^2, \\
    m_A ^2 &= g^2\expval{\phi}^2.
\end{align}
Our theory now has a concrete prediction,
\begin{align}
    \frac{m_\phi ^2}{m_A ^2} = \frac{3g^2}{8\pi^2},
\end{align}
which establishes a hierarchy between masses. In a reverse argument, the equation above shows that perturbative expansion in theories with RSB requires the gauge field to be heavier than the scalar. If it were the other way around, then the gauge coupling would have to be large, thus invalidating the perturbation theory.\footnote{ 
We have calculated the LO effective potential using a particular choice of gauge. However, it can be shown that while the effective potential itself is gauge-dependent, its physical predictions are gauge-invariant in perturbation theory \cite{Andreassen_2015}. Therefore, the above ratio of masses is in fact gauge-invariant \cite{Schwartz:2014sze}.
}
Since we started with massless theory, at the beginning, there were no dimensionful parameters in the classical Lagrangian. Such theories are often called \textit{conformal} or theories with \textit{classical scale invariance}. We have seen, however, that quantum corrections introduce an explicit mass scale which breaks the classical scale invariance. This method of dynamical scale generation cannot be applied to the Standard Model, as it is not possible to obtain the measured mass of the Higgs boson. However, such theories are still popular in the context of extension of the Standard Model, see e.g. \cite{PhysRevD.99.015026} for a review.

Lastly, note that the effective potential in eq.~\eqref{eq:VLO_CWmodel_final} is valid for the particular relation of couplings. Such a relation is indeed satisfied at the global minimum of the theory; however, it may not be satisfied for smaller field values, i.e. closer to the origin of field space. This can be important, e.g. in the studies of vacuum decay \cite{Andreassen:2016cvx, Coleman:1977py, PhysRevD.16.1762} or thermal phase transitions. In fact, usually there is a large separation between the field values near the origin and close to the global minimum. Thus, to have an effective potential valid over a large field range, it is necessary to obtain an RG-improved potential \cite{Kierkla:2022odc}. We will use such potential in the context of the classically scale-invariant extension of the Standard Model studied in this thesis.

\chapter{Thermal field theory and phase transitions}
\label{chapter:TFT}

In this chapter, we will review the basics of thermal field theory, a framework that will allow us to study the cosmological phase transition in the hot early Universe. 
We will begin with a general description of the first-order phase transition and its dynamics.
Then we will move on to theoretical formalism. 
We will introduce imaginary-time formalism \cite{Matsubara:1955ws, Laine_2016} and study the relation between the free energy and the effective potential at finite temperature. Then we will introduce \textit{high-temperature dimensional reduction} \cite{GINSPARG1980388, Appelquist:PhysRevD.23.2305, Nadkarni:PhysRevD.27.917, LANDSMAN1989498}, a powerful tool based on the effective field theory framework, which allows us to neatly organise perturbative computations at finite temperature \cite{Kajantie:1995dw, Farakos_1994, Braaten:1995cm, Braaten:PhysRevD.53.3421}. 
We will illustrate the method with a simple example and then finally show how it can be used to perturbatively compute the thermal nucleation rate, which is the main quantity to describe the phase transitions. 

This chapter does not aim to be a detailed review, but rather serves as a basic introduction to the problems studied in this thesis and the tools used to handle them, illustrated within a simple toy model. 
The content is based mostly on textbooks \cite{Laine_2016, Lancaster:2014pza}, and articles \cite{Farakos_1994, Kajantie:1995dw, Gould:2021, Gould:2023ovu, Lofgren:2021ogg, Ekstedt:2021kyx, Ekstedt:2022tqk, Ekstedt:2022ceo, Gould:2021dzl, Gould:2021ccf, Ekstedt:2023sqc}. 

\section{First-order phase transitions}
\begin{figure}
    \centering
    \includegraphics[width=.5\textwidth]{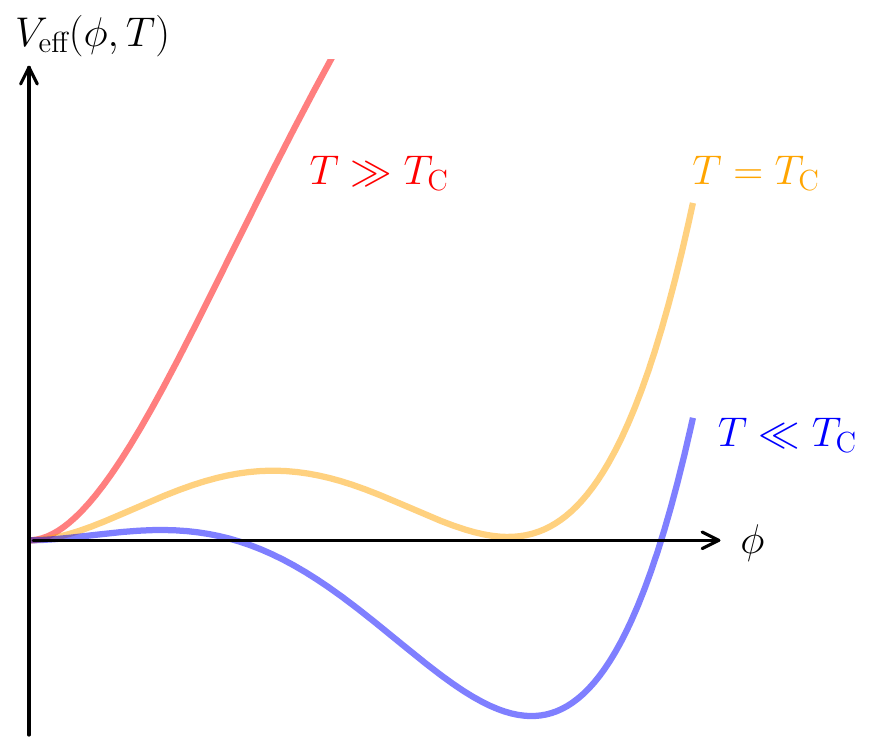}
    \caption{Potential exhibiting first-order phase transition. The value of global minimum changes at the temperature $\Tc$, hence providing discontinuity in $\dd f / \dd T.$}
    \label{fig:VPT}
\end{figure}

Phase transitions in a given system can be studied by considering the analyticity of the free energy density $f(T,\mu)$ on a line in $(T, \mu)$-space, where $T$ is temperature and $\mu$ is the chemical potential. 
In the context of cosmological phase transitions, the free energy is equal to the value of the effective potential at the global minimum  $\phi_{\rm min}$ at a given temperature (see e.g. \cite{Laine_2016} for a pedagogical review):
\begin{align}
    f(T) = V_{\rm eff}(\phi_{\rm min}(T), T).
\end{align}
The order of a phase transition is related to the discontinuity in a given order of derivative of the free energy. For example, if $\pdv{f}{T}$ or $\pdv{f}{\mu}$ has discontinuity, then the transition is of \textit{first-order}. 
Now, if we write the temperature derivative of the effective potential, we get: 
\begin{align}
    \dv{ f(T)}{T} =
    \dv{ V_{\rm eff}}{T} =
    \eval{\qty(\pdv{V_{\rm eff}}{\phi} \pdv{\phi_{\rm min}}{T} + \pdv{V_{\rm eff} }{T})}_{\phi=\phi_{\rm min}} = 
    \eval{\pdv{V_{\rm eff} }{T}}_{\phi=\phi_{\rm min}}.
\end{align}
We see that the discontinuity can appear at some $T=\Tc$, if the following is true:
\begin{align}
    \lim_{T\rightarrow T_c^+} \dv{f}{T} \neq \lim_{T\rightarrow T_c^-} \dv{f}{T}.
\end{align}
We are going to refer to $\Tc$ as the \textit{critical temperature}.
In terms of the effective potential, the above expression translates to,
\begin{align}
    \lim_{T\rightarrow T_c^+} \phi_{\rm min} \neq \lim_{T\rightarrow T_c^-} \phi_{\rm min},
\end{align}
i.e. there must be a discontinuity in the position of the global minimum in the effective potential. A situation like this occurs whenever the effective potential possesses a barrier between minima, as shown in figure~\ref{fig:VPT}.
A first-order phase transition possesses a non-trivial dynamics. This is due to the discontinuity in the energy density -- a latent heat. This energy, stored in the old phase, needs to be transported or dissipated. 
As the temperature drops below the critical temperature, the system will be supercooled, and the transition will proceed via nucleation of bubbles of the new phase. This new phase corresponds to the global minimum of the effective potential.  Then, these bubbles will expand fast until they fill the entire volume (such as the whole Universe) and complete the phase transition. Thus, the first-order transition takes place at some lower temperature, $T<\Tc$, which we will denote as the nucleation temperature. Then, we can also consider the temperature of percolation of the bubbles, which can be identified as the completion of the transition. 

The overall goal of the next section is to determine the probability of bubble nucleation per unit time and volume at a given temperature. It is also often called the \textit{bubble nucleation rate}. It will be the key quantity that will allow us 
to study thermal phase transitions. For example, it will allow us to estimate the nucleation temperature, as the temperature at which at least one bubble nucleates in a considered volume.
Then, in the next chapters of this thesis, we will apply this framework in the context of cosmological phase transitions, taking place in the early Universe.




\section{Imaginary time formalism} 
\label{sec:matsubara_formalism}

\subsection{Partition function and Matsubara modes}

Thermal (or statistical) field theory can be formulated via so-called \textit{imaginary-time formalism}. Sometimes it is also called \textit{Matsubara formalism} \cite{Matsubara:1955ws}. This formalism allows for studying equilibrium quantities, and it captures equal-time correlation functions of a QFT at finite temperature. In this framework, time is imaginary and periodic. Therefore, one can imagine thermal field theory as a quantum field theory with 3 dimensions and a warped temporal dimension, i.e. fields live in a $\mathbb{E}^3 \times \mathbb{S}^1$ manifold, where $\mathbb{E}^3$ is a three-dimensional (3d) Euclidean space and $\mathbb{S}^1$ is a circle. The theory can be defined via a partition function
\begin{align}
    \mathcal{Z} = \int \mathcal{D}\Phi e^{-S_\rmii{E}[\Phi]},
\end{align}
where ${S_\rmii{E}}$ is the Euclidean action given by
\begin{align}
    S_\rmii{E} \equiv \int_0^\beta \dd{\tau} \int\dd[3]{x} \mathcal{L}_\rmii{E},
\end{align}
and $\mathcal{L}_\rmii{E}$ is a 3d Lagrangian density. The first integral here corresponds to the ``time-circle'' we have mentioned before. The imaginary time $\tau$ is related to real time by $\tau=-it$. The period of temporal dimension is given by $\beta \equiv 1/k_\rmii{B} T$, where  $k_\rmii{B}$ is the Boltzmann constant and $T$ is the temperature. From now on, we will work in units $k_\rmii{B} =1$. Note that setting $T=0 $ results in an infinite period, which just corresponds to a regular, non-warped temporal dimension. Thus, we would recover vacuum 4d Euclidean integration.
Fields governed by the partition function are then decomposed as 
\begin{align}
    \Phi(\tau, \vec{x}) = 
    T \sum_{\omega_n} \int_{k} \phi_n(k) e^{\omega_n \tau - \vec{k} \cdot \vec{x}},
\end{align}
where $\int_k = \int \frac{\dd[3]k}{(2\pi)^3}$.
The modes $\phi(k)$ are called \textit{Matsubara modes} and 
$K^2 = \omega_n^2 + k^2$, where $ \omega_n$ are \textit{Matsubara frequencies} defined as
\begin{align}
    \omega_n &\equiv 2n\pi T \quad \mbox{for bosons,}\\
    \omega_n &\equiv (2n+1)\pi T \quad \mbox{for fermions,}
\end{align}
and $n=0,1,2,\dots$ are positive integers. These modes are a result of different periodic boundary conditions for the quantum fields. Boson fields are periodic in the imaginary time direction, leading to even-integer numbers, while fermion fields are anti-periodic, which gives odd numbers in the Matsubara frequency. 
For our purposes, we can focus on bosonic fields only, but we want to stress that the formalism summarised above can also handle fermionic degrees of freedom. 

It is worth investigating a bosonic propagator \cite{Laine_2016} 
\begin{align} 
\label{eq:matsubara_boson_propagator}
    \sumint_{~K} \frac{1}{P^2 + m^2} = 
    \sumint_{~K} \frac{1}{\omega_n^2 + m^2 + k^2 }, 
     \mbox{ where } \sumint_{~K} \equiv T \sum^\infty _{n=-\infty} \int_k,
\end{align}
and $m$ is the mass of a boson. We can observe that the Matsubara frequency acts as an effective, ``thermal'' mass. At high temperatures, these contributions can even dominate the total ``effective'' mass. This will be crucial for the studies of thermal phase transitions, as then we are interested precisely in the regime where temperature fluctuations dominate over quantum fluctuations, and we can expect $T\gg m$.

\subsection{Finite temperature effective potential}
Now, let us discuss the effective potential at finite temperature, which will be our main tool to study possible phases in a given theory. 
Let us then consider a static background $\phib$, which corresponds to a Matsubara zero mode with $\omega_n=0$ and $k=0$. It can be shown then that the total effective potential can be divided into a vacuum ($T=0$) and thermal part, see e.g.~ref.~\cite{Laine_2016, Quiros:1999jp} for a pedagogical derivation:
\begin{align}
\label{eq:Veff_T_nodaisy}
    V_\rmii{eff}(\phib, T) = V_\rmii{tree}(\phib) + V_\rmii{CW}(\phib) + V_\rmii{$T$}(\phib, T).
\end{align}
Vacuum part is the effective potential at zero temperature, which we have described in the previous chapter; the thermal part, on the other hand, at one-loop order is given via the so-called \textit{thermal functions}
\begin{align}
    V_T(\phib, T) &\equiv \frac{T^4}{2\pi^2}  \sum_a n_a J_\rmii{b/f}\qty(\frac{m_a}{T}),
\end{align}
where $n_a$ denote degrees of freedom of a $a$-th particle species and
\begin{align}
    J_\rmii{b/f}\qty(y) &\equiv \int_0^\infty \dd{x} x^2 
    \log
    \qty(
        1\pm e^{-\sqrt{x^2+y^2}}  
    ),
\end{align}
with the ``$+$'' sign for fermions and ``$-$'' sign for bosons~\cite{Laine_2016}. It is important to emphasise that $m_a$ corresponds to the \textit{effective} mass of a particle, which is given by a second derivative of the zero-temperature potential. 
These functions cannot be expressed in terms of elementary functions, although they can be evaluated numerically, see e.g.~\cite{Fowlie:2018eiu}. Moreover, there exist both high-$T$ (small $y$) and low-$T$ (large $y$) expansions. When the temperature is high with respect to the mass scale, $m_{a}/T\ll 1$, the thermal functions can be expanded as 
\begin{align}
\label{eq:T-master-high-T}
J^{y \ll 1 }_{T,b}(y)= &\ -\frac{\pi^4}{45} + \frac{\pi^2}{12} y^2 - \frac{\pi}{6} y^{3} - 
\frac{1}{32} y^4\left( \log\frac{y^2}{16 \pi^2} -\frac{3}{2} + 2 \gamma_E \right) + \nonumber\\
&+ \pi^2 y^2 \sum_{i=2}^{\infty}\left(-\frac{1}{4 \pi^2} y^2 \right)^
i \frac{(2 i - 3)!! \zeta(2 i - 1)}{(2 i)!! (i + 1)}, \nonumber \\
J^{y\ll 1}_{T,f}(y)=&\   \frac{7\pi^4}{360} - \frac{\pi^2}{24} y^2 - \frac{1}{32} y^4 \left(\log\frac{y^2}{\pi^2}-\frac{3}{2} + 2 \gamma_E\right) + \nonumber\\
&+\pi^2 y^2 \sum_{i=2}^{\infty}\left(\frac{-1}{4 \pi^2} y^2\right)^i \frac{(2 i - 3)!! \zeta(2 i - 1)}{(2 i)!! (i + 1)} \left(2^{2 i - 1} - 1\right),
\end{align}
and in the opposite regime $m _a/T\gg 1$, the expansion is
\begin{align}
J_{T}^{y\gg 1}(y)= &-e^{-y} \left(\frac{\pi}{2} y^{3}\right)^{\frac{1}{2}}
\sum_{i=0}^{\infty}\frac{1}{2^i i!} \frac{\Gamma(\frac{5}{2} + i)}{\Gamma(\frac{5}{2} - i)} y^{-i},\label{eq:low-T}
\end{align}
which is the same for bosons and fermions. Above, $\gamma_E$ denotes the Euler-Mascheroni constant, $\zeta$ is the Riemann zeta function, and $\Gamma$ is the gamma function.

\subsection{Infrared problems at high temperatures}
Unfortunately, the effective potential at finite temperature does not admit a straightforward loop expansion. Naive computations in an interacting theory lead to infrared divergences in loop corrections. In particular, corrections containing odd-powers of mass $m$ are characterised by an effective expansion parameter  $\sim \frac{\lambda T^2}{m^2}$ (see e.g. \cite{Laine_2016}). We can thus immediately see that for small masses, or for high temperatures, i.e., $T\gg m$, the expansion is no longer under control -- higher loops can contribute at the same order as tree level. 

\begin{figure}
    \centering
    \begin{tikzpicture}[very thick]
        \begin{feynman}
            \def\radius{1.3}  

            \draw[dashed] (0,0) circle (\radius);
            
            \draw[] (-0.707*1.3*\radius, -0.707*1.3*\radius) circle (\radius*0.3);
            \draw[] (-1.3*\radius, 0) circle (\radius*0.3);
            \draw[] (0, 1.3*\radius) circle (\radius*0.3);
            \draw[] (-0.707*1.3*\radius, 0.707*1.3*\radius) circle (\radius*0.3);
            \draw[] (+0.707*1.3*\radius, 0.707*1.3*\radius) circle (\radius*0.3);
            \filldraw[black] (1.3*\radius,0) circle (2pt);
            \filldraw[black] (0.9659 * 1.3*\radius,0.2588* 1.3*\radius) circle (2pt);
            \filldraw[black] (0.9659 * 1.3*\radius,-0.2588* 1.3*\radius) circle (2pt);
            \draw[] (+0.707*1.3*\radius, -0.707*1.3*\radius) circle (\radius*0.3);
        \end{feynman}
    \end{tikzpicture}
    
    \caption{Daisy diagram. Dashed lines correspond to the Matsubara zero mode, while solid lines are propagators of non-zero modes.}
    \label{fig:daisy_diagram}
\end{figure}
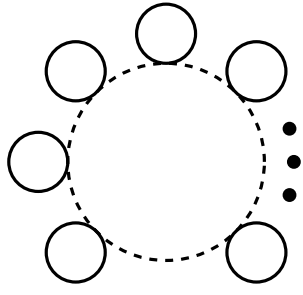

This issue can be partially addressed by resummations of leading-order diverging terms \cite{Arnold:1992, Lofgren:2023sep, Parwani:PhysRevD.45.4695}. These terms can be identified as so-called \textit{daisy diagrams} as shown in figure~\ref{fig:daisy_diagram}. Such a diagram contains a zero-mode loop that gets corrections from $N$-loops of non-zero modes. These contributions have a physical interpretation: at high temperature, the zero Matsubara mode is effectively screened by non-zero modes. Fortunately, these contributions can be summed, leading to a finite contribution --  an effective ``thermal mass'' of a form $\sim \lambda^{\frac{3}{2}}T^2$ in scalar theory.\footnote{This mass should not be mistaken with the Matsubara frequency term in the bosonic propagator. Moreover, for the zero Matsubara mode $\omega_n =0$ anyway.} 
This scaling indeed modifies coupling expansion for the effective potential in comparison to the vacuum case, as now the most relevant correction to the tree level is not of order $\mathcal{O}(\lambda^2)$ but rather $\mathcal{O}(\lambda^{\frac{3}{2}})$. 
Incorporating this resummation into the effective potential, we arrive at the schematic expression
\begin{align}
    V_\rmii{eff}(\phib, T) = V_\rmii{tree}(\phib)+ V_\rmii{CW}(\phib) + V_\rmii{$T$}(\phib, T) + V_\rmii{daisy}(\phib, T).
\end{align}
The exact form of the contribution from resumming IR divergent terms is, of course, model dependent; thus, we will now illustrate it with a simple example. 

\subsection{Scalar theory at finite temperature}

Consider a theory of a real, interacting scalar field, described by a Lagrangian
\begin{align}
\label{eq:LagrangianE_lamhi4}
    L_\rmii{E} = \frac{1}{2} (\partial_\mu \phi)^2
    +\frac{1}{2} m^2 \phi^2 + \frac{1}{4!}\lambda\phi^4.
\end{align}
Following \cite{Laine_2016}, we can construct the effective potential by decomposing the field into zero Matsubara mode (static background), which we will denote as $\phib$, and non-zero modes denoted as $\phi^\prime$
\begin{align}
    \phi = \phib + \phi^\prime.
\end{align}
Then, following the recipe in eq.~\eqref{eq:Veff_T_nodaisy} and the results from the previous chapter, we immediately obtain
\begin{align}
    V_\rmii{tree}(\phib) &= \frac{1}{2} m^2 \phib^2 
    + \frac{1}{4!}\lambda\phib^4, \\
     V_\rmii{CW}(\phib) &= 
        \frac{m^4_\rmii{eff}}{4(4\pi)^2}\qty(\log{\frac{m^2_\rmii{eff}}{\mu^2}} -\frac32), \\
    V_\rmii{$T$}(\phib, T) &= \frac{T^4}{2\pi^2} 
    J_\rmii{b}\qty(
        \frac{m_\rmii{eff}}{T} 
        ),
\end{align}
where $m^2_\rmii{eff} = (m^2 + \frac12 \lambda \phib^2)$. 
As we are interested in the high-temperature regime, then in high-T expansion the term $ V_\rmii{$T$} $ can be written as:
\begin{align}
\label{eq:VT_lambda_phi4}
     V_\rmii{$T$}(\phib, T) \simeq 
     -\frac{\pi^2 T^4}{90} 
     +\frac{m^2_\rmii{eff}T^2}{24} 
     -\frac{(m^2_\rmii{eff})^{3/2} T}{12\pi}
     + \frac{m^4_\rmii{eff}}{4(4\pi)^2} \qty(
        \log\frac{m^2_\rmii{eff}}{(4\pi T)^2}
        -\frac32 +2\gammaE
     ).
\end{align}
Looking at the quadratic terms in $V_\rmii{tree} +  V_\rmii{$T$} $, we see that the scalar field receives a thermal contribution to its mass
\begin{align}
    V_\rmii{tree} +  V_\rmii{$T$} &\supset 
    \frac12 \qty(m^2 + \Pi_T)\phi^2, \\
    \Pi_T &\equiv \frac{\lambda T^2}{24}.
\end{align}
We see that such a correction directly affects the tree-level potential despite the fact that it is a one-loop effect. Now, let us investigate the influence of a set of leading higher-order loop diagrams. In particular, we need to resum daisy diagrams, see figure~\ref{fig:daisy_diagram}. Such a diagram consists of a main loop containing zero Matsubara modes of the scalar field, and other $N=0,1,2,\dots$ loops of non-zero modes attached to it. In general, we need to consider infinitely many loops attached. Fortunately, the sum of such diagrams converges.  
For this theory, the resummation of all daisy diagrams is given by \cite{Laine_2016}
\begin{align}
    \sum_{N=0}^\infty
    \frac{1}{N!} 
    \qty(\frac{\lambda T^2}{24} )^N
    \qty(\dv{ }{m_\rmii{eff} ^2}  )^N
    \qty(- \frac{-m_\rmii{eff} ^3 T}{12\pi}) 
    =
    -\frac{T}{12\pi} 
    m_{\rm th} ^{\frac{3}{2}},
\end{align}
where, 
\begin{align}
     m_{\rm th} ^{\frac{3}{2}} &\equiv \qty(m_\rmii{eff}^2 + \frac{\lambda T^2}{24})^{\frac{3}{2}}. 
\end{align}
Since we have resummed all $N$-daisy-loops, this expression also contains the contribution of the ``tree-level'', $N=0$ term, i.e, the cubic term in eq.~\eqref{eq:VT_lambda_phi4}. Therefore, to avoid double-counting, we can express the daisy correction term in a form
\begin{align}
    V_{\rm daisy}(\phib,T) \equiv \frac{T}{12\pi}
    \qty[
        (m_\rmii{eff}^2)^{\frac{3}{2}} - (m_{\rm th} ^2)^{\frac{3}{2}}
    ],
\end{align}
which in the literature is often referred to as \textit{Arnold-Espinosa resummation scheme} \cite{Arnold:1992}. 
The daisy-correction term in such form allows us to just add it to the rest of the effective potential ingredients, which will result in the replacement of the previous cubic term with the one containing a daisy resummed thermal mass
\begin{align}
\label{eq:Veff_daisy_phi4}
    V_{\rm eff}(\phib, T) =& 
    V_\rmii{tree}(\phib) + V_\rmii{CW}(\phib) + V_\rmii{$T$}(\phib, T) + V_\rmii{daisy}(\phib, T) \notag \\
    =&
    -\frac{\pi^2 T^4}{90} 
    +\frac{1}{2} m^2 \phib^2 + \frac{1}{4}\lambda\phib^4 
     +\frac{m^2_\rmii{eff} T^2}{24} 
     -\frac{T}{12\pi} (m_{\rm th} ^2)^{\frac{3}{2}} 
     -\frac{m^4_\rmii{eff}}{4(4\pi)^2} L_b(\mu) ,
\end{align}
where, following refs.~\cite{Kajantie:1995dw, Braaten:1995cm} we introduce
\begin{align}
    L_b \equiv \log\frac{e^{2\gammaE} \mu^2}{(4\pi T)^2}.
\end{align}
We see now that this form of effective potential is completely IR-safe, as taking the $m\rightarrow0$ limit results in a finite expression.   

\subsection{Why daisy-resummation is not enough}
This form of the effective potential at finite temperature is arguably the most commonly used in the literature for the study of cosmological phase transitions. This approach, while relatively simple, has many drawbacks and, in general, is not sufficient for reliable predictions regarding phase transition physics. While the daisy resummation approach is capable of capturing the qualitative behaviour of the system, it suffers from many theoretical uncertainties \cite{Croon:2020cgk, Gould:2021,  Gould:2021ccf, Lofgren:2023sep}, such as 
\begin{itemize}
    \item gauge-dependence,
    \item imaginary parts in the effective potential,
    \item uncancelled RG-scale dependence.
\end{itemize}
Arguably, the last issue causes the most trouble \cite{Lewicki:2024xan, Croon:2020cgk} as it can affect the phenomenological predictions, e.g., the amplitude of the gravitational wave signal by a few orders of magnitude. Here, following \cite{Gould:2021, Gould:2023ovu}, we will quickly review the core of problematic RG-scale dependence in the daisy-resummation approach. Let us consider the effective potential for the real scalar theory given by eq.~\eqref{eq:Veff_daisy_phi4}. For the following discussion, it is convenient to separate the potential into powers of the couplings
\begin{align}
V_\rmii{eff} = 
\underbrace{
    -\frac{\pi^2 T^4}{90} 
    }_{\rm const.} 
    + \underbrace{
        \frac{1}{2} m^2 \phib^2 
        + \frac{1}{4!}\lambda\phib^4 
        +\frac{m^2_\rmii{eff} T^2}{24}
    }_{   \equiv V^{(\lambda)}   }
     -\underbrace{
        \frac{T}{12\pi} (m_{\rm th} ^2)^{\frac{3}{2}}
     }_{ \equiv V^{(\lambda^{3/2})} }
     -\underbrace{
       \frac{m^4_\rmii{eff}}{(4\pi)^2} \frac{L_b(\mu)}{4} 
     }_{ \equiv V^{(\lambda^{2})} _\rmii{ log} }.
\end{align}
While the beta functions of the couplings are unaffected by finite-temperature corrections, i.e., they are the same as in eq.~\eqref{eq:beta_funcs_phi4}, we see that the running of the leading-order effective potential will actually be affected
\begin{align}
\label{eq:Vlambda_beta}
    \mu\dv{}{\mu} V^{(\lambda)} 
    &= \frac12 \mu \dv{m^2}{\mu} \phib^2
    + \frac{1}{4!}\mu \dv[]{\lambda}{\mu}  \phib^4
    + \frac{T^2}{24} \mu \dv{m^2_\rmii{eff}}{\mu}\\ 
    &= \frac{1}{(4\pi)^2}\qty(
        \frac12 \lambda m^2 \phib^2 
        +\frac18 \lambda^2 \phib^4
        +\frac{1}{24} \lambda T^2 m^2 
        +\frac{1}{16}\lambda^2 T^2 \phib^2
    ).
\end{align}
Note that the constant term could be dropped, as in practice one is interested in a normalised effective potential. Then, at subleading order $\mathcal{O}(\lambda^{3/2})$ of the effective potential, its running is of higher order
\begin{align}
    \mu\dv{}{\mu} V^{(\lambda^{3/2})} 
    &= \frac{T}{12\pi} \mu \dv{}{\mu} m_{\rm th} ^3 (\phib) 
    = \frac{T}{12\pi}  \qty(
        \frac32 m_{\rm th}(\phib) \qty(
            \mu\dv{m_\rmii{eff}^2}{\mu} + \frac{T^2}{24} \mu\dv{\lambda}{\mu}
        )  
    ) \sim \mathcal{O}(\lambda^{5/2}).
\end{align}
Finally, the running of the $\mathcal{O}(\lambda^{2})$ part of the effective potential is similar to the vacuum case
\begin{align}
\label{eq:Vlambda2_beta}
    \mu\dv{}{\mu} V^{(\lambda^2)}_\rmii{ log}  
    &= 
    \mu\dv{}{\mu} \qty(-
        \frac{m^4_\rmii{eff}}{(4\pi)^2} \frac{L_b(\mu)}{4} 
    )
    = 
    \mu\dv{}{\mu} V_\rmii{CW},
\end{align}
where $V_\rmii{CW}$ is the Coleman Weinberg term, i.e. scalar one-loop correction at zero temperature.
Then, in analogy to the vacuum effective potential, one could expect that implicit running of tree-level parameters will cancel the logarithms from one-loop corrections. Unfortunately, this is not the case, as by looking at equations \eqref{eq:Vlambda_beta} and \eqref{eq:Vlambda2_beta} we see that there will be some leftover dependence. More specifically, by investigating the running of the total effective potential, we get  
\begin{align}
    \mu\dv{}{\mu} V_\rmii{eff} &= 
    \mu\dv{}{\mu} V^{(\lambda)} 
    + \mu\dv{}{\mu} V^{(\lambda^{3/2})} 
    + \mu\dv{}{\mu} V^{(\lambda^2)}_\rmii{ log} \\
    &= 
    \underbrace{
        \mu\dv{}{\mu} V_\rmii{tree}
        + \mu\dv{}{\mu} V_\rmii{CW}
    }_{=0}
    + \underbrace{
         \frac{T^2}{24} \mu \dv{m^2_\rmii{eff}}{\mu} 
    }_{\rm uncancelled}
    + \underbrace{
        \mu\dv{}{\mu} V^{(\lambda^{3/2})} 
    }_{\sim~\mathcal{O}(\lambda^{5/2})}.
\end{align}
Thus, the effective potential at $\mathcal{O}(\lambda^2)$ is not RG-scale invariant. The reason is the running of the thermal contribution to the scalar mass, i.e. 
\begin{align}
\label{eq:selfE_phi4_4d_running}
    \mu\dv{}{\mu} V_\rmii{eff} = 
    \mu\dv{}{\mu} \Pi_T = 
    \mu\dv{}{\mu} \qty(\frac{\lambda T^2}{24})
    = \frac{1}{(4\pi)^2}\qty(\frac{1}{16}\lambda^2 T^2),
\end{align}
where we dropped the constant term. This is a general feature of the daisy resummation approach for computing the effective potential. To see why it can cause problems, let us consider more general theories. There, the above running of thermal contribution to the self-energy will be affected by other couplings, which can be large. Such situations can occur, for example, in BSM theories where large couplings are necessary to change the order of the electroweak phase transition. In such a situation, the daisy-resummation approach to the effective potential will cause a large RG-scale dependence of the phenomenological predictions \cite{Croon:2020cgk, Gould:2021}. 

Even if there might be theories where the daisy term alone gives good convergence in some perturbative expansion, the list of problems listed at the beginning of this subsection prevails. The only and main advantage of this approach is its relative simplicity. It is also no surprise that such an easy fix is not a general and robust framework. Instead, the main lesson of calculations at finite temperature is that each theory under consideration should first be carefully analysed in terms of mass and coupling hierarchies. Then, the calculation of observables should be done in a strict perturbative way \cite{Lofgren:2023sep}. There exists a robust framework for that, based on the grounds of high-temperature \textit{effective field theories}.



\section{Dimensional reduction}
Looking at the form of the bosonic propagator in eq.~\eqref{eq:matsubara_boson_propagator}, we see that if the mass parameter is smaller than the lightest Matsubara frequency, i.e. $m \ll 2\pi T$, then there exists a strict scale hierarchy between the mass of the (light) zero mode and the (heavy) non-zero modes. We can also see that from the point of view of the Lagrangian. Writing the action for a bosonic degree of freedom leads to 
\begin{align}
S = 
\frac{1}{T} 
\int d^3x { \sum_{n=0}^\infty
    \left[ 
        \frac12(\partial_i \phi_n)^2 
        + \frac12 (2\pi n T)^2\phi_n ^2
        + \frac12 m^2 \phi_n ^2
    \right]
}. 
\end{align}
We see that $n$-th Matsubara mode carries a mass 
\begin{align}
    m_n ^2 = m^2 + (2\pi nT)^2 \xrightarrow{T\gg m} (2\pi nT)^2,
\end{align}
then, at high temperatures, we have the explicit hierarchy
\begin{align}
    m_0^2 \ll m_1^2 \ll m_2^2 \ll \dots \ll m_n^2.
\end{align}
Such a scale hierarchy allows for the construction of an effective field theory (EFT) for the lightest, zero mode. In such a framework, the heavy (or UV) modes are integrated out, and their effects on the dynamics of the zero mode (IR) are encoded in the effective parameters of the EFT. For a review of the general concept of EFTs, see e.g.~\cite{Cohen:2019wxr, Burgess_2020}. 

In this section, we will discuss the most important concepts of EFT at high temperature, called \textit{dimensional reduction}, and illustrate the exact construction for a simple toy model. Finally, we will discuss the application to the gauge theory, which will be relevant for the main results of this thesis. 

\subsection{Thermal scales hierarchy}
The root of all problems in perturbation theory of thermal field theory is the infrared physics of bosonic modes. We saw that for small masses, or equivalently for large temperatures, the perturbative expansion in terms of loops may be insufficient to correctly calculate free energy. To see that in more general terms, let us denote the expansion parameter for bosons at high
temperatures, i.e. $\pi T \gg m$ as \cite{Laine_2016}
\begin{align}
    \epsilon_\rmii{b} \sim \frac{g^2 T}{\pi m},
\end{align}
where $g$ is the largest coupling in the theory, e.g. it can be the gauge coupling.\footnote{It may help to assume scaling $g^2\sim \lambda$, which allows for direct analogy with the real scalar model we have discussed previously.} 
The non-zero Matsubara modes are protected from making the expansion parameter large, since the lightest of these modes has mass $\mathcal{O}(\pi T)$. This scale is called \textit{hard scale}, and the modes above this scale are called hard modes. For the hard modes, the expansion parameter becomes 
\begin{align}
    \epsilon_\rmii{b} ^\rmii{hard} \sim \frac{g^2 T}{\pi^2 T} = \frac{g}{\pi^2},
\end{align}
On the other hand, if we look at the zero Matsubara mode, in the limit of vanishing mass, the expansion parameter diverges, and the naive perturbation theory seems to break down. However, as we have seen in the previous sections, it is possible to resum the leading set of higher-loop diagrams which modifies the mass of the zero mode. For a gauge mode, such a correction leads to an effective mass $m^2 _\rmii{eff} \sim g^2 T^2/4$. Physically, this effect can be understood as the screening of low-energy modes by hard scale fluctuations, which induces an effective mass. The scale $\mathcal{O}(gT)$ is called the \textit{soft scale}.
In the case of soft modes, the expansion parameter becomes, 
\begin{align}
    \epsilon_\rmii{b} ^\rmii{soft} \sim \frac{g^2 T}{4\pi g^2 T^2 } = \frac{g}{4\pi},
\end{align}
which allows for perturbative calculations as long as $g\ll \pi$. We see that such a perturbative expansion differs from the vacuum case, as odd powers of the coupling appear, cf. eq.~\eqref{eq:Veff_daisy_phi4}. 
Moreover, gauge fields have $m_\rmii{eff} \lesssim g^2T/4\pi$ which results in 
\begin{align}
    \epsilon_\rmii{b} ^\rmii{ultrasoft} \gtrsim \frac{g^2T}{g^2 T} = 1, 
\end{align}
which leads to explicit breakdown of perturbation theory \cite{Laine_2016}. This phenomenon is called \textit{Linde problem} of thermal gauge theory. The scale where perturbation theory breaks down $\mathcal{O}(g^2T/4\pi)$ is called the \textit{ultrasoft scale}.

In summary, the general thermal scale hierarchy at high temperature is 
\begin{align}
    \underbrace{\frac{g^2T}{4\pi}}_{\text{ultrasoft scale} } \ll
    \underbrace{\frac{gT}{4}}_{\text{soft scale} } \ll
    \underbrace{\pi T}_{\text{hard scale} }. 
\end{align}

\subsection{Effective field theory at high temperature }
\begin{figure}
    \centering
    \includegraphics[width=1\linewidth]{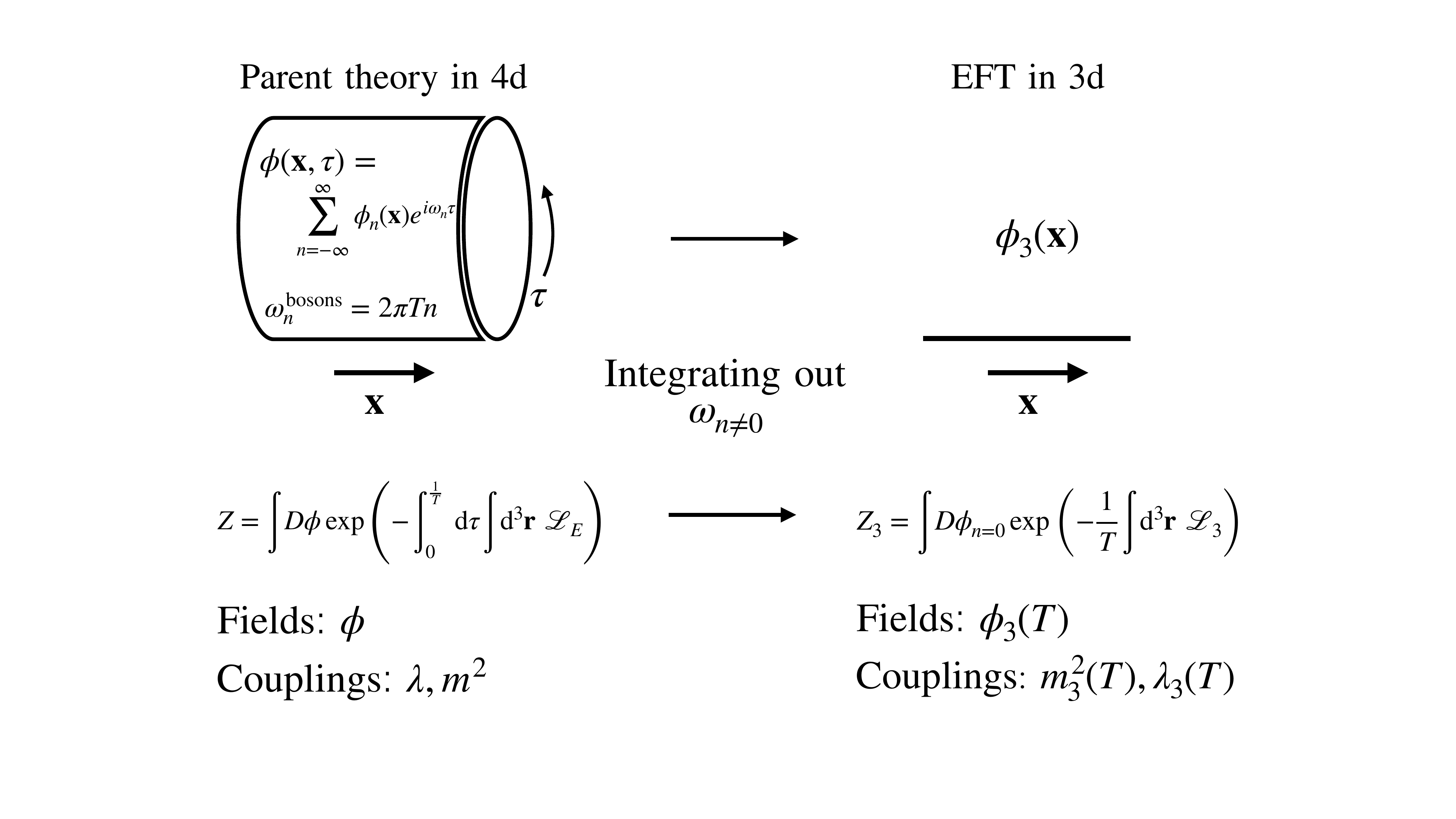}
    \caption{Schematic depiction of dimensional reduction in a real scalar theory.}
    \label{fig:DR_scheme_phi4}
\end{figure}

The EFT construction between the light zero modes at the soft scale and the heavy hard modes is called \textit{dimensional reduction} (DR). The EFT describes the physics of zero modes and is obtained by integrating out the heavy, hard modes, i.e. modes with energies  $E > \pi T$. The zero mode is static,\footnote{ Since for the zero mode the field is  $\phi(\mathbf{x},\tau) = \phi_0(\mathbf{x}) e^{\omega_0 \tau} = \phi_0(\mathbf{x})$.} thus the resulting EFT is three-dimensional. Hence, the name is dimensional reduction. Figure~\ref {fig:DR_scheme_phi4} shows a schematic depiction of DR for a real scalar theory. In practice, dimensional reduction can be achieved by performing the following steps:
\begin{enumerate}
    \item Write down a 3d theory for the soft mode obeying the same symmetries and containing the same degrees of freedom as the original, parent 4d theory.
    \item Calculate the static correlation functions for both theories using strict perturbative expansion, i.e. treating masses as small interactions.
    \item Determine the effective parameters of the 3d theory by matching the correlation functions.
\end{enumerate}
We will now illustrate this procedure using the scalar theory discussed in previous sections.\footnote{
This approach is based on the matching of correlation functions in both 3d and 4d theories. However, there exists a different approach \cite{Hirvonen:2022jba}, where the 3d EFT is a direct result of performing a path integral over the heavy modes, which is in closer resemblance to the background field method for calculating effective actions. Two methods are completely equivalent.
}

\subsection{Dimensional reduction in real scalar theory}
\label{sec:dr_phi4}
Let us consider again the simple model of a $Z_2$-symmetric, real scalar field defined by the Lagrangian in eq.~\eqref{eq:LagrangianE_lamhi4}. Following ref.~\cite{Gould:2021dzl}, we will now calculate the 3d effective theory for it.
First, we will write the most general 3d (Euclidean) Lagrangian obeying the same symmetries as the ``parent'' 4d theory
\begin{align}
    \mathcal{L}_\rmii{3} = 
    \frac{1}{2} (\partial_i \phithree)^2
    +\frac{1}{2}\mthree ^2 \phithree ^2
    + \frac{1}{4!} \lambda_\rmii{3} \phithree ^4.
\end{align}
Now, for the 4d theory, we want to use strict perturbation theory, i.e. treat mass terms as interactions
\begin{align}
    \mathcal{L}_\rmii{free} &= 
    \frac{1}{2} (\partial_\tau \phi)^2 + (\partial_i \phi)^2, \\
    \mathcal{L}_\rmii{int} &= 
    \frac12(m^2 +\delta m^2) \phi^2 
    + \frac14(\lambda+\delta\lambda)\phi^4,
\end{align}
where we have explicitly written the relevant counterterms. Note that $\phi$ here is the background field, but we have dropped the subscript ``b''. Now, performing the split for the 3d Lagrangian in the manner of strict perturbation theory, we analogously obtain
\begin{align}
    \mathcal{L}_\rmii{3,free} &= 
        (\partial_i \phithree)^2, \\
    \mathcal{L}_\rmii{3,int} &= 
        \frac12(\mthree^2 +\delta \mthree^2) \phithree^2 
        + \frac{1}{4!} \lamthree \phithree^4.
\end{align}
The lack of a counterterm for the 3d quartic coupling follows from the superrenormalisability of the 3d EFT. In fact, there is only a finite number of divergent diagrams, and the mass counterterm written here is needed scarcely at two-loop order. 

Now the main goal is to obtain the 3d effective parameters in the EFT. We can do that by writing certain \textit{matching conditions}, i.e. match observables calculated in both theories. One of the most suitable choices is matching one-particle irreducible correlation functions (see e.g. \cite{Kajantie:1995dw, Gould:2021dzl}). We will denote them as $\Gamma^{(k)}$ for $k$-point functions in the 4d theory, and $\Gamma^{(k)}_3$ for their counterparts within 3d EFT. These functions can be obtained in a diagrammatic expansion as a (minus) sum of all possible connected 1PI Feynman diagrams with $k$ legs, as was demonstrated in the first chapter for the vacuum case. We want to achieve $\mathcal{O}(\lambda^2)$ accuracy, which will also require computing two-loop diagrams.

\paragraph{Four-point correlation function}
%
%
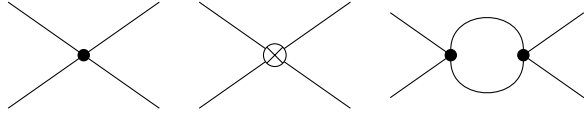
\begin{figure}
    \centering
    \begin{tikzpicture}[baseline=(A.base)]
        \begin{feynman}
            \vertex      (i1) at (-1, 0.7);
            \vertex      (i2) at (-1,-0.7);
            \vertex[dot]  (A) at (0,0) {};
            \vertex      (f1) at (1, 0.7);
            \vertex      (f2) at (1,-0.7);
            \diagram* {
                (i1) -- [plain] (A) -- [plain] (i2),
                (f1) -- [plain] (A) -- [plain] (f2),
            };
        \end{feynman}
    \end{tikzpicture}
    \quad
    \begin{tikzpicture}[baseline=(A.base)]
        \begin{feynman}
            \vertex             (i1) at (-1, 0.7);
            \vertex             (i2) at (-1,-0.7);
            \vertex[crossed dot] (A) at (0,0) {};
            \vertex             (f1) at (1, 0.7);
            \vertex             (f2) at (1,-0.7);
            \diagram* {
                (i1) -- [plain] (A) -- [plain] (i2),
                (f1) -- [plain] (A) -- [plain] (f2),
            };
        \end{feynman}
    \end{tikzpicture}
    \quad
    \begin{tikzpicture}[scale=0.8, baseline=(A.base)]
        \begin{feynman}
            \vertex      (i1) at (-1, 0.7);
            \vertex      (i2) at (-1,-0.7);
            \vertex[dot]  (A) at (0,0) {};
            \vertex[dot]  (B) at (1.2,0) {};
            \vertex      (f1) at (2.2, 0.7);
            \vertex      (f2) at (2.2,-0.7);
            \diagram* {
                (i1) -- [plain] (A),
                (i2) -- [plain] (A),
                (A) -- [plain, half left] (B) -- [plain, half left] (A),
                (B) -- [plain] (f1),
                (B) -- [plain] (f2),
            };
        \end{feynman}
    \end{tikzpicture}
    \caption{Diagrams contributing to the four-point correlation function. Crossed-dot vertex denotes the insertion of a relevant counterterm. }
    \label{fig:Gammak4_diagrams}
\end{figure}
First, we will obtain four-point correlation functions, as it requires diagrams only up to 1-loop in order to achieve the desired accuracy. The relevant diagrams are collected in figure~\ref{fig:Gammak4_diagrams}.
We will start with computing these diagrams in the 4d theory, including the relevant counterterm $\delta \lambda$. The sum of diagrams leads to the following integrals, where their ordering corresponds to the ordering in  figure~\ref{fig:Gammak4_diagrams} 
\begin{align}
\Gamma^{(4)}(\mathbf{p}, \mathbf{q}, \mathbf{r},-\mathbf{p}-\mathbf{q}-\mathbf{r}) & \approx 
\lambda+\delta \lambda
-\frac{3}{2} \lambda^2 \sumint_Q \frac{1}{Q^4}
\\
\label{eq:Gamma4_4d_phi4}
& \approx 
\lambda
-\frac{1}{(4 \pi)^2}\left(
    \frac{3}{2} \lambda^2 L_b(\Lambdam)
\right),
\end{align}
where $Q$ denotes the momenta in the loop.
The approximation sign denotes evaluating diagrams in strict perturbation theory, i.e. expanding in $m\rightarrow 0$.  
The scale $\Lambdam$ is the RG-scale from the evaluation of a one-loop diagram in dimensional regularisation and $\overline{\rm{MS}}$ scheme. This scale is going to correspond to the \textit{matching scale} of the EFT. 
Now, evaluating the same diagrams in 3d theory leads to a much simpler expression. Since the diagrams contain the zero Matsubara mode with $\omega_0 =0 $, all the loop diagrams vanish in dimensional regularisation, yielding a simple result
\begin{align}
\label{eq:Gamma4_3d_phi4}
    \Gamma^{(4)}_3 (\mathbf{p}, \mathbf{q}, \mathbf{r},-\mathbf{p}-\mathbf{q}-\mathbf{r}) & \approx 
    \lamthree + \delta\lamthree.
\end{align}

\paragraph{Two-point correlation function}
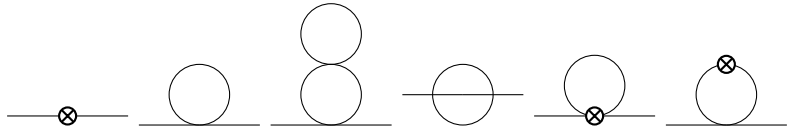
\begin{figure}[ht]
    \centering
    %
    \begin{tikzpicture}[scale=0.8]
    \begin{feynman}
        \vertex (A) at (0,0);
        \vertex (C) at (1,0);
        \vertex (B) at (2,0);
        \diagram* {
            (A) -- [plain] (C) -- [plain] (B),
        };
        \draw[fill=white,cross,thick] (C) circle (4pt);
    \end{feynman}
    \end{tikzpicture}
    \begin{tikzpicture}[scale=0.8]
        \begin{feynman}
            \vertex (a) at (0,0);
            \vertex (b) at (1,0);
            \vertex (c) at (2,0);
            \diagram* {
              (a) -- [plain] (b) -- [plain] (c)
            };
            \path (b)--++(90:0.5) coordinate (A);
            \draw [solid] (A) circle(0.5) ;
        \end{feynman}
    \end{tikzpicture}
    %
    \begin{tikzpicture}[scale=0.8]
        \begin{feynman}
            \vertex (a) at (0,0);
            \vertex (b) at (1,0);
            \vertex (c) at (2,0);
            \diagram* {
              (a) -- [plain] (b) -- [plain] (c)
            };
            \path (b)--++(90:0.5) coordinate (A);
            \draw [solid] (A) circle(0.5);
            \path (b)--++(90:1.5) coordinate (B);
            \draw [solid] (B) circle(0.5);
        \end{feynman}
    \end{tikzpicture}
    %
    \begin{tikzpicture}[scale=0.8]
        \begin{feynman}
            \vertex (a) at (0,0);
            \vertex (b) at (1,0);
            \vertex (c) at (2,0);
            \diagram* {
              (a) -- [plain] (b) -- [plain] (c)
            };
            \draw [solid] (b) circle(0.5);
        \end{feynman}
    \end{tikzpicture}
    \begin{tikzpicture}[scale=0.8]
        \begin{feynman}
            \vertex (A) at (0,0);
            \vertex (C) at (1,0);
            \vertex (B) at (2,0);
            \diagram* {
                (A) -- [plain] (C) -- [plain] (B),
            };
            \path (C)--++(90:0.5) coordinate (D);
            \draw [solid] (D) circle(0.5);
            \draw[fill=white,cross,thick] (C) circle (4pt);
        \end{feynman}
    \end{tikzpicture}
    \begin{tikzpicture}[scale=0.8]
        \begin{feynman}
            \vertex (A) at (0,0);
            \vertex (C) at (1,0);
            \vertex (B) at (2,0);
            \diagram* {
                (A) -- [plain] (C) -- [plain] (B),
            };
            \path (C)--++(90:0.5) coordinate (D);
            \draw [solid] (D) circle(0.5);
            \draw[fill=white,cross,thick] (1,1) circle (4pt);
        \end{feynman}
    \end{tikzpicture}
    
    \caption{Diagrams contributing to two-point correlation functions up to the 2-loop level.}
    \label{fig:Gammak2_diagrams}
\end{figure}
%
Calculation of the two-point correlation function is more involved as it requires two-loop diagrams to achieve $\mathcal{O}(\lambda^2)$. This is characteristic of thermal field theory at large temperatures. The required diagrams are collected in figure~\ref{fig:Gammak2_diagrams}. 
Evaluating the diagrams in the 4d theory leads to the following expression for the two-point correlation function, where the ordering of integrals corresponds to the ordering of diagrams in figure~\ref{fig:Gammak2_diagrams} 
\begin{align}
\Gamma^{(2)}(-\boldsymbol{p}, \boldsymbol{p}) =&
p^2  +m^2+\delta m^2
+\frac{\lambda}{2} \sumint_Q \frac{1}{Q^2}
-\frac{\lambda^2}{4} \sumint_{Q R} \frac{1}{Q^4 R^2}\notag\\
&-\frac{\lambda^2}{6} \sumint_{Q R} \frac{1}{Q^2 R^2(Q+R)^2} 
 +\frac{\delta \lambda}{2} \sumint_Q \frac{1}{Q^2}
-\frac{\lambda}{2}\left(m^2
+\delta m^2\right) \sumint_Q \frac{1}{Q^4} \notag\\
\label{eq:Gamma2_4d_phi4}
\approx& 
p^2  +m^2 + \frac{\lambda T^2}{24} \notag
\\
&+ \frac{1}{\fpisq}\qty[
     \frac{\lambda^2 T^2}{12} \qty(
        \frac{1}{2\epsilon} +\frac14 L_b(\Lambdam) -\gammaE + 12\log{A}
     )
     -\lambda m^2 \frac{L_b(\Lambdam)}{2}  
],
\end{align}
where $Q,R$ denote momenta in relevant loops, and $A\simeq 1.282$ denotes the Glaisher-Kinkelin constant and the integrals were expanded assuming $p\sim \lambda^\frac12 T$, see \cite{Gould:2021dzl} for details.
For the 3d theory, the computation again becomes trivial due to the lack of a mass scale in the integrals. The correlation function in the 3d EFT becomes
\begin{align}
\label{eq:Gamma2_3d_phi4}
    \Gamma^{(2)}(-\boldsymbol{p}, \boldsymbol{p}) &\approx
p^2  +\mthree^2+\delta \mthree^2
\end{align}

\paragraph{Matching relations}
Finally, we can express the effective parameters of the 3d EFT in terms of the 4d ones. In the theory under consideration, we have three parameters $\phithree, \mthree, \lamthree$. Since we can equate two correlation functions, we are missing one more equation that would allow us to find the last parameter. This condition corresponds to matching the kinetic operators in both theories, and will allow us to find $\phithree$, while equating correlation functions allows for finding the effective parameters in the potential. 

We can achieve matching of kinetic operators by matching the momentum dependence of the zero Matsubara mode in 4d theory $\phi_0$ and the field operator within 3d EFT: 
\begin{align}
    \phithree^2 = 
    \frac{1}{T} 
    \eval{
        \pdv{\Gamma^{(2)}(-\boldsymbol{p}, \boldsymbol{p})}{p^2}
    }_{p^2=0} \phi_0 ^2
    =
    \frac{1}{T} \qty(
        1 + \eval{\Pi^\prime (-\boldsymbol{p}, \boldsymbol{p})}_{p^2 =0}
    ) \phi_0 ^2,
\end{align}
where $\Pi$ is the self-energy of $\phi_0$ and following \cite{Gould:2021dzl} we used a notation $\empty^\prime \equiv \pdv{}{p^2}$. Note that for the 3d field $\phithree$ to be canonically normalised in 3d EFT, we had to include a factor of the inverse of $T$.
Consequently, the rest of the matching relation can be obtained by equating the correlation functions at zero momentum and normalising as well by powers of $T$
\begin{align}
    \Gamma^{(2)}_3 (\vb{0}, \vb{0}) &= 
    T^0 \qty(
        1 - \eval{ \Pi^\prime (\vb{p}, -\vb{p}) }_{p^2 =0} 
    )
    \Gamma^{(2)}(\vb{0}, \vb{0}), \\
    \label{eq:Gamma4_matching_phi4}
    \Gamma^{(4)}_3 (\vb{0},\vb{0},\vb{0},\vb{0}) &= 
    T \qty(
        1 - 2\eval{ \Pi^\prime (\vb{p}, -\vb{p}) }_{p^2 =0}
    )
    \Gamma^{(4)} (\vb{0},\vb{0},\vb{0},\vb{0}).
\end{align}
Now we can use the explicit form of calculated correlation functions to write the matching relations. All of the relations contain a derivative of self-energy with respect to momentum squared. Looking closer at the two-point function, we see that such a derivative will lead to a null result. It is not a general feature, as a real scalar model without $Z_2$-symmetry would lead to a non-zero $\Pi^\prime$. In our case, though, the relations are thus simpler. Starting with the field operator, we simply have
\begin{align}
    \phi_3 ^2 = \frac{1}{T} \phi_0 ^2,
\end{align}
then by comparing static four point functions, i.e. substituting  eq.~\eqref{eq:Gamma4_4d_phi4} and eq.~\eqref{eq:Gamma4_3d_phi4} in eq.~\eqref{eq:Gamma4_matching_phi4} at zero momenta, we arrive at
\begin{align}
    \lamthree + \delta\lamthree &= 
    T\qty[
        \lambda - \frac{1}{\fpisq}\qty(\frac32 \lambda^2 L_b(\Lambdam) )
    ] \\
    &=
    T\underbrace{
    \qty[
        \lambda - \frac{1}{\fpisq}\qty(\frac{\beta_\lambda}{2} L_b(\Lambdam) )
    ]
    }_{\equiv \Bar\lambda}
\end{align}
In the 3d EFT, there are no divergences that should be cancelled by $\delta\lamthree$, thus it is simply $\delta\lamthree = 0$. Then, note that the expression in square brackets is 4d RG-scale invariant as the coefficient in front of $L_b/2$ is the beta function of the quartic coupling.  Finally, the matching relation for $\lamthree$ becomes
\begin{align}
\label{eq:lambda3_matching_phi4}
    \lamthree = T \Bar\lambda.
\end{align}
Now, let us move to the 3d mass parameter. Again, we need to compare static correlation functions in eqs.~\eqref{eq:Gamma2_4d_phi4} and \eqref{eq:Gamma2_3d_phi4}. We will rewrite the relation such that the barred couplings are more visible
\begin{align}
    \mthree^2 + \delta \mthree^2 =&
    \underbrace{
        \qty[m^2 - \frac{\lambda m^2}{\fpisq} \frac{L_b}{2} 
    ]}_{\equiv \bar{m} ^2}
    + \underbrace{
    \frac{ T^2}{24} \qty[ \lambda -  \frac{3 \lambda^2}{\fpisq}    \frac{L_b}{2}] }_{\frac{\bar{\lambda} T^2}{24} }
    +\frac{\lambda^2 T^2}{6 \fpisq} \qty[
        L_b - \frac{\gammaE}{2} + \log{A^6} + \frac{1}{4\epsilon}
    ] \notag \\
    \label{eq:m3Sq_matching_4dpole_phi4}
    =& 
    \bar{m} ^2 + \frac{\bar{\lambda} T^2}{24}
    +\frac{\lambda^2 T^2}{6 \fpisq} \qty[
        \log\frac{\Lambdam}{3T} - c + \frac{1}{4\epsilon}
    ],
\end{align}
where in the second line we have simplified the expression inside square brackets by introducing a constant \cite{Kajantie:1995dw}
\begin{align}
    c \equiv -\log\frac{ 
        3 e^{\gammaE/2} A^6
    }{
        4\pi
    } \simeq - 0.348723.
\end{align}
Let us make two comments now. First, note that the second term here, i.e. $\bar\lambda T^2/24$, is RG-scale invariant due to the log term inside the square bracket. This term originates from the two-loop ``snowman'' diagram, and is the $\mathcal{O}(\lambda^2)$ contribution that was missing in the daisy-resummation approach, cf. eq.~\eqref{eq:selfE_phi4_4d_running}. In fact, daisy-resummation is equivalent to dimensional reduction if we truncate the matching relations up to $\mathcal{O}(\lambda)$ order. This is arguably the biggest virtue of the EFT approach -- it allows one to systematically include all relevant terms at each order in the chosen perturbative expansion. 

Now, for the second comment, let us discuss the divergent term in the above matching relation. This temperature pole should be cancelled by the analogous pole in the 3d EFT. More specifically, computation of the two-loop effective potential in the EFT will require computing a 3d scalar sunset. This will result in the following form of the 3d mass counterterm   
\begin{align}
    \delta \mthree^2 = \frac{\lamthree ^2}{24 \fpisq \epsilon}.
\end{align}
We see that such a counterterm would exactly remove the pole in the matching relation. The only difference between this counterterm and the corresponding term in the matching relation is in the quartic couplings (3d here and 4d in the matching relations). We can thus use the matching relation for quartic coupling, eq.~\eqref{eq:lambda3_matching_phi4}, to express the 3d counterterm in terms of 4d coupling
\begin{align}
    \delta \mthree^2 = \frac{1}{24 \fpisq \epsilon} \lambda ^2 T^2 +\mathcal{O}(\lambda^3).
\end{align}
Plugging that into the eq.~\eqref{eq:m3Sq_matching_4dpole_phi4} results in a 
finite expression in the matching relation for the mass parameter up to $\mathcal{O}(\lambda^2)$ order
\begin{align}
\label{eq:mthree_4dmatching_nopole_phi4}
\mthree^2 = 
    \bar{m} ^2 + \frac{\bar{\lambda} T^2}{24}
    +\frac{\lambda^2 T^2}{6 \fpisq} \qty[
        \log\frac{\Lambdam}{3T} - c 
    ].
\end{align}
This is the basic logic behind the procedure of obtaining parameters in the 3d EFT. However, for a completely RG-scale invariant calculation, there is one more modification needed for the matching relation of the 3d mass term. This modification will result in expressions found in the literature, see e.g. \cite{Gould:2021, Gould:2021dzl, Kajantie:1995dw} and is described in detail in the next paragraph.

\paragraph{RG-scale dependence of parameters}

Looking at eq.~\eqref{eq:lambda3_matching_phi4}, we can conclude that the resulting 3d quartic coupling is independent of the RG-scale $\Lambdam$. However, the parameters of 4d theory still formally run, thus we can use it to run them from the matching scale $\Lambdam$ to some new renormalisation scale $\mu_4$. This running can be used to minimise large logarithms in the perturbation theory, as well as to investigate any scale dependence of the computed observables. Then we see that the 3d mass parameter still retains some logarithmic $\Lambdam$ scale dependence, while the rest of the terms are RG-scale invariant. This dependence can be cancelled by computing two-loop diagrams contributing to the effective potential within 3d theory. Then, one needs to expand these terms in 4d parameters, which will result in cancellation of leftover $\Lambdam$ dependence. We will now demonstrate it explicitly.

Let us, for now, partially forget about the 4d theory and focus instead on the 3d EFT. First, we want to investigate the running of the 3d parameters. The quartic coupling is independent of the $\Lambdam $ scale, thus in the 3d EFT it is not running with the RG-scale of 3d EFT, which we will denote as $\mu_3$
\begin{align}
    \dv{\lamthree }{\log\Lambdam} &= 0 \\
    \Rightarrow \beta_{\lamthree} \equiv \mu_3 \dv{\lamthree(\mu_3) }{\mu_3} &= 0.
\end{align}
On the other hand, though, similar arguments suggest that the mass parameter will be running. Indeed, the associated beta function reads (see \cite{Gould:2021dzl} and references therein) 
\begin{align}
\label{eq:betamthree_phi4}
    \beta_{\mthree^2} \equiv \mu_3 \dv{\mthree^2}{\mu_3} 
    = \frac{\lamthree ^2}{6(4\pi)^2}
\end{align}
Within the 3d EFT, this is the exact beta function up to all orders in perturbation theory. We can analytically solve the RGE for the mass parameter, which results in 
\begin{align}
    \label{eq:mthree_mu3_dependence}
    \mthree ^2 (\mu_3) = 
    \mthree ^2 (\Lambdam) 
    + \frac{\lamthree ^2}{6 \fpisq} \log\frac{\mu_3}{\Lambdam},
\end{align}
where we have used eq.~\eqref{eq:mthree_4dmatching_nopole_phi4} for the initial condition of RGE.  

We may expect, then, that similarly to the 4d vacuum case, the running will cancel against some logarithm coming from loop computations. In the 3d EFT, the logarithm divergence appears at two two-loop level. Let us thus write here the effective potential at two-loop order in the 3d theory \cite{Gould:2021}:
\begin{align}
\label{eq:Veff_3d_phi4}
    V_\rmii{3,eff}(\phithree) = 
    V_\rmii{3} ^\rmii{tree}(\phithree) 
    + V_\rmii{3} ^\rmii{1-loop}(\phithree)
    + V_\rmii{3} ^\rmii{2-loop}(\phithree), 
\end{align}
where 
\begin{align}
    V_\rmii{3} ^\rmii{tree}(\phithree) &\equiv 
    \frac{1}{2}\mthree^2 \phithree^2 
    + \frac{1}{4!} \lamthree \phithree^4, \\
    V_\rmii{3} ^\rmii{1-loop}(\phithree) &\equiv 
    - \frac{1}{12\pi} m_\rmii{3,eff} ^3,\\
    V_\rmii{3} ^\rmii{2-loop}(\phithree) &\equiv
    \frac{1}{(4\pi)^2}\qty[
        \frac{\lamthree}{8} m_\rmii{3,eff} ^2
        + \frac{\lamthree^2 \phithree^2}{24}\qty(2\log\frac{3 m_\rmii{3,eff} }{\mu_3} -1 )
    ],
\end{align}
and we have defined $ m_\rmii{3,eff} ^2 \equiv \qty(\mthree^2 + \frac12 \lamthree \phithree^2 ) $. We may now investigate the 3d scale dependence by computing 
\begin{align}
    \beta_{\mthree^2} \pdv{(V_\rmii{3} ^\rmii{tree} +  V_\rmii{3} ^\rmii{1-loop})}{\mthree^2} 
    + \pdv{V_\rmii{3} ^\rmii{2-loop}(\phithree)}{\log{\mu_3}} 
    &= \frac{\lamthree ^2}{6(4\pi)^2} \qty (
            \frac12 \phithree ^2 
             -\frac{1}{12\pi} \frac{1}{\mthree^2} m_\rmii{3,eff} ^3
    )
    - \frac{1}{(4\pi)^2} \frac{\lamthree^2 \phithree^2}{12} \\
    &= -\frac{\lamthree ^2}{6(4\pi)^2} 
    \frac{1}{12\pi} \frac{1}{\mthree^2} m_\rmii{3,eff} ^3 \\
    &= \mathcal{O}(\lamthree^3).
\end{align}
Thus, the effective potential computed within 3d theory is $\mu_3$-scale invariant to the order under consideration, as the leftover dependence is of higher order.  

Lastly, let us now address the issue of leftover logarithmic $\Lambdam$ scale dependence in eq.~\eqref{eq:mthree_4dmatching_nopole_phi4}. We can now show  explicitly the cancellation by expanding the logarithmic part of the two-loop potential $V_\rmii{3} ^\rmii{2-loop}$ in 4d parameters up to $\order{\lambda^2}$ order,
\begin{align}
    V_\rmii{3} ^\rmii{tree} + V_\rmii{3} ^\rmii{2-loop} &= 
    \frac12 \mthree ^2 \phithree^2 + \frac1\fpisq\qty[\frac{\lamthree^2 \phithree^2}{24}\qty(2\log\frac{3 m_\rmii{3,eff} }{\mu_3} -1 )] 
    + \dots \\
    &= 
    \frac{\lambda^2 T^2 \phithree^2}{12\fpisq}\qty(
        \log\frac{\Lambdam}{3T} 
        +
        \log\frac{3 m_\rmii{3,eff}}{ \mu_3} -c -\frac12
    ) +  \order{\lambda^3} + \dots. 
\end{align}
We see now that setting $\mu_3 = \Lambdam$ results in cancellation of $\Lambdam$-scale dependence. In practice, we can conclude that matching scale is not physically relevant, but rather is replaced by two different scales: $\mu_4$ associated with running of 4d parameters, and the $\mu_3$, which is the RG-scale of the 3d EFT. Both of these scales can be chosen independently.\footnote{
    All of this conclusion could also have been deduced from eq.~\eqref{eq:mthree_mu3_dependence}. Expanding this in $\lambda$ leads to
    $$\mthree^2 = 
    \bar{m} ^2 + \frac{\bar{\lambda} T^2}{24}
    +\frac{\lambda^2 T^2}{6 \fpisq} \qty[
        \log\frac{\Lambdam}{3T} + \log\frac{\mu_3}{\Lambdam} - c ] +\order{\lambda^3},$$
        where the matching scale is immediately replaced by $\mu_3$, the scale of 3d EFT. The author thanks Oliver Gould for pointing out this trick.
}
This is a general feature of EFTs; the matching scale always disappears, and large logarithms can be minimised by running parameters in the parent and effective theory. Thus, it is much more convenient to rewrite the matching relations in terms of the 3d RG-scale, i.e. swapping $\Lambdam \rightarrow \mu_3$. Moreover, since the 3d EFT is superrenormalisable, we may include the 3d quartic coupling in the matching relation for the mass parameter, which will effectively result in higher-order resummations and will extend the cancellation of the matching scale to all orders. Taking all of these remarks into account results in a final form for the matching relations
\begin{align}
    \label{eq:matching_phi3}
    \phithree^2 &= \frac{1}{T} \phi^2, \\
    \lamthree &= T\bar{\lambda}, \\
    \label{eq:matching_m3_final_phi4}
    \mthree ^2 &= \bar{m} ^2 + \frac{\bar{\lambda} T^2}{24}
    +\frac{\lamthree^2}{6 \fpisq} \qty[
        \log\frac{\mu_3}{3T} - c 
    ].
\end{align}
To conclude, let us comment on the calculation of the effective potential to higher orders. 
Note that having matching relations in such form, and since the beta functions are exact in the 3d EFT, going to $\order{\lambda^{\frac52}}$ for the effective potential will require only computations within the EFT, i.e. computing three-loop contributions. 
At the same time, no new sum-integrals are needed to improve matching relations. In other words, one would improve potential in eq.~\eqref{eq:Veff_3d_phi4} but not the eqs.~\eqref{eq:matching_phi3}-\eqref{eq:matching_m3_final_phi4}.
\paragraph{Conclusions}
The effective field theory approach, known as dimensional reduction, leads to a perturbative and consistent description of field theory in thermal equilibrium. It is free of the theoretical inconsistencies that plague the naive 4d computations based on the resummation of daisy diagrams. 
In the end, of course, the root of all problems in the daisy approach lies in omitting contributions relevant at a given order of perturbative expansion in powers of coupling. Or stubbornly relying on loop-expansion, which does not work at large temperatures.\footnote{Technically, the daisy-resummation approach includes higher-loop effects. However, the rest of the potential is just tree-level plus one-loop corrections. Therefore, we consider this approach as a slightly ``augmented'' loop expansion.} 
It would be possible, however, to perform all the computations purely within 4d theory, see e.g. ref.~\cite{Lofgren:2021ogg}, or even combine the two approaches \cite{Navarrete:2025yxy}. The strength of EFT lies in the fact that it allows for much easier bookkeeping of relevant contributions.
Thus, the author would like to conclude this rather involved section with a simple rule for computing \textit{anything} in quantum field theory at zero or finite temperature. 
\begin{quote}
    Respect perturbation theory. 
\end{quote}


\subsection{Blueprint of dimensional reduction in gauge theories}
\label{sec:DR_SU(2)_sketch}
Dimensional reduction has been used to describe the nature of the electroweak phase transition in the Standard Model, and is most suitable for studying BSM physics at finite temperature. 
Here, we will review the construction of the 3d EFT for a gauge theory, which is relevant for many BSM theories. Here we list some studies of popular theories in the context of thermal phase transitions:
\begin{itemize}
    \item SM-like, SU(2)+Higgs e.g.  \cite{Kajantie:1995dw, Farakos_1994, Kajantie:1996mn, Ekstedt:2021kyx, Ekstedt:2024etx},
    \item Abelian Higgs e.g. \cite{Hirvonen:2021zej, Lofgren:2021ogg, Bernardo:2025vkz, Ekstedt:2024etx},
    \item triplet extension e.g. \cite{Gould:2021ccf, Niemi:2018asa, Niemi:2022bjg},
    \item singlet extension e.g \cite{ Croon:2020cgk, Niemi:2018asa, Schicho:2021gca, Niemi:2021qvp, Gould:2024jjt},
    \item conformal extensions \cite{Kierkla:2023von, Kierkla:2025qyz, Kierkla:2025vwp},
\end{itemize} 
We are not going to derive all the specific matching relations, as was done previously in the literature. 
Moreover, derivation of matching relations has been automatized in the \texttt{DRalgo} Mathematica package \cite{Ekstedt:2021kyx}. Instead, we will focus on the general construction of 3d EFT for gauge theories. Now, let us review the dimensional reduction for a case of SU(2)+Higgs, which will be relevant for the rest of the thesis.  

Consider the following theory described by the 4d Euclidean action
\begin{align}
    S_4 = \int \dd[4]{x} \qty[
    -\frac14 F_{\mu\nu}^a F^{\mu\nu}_a
    + \frac12 (D_\mu \Phi)^\dagger (D^\mu \Phi)
    + \frac{1}{2}m^2 \Phi^\dagger \Phi
    + \frac{1}{4}\lambda ( \Phi^\dagger \Phi)^2
    ],
\end{align}
where the gauge field strength tensor and covariant derivative are defined in a standard fashion as 
\begin{align}
    F_{\mu\nu}^a(x) &\equiv \partial_\mu A_\nu ^a(x) - \partial_\nu A_\mu ^a (x) +gf^{abc}A^b_\mu A^c_\nu\\
    D_\mu \Phi &\equiv \partial_\mu \Phi - ig \tau^a A_\mu^a \Phi,
\end{align}
where $f^{abc}$ is the SU(2) structure constant tensor and $\tau^a$ are generators of the corresponding Lie algebra (for more details see e.g. the textbooks \cite{Schwartz:2014sze, Peskin:1995ev}). Following the procedure of dimensional reduction, we can integrate out non-zero Matsubara modes (hard scale), which leads to the following 3d theory \cite{Kajantie:1995dw}:
\begin{align}
\label{eq:SU(2)_soft_Seff}
    \SthreeSoft  [A_i^a, A_0^a, \Phi_3] =
    \int \dd[3]{x} \biggl\{
        &\frac14 F_{ij}^a F^{ij}_a 
        +\frac12 (D_i A_0 ^a)^2 
        +\frac12 (D_i \Phi_3)^\dagger (D^i \Phi_3)  \notag \\
        &+\frac12 \mD^2 A_0^a A_{0}^a
        +\frac14 \kappa_3 ( A_0^a A_{0}^a)^2 \notag \\
        &+\frac12 \mthree^2 \Phi_3 ^\dagger \Phi_3
        + \lamthree (\Phi_3 ^\dagger \Phi_3)^2 
        + h_3 \Phi_3 ^\dagger \Phi_3 A_0^a A_{0}^a
    \biggr\}.
\end{align}
Matching relations up to $\order{g^4}$ for the 3d parameters can be found in refs. \cite{Kierkla:2022odc, Kierkla:2025vwp, Kajantie:1995dw}.
First, we can see that in this theory, the temporal component of the 4d gauge field has become a scalar triplet $A_{0}^a$ within the EFT. Moreover, it is a massive field. Physically, this is again a consequence of heavy modes screening the zero mode of the temporal part of the gauge field. Its mass in the EFT is denoted as \textit{Debye mass}, and to leading order it is given by 
\begin{align}
    \mD^2 \sim g^2T^2.
\end{align}
Appearance of Debye mass is often also described as a consequence of breaking Lorentz symmetry, as the thermal bath of particles becomes a preferred frame of reference. 

Now we need to make an important remark. Since in the parent theory there were two couplings $\lambda$ and $g$, we need to establish a hierarchy between them, which will allow for organising the perturbative expansion. The usual scaling which is relevant for phase transitions is \cite{Gould:2023ovu, Camargo-Molina:2024sde}
\begin{align}
    \lambda \sim g^3.
\end{align}
Then, the perturbation theory will be organised in powers of $g$. Such expansion is similar to the case of the Coleman-Weinberg model \cite{PhysRevD.7.1888} that we have discussed in the previous chapter.\footnote{Note that in this case the scaling was $\lambda\sim g^4 $. However, this scaling is valid only in the vicinity of the global minimum. Near the origin of field space (in particular, near the barrier), the scaling can be $\lambda\sim g^3$ instead.} 

Now, moving to the scalar field $\phithree$, its mass parameter is given at leading order as 
\begin{align}
    \mthree^2 \sim - \frac12 |m|^2 + \#g^2T^2 ,
\end{align}
where $\#$ denotes some numerical constant. As we can see, there is a partial cancellation between the thermally induced mass and the 4d mass. Therefore, the scalar field belongs to a different, lower energy scale than the \textit{soft scale} of the effective theory given above. This next scale we shall denote as \textit{supersoft}, and its expansion parameter is given as 
\begin{align}
    \epsilon_\rmii{b}^\rmii{supersoft} \sim \qty(\frac{g }{4\pi })^\frac{3}{2}.
\end{align}
Then, for gauge theories, we can establish a general hierarchy between thermal scales \cite{Gould:2023ovu}
\begin{align}
    \underbrace{\frac{g^2T}{4\pi}}_{\text{ultrasoft scale} } \ll
    \underbrace{ \frac{\qty(\frac{1}{4}g)^\frac32 T}{ \sqrt \pi} }_{\text{supersoft scale} } \ll
    \underbrace{\frac{gT}{4}}_{\text{soft scale} } \ll
    \underbrace{\pi T}_{\text{hard scale} }. 
\end{align}
Thus, to describe the dynamics of the scalar field driving the transition, we need to integrate out the modes heavier than the supersoft scale. This formally results in a new EFT. 
The new effective action can be obtained by integrating out the leftover spatial and temporal gauge modes. Note that the derivation of this new EFT is more analogous to calculating effective action in the vacuum 3d case than high-temperature dimensional reduction. We can, thus, use similar techniques as in the vacuum case, i.e. compute the effective action using the background field method. We assign a background to the 3d scalar field as
\begin{align}
    \Phi_3 = \frac{1}{\sqrt2}\mqty(0 \\ \phi_3).
\end{align}
Now, by integrating out the gauge modes at one-loop order and keeping terms up to $\mathcal{O}(g^3)$, we arrive at the leading-order supersoft EFT
\begin{align}
\label{eq:SU(2)_supersoft_Seff}
    S_\rmii{LO}^\rmii{(supersoft)}[\phi_3] &= 
    \int \dd[3]{x}
    \qty[
        \frac12(\partial_i \phi_3)^2 
        + V_\rmii{LO}^\rmii{(supersoft)}[\phi_3]
    ], \\
\label{eq:SU(2)_supersoft_VLO}
    V_\rmii{LO}^\rmii{(supersoft)}[\phi_3] &= 
         \frac12 \mthree^2 \phi_3 ^2 
        +\frac14 \lamthree \phi_3 ^4
        + \frac{1}{12\pi} \qty(
        6m_A ^3 + 3m_{A_0}^3
        ),
\end{align}
where the background-dependent gauge masses are given by
\begin{align}
    m_A^2 = \frac{1}{4}g_3\phi_3 ^2,
    \quad
    m_{A_0}^2 = \mD^2 + \frac{1}{2} h_3 \phi_3 ^2.
\end{align}
We can see that these terms give a cubic term in the effective potential.\footnote{ 
In fact, this potential can be translated back into the 4d potential with daisy correction, if we truncate the matching relations. }
This means that the supersoft potential here has a barrier.
Thus, supersoft EFT describes a potential with a first-order phase transition. Let us emphasise that these conclusions hold only if we can trust perturbative expansion, i.e. there must be hierarchy between the masses of scalar and gauge modes. 
Similarly to models where symmetry is broken by radiative corrections, as we have described in chapter \ref{chapter:Seff_vacuum}, models where the barrier is generated due to integrating out ``quantum'' fluctuations are described as models with \textit{radiative barrier}. 

To acquire RG-scale invariant predictions, one can also compute the two-loop corrections to the effective potential. Then the argument follows the same logic as in the case of the real scalar model. Effective potential computed in perturbative expansion is also gauge invariant, see e.g. \cite{Baacke:1999sc}. 


\section{Thermal bubble nucleation}
\label{sec:T_bubble_nucl}

The thermal first-order phase transition proceeds due to thermal fluctuations of long-range modes of the fields.\footnote{In principle, the phase transition can proceed via vacuum decay, i.e. due to quantum tunnelling \cite{Coleman:1977py}. However, at finite temperatures these effects are always negligible, and the transition indeed always proceeds do to the thermal fluctuations.} 
Physically, it is driven by nucleation of bubbles containing the new phase in the volume filled with the old, metastable phase. In its nature, this is a classical process, not a quantum one; therefore, a strategy for describing such a transition for a given quantum theory is to establish an effective theory which can be matched into the classical formalism of \textit{bubble nucleation} developed by Langer \cite{Langer:1969bc, Gould:2021ccf}. In its essence, thermal bubble nucleation rate can be factorised into a product of two parts,
\begin{align}
    \Gamma_T = \Adyn \cross  \Astat,
\end{align}
where $\Astat$ is the statistical part which contains contributions from in-equilibrium physics and is directly associated with the effective free energy of the system under consideration. On the other hand, the dynamical part $\Adyn$ contains out-of-equilibrium dynamics and physically corresponds to the exponential growth of the critical bubble. 

Let us now formalise the nucleation rate. We will be basing our discussion mostly on refs.~\cite{Gould:2021ccf, Ekstedt:2022tqk, Ekstedt:2021kyx, Ekstedt:2023sqc, Andreassen:2016cvx, Langer:1969bc} and references therein.
Let us assume that we know the effective free energy $F_\rmii{eff}[\phi]$ of the system. Then we also will consider two phases of such a system -- the metastable phase denoted by $\phi_\rmii{meta}$ and the stable phase $\phi_\rmii{stable}$. One can then show that in the configuration space exists a transition surface which is a separatrix between gradient flows in $F_\rmii{eff}[\phi]$, i.e. it separates two phases (for details see ref.~\cite{Gould:2021ccf}). 
Every field configuration which crosses the transition surface corresponds to a nucleated bubble. One can imagine a whole system in the metastable phase $\phi_\rmii{meta}$, and then each of such nucleated bubbles would contain a region of stable phase $\phi_\rmii{stable}$ inside itself. 
Bubble-configurations are (exponentially) Boltzmann suppressed, and we can then denote a so-called \textit{critical bubble,} $\phi_\rmii{CB}$ which is the least suppressed field configuration on the transition surface. Such a configuration is a saddle point of the effective free energy, 
\begin{align}
    \eval{
        \frac{\delta F_\rmii{eff}[\phi]}{\delta \phi} 
    }_{\phi=\phi_\rmii{CB}} 
    =0,
\end{align}
where the boundary conditions are such that the field configuration at large spatial distances from the centre of the bubble is in the metastable phase. As any other field configuration that allows for transition should be much more suppressed, the contributions of all ``bubble-like'' configurations to the free energy could be expressed solely by the critical bubble and fluctuations on top of it. Mathematically, it simply means using the saddle-point approximation for the effective free energy 
\begin{align}
\label{eq:Feff_saddle_expn}
    F_\rmii{eff}[\phi_\rmii{CB} + \Tilde{\phi} ] &\simeq
    F_\rmii{eff}[\phi_\rmii{CB}]
    +\frac12 \int_{\mathbf{x}} \int_{\mathbf{y}}
       \Tilde{\phi}(\mathbf{x}) 
       F_\rmii{eff} ^{\prime\prime }[\phi_\rmii{CB}] (\mathbf{x}, \mathbf{y})
       \Tilde{\phi}(\mathbf{y}),
\end{align}
where $\Tilde{\phi}$ denote fluctuations over the critical bubble background, and we have introduced a \textit{fluctuation operator}\footnote{Note, that this operator is analogous to the scalar field fluctuation operator we have encountered while deriving 1PI effective action. Thus, in this analogy, the first term in the saddle-point expansion in eq.~\eqref{eq:Feff_saddle_expn} would correspond to the ``tree-level'' action and the second term would describe one-loop ``quantum'' corrections. }
\begin{align}
     F_\rmii{eff} ^{\prime\prime }[\phi_\rmii{CB}] (\mathbf{x}, \mathbf{y})
     &\equiv
     \eval{
        \frac{\delta^2 F_\rmii{eff}}{\delta\phi(\mathbf{x}) \delta\phi(\mathbf{y})}
        }_{\phi = \phi_\rmii{CB}}.
\end{align}
This operator has one negative eigenvalue, $\lambda_-$, which corresponds to the direction in the configuration space where the free energy decreases. Note that this direction is always orthogonal to the transition surface. In other words, it describes the growth of the critical bubble. Additionally, the bubble background
can break symmetries, and each broken symmetry will result in the appearance of a zero eigenvalue.
In particular, a bubble will break translation symmetry, which in 3d leads to three zero eigenvalues. Then, all the remaining eigenvalues are positive. 

Finally, with the above definitions and following \cite{Gould:2021ccf}, we can show a general formula for the statistical and dynamical parts in the form of the effective free energy 
\begin{align}
    \label{eq:Adyn_Feff}
    \Adyn &\equiv \frac{\kappa}{2 \pi}
    \\
    \label{eq:Astat_Feff}
    \Astat &\equiv 
        \frac{\mathcal{N}}{Z_\rmii{meta}}
        \int \mathcal{D}\phi ~\delta(\phi_-) e^{-\beta F_\rmii{eff}[\phi]},
\end{align}
where inverse of $Z_\rmii{meta}$ gives the normalization to the metastable phase. Note that the integral in the statistical part should be performed only over the transition surface. This is ensured by the delta function term, where $\phi_-$ denotes the coefficient of the negative eigenmode in the expansion $\phi=\sum_n \phi_n f_n$, where the functions $f_n$ are the normalised eigenmodes of the fluctuation operator. Then, following \cite{Gould:2021ccf} we introduce a normalization factor 
\begin{align}
    \mathcal{N} \equiv \sqrt{\beta \frac{|\lambda_-|}{2\pi}},
\end{align}
which will compensate for the removed negative mode in the path integral in the saddle-point approximation. We will comment on that soon. 
The dynamical part now has a concrete interpretation as the exponential growth of the critical bubble denoted as $\kappa$, which can be computed in terms of the damping coefficient $\eta$
\begin{align}
\label{eq:kappa_eta}
\kappa \equiv \sqrt{|\lambda_-|+\frac{\eta^2}{4}}-\frac{\eta}{2}.
\end{align}
The damping coefficients are the terms that contain all the effects of non-equilibrium dynamics in the nucleation rate. In this thesis, we will work in the ``no-damping'' approximation where $\eta\rightarrow 0$. For more general calculations, see refs.~\cite{Ekstedt:2022tqk, Hirvonen:2024rfg}.

The last step is performing the path integral in eq.~\eqref{eq:Astat_Feff} using the saddle-point approximation. Using the expansion in eq.~\eqref{eq:Feff_saddle_expn} in the integral leads to
\begin{align}
    \Astat \simeq
    \mathcal{V}_0 
    \qty(
    \qty|
        \frac{\det^\prime \qty(\frac{\beta}{2\pi} F_\rmii{eff} ^{\prime\prime }[\phi_\rmii{CB}])}{\det\qty(\frac{\beta}{2\pi} F_\rmii{eff} ^{\prime\prime }[\phi_\rmii{meta}])}
    |
    )^{-\frac{1}{2}}
    e^{-\beta \Delta F_\rmii{eff}[\phi_\rmii{CB}]},
\end{align}
where $\Delta F_\rmii{eff}[\phi_\rmii{CB}] \equiv F_\rmii{eff}[\phi_\rmii{CB}]-F_\rmii{eff}[\phi_\rmii{meta}]$. The prime on the determinant denotes removal of the zero modes. Note that because of that, the ratio of determinants is going to be a dimensionful quantity. For thermal transitions, in $d=3$ spatial dimensions, it will have then mass-dimension three.
Instead, the volume factor $\mathcal{V}_0$ contains the contribution of the zero modes as well as deviations of the critical bubble, which leave the effective free energy unchanged. More concretely, in three spatial dimensions, there will be three zero modes due to broken translation invariance. It can be shown that such zero modes are simply spatial derivatives of the field configuration
(see e.g. \cite{Gould:2021ccf, Coleman:1985rnk}). 
The volume factor can then be written as
\begin{align}
\label{eq:Vol_zeromodes}
    \mathcal{V}_0 = \mathcal{V}_\rmii{space} 
    \prod^3_{i=1}\qty[
        \int_\mathbf{x}(\partial_i \phi)^2
    ]^\frac{1}{2}
    = 
    \mathcal{V}_\rmii{space} \Delta  (F_\rmii{eff}[\phi_\rmii{CB}])^{\frac{3}{2}}.
\end{align}
Here $\mathcal{V}_\rmii{space}$ is the volume of space, and in the second step we have assumed canonical normalisation of the kinetic term of the field.
In general, there may be additional zero modes related to the breaking of internal symmetries. A more general prescription for the removal of the zero modes is the method of collective coordinates. Then the volume factor of zero modes is simply a Jacobian associated with the change to collective coordinates \cite{Ekstedt:2023sqc, Bhattacharya:2024chz}.
By performing the saddle-point approximation, the normalisation factor $\mathcal{N}$ in eq.~\eqref{eq:Astat_Feff} is absorbed into the determinant.\footnote{To understand this, consider the fact that the path integral without a delta function, in the saddle-point approximation, would result in a Gaussian integration. Since we remove the negative eigenfunction part of the Gaussian integral, it must be compensated by, which is what the $\mathcal{N}$ factor does exactly. In other words, only by including $\mathcal{N}$ with the path integral containing a delta function, we can perform a Gaussian integration.}

Taking all the pieces together, the resulting statistical part, or more specifically, $-\log\Astat$, can be connected to the effective action of a given theory by an analytic continuation \cite{Gould:2021ccf}. More specifically, it is possible to express the nucleation rate in the framework of the 3d EFT we have introduced in the previous section. The general equivalence can be written as,
\begin{align}
    \beta F_\rmii{eff}[\phi_\rmii{CB}] 
    \equiv 
    S_\rmii{nucl}[\phi_\rmii{CB}],
\end{align}
where $\beta=1/T$, while $S_\rmii{nucl}$ is the action of \textit{nucleation EFT} and $\phi_\rmii{CB}$ is its stationary configuration. This effective theory is associated with the nucleation scale $\Lambda_\rmii{nucl}$, which in the usual scenario corresponds to the mass of the nucleating scalar field. Thus, given a concrete theory, one should identify the heavy d.o.f.s and integrate out all the fluctuations above the nucleation scale. At the one-loop level, this can be schematically written as
\begin{align}
\label{eq:Astat_Snucl}
    \Astat = 
    \underbrace{
    \mathcal{V}_0 
    \qty(
    \qty|
        \frac{
                \det^\prime \qty( S_\rmii{eff} ^{\prime\prime }[\phi_\rmii{CB}])
            }{
                \det\qty( S_\rmii{eff} ^{\prime\prime }[\phi_\rmii{meta}])
            }
    |
    )^{-\frac{1}{2}}
    }_{\text{modes } E < \Lambda_\rmii{nucl} }
    \underbrace{
    e^{- \Delta S_\rmii{nucl}[\phi_\rmii{CB}]}
    }_{\text{modes } E > \Lambda_\rmii{nucl} },
\end{align}
where again $\Delta S_\rmii{nucl}[\phi_\rmii{CB}] = S_\rmii{nucl}[\phi_\rmii{CB}] - S_\rmii{nucl}[\phi_\rmii{meta}]$. 

To sum up, to compute the nucleation rate, one should follow the steps listed below
\begin{enumerate}
    \item Construct nucleation EFT $S_\rmii{nucl}$ by integrating out the fluctuations with energies $E > \Lambda_\rmii{nucl}$.
    \item Find the critical bubble field configuration by solving for the saddle point of $S_\rmii{nucl}$, i.e.     
    $\eval{
        \frac{\delta S_\rmii{nucl}[\phi]}{\delta \phi} 
    }_{\phi=\phi_\rmii{CB}} 
    =0$.
    \item Integrate over the scalar fluctuation on the critical bubble background, i.e. modes with energies $E < \Lambda_\rmii{nucl}$ to obtain the fluctuation determinant. 
    \item Evaluate the determinant and the exponent on the critical bubble background and normalise to the metastable phase field configuration.
\end{enumerate}
The last missing piece is the dynamical factor. As it contains non-equilibrium effects, it cannot be obtained solely in terms of nucleation EFT. However, in a non-damping approximation, it can be obtained using equations \eqref{eq:Adyn_Feff} and \eqref{eq:kappa_eta}, and employing there the negative eigenvalue associated with the fluctuation operator $S_\rmii{eff} ^{\prime\prime }[\phi_\rmii{CB}]$.

As a result, the nucleation rate computed this way inherits all the desired qualities of the EFT. If computed properly within a strict perturbative expansion, $\GammaT$ is real, and both RG-scale and gauge invariant to all orders and properly incorporates all the necessary contributions of heavy and light modes in the theory \cite{Ekstedt:2022tqk, Gould:2021ccf, Ekstedt:2021kyx}. 
We will now show how to calculate nucleation rate explicitly using an example 3d EFT from previous sections. 

\section{Nucleation rate in a model with radiative barrier}
\label{sec:nucl_rate_model_radiativebarrier}
Following refs.~\cite{Gould:2021ccf, Ekstedt:2022tqk}, we will now show how to compute both factors within the framework of 3d effective theory introduced in previous sections. In particular, we will discuss the SU(2) gauge theory, whose EFT  represents a broader class of models with a radiative barrier.

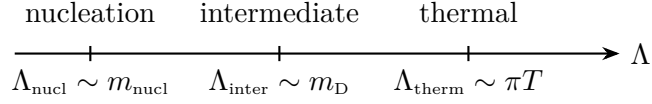
\begin{figure}
    \centering
    \begin{tikzpicture}[>=Stealth,thick]

      \draw[->] (0,0) -- (8,0) node[anchor=west] {$\Lambda$};
    
      \draw (1,-0.1) -- (1,0.1)
      node[above=5pt] {nucleation}
      node[below=5pt] {$\Lambda_\rmii{nucl}\sim m_\rmii{nucl}$};
      \draw (3.5,-0.1) -- (3.5,0.1) 
      node[above=5pt] {intermediate}
      node[below=5pt] {$ \Lambda_\rmii{inter}\sim \mD$};
      \draw (6,-0.1) -- (6,0.1) 
      node[above=5pt] {thermal}
      node[below=5pt] {$ \Lambda_\rmii{therm}\sim \pi T$};
    
    \end{tikzpicture}
    \caption{Thermal scale hierarchy for bubble nucleation.}
    \label{fig:scales_su2_higgs}
\end{figure}

\paragraph{Step 1}
From the discussion in section \ref{sec:DR_SU(2)_sketch} we can conclude that SU(2)+Higgs theory exhibits the following scales: 
\begin{itemize}
    \item thermal (hard) scale associated with the lightest non-zero Matsubara mode $\Lambda_\rmii{therm} = \Lambda_\rmii{hard} \sim \pi T$,
    \item intermediate (soft) scale associated with the Debye mass $\Lambda_\rmii{inter} = \Lambda_\rmii{soft} \sim \mD \sim gT$,
    \item nucleation (supersoft) scale associated with the mass of scalar field $\Lambda_\rmii{nucl} = \Lambda_\rmii{supersoft} \sim \mthree \sim \frac{g^{\frac32}}{\sqrt{\pi}}T$.
\end{itemize}
See also figure~\ref{fig:scales_su2_higgs} for a schematic depiction of these scales.
In order to calculate nucleation rate in such a scenario, one needs first to integrate out the thermal scale via dimensional reduction as in eq.~\eqref{eq:SU(2)_soft_Seff}, which leads to the soft EFT. Then we can integrate out the gauge modes by constructing an effective action within the soft EFT. Resulting supersoft effective action as shown in eq.~\eqref{eq:SU(2)_supersoft_Seff} is the leading-order nucleation EFT we need
\begin{align}
    S_\rmii{nucl} = S^\rmii{(supersoft)}_\rmii{LO}
\end{align}

\begin{figure}
    \centering
    \includegraphics[width=0.8\linewidth]{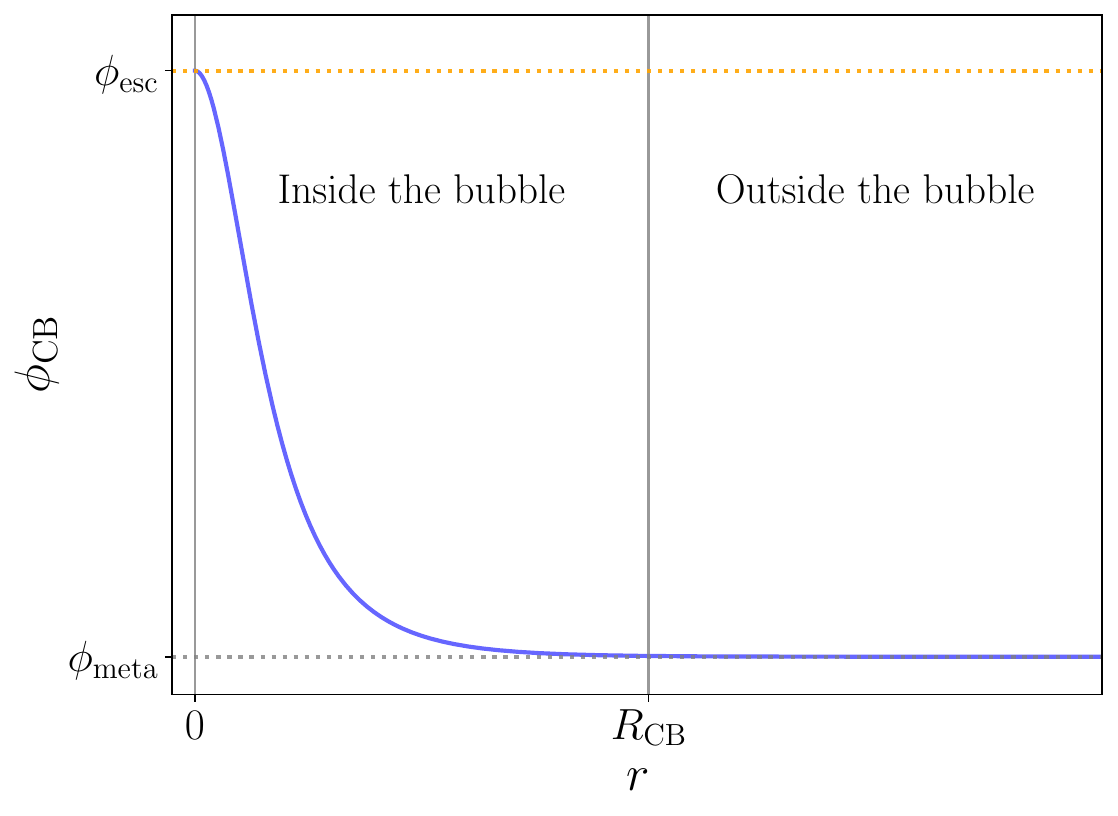}
    \caption{
        Example plot of a critical bubble (or bounce solution). Vertical grey line on the left denotes the centre of the bubble, while the grey line on the right corresponds to the approximate total size of the critical bubble, $R_{\rm{CB}}$, which separates the inside of the bubble (stable phase) from the outside (metastable phase). The orange dotted line denotes the field value at the centre of the bubble, while the grey dotted line denotes the metastable phase.
    }
    \label{fig:toy_bounce}
\end{figure}

\paragraph{Step 2}
We can find the critical bubble by solving for the saddle point of the leading-order action. First, let us note that the bubble solution is $O(3)$-symmetric, thus the critical bubble field configuration will be a function of the radial coordinate $r$. 
The Euclidean nucleation action can then be rewritten in terms of such a background, in radial coordinates, as 
\begin{align}
\label{eq:Snucl_radial}
    S_\rmii{nucl}[\phi_b(r)] &= 
    \int_\mathbf{r} 
    \qty{
        \frac12(\partial_r \phi_b)^2
        + V _\rmii{nucl}[\phi_b(r)]
    },\\
    V _\rmii{nucl} [\phi_b(r)] &= 
        \frac12 \mthree^2 \phi_b ^2(r) 
        +\frac14 \lamthree \phi_b ^4(r) 
        +\frac{1}{12\pi} \qty( 
                6m_A ^3[\phi_b(r)] + 3m_{A_0}^3 [\phi_b(r)]
            ),
\end{align}
where we have introduced notation $\int_\mathbf{r} \equiv 4\pi^2 \int\dd{r} r^2$.
Varying the action $S_\rmii{nucl}$ with respect to the background field leads to so-called \textit{bounce equation} \cite{Coleman:1977py, LINDE1983421, LINDE198137} 
\begin{align}
    \dv[2]{\phi_\rmii{CB}}{r} 
    + \frac{2}{r} \dv[]{\phi_\rmii{CB}}{r} 
    = \dv[]{V _\rmii{nucl}}{\phi_\rmii{CB}},
\end{align}
which determines the critical bubble -- or a bounce solution -- that obeys the following boundary conditions
\begin{align}
    \eval{\dv[]{\phi_\rmii{CB}}{r} }_{r=0} = 0, \quad 
    \phi_\rmii{CB}(r\rightarrow \infty) = \phi_\rmii{meta}.
\end{align}
First condition ensures the solution is analytical, which is crucial, especially in 4d, as then the solution needs to be translatable between Euclidean and Minkowski space. The second condition means that for large values of radial coordinates, we should be outside the bubble, where the system is still in the metastable phase.
%
%
We will denote the field value at the centre of the bubble as $\phi_\rmii{esc} \equiv \phi_\rmii{CB}(0)$, as it corresponds to the escape point to which the field transitions from the metastable phase.
A schematic solution is shown in figure~\ref{fig:toy_bounce}. 

\paragraph{Step 3}
Knowing the critical bubble and the nucleation EFT action, we can integrate out the scalar fluctuations. Let us expand the action on the bubble background, 
\begin{align}
    S_\rmii{nucl}[\phi_\rmii{CB}(r) + \Tilde{\phi}(\mathbf{x})] = 
    S_\rmii{nucl}[\phi_\rmii{CB}] 
     +\frac12 \int_{\mathbf{x}} \int_{\mathbf{y}}
       \Tilde{\phi}(\mathbf{x}) 
       S_\rmii{nucl} ^{\prime\prime }[\phi_\rmii{CB}] (\mathbf{x}, \mathbf{y})
       \Tilde{\phi}(\mathbf{y}),
\end{align}
where the linear term disappears as the critical bubble is a stationary point of the action. The fluctuation operator can be obtained by taking a double functional derivative of the LO action:
\begin{align}
    S_\rmii{nucl} ^{\prime\prime} [\phi_\rmii{CB}] =
    \partial_r ^2  
    + V^{\prime\prime}_\rmii{nucl}[\phi_\rmii{CB}].
\end{align}
Performing the Gaussian path integral leads to the fluctuation determinant 
\begin{align}
\label{eq:scalar_det_toy}
    \int \mathcal D \Tilde \phi
    ~e^{-\frac12 \int_{\mathbf{x}} \int_{\mathbf{y}}
       \Tilde{\phi}(\mathbf{x}) 
       S_\rmii{nucl} ^{\prime\prime }[\phi_\rmii{CB}] (\mathbf{x}, \mathbf{y})
       \Tilde{\phi}(\mathbf{y})}
    = \qty(\det S_\rmii{nucl} ^{\prime\prime} [\phi_\rmii{CB}])^{-\frac{1}{2}}.
\end{align}

\paragraph{Step 4}
Taking all the pieces together, we can now evaluate the exponential containing the effective action as well as the prefactor containing the fluctuation determinant. We need to evaluate them on the critical bubble and also the metastable phase, which serves as a normalisation of the nucleation rate. We can also use eq.~\eqref{eq:Vol_zeromodes} to write the volume factor explicitly in terms of the action. The statistical part of the thermal nucleation rate thus becomes
\begin{align}
\Astat =
    V_\rmii{space}
    \qty(\frac{S_\rmii{nucl}[\phi_\rmii{CB}]}{2\pi})^\frac{3}{2}
    \qty(\qty|
        \frac{\det^\prime (\partial_r ^2  
            + V^{\prime\prime}_\rmii{nucl}[\phi_\rmii{CB}]) 
        }{\det (\partial_r ^2  
            + V^{\prime\prime}_\rmii{nucl}[\phi_\rmii{meta}]) 
        }
    |)^{-\frac12}
    e^{-\Delta S_\rmii{nucl}[\phi_\rmii{CB}]}.
\end{align}
The $(2\pi)^{-3/2}$ term in the zero modes volume factor comes from the fact that we deal with gauge theory.
Computing the fluctuation determinant can be quite involved, especially if there is some mixing between fluctuating modes in a given theory. Nevertheless, it is possible by using Gelfand-Yaglom theory and its generalisations. Moreover, there exists a public code \texttt{BubbleDet} \cite{Ekstedt:2023sqc} written in Python, which allows for numerical calculation of the one-dimensional fluctuation determinants. 

\paragraph{Dynamical factor}
We can also compute the dynamical part in the no-damping approximation. To do that, we need to find the negative eigenvalue $\lambda_-$ of the fluctuation operator. We can use the spherical symmetry of the critical bubble to decompose the eigenfunctions of the fluctuation operator into spherical harmonics. The operator becomes then 
\begin{align}
    (\partial_r ^2 + \frac{d-1}{r} \partial_r 
    - \frac{l(l+d-2)}{r^2}
            + V^{\prime\prime}_\rmii{nucl}[\phi_\rmii{meta}]).
\end{align}
Here $l$ is the orbital number. In our case, $d=3$ and since the negative eigenmode $f_- (r)$ is spherically symmetric, then $l=0$. The eigenvalue is thus defined by the equation
\begin{align}
    (\partial_r ^2 + \frac{2}{r} \partial_r 
            + V^{\prime\prime}_\rmii{nucl}[\phi_\rmii{meta}])
    f_- (r) 
    =\lambda_- f_- (r), 
\end{align}
with boundary conditions
\begin{align}
    \partial_r f_-(r\rightarrow 0^+) = 0, \quad 
    f_-(r\rightarrow \infty) = 0.
\end{align}
For practical purposes, in numerical studies, it is convenient to use public code \texttt{BubbleDet} \cite{Ekstedt:2023sqc} and a function \texttt{findNegativeEigenvalue()} therein.

\paragraph{Final formula}
Finally, we can write the complete formula for the leading-order thermal nucleation rate per unit volume, in our toy SU(2)+Higgs theory,
\begin{align}
    \Gamma_T = 
    \sqrt{\frac{\lambda_-}{2\pi}}
    \qty(\frac{S_\rmii{nucl}[\phi_\rmii{CB}]}{2\pi})^\frac{3}{2}
    \qty(\qty|
        \frac{\det^\prime (\partial_r ^2  
            + V^{\prime\prime}_\rmii{nucl}[\phi_\rmii{CB}]) 
        }{\det (\partial_r ^2  
            + V^{\prime\prime}_\rmii{nucl}[\phi_\rmii{meta}]) 
        }
    |)^{-\frac12}
    e^{-\Delta S_\rmii{nucl}[\phi_\rmii{CB}]}.
\end{align}
One last thing to note is the possibility of including higher orders in the nucleation rate. Moreover, in previous sections, we have seen that the RG-scale invariance of the EFT required computing the NLO effective potential within the 3d EFT. Indeed, the LO nucleation rate written above is computed up to $\mathcal{O}(g^3)$ in the coupling constant, and to inherit the RG-scale independence of the 3d EFT, one would have to go to the NLO, i.e. include terms of order $\mathcal{O}(g^4)$. 
One can improve the accuracy even higher, as in general the effective action $\Seff$ can be expanded on a general background $\phib$ as
\begin{align}
    \Seff [\phib]  
    &= \SLO[\phib] + \SNLO[\phib] + \dots\\
    &= \SLO[\phib ^\rmii{LO}] + \SNLO[\phib ^\rmii{LO}]
    - \frac12 (\phib ^\rmii{NLO})^2 \delta^2 \SLO[\phib ^\rmii{LO}]
    + \dots
\end{align}
We see that at higher orders, the background configuration also receives corrections. Such corrections are formally the effect of incorporating all tadpole diagrams, for details see \cite{Ekstedt:2021kyx, Ekstedt:2022tqk, Gould:2021ccf}. 

In the next parts of this thesis, we will apply the framework described here to obtain the NLO nucleation rate in a classically scale-invariant extension of the Standard Model. The details are going to be described in chapter \ref{chapter:supercoolEFT}.

\chapter{Conformal extension of the Standard Model}
\label{chapter:su2csm}

Now we will use the formalism described in previous chapters to study the cosmological phase transition in a classically scale-invariant extension of the Standard Model. First, we will introduce a concrete model called SU(2)cSM, and then we will describe its general features. We will obtain the effective potential and show how to treat the renormalisation scale dependence carefully. 
This chapter contains the original results, mainly the ones described in ref.~\cite{Kierkla:2022odc}.

\section{Introducing the model}

A specific BSM scenario, which we will study here, is a model called SU(2)cSM.
It is a classically scale-invariant version\footnote{Or in other words ``conformal'', hence the ``c'' in SU(2)cSM.} of the Standard Model, but extended with a new SU(2)$_X$ gauge (dark) sector.
The total field content is thus Standard Model fields plus the new SU(2)$_X$ fields: a scalar field, $\Phi$, and a gauge boson, $X$. The new scalar field $\Phi$ is a doublet under the new SU(2)$_X$, whilst a singlet of the SM gauge group. The Lagrangian of the SU(2)cSM is given as
\begin{align}
    \mathcal{L}_{\rmii{SU(2)cSM}} = 
    \mathcal{L}_{\rmii{cSM}} + 
    \frac{1}{4} F^a_{\mu\nu} F^{a\mu\nu} 
    + (D_\mu \Phi)^\dagger (D^\mu \Phi)
    - V(\Phi, H),
\end{align}
where $F^a_{\mu\nu}$ is the SU(2) field strength tensor and index $a=1,2,3$ corresponds to SU(2) gauge group index. Then, the covariant derivative of dark SU(2) is given as
\begin{align}
    D_\mu \phi &= \partial_\mu \phi -i\gx \frac{\tau^a}{2} X^a_\mu \phi,
\end{align}
where $\gx$ is the associated gauge coupling.
The term $\mathcal{L}_{\rmii{cSM}}$ corresponds to the Lagrangian of the SM without the Higgs potential. Instead, there is a new potential introduced where the SM sector and the SU(2)$_X$ sector communicate only via the Higgs-portal term
\begin{align} 
    V(\Phi, H) = \lambda_h (H^\dagger H)^2 + \lambda_{h\varphi} (\Phi^\dagger \Phi)(H^\dagger H)^2 + \lambda_\varphi  (\Phi^\dagger \Phi)^2,
\end{align}
where $H$ is the SM scalar field. The tree-level potential is classically scale-invariant, as there are no dimensionful parameters present. Moreover, gauge bosons $X$ do not couple to the SM directly, but only through the mixing of $H$ and $\Phi$. 
We will discuss symmetry-breaking patterns in SU(2)cSM in more detail in the following sections.

This model is one of the possible choices in the class of classically scale-invariant extensions of the SM. Such models are very promising in the sense that they are highly predictive and perturbative. Among other popular choices of the extra gauge sector is the ``pure'' U(1) version a'la Coleman-Weinberg (see e.g., \cite{Lewicki:2021xku, Das:2016zue}) or a U(1) variant with a gauged $B-L$ (see e.g., \cite{Jinno:2016, Oda:2017kwl, Ellis:2020, Marzo:2018nov}) quantum number. For a review of classically scale-invariant models with different gauge groups, see, e.g., \cite{Khoze:2014xha}. 

The choice of the SU(2) gauge group in our case is motivated by several factors. It does not possess a UV Landau pole, as the theory is asymptotically free. On the other hand, there exists a region in parameter space with possible strong dynamics \cite{Kierkla:2022odc}, which broadens the phenomenology of the model. Then, this model provides a candidate for vector dark matter, as the dark gauge boson is stabilised after the phase transition by a residual symmetry \cite{Hambye:2018qjv, Hambye:2013}. 
In this thesis, however, we will focus on the accurate description and phenomenology of the supercooled phase transition in SU(2)cSM. For more general study, including the interplay of dark matter and gravitational wave signals, see e.g. refs.~\cite{Kierkla:2022odc, Marfatia:2020, Baldes:2018}.
It should be emphasised, though, that the methods and their applications that are going to be described in the following chapters (such as derivation of effective action, calculation of the thermal nucleation rate and the description of phase transition) can be applied to any kind of model exhibiting a supercooled phase transition or possessing a classical scale invariance.

\section{Effective potential at zero temperature}
To begin the analysis of the theory at zero temperature, let us begin with the derivation of the effective potential. Note that each of the sectors has its own SU(2) symmetry. Thus, analogously to the usual SM practice, we can write the radial components squared of both doublets as:
\begin{align}
    H^\dagger H = \frac{1}{2}h^2, \quad
    \Phi^\dagger \Phi = \frac{1}{2} \varphi^2.
\end{align}
Then the tree-level contributions to the effective potential can be written as
\begin{align}
     V^{(0)}(h, \varphi) \equiv 
     \frac{1}{4} \lambda_h h^4
     + \frac{1}{4} \lambda_{\varphi h}  h^2 \varphi^2
    + \frac{1}{4} \lambda_\varphi \varphi^4,
\end{align}
It should be noted that here the radial components $h,\varphi$ are now identified with the classical background fields, as they are arguments of the effective potential. From now on, every time we write the effective action, we will adopt this implicit notation, unless specified otherwise.
One-loop quantum corrections are given by the standard formula (in the $\overline{\rm MS}$ scheme\footnote{For a discussion of CSI models in different renormalisation schemes see \cite{Sojka:2024btp}} and Landau gauge)~\cite{PhysRevD.7.1888}:
\begin{align}
    V^{(1)}(h, \varphi)\equiv
    \frac{1}{64 \pi^2} \sum_a n_a M_a^4(h, \varphi)\left(\log \frac{M_a^2(h, \varphi)}{\mu^2}-C_a\right),
\end{align}
where the sum goes over all particle content. 
The parameter $n_a$ corresponds to the number of degrees of freedom, and it is given as:
\begin{align}
    n_a = (-1)^{2s_a}Q_a N_a (2s_a +1),
\end{align}
where $s_a $ is the spin, then $Q_a = 1$ for uncharged particles, while $Q_a = 2$ for charged ones (because charged particles have their antiparticles), $N_a=1$ for colour singlets and $N_a=3$ for particles carrying the colour charge.
In dimensional regularisation, the constant $C_a$ is equal to $C_a = \frac{5}{6}$ for vector bosons and $C_a=\frac{3}{2}$ for all the other particles, while $\mu$ is the RG-scale.

Finally, the parameter $M_a(h,\varphi)$ is the field-dependent mass of a particle labelled by $a$. The most relevant particles from the SM sector (aside from the Higgs boson) that contribute to the effective potential are the top quark and the weak gauge bosons, as they are the heaviest. 
The tree-level field-dependent mass squared for the scalars is given by the second derivative of the $V^{(0)}$, i.e. the Hessian matrix
\begin{align}
M^2(h, \varphi)=
\left(\begin{array}{cc}
    3 \lambda_1 h^2+\frac{\lambda_2}{2} \varphi^2 & \lambda_2 h \varphi \\
    \lambda_2 h \varphi & 3 \lambda_3 \varphi^2+\frac{\lambda_2}{2} h^2
\end{array}\right),
\end{align}
from which one can obtain tree-level mass eigenvalues 
\begin{align}
\begin{aligned}
M_{ \pm}^2(h, \varphi)= & \frac{1}{2}\left(3 \lambda_1+\frac{\lambda_2}{2}\right) h^2+\frac{1}{2}\left(\frac{\lambda_2}{2}+3 \lambda_3\right) \varphi^2 \\
& \pm \frac{1}{2} \sqrt{\left[\left(3 \lambda_1-\frac{\lambda_2}{2}\right) h^2-\left(3 \lambda_3-\frac{\lambda_2}{2}\right) \varphi^2\right]^2+4 \lambda_2^2 h^2 \varphi^2} .
\end{aligned}
\end{align}
In most parts of the parameter space, the heavier state can be associated with a new scalar particle; however, as we will show, this is not always the case. 
Then, the associated field-dependent Goldstone masses read
\begin{align}
\begin{gathered}
M_G^2(h, \varphi)=M_{G^{ \pm}}^2(h, \varphi)=\lambda_1 h^2+\frac{1}{2} \lambda_2 \varphi^2, \\
M_{G_X}^2(h, \varphi)=M_{G_X^{ \pm}}^2(h, \varphi)=\lambda_3 \varphi^2+\frac{1}{2} \lambda_2 h^2 .
\end{gathered}
\end{align}
Finally, the tree-level vector boson and top quark masses are given in the usual way,
\begin{align}
  M_{\rmii{$W$}}(h)&=\frac{1}{2} g_2 h, \quad 
  M_\rmii{$Z$}(h)=\frac{1}{2} \sqrt{g_2^2+g_Y^2} h, 
  \quad M_\rmii{$t$}(h)=\frac{1}{\sqrt{2}} y_t h,\quad 
\MX( \varphi)=\frac{1}{2} g_X \varphi,  
\end{align}
where $g_1, g_2$ are the SM gauge electroweak couplings, $y_t$ is the top quark Yukawa coupling and $g_X$ is the gauge coupling of $\mathrm{SU}(2)_X$.


\subsection{Perturbative expansion and power counting}
\label{sec:power_counting_T0}
As mentioned before, the symmetry breaking in the dark sector proceeds via the Coleman-Weinberg mechanism \cite{PhysRevD.7.1888}. 
In order for it to work, one-loop corrections should change the potential qualitatively, which seems like an explicit breakdown of perturbation theory. And it is, but only if one organises the perturbative expansion in terms of loops. Indeed, for radiative symmetry breaking to work, one needs to reorganise perturbation theory in some new expansion parameter $g$. As we have shown in chapter~\ref{chapter:Seff_vacuum}, first we need to establish some hierarchy between the couplings to distinguish between terms that contribute to LO or to NLO.
For the SU(2)cSM, we shall adopt the power-counting scheme from \cite{Kierkla:2022odc}. It is worth emphasising here that this power-counting is valid only in the global minimum of the theory. Thus, the tree-level terms scale as $\sim g^2$, while the one-loop contributions (such as from the gauge bosons $X$) scale as $\sim g^4$. Moreover, we assume that the SM couplings, $g_2 ^2, g_Y^2, y_t ^2$ all scale as $\sim g^2$. Another crucial thing to note here is that scalar one-loop corrections are not included at LO in such power counting, since they contribute as $\lambda^2\sim g^8$.
As for the scalar sector, the portal coupling $\lamphih$ is small (usually $\lamphih < 10^{3}$) and should be of order $\order{g^4}$. This magnitude of portal coupling is required to reproduce the correct EW minimum, as the vev of the new scalar is usually much larger than the EW scale, and the portal coupling must compensate for it.  
Similarly, the new scalar coupling scales as $\lamphi \sim g^4$, which is required by the radiative symmetry breaking. 
Then, Higgs quartic coupling scales as $\lamh \sim g^2$ just like the rest of the SM couplings. This is required to reproduce correct SM physics.


From this discussion, it follows that the scalar one-loop contributions to the effective potential scale as $\lambda^2 \sim g^8$ and thus belong to higher orders in $g$-expansion and should be neglected at $g^4$. See figure \ref{fig:VLO_RSB} for the scheme of the LO effective potential. For further details on the scalings of couplings in SU(2)cSM, see ref.~\cite{Kierkla:2022odc}.
\begin{figure}
    \centering
    \begin{equation*}
    V_\rmii{eff} =  
    \overbrace{
    \begin{tikzpicture}[scale=0.8, baseline=-0.5ex]
        \begin{feynman}
            \vertex[dot] (a) at (0,0) {};
            \node[below] (a) at (0,-0.5) {$\sim \order{g^2}$};
        \end{feynman}
    \end{tikzpicture}
    \;+\;
    \begin{tikzpicture}[scale=0.8, baseline=+1.5ex]
        \begin{feynman}
            \vertex (b) at (0,0);
            \path (b)--++(90:0.5) coordinate (A);
            \draw [photon] (A) circle(0.5);
            \node[below] (b) at (0,-0.05) {\(\sim \order{g^4}\)};
        \end{feynman}
    \end{tikzpicture}
    }^{V_\rmii{eff} ^\rmii{(LO)} }
    \;+\;
    \begin{tikzpicture}[scale=0.8, baseline=+1.5ex]
        \begin{feynman}
            \vertex (b) at (0,0);
            \path (b)--++(90:0.5) coordinate (A);
            \draw [plain] (A) circle(0.5);
            \node[below] (b) at (0,-0.05) {\(\sim \order{g^8}\)};
        \end{feynman}
    \end{tikzpicture}
    .
    \end{equation*}
    \caption{Schematic representation of the effective potential and power counting in SU(2)cSM. Dot denotes a tree-level potential, the wiggly-loop corresponds to gauge one-loop correction, while the solid-line loop is the one-loop scalar correction.}
    \label{fig:VLO_RSB}
\end{figure}
Thus, we conclude that the only relevant 1-loop contributions to the effective potential come from gauge bosons and the top quark. This is fortunate, as it makes the derived effective potential explicitly gauge invariant. All the gauge-fixing parameter $\xi$-dependence in $R_\xi$-gauges is associated with mixed Goldstone-gauge terms in the action (see, e.g. \cite{PhysRevD.99.015026, Andreassen_2015}). 
Explicitly, the LO effective potential becomes:
\begin{align}
    V_\rmii{eff} ^\rmii{(LO)} = 
    &\frac{1}{4} \lambda_h h^4
     + \frac{1}{4} \lambda_{\varphi h}  h^2 \varphi^2
    + \frac{1}{4} \lambda_\varphi \varphi^4 
    +\frac{9\MX^4}{64 \pi^2}\left( 
        \log \frac{\MX^2}{\mu^2}-\frac56 
    \right)
    -\frac{12 M_\rmii{$t$}^4}{64 \pi^2}\left( 
        \log \frac{M_\rmii{$t$}^2}{\mu^2}-\frac32 
    \right) \notag \\
    &+\frac{6 M_\rmii{$W$}^4}{64 \pi^2}\left( 
        \log \frac{M_\rmii{$W$}^2}{\mu^2}-\frac56 
    \right)
    +\frac{3 M_\rmii{$Z$}^4}{64 \pi^2}\left( 
        \log \frac{M_\rmii{$Z$}^2}{\mu^2}-\frac56 
    \right).
\end{align}

\subsection{Symmetry breaking patterns}
Now we will investigate symmetry breaking in the SU(2)cSM model. To do that, we need to minimise the potential. 
The stationary-point equations (divided by vevs) for the LO potential are given as:
\begin{align}
\frac{1}{v^3}\frac{\partial V}{\partial h}=
\lambda_h&  
+\frac{1}{2} \lambda_{\varphi h} \left(\frac{w}{v}\right)^2 
+\frac{1}{v^3}\left.\frac{\partial V^{(1)}}{\partial h}\right|_{h=v, \varphi=w}=0,\label{eq:min1}\\
\frac{1}{w^3}\frac{\partial V}{\partial \varphi}=
\lambda_\varphi& 
+\frac{1}{2} \lambda_{\varphi h} \left(\frac{v}{w}\right)^2 
+\frac{1}{w^3}\left.\frac{\partial V^{(1)}}{\partial \varphi}\right|_{h=v, \varphi=w}=0.
\label{eq:min2}
\end{align}
Let us look first at the right side of eq.~\eqref{eq:min2}. The term proportional to $\lambda_{\varphi h} $ is suppressed by the ratio $(v/w)^2$, so we can neglect it, as it is going to be negligible with respect to $\lamphih$ and the derivative term.
As explained before, at $g^4$-order, we include only one-loop terms coming from the $X$ gauge boson (note that there are only $\varphi$ derivatives in eq.~\eqref{eq:min2}, so only dark sector terms matter here).  
Therefore, writing explicitly the Coleman--Weinberg term,  eq.~\eqref{eq:min2} can be expressed as
\be
\lambda_\varphi= - \frac{9}{256\pi^2}
\gX^4\left[2\log\left(\frac{\frac{1}{2} \gx w}{\mu}\right) -\frac{1}{3}\right].\label{eq:min-phi}
\ee
This confirms our earlier claim that $\lambda_{\varphi} \sim \gx^4 \sim g^4$.
It is a typical Coleman--Weinberg relation between the scalar and the gauge couplings, and this relation is the exact reason that radiative symmetry breaking is possible since it ensures that the gauge one-loop term is of the same order as the tree-level terms in the $g$-expansion. 

\begin{figure} 
    \centering
    \includegraphics[width=0.5\linewidth]{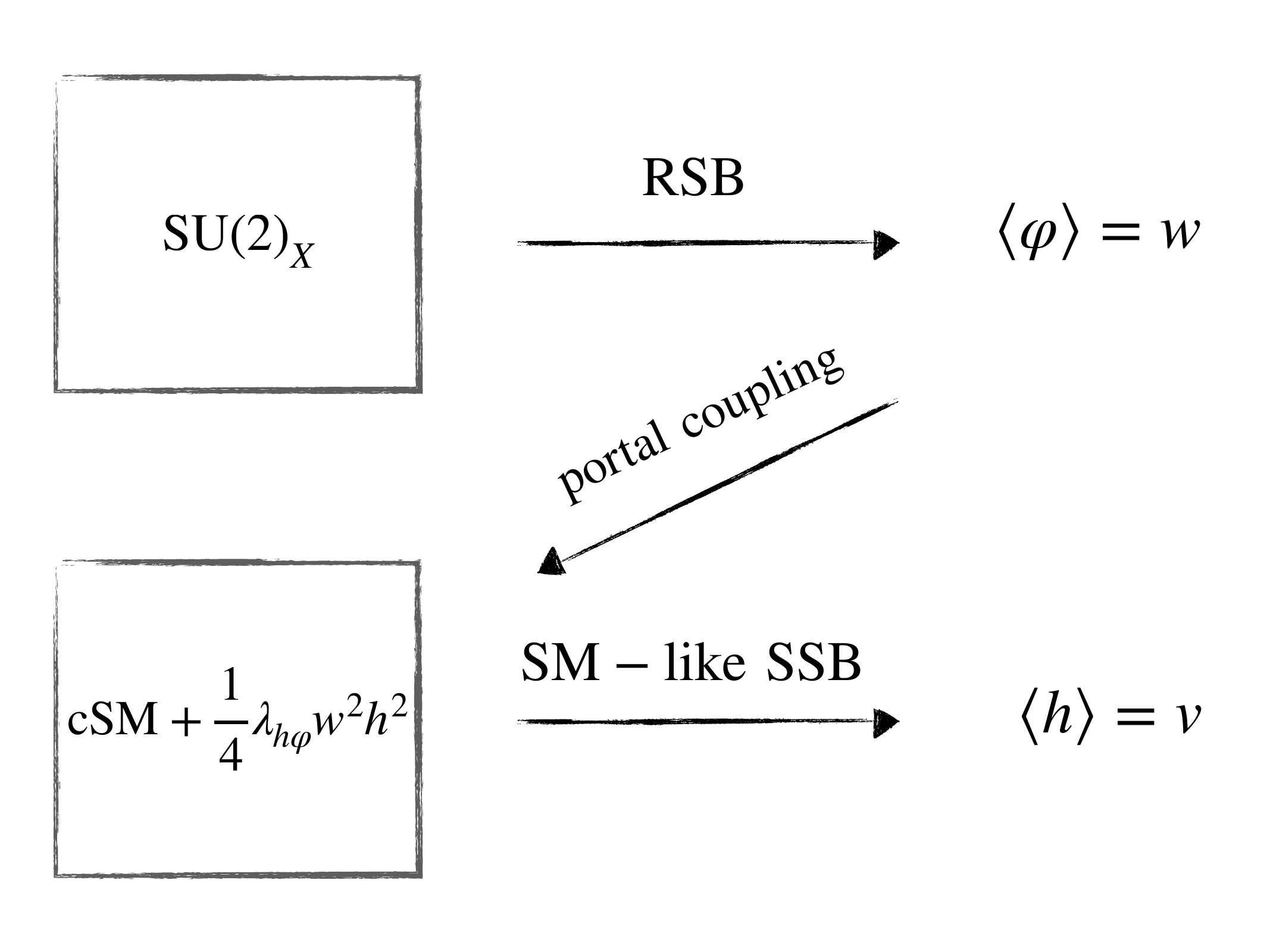}
    \caption{Scheme of symmetry breaking pattern in SU(2)cSM. First, the vev $w$ is generated in the dark SU(2) sector via RSB, then via the portal coupling, a mass term of the SM Higgs is generated, allowing for regular, "SM-like" SSB and generation of Higgs vev $v$.}
    \label{fig:ssb_patterns}
\end{figure}

Now let us discuss the other stationary-point condition, eq.~\eqref{eq:min1}. The term proportional to $\lambda_{\varphi h}$ is enhanced by the ratio $\left(\frac{w}{v}\right)^2$ instead. Moreover, we can use the fact that the SM tree-level couplings scale as $g^2$, which makes them larger than 1-loop corrections that belong to $g^4$ order. Therefore, by neglecting the subleading terms, we arrive at a simple relation:
\be
\lamh +\frac{1}{2} \lamphih \left(\frac{w}{v}\right)^2=0.
\label{eq:tree-level-min}
\ee
This relation indicates the ``SM-like'' symmetry-breaking pattern. The second term here can be identified with the ``SM-like'' mass term for the Higgs boson.\footnote{For completeness, we can also consider the $g^4$ corrections, which correspond to the one-loop contributions from the SM particles (in the eq.~\eqref{eq:min1}, there are only $h$ derivatives, so we can focus on the SM sector only). Thus, the corrected stationary point condition becomes
\be
\lamh +\frac{1}{2} \lamphih \left(\frac{w}{v}\right)^2 +\frac{1}{16\pi^2}\sum_{W^{\pm},Z,t}n_a \frac{M_a^4(h,\f)}{v^4}\left(\log\frac{M_a^2(h,\f)}{\mu^2}-C_a+\frac{1}{2}\right)=0. \label{eq:condition-lambda2}
\ee
}
The summary of symmetry-breaking patterns described in this section can be written as (see also figure\ref{fig:ssb_patterns}):  
\begin{enumerate}
    \item First radiative symmetry breaking occurs in the dark sector, i.e. along $\varphi$ direction, and the vev, $w$, is generated.
    \item Then, in the SM sector, symmetry is broken in a ``regular'' way, as the $h$ field now has a negative ``tree-level mass term'' generated by the portal coupling $\lamphih$ and new scalar vev $w$.
\end{enumerate}
This pattern is also present in other models with classical scale invariance and dark gauge sectors, see e.g. \cite{Khoze:2014xha, Kierkla:2025vwp, Schmitt:2024pby}.

\subsection{Masses and mixing of the scalars}
Let us now discuss further the masses of scalars.
The easiest way to compute the one-loop corrected masses of scalar particles is to obtain eigenvalues of the Hessian of the LO effective potential. Such ``running'' masses (as their value depends on the chosen RG-scale $\mu$) are evaluated then at zero momentum, as the effective potential is the zero-momentum part of the effective action. However, the actual physical masses $M^2_{\rm pole}$ are associated with the pole of the propagator: 
\begin{align}
    M^2_{\rm pole} = m^2_{\rm tree} + \Re[\Sigma(p^2=M^2_{\rm pole})],
\end{align} 
where $\Sigma$ is the one-loop self-energy of the scalar field.
Fortunately, the correction from the non-zero momentum is rather small, as it was shown in \cite{Kierkla:2022odc}; therefore, here we shall use the running masses, which we can formally define as:
\begin{align}
    M^2_{\rm } \equiv m^2_{\rm tree} + \Re[\Sigma(p^2=0)],
\end{align}
where $\Re[\Sigma(p^2=0)] = \partial^2 V^{(1)}(\varphi, h)$ evaluated in the vev of the theory.
In the case of SU(2)cSM, the mass matrix is then given by
\begin{align}
    M^2 = 
    \mqty(
        3\lambda_h v^2 + \frac{1}{2} \lambda_{\varphi h} w^2 &  \lambda_{\varphi h} w^2 v w \\
        \lambda_{\varphi h} w^2 v w & 3\lambda_\varphi w^2 + \frac{1}{2} \lambda_{\varphi h} v^2
    )
    + \mqty(
        \Sigma_{hh}(0) & \Sigma_{h\varphi}(0) \\
        \Sigma_{h\varphi}(0) & \Sigma_{\varphi\varphi}(0)
    ),
\end{align}
where $\Sigma(0)=\left.\frac{\partial^2 V^{(1)}}{\partial \varphi_i \partial \phi_j}\right|_{h=v, \varphi=w}$. The running masses are finally obtained as the eigenvalues, 
\begin{align} \label{su2csm:eq:M_eigenvals}
\begin{aligned}
    M_{ \pm}^2\left(p^2\right)= & \frac{1}{2}\left\{\left(3 \lambda_1+\frac{\lambda_2}{2}\right) v^2+\frac{1}{2}\left(\frac{\lambda_2}{2}+3 \lambda_3\right) w^2+\Sigma_{h h}\left(0\right)+\Sigma_{\varphi \varphi}\left(0\right)\right. \\
    & \left. \pm \sqrt{\left[\left(3 \lambda_1-\frac{\lambda_2}{2}\right) v^2-\left(3 \lambda_3-\frac{\lambda_2}{2}\right) w^2+\Sigma_{h h}\left(0\right)-\Sigma_{\varphi \varphi}\left(0\right)\right]^2+4 \lambda_2^2 v^2 w^2}\right\} .
\end{aligned}
\end{align}
It is not clear at first glance which mass eigenvalue corresponds to the Higgs particle and which to the new scalar. We can simplify the expression in eq.~\eqref{su2csm:eq:M_eigenvals} by neglecting terms suppressed by the $\lambda_{\varphi h}, \lambda_\varphi$ couplings, which results in: 
\begin{align}
M_{+}^2(h, \varphi) & =3 \lambda_\varphi \varphi^2+\Sigma_{\varphi \varphi}\left(0\right) \label{eq:M_S}\\
M_{-}^2(h, \varphi) & =3 \lambda_{ h} h^2+\frac{1}{2} \lambda_{\varphi h} \varphi^2+\Sigma_{h h}\left(0\right) \label{eq:M_H}
\end{align}
in the case of 
$\Delta \equiv 3\lambda_h +v^2 - 3\lambda_{\varphi h}\varphi^2 + \Sigma_{h h}\left(0\right) - \Sigma_{\varphi \varphi}\left(0\right) < 0 $. If this $\Delta$ is positive, then $M_{+}, M_{-}$ swap places. 
The form of eq.~\eqref{eq:M_S} suggests that it corresponds to the mass eigenstate $S$, i.e. the new scalar (it only depends on the $\varphi$--related terms), while eq.~\eqref{eq:M_H} is the mass of SM-like mass eigenstate $H$. Thus, for $\Delta < 0$ the new scalar $S$ is heavier, and the mass--order indeed reverses for $\Delta >0$.
Nevertheless, in numerical calculations related to the phase transition in SU(2)cSM, we incorporate the momentum-corrected masses to maximise the diligence of our computations.
The details on treatment of the masses and corrections from non-zero momenta can be found in \cite{Kierkla:2022odc}. 

Then, the mass eigenstates $(\phi^-, \phi^+)$ are a result of rotation of the gauge interaction eigenstates $(h, \varphi)$ by the matrix: 
\begin{align}
    \mqty(\phi^- \\ \phi^+) = 
    \mqty(
        \cos\theta & \sin\theta \\
        -\sin\theta & \cos\theta 
    )
    \mqty(h \\ \varphi)
\end{align}
where for most of the parameter space $(\phi^-, \phi^+) = (H, S)$. The mixing angle $\theta$ is in the range $\{ \frac{-\pi}{2}, \frac{\pi}{2}\}$. Moreover, following \cite{Kierkla:2022odc}, we define the mixing parameters $\xi_H$ and $\xi_S,$ which represent rescaling of scalar couplings with respect to the SM as
\begin{align}
    \xi_H=\left\{\begin{array}{rll}
\cos \theta & \text { for } & M_H \leqslant M_S \\
-\sin \theta & \text { for } & M_H>M_S
\end{array}, \quad \xi_S=\left\{\begin{array}{rll}
-\sin \theta & \text { for } & M_H \leqslant M_S \\
\cos \theta & \text { for } & M_H>M_S
\end{array} .\right.\right.
\end{align}

\subsection{Free parameters of SU(2)cSM and RG running}
\label{sec:su2csm_thermal_scale_running}

The Lagrangian of SU(2)cSM contains all SM couplings (except the SM Higgs potential terms) plus four new ones:
\begin{align}
    \lambda_h &\quad \mbox{- ''Higgs" self-coupling},\\
    \lambda_{\varphi h} &\quad \mbox{- portal coupling},\\
    \lambda_{\varphi } &\quad \mbox{- new scalar self-coupling},\\
    \gX &\quad \mbox{- SU(2)$_X$ gauge coupling }.
\end{align}
In principle, these are the free parameters of the model. However, it is convenient to use the measured values of Higgs vev, $v$, and Higgs mass, $\MH$, to find the values of two parameters, thus leaving only two free parameters. Following the procedure described in \cite{Kierkla:2022odc}, we do that and choose the free parameters to be $\gX$ and $\MX$. The procedure can be summarised as
\begin{enumerate}
    \item  We set the values of $\gx$ and $\MX$ at the scale $\mu=\MX$, as they are free parameters.
    \item We minimise the one-loop potential, which fixes the value of scalar quartic coupling $\lambda_\varphi$.
    \item We use the running couplings to evaluate the vev at the scale $\mu=\MZ$ and then also match the portal coupling $\lamphih$ and Higgs quartic coupling $\lamh$ so it is possible to generate the correct EW vev and Higgs mass.
\end{enumerate}

\begin{figure} 
    \centering
    \includegraphics[width=0.5\linewidth]{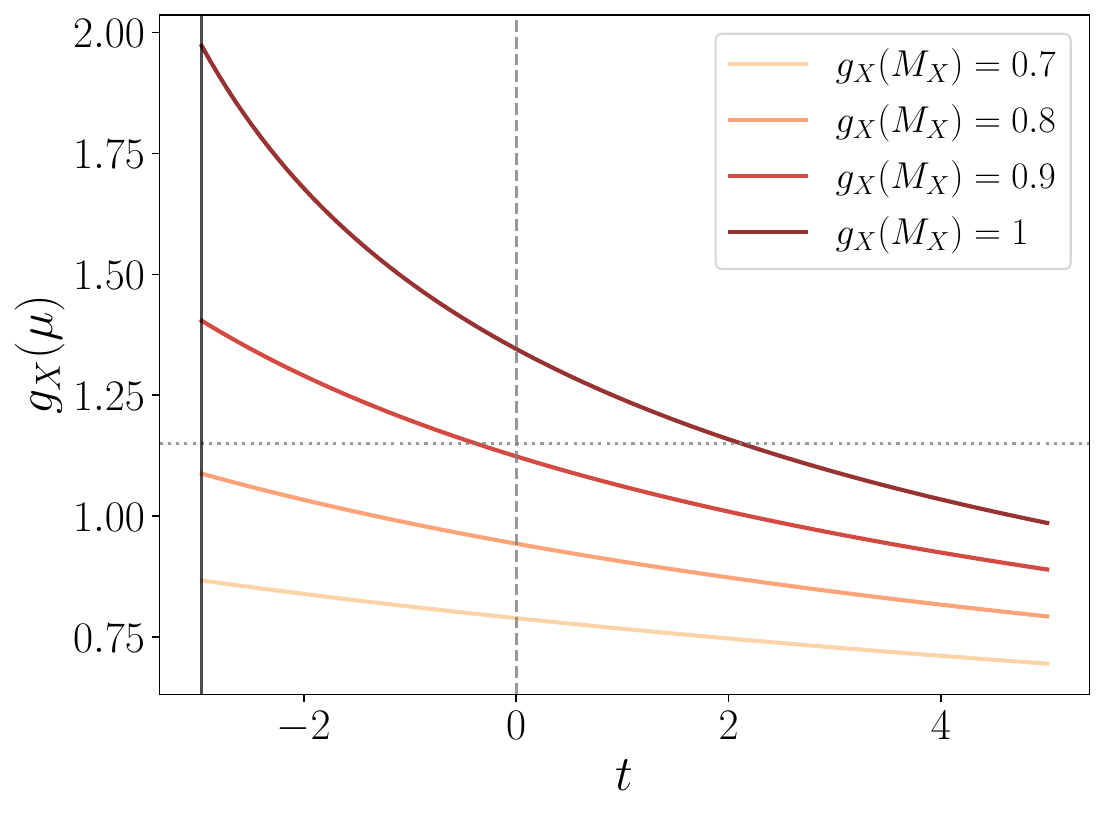}
    \caption{
        Example running of gauge coupling $\gx$. For the curves, we kept a constant mass $\MX = 10000$ GeV, and changed the value of $\gx$ defined at $\MX$ scale. 
        The argument on the $x$-axis is defined as $t\equiv\log\frac{\mu}{\MZ}$. 
        Dashed grey line denotes the $\mu=\MZ$. Dotted line denotes $\gx(\mu)=1.15$. The black solid line in the left part of the plot denotes QCD scale $\mu=\Lambda_{\rm{QCD}} \simeq 0.1$ GeV.
    } 
    \label{su2csm:fig:gx_running_scheme}
\end{figure}

In our setup, the free parameters are defined at the renormalisation scale $\mu= M_X$. Then we use the one-loop $\beta$-functions to evolve all the relevant couplings to the electroweak scale, $\mu=m_Z$, where the masses of scalars are computed. 
Using the running couplings allows us to perform the computations at the scales relevant for the given problem (e.g., scale of thermal fluctuations for the tunnelling, or the vev for the energy budget of phase transition). 
Despite its usefulness, this method must be used with caution. Since the dark sector contains a non-Abelian gauge group, one can expect strong dynamics in the IR energy scales. Therefore, even if the $\gX$ coupling is perturbative at the scale of $\mu=M_X$ (which roughly corresponds to the vev $w$), it may grow large for small energy scales, thus signalling the breakdown of perturbation theory. To avoid this problem, we set a heuristic constraint on $g_X$:
\begin{align} \label{eq:pert-gx}
   \eval{ \gX }_{\mu=\MZ} \leq 1.15,
\end{align}
which is schematically depicted on figure~\ref{su2csm:fig:gx_running_scheme}. Using this constraint\footnote{The value $1.15$ here is not motivated by ``physics'' of this model, but rather it is an estimate that was proved to work well by performing various scans of the model's parameter space.} 
allowed us to study the perturbative part of parameter space using RG-improved effective potential.

\subsection{Parameter space of SU(2)cSM} 
\label{sec:zero-temp}
\begin{figure} 
\center
\includegraphics[width=.4\textwidth]{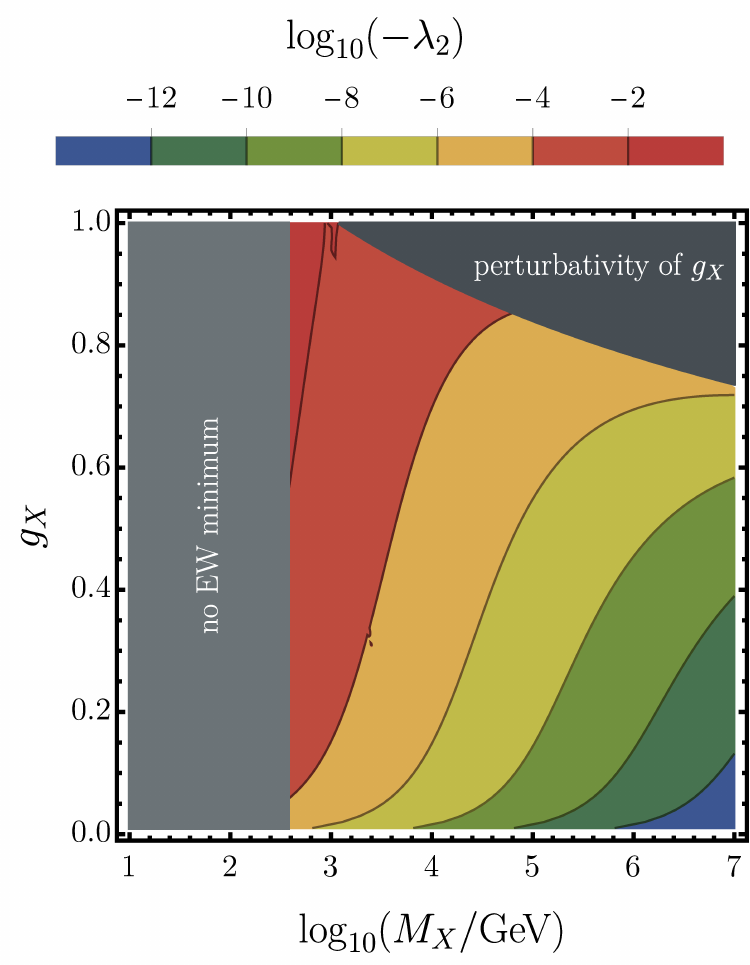}\hspace{20pt}
\includegraphics[width=.4\textwidth]{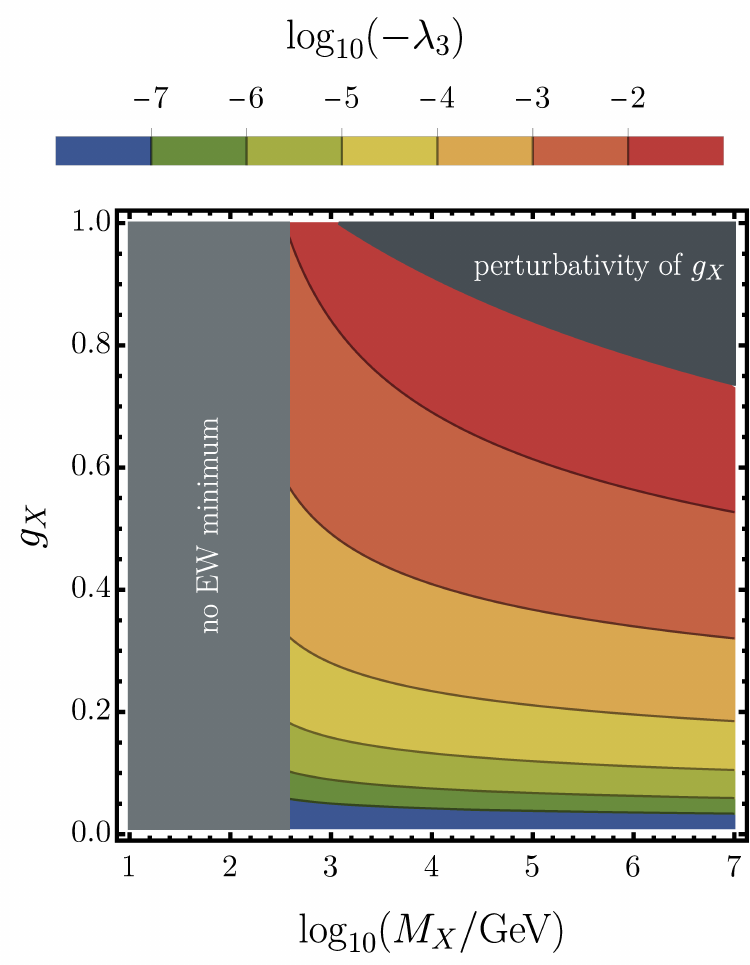}
\caption{Values of the scalar couplings $\lamphih$ (denotes as $\lb$ on left panel) and $\lamphi$  (denoted as $\lc$ on right panel) evaluated at the electroweak scale. Grey shaded regions are excluded, from left to right: no electroweak minimum with correct mass and vev of the Higgs exists, perturbativity of $\gX$ (see eq.~\eqref{eq:pert-gx}). 
}
\label{fig:scalar-couplings}
\end{figure}

Using the procedure we have just outlined, we are able to study the available parameter space of the model. We will first investigate the possible values of the scalar couplings. 
Figure~\ref{fig:scalar-couplings} shows the scalar couplings — the portal coupling, $\lamphih$, on the left and the $\varphi$ self-coupling $\lamphi$ on the right.
The results confirm the scaling we have previously assumed, i.e. they are small (in comparison to SM couplings) and scale as $\sim g^4$. The Higgs self-coupling $\lamh$ is very close to $\lamh = 0.14$, i.e. the SM value at $\mu=\MZ$. 
This behaviour of the portal coupling is especially visible in the case of $M_S > M_H$ mass ordering, as it corresponds to a scenario where the SM sector is very weakly coupled to a SU(2) dark sector, i.e., the portal coupling is much smaller in that region. 
The small deviations from this value of $\lamh$  are visible in the region where $M_S < M_H$.
The excluded parts of the parameter space correspond to the perturbativity constraint that we have introduced before, i.e. eq.~\eqref{eq:pert-gx} (upper-right corner, darker grey) and the region with low $\MX$ masses, where it is not possible to reproduce a stable SM minimum with correct Higgs mass and vev (left, lighter grey).  

\begin{figure}
\center
\includegraphics[width=.4\textwidth]{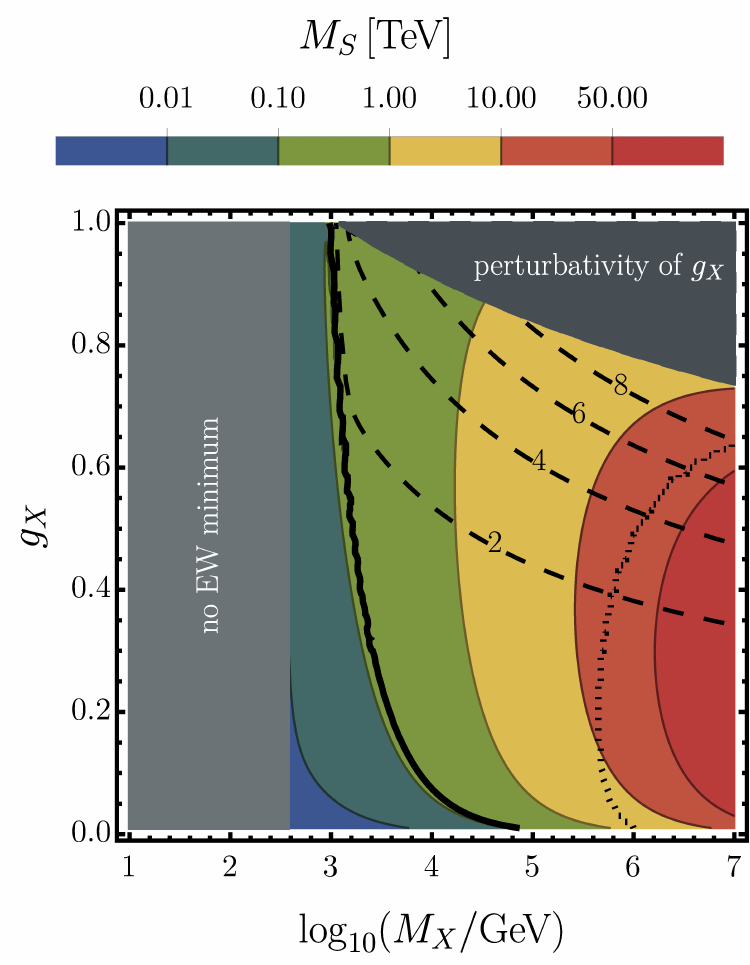}\hspace{20pt}
\includegraphics[width=.4\textwidth]{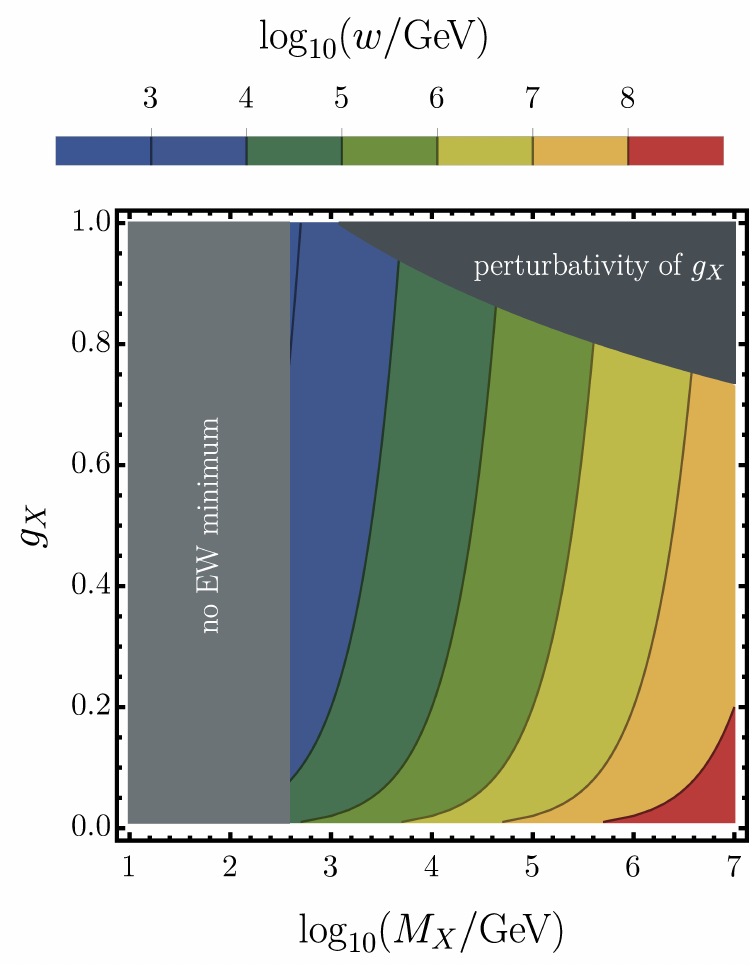}
\caption{Values of the new scalar mass $M_S$ (left panel) and the vev $w$ (evaluated at $\mu=\MX$) (right panel). In the left panel, the thick black line indicates where $M_S=M_H=125\g$ and across this line, mass ordering between $S$ and $H$ changes (to the left of the line $M_S<M_H$, and to the right $M_H<M_S$). To the right of the dotted line, $\xi_H$ becomes numerically equal to 1. The dashed lines indicate a discrepancy between the running and the pole mass (in percent). Grey shaded regions are excluded, see caption of figure~\ref{fig:scalar-couplings}. 
}
\label{fig:MS-w}
\end{figure}

Figure~\ref{fig:MS-w} shows the results of the scans for the new scalar mass, $M_S$, and its vev, $w$. The new scalar is indeed heavier than the Higgs boson for most of the parameter space, in the region on the right of the thick black line. However, on the left of the line, the mass ordering is reversed, and the new scalar is lighter than the Higgs boson. Note that since the self-coupling $\lambda_\varphi$ is small, the new scalar is going to be always lighter than the gauge boson $X$, and we have,
\begin{align}
    \frac{M_S}{\MX} \sim g^2,
\end{align}
which is a general feature of models with radiative symmetry breaking \cite{PhysRevD.7.1888}. 
%
The mixing between states $H$ and $S$ as parametrised by $\xi_H$ is weak, as for the majority of the parameter space $\xi_H = 0.99$ except for the vicinity of the thick black line where the mass ordering changes. Therefore, the experimental constraint on the scalar mixing with Higgs, i.e. $\xi_H >0.95,$ is always fulfilled and does not affect the viable parameter space of SU(2)cSM (see e.g. ref.~\cite{Robens:2021} for recent constraints on the mixing angle in the singlet-extended SM, which apply also to our model).

\subsection{Renormalisation-group improvement of effective potential}
\label{sec:RGimproved_Veff_su2csm}
As we have discussed in the previous chapter, the RG-scale dependence is one of the most important theoretical uncertainties in the calculation of the effective potential. This uncertainty can also impact the description of cosmological phase transitions (see \cite{Croon:2020cgk, Lewicki:2024xan} for a review). 
In principle, this problem is even more pronounced in models with classical scale invariance. 
The reason is that due to the $\lambda \sim g^4$ scaling, there is usually a few orders of magnitude difference between the field values close to the symmetric phase and the global minimum. However, in order to study phase transition, we need to have a good description of the potential along the field ranges in the vicinity of both phases. For this reason, there is no good single choice for a value of the renormalisation scale present in the effective potential if we are interested in transition dynamics. 

To address this issue, following \cite{Kierkla:2022odc} and chapter~\ref{chapter:Seff_vacuum}, we are going to use the RG-improved effective potential. In such a setup, the renormalisation scale dynamically follows the value of the field, and at the same time the couplings run with the same scale -- they become functions of the field as well. Schematically, we now have 
\begin{align}
    g \rightarrow g(t), \quad t\equiv\log\frac{\mu}{\mu_0},
\end{align}
where $\mu_0$ is a reference scale for which we choose the physical mass of the $Z$-boson at the electroweak scale, $\mu_0 = M_Z \simeq 91.18$ GeV \cite{ParticleDataGroup:2024cfk}. The fields are multiplied by the field-renormalisation factors, $Z_h$ and $Z_\varphi$ which also depend on the scale $\mu$,
\begin{align}
    \varphi &\rightarrow \sqrt{Z_{\varphi}(t)}\varphi,\\
    h &\rightarrow \sqrt{Z_{h}(t)} h.
\end{align}
The relevant $\beta$- and $\gamma$-functions needed to obtain the running couplings by solving the RGEs are listed in appendix \ref{app:betafunc_su2csm}. 

The most important part of the effective potential for the phase transition in SU(2)cSM is going to be associated with the $\varphi$ direction. 
Indeed, as it was shown in \cite{Kierkla:2022odc}, the phase transition proceeds in the dark sector and the new scalar alone drives the thermal nucleation.
Therefore, in practice, when studying the dynamics of phase transition in SU(2)cSM, one deals with a one-dimensional effective potential. In such a case, it is convenient to define the renormalisation scale in the following way
\begin{align}
    \mufour \equiv 
    \max\qty(\mxbar(\varphi), \mu_\rmii{IR}) \equiv 
    \max\qty(\frac{1}{2}\gxbar \varphi, \mu_\rmii{IR}),
\end{align}
where $\mu_\rmii{IR}$ is an infrared cut-off, which for the discussion relevant for phase transition is going to be related to the temperature, we will discuss this matter in the next section. Moreover, we have introduced, 
\begin{align}
    \mxbar \equiv \frac{1}{2} \gxbar \varphi = 
    \frac{1}{2}\gx(\MX) \varphi, 
\end{align}
which contains ``frozen'' coupling $\gX$ evaluated at the $\mu=\MX$ scale. The running value of the coupling would itself be $\mu$-dependent, which creates a kind of ``recurrence'' problem. 
To summarise, we will now write the full form of the RG-improved LO effective potential in the $\varphi$ direction, which we will denote as 
\begin{align}
    V_\rmii{RG}(\varphi) \equiv
    \frac{1}{4}\lambda_{\varphi}(t) Z_\varphi(t)^2 \varphi^4 
    + \frac{9 M_X(\varphi,t)^4}{64\pi^2} \qty(
        \log \frac{M_X(\varphi,t)^2}{\mu_4^2} -\frac56
    ),
\end{align}
where $t=\log\frac{\mu_4}{M_X}$, and $M_X(\varphi,t) = \frac12 \gX(t) Z_\varphi ^{\frac12}(t) \varphi$. Effective potential in this form should be used to study the zero-temperature behaviour of the dark sector in this model.

\chapter{Effective theory for supercooled phase transitions}
\label{chapter:supercoolEFT}

In this chapter, we will study the SU(2)cSM at finite temperature. First, we will obtain the effective potential at finite temperature for the SU(2)cSM. We will review the high- and low-temperature regimes of the theory, and finally show the effective field theory approach to the dark SU(2)$_X$ sector. 
Matsbura formalism allows only to capture equilibrium effects, which hints at the fact that the effective potential alone is not an object capable of fully describing bubble nucleation -- a process that is out of equilibrium in its nature. Therefore, in the later sections, we will review the computation of the thermal nucleation rate in SU(2)cSM. We will discuss various approaches and explicitly show how to include the relevant corrections up to NLO.  
This chapter contains original results of this thesis that were also described in refs.~\cite{Kierkla:2023von, Kierkla:2025qyz, Kierkla:2025vwp}

\section{Finite temperature SU(2)cSM}

\subsection{Thermal corrections to the 4d effective potential}
The effective potential is a function of a static, homogenous background field. One can introduce thermal effects to the potential by using the Matsubara imaginary time formalism. The effective potential, in its global minimum, corresponds then to the free energy of a field coupled to a thermal bath of particles, see section \ref{sec:matsubara_formalism}. 
As described in previous sections, the temperature dependence in the effective potential is introduced by the term
\begin{align}
    V_T = \frac{T^4}{2\pi^2}\sum_a n_a J_{b,f} \left(\frac{M_a}{T} \right),
\end{align}
where $n_a$ is the number of degrees of freedom, and the index $a$ goes over particles coupled to the scalar field. So-called thermal functions $J_{b,f}$ are given by:
\begin{align}
    J_{b,f} \left(y \right) = \int^\infty_0 \text{d}x ~x^2 \log\left(
        1\pm \exp\left(-\sqrt{x^2+y^2}\right)
    \right),
\end{align}
where the plus sign corresponds to fermions while the minus sign corresponds to bosons. 
Here, we are interested in the $\SUTwoX$ dark sector only, i.e., a scalar coupled to a gauge boson. The thermal correction term then becomes
\begin{align} \label{eq:VT_Xboson}
    V_T(\varphi) = \frac{9 T^4}{2\pi^2}  J_b\left( \frac{\MX(\varphi, t)}{T} \right).
\end{align}
Note that the mass term here is also RG-scale dependent, due to the running couplings. This expression is valid across all temperatures and field values in the perturbative regime. 
However, as described in previous sections, in the high-temperature regime, i.e., when the temperature becomes larger than the mass scale, perturbation theory suffers from the Linde problem. Therefore, due to the slower convergence of perturbative expansion, a more careful treatment is necessary, for example, by employing various resummations or using high-temperature dimensional reduction.
Here we will start by discussing the daisy-resummation approach (see chapter \ref{chapter:TFT}), which we will later compare against a more advanced framework. 
The ``daisy resummation'' term in SU(2)cSM is given by: 
\begin{align}
\label{eq:daisy}
    V_\text{daisy}(\varphi, T) =
    -\frac{3T}{12\pi} \qty( 
         m_\rmii{$X$,th} ^3(\varphi, t) - \MX^3(\varphi, t)
    ).
\end{align}
Physically, this term affects the temporal gauge mode that has now gained a ``thermal mass''. This contribution is called the Debye mass, $\mD$,
\begin{align}
    m_\rmii{$X$,th}^2 = \MX(\varphi)^2 + \mD^2(T), \quad
    \mD^2(T) = \frac{5}{6} \gx^2 T^2.
\end{align}
Note that in our power counting (see section \ref{sec:power_counting_T0}), the diagrams involving scalar loops are beyond the LO, and thus only the resummation of the non-zero Matsubara modes of the gauge temporal mode is necessary.  
Following chapters \ref{chapter:TFT} and \ref{chapter:su2csm}, the temperature-dependent, RG-improved effective potential with daisy-resummation term takes the following form
\begin{align}\label{eq:V4_T}
\begin{split}
    V_\rmii{eff}^\rmii{daisy} &\equiv 
    V_\rmii{eff}^\rmii{LO}(\varphi, T=0) + V_T(\varphi, T) + V_\text{daisy}(\varphi, T) \\
     &=\frac{1}{4}\lambda_{\varphi}(t) Z_\varphi(t)^2 \varphi^4 
    + \frac{9 \MX(\varphi,t)^4}{64\pi^2} \qty(
        \log \frac{\MX(\varphi,t)^2}{\mufour^2} -\frac56
    ) \\
    &\quad+\frac{9 T^4}{2\pi^2}  J_b\left( \frac{\MX (\varphi, t)}{T} \right)
    -\frac{3T}{12\pi} \qty( 
        m_\rmii{$X$,th}^3(\varphi, t) - \MX^3(\varphi, t)
    ),
\end{split}
\end{align}
where the daisy term should be included only inside the high-temperature regime, as we will discuss in more detail in the next section. 
Such a form of the effective potential with the daisy term included is the most commonly used approach in the phenomenological studies of phase transitions. It can capture the general physical behaviour of the system; however, it suffers from many theoretical uncertainties (see e.g.~\cite{Croon:2020cgk} for a review). The reason is that it does not capture all relevant corrections in the expansion in powers of the gauge coupling. 
In this thesis, we will demonstrate that this approach is also not sufficient for a robust study of supercooled transition dynamics and its phenomenological predictions \cite{Kierkla:2023von, Kierkla:2025qyz, Kierkla:2025vwp}. 

\subsection{High- and low-temperature regimes and IR cut-off}
\label{sec:HT-LT-regimes}

\begin{figure}
    \centering
    \includegraphics[width=0.5\linewidth]{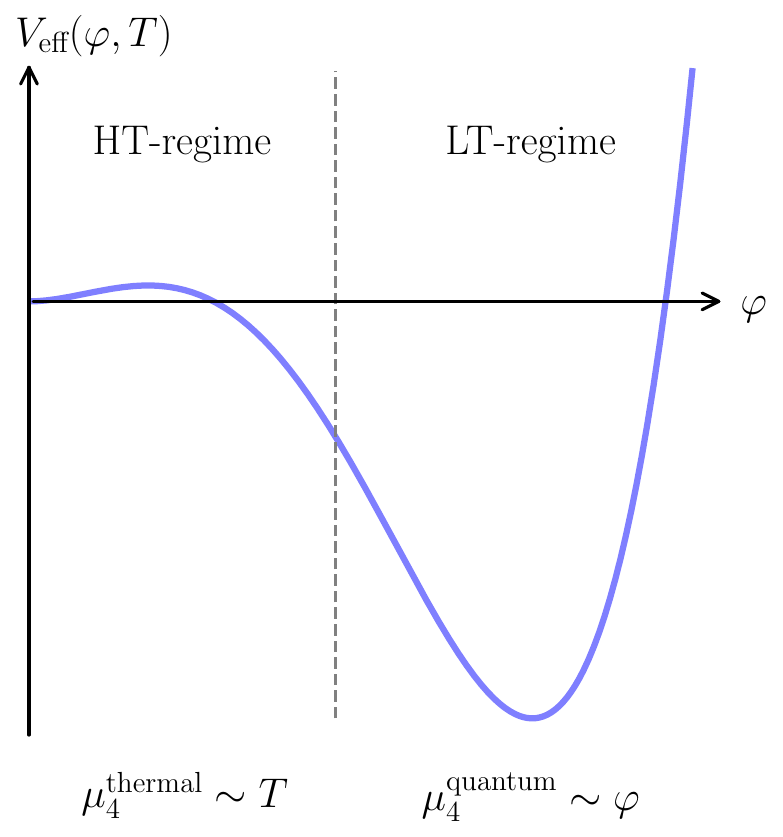}
    \caption{Schematic depiction of high- and low-temperature regimes in a model with classical scale invariance.}
    \label{fig:HTLT_regimes}
\end{figure}

In the previous section, we mentioned that daisy-resummation should be added only for the field values in the high-temperature regime. 
In general, the ratios of the field-dependent masses to the temperature -- arguments of thermal functions -- determine whether the high-temperature (HT) or low-temperature (LT) limit should be considered for constructing the effective potential. 
Large field values correspond to large masses, and thus if we have $M_a > T$, 
then the potential should be computed using the LT limit (and without daisy corrections), while small field values can correspond to the HT limit if $M_a < T$. 
In models with classical scale invariance, which feature supercooled phase transitions, the scales associated with the global minimum of the potential (the location of which ultimately determines the energy difference between two phases) and with the location of the barrier (relevant for nucleation) are widely spread (see e.g.~\cite{Kierkla:2023von, Kierkla:2025qyz, Kierkla:2025vwp}). 
Therefore, we cannot naively construct the effective potential in just one of the limits. To have the full description of the supercooled transition, we need to consider two regimes, see fig.~\ref{fig:HTLT_regimes}.
In classically scale-invariant models, such as SU(2)cSM, the potential around the global minimum, for temperatures much below the critical temperature, is indeed in the low-temperature regime. In practice, this means that we can use the one-loop thermally corrected potential without worrying about the Linde problem. 
Specifically, we do not need to include any resummations to compute, e.g. the energy stored in metastable vacuum \cite{Kierkla:2023von}. We will discuss the explicit calculations of phase transition parameters in detail in the next chapters. 

In principle, in the presence of large field ranges, we should use the RG-improved effective potential, which effectively resums the large logarithms, and keeps the effective potential perturbative over a wide range of field values, as we have shown in chapter \ref{chapter:Seff_vacuum}. 
For theories in which the one-loop radiative corrections to the potential are dominated by a single mass scale $M(\field)$ -- such as for gauge mass in the case of radiative symmetry breaking -- the construction of the RG-improved potential is straightforward. We can achieve that by making the RG scale field-dependent, $\mu=M(\field)$, while ensuring that all couplings also run with this scale. 
However, going from the large field values to lower field values, at some point the ratio $M(\field)/T$ becomes small, which signals the onset of the high-temperature regime. Then the choice of RG-scale becomes modified \cite{Kierkla:2023von}. Let us now demonstrate this feature explicitly for the SU(2)cSM. 

First, note that inside the high-temperature regime, we may use the HT expansion of the bosonic thermal function, 
\begin{align}
\label{eq:Jb_X_HT}
J^\rmii{HT}_{b}\qty(\frac{\MX}{T})= 
&-\frac{\pi^4}{45} 
+ \frac{\pi^2}{12} \qty(\frac{\MX}{T})^2 
- \frac{\pi}{6} \qty(\frac{\MX}{T})^{3} 
- \frac{1}{32} \qty(\frac{\MX}{T})^4 \left( 
    \log\frac{\MX^2}{16 \pi^2 T^2} -\frac{3}{2} + 2 \gamma_E \right) + \nonumber\\
&+ \pi^2 \qty(\frac{\MX}{T})^2 \sum_{i=2}^{\infty}\left(-\frac{1}{4 \pi^2} \qty(\frac{\MX}{T})^2 \right)^
i \frac{(2 i - 3)!! \zeta(2 i - 1)}{(2 i)!! (i + 1)}, 
\end{align}
which allows us to write the thermal correction term for the $X$-boson as
\begin{align}
\label{eq:VTHT}
\begin{split}
    V_T ^{\rmii{HT}}({\varphi, T}) \simeq 
    &- \frac{9 \pi^2 T^4}{90}
    +\frac{9 T^2}{24}  \MX(\varphi, t)^2
    -\frac{9  T}{12\pi} \MX(\varphi, t)^3\\
    &-\frac{9 \MX(\varphi, t)^4}{64\pi^2 T^4} \left[
        \log\qty(\frac{\MX(\varphi, t)^2}{16\pi^2 T^2})  -\frac32 +2\gamma_E
    \right],
\end{split}
\end{align}
where we have dropped higher-order terms containing field operators of dimension larger than four. This approximation is well behaved as long as the HT expansion works well, since these higher-order operators are suppressed by powers of temperature.\footnote{Recently, there were studies of the effect of higher order operators on the 3d EFT and the phase transition, see refs.~\cite{Bernardo:2025vkz, Chala:2024xll}.} 
Keeping that in mind, the total effective potential in the HT regime can now be written as,
\begin{align}
\label{eq:V4HT}
\begin{split}
    V_4 ^\rmii{HT}(\varphi, T) &\equiv 
    V_4(\varphi, T=0) + V_T^\rmii{HT}(\varphi, T) + V_\text{daisy}(\varphi, T) \\
     &=
     - \frac{9 \pi^2 T^4}{90}
     +\frac{9 T^2}{24}  \MX(\varphi, t)^2
     -\frac{9  T}{12\pi} \MX(\varphi, t)^3
     +\frac{1}{4}\lambda_{\varphi}(t) Z_\varphi(t)^2 \varphi^4
     \\
    &\quad + \frac{9 \MX(\varphi,t)^4}{64\pi^2} 
    \qty(
        L_b(\mu_4) 
         +\frac23 
    )
    -\frac{3T}{12\pi} \qty( 
        m_\rmii{$X$,th}^3(\varphi, t) - \MX^3(\varphi, t)
    ).
\end{split}    
\end{align}
Note the similarity to eq.~\eqref{eq:Veff_daisy_phi4}. 
We can observe here that the resulting logarithm now contains a ratio of temperature to the RG-scale. Therefore, to minimise this logarithm, we no longer should make the RG-scale field-dependent, but rather set it as 
\begin{align}
    \mu_4 = \kappa T,
\end{align}
where $\kappa$ is some $\order{1}$ constant. The logarithmic term becomes then
\begin{align}
    \log\frac{16\pi^2}{\kappa^2}.
\end{align}
Then, a convenient choice for the RG-scale in the HT regime is the physical scale associated with the heaviest non-zero Matsubara mode, i.e. the hard scale, $\mu_4 =\pi T$. Therefore, we choose $\pi T$ for the value of the IR cut-off scale $\mu_{\rm IR}$ we have introduced in \ref{sec:RGimproved_Veff_su2csm}. The running of the renormalisation scale at finite temperature in SU(2)cSM thus becomes,
\begin{align} \label{eq:mu4_final}
    \mu_4(\varphi) = 
    \max\qty(\mxbar(\varphi), \pi T).
\end{align}
Finally, let us discuss the estimate of the field value where we enter the HT regime. 
Indeed, the HT expansion of the thermal functions works well for a small ratio $\MX /T \ll 1$. However, the total effective potential can still be approximated well in the HT expansion even when $\MX/T \sim \mathcal{O}(1)$ \cite{Kierkla:2023von}. 
Thus, we may use the RG thermal cut-off, i.e. the hard scale, also as a criterion for the high-temperature regime,
\begin{align}
\frac{\MX(\varphi_\rmii{HT})}{\pi T} &= 1, \\
\Rightarrow \quad
\varphi_\rmii{HT} 
&\simeq \frac{2}{\gx} \pi T,
\end{align}
where we have assumed $\sqrt{Z(\varphi_\rmii{HT})} \simeq 1$. For the discussion of the field renormalisation, see appendix \ref{sec:rescaling-field-4D}.

The goal of the discussion in this section was to scrutinise whether high-temperature effects may also affect models with CSI (or with a supercooled transition in more general terms). Our analysis shows that indeed, such an effect can affect the description of thermal nucleation, which is crucial for the correct calculation of phase transition parameters and resulting phenomenological observables. Moreover, a watchful reader might have noticed that the discussion here was similar to the description of dimensional reduction in chapter \ref{chapter:TFT}. This is not a coincidence, as now our goal is to investigate the higher-order thermal effects using the framework of effective field theory.


\subsection{High-temperature effective theory}
As discussed previously, thermal field theory suffers from poor convergence of the perturbative expansion, which is now not simply organised in the number of loops. This issue impacts the precision with which the parameters of the phase transition are calculated. 
Even though daisy resummation correction, see eq.~\eqref{eq:daisy}, allows for taking into account a leading set of IR-sensitive diagrams, and is correct at $\mathcal{O}(\gx^3)$, a problem persists: parametrically large $\mathcal{O}(\gx^4)$ contributions are still missing \cite{Kierkla:2023von}. 
These corrections affect calculations of phase transition parameters drastically, even in the case of supercooled phase transitions as we will demonstrate (see also \cite{Croon:2020cgk, Kierkla:2023von, Kierkla:2025qyz, Kierkla:2025vwp}). 
In particular, stopping at $\mathcal{O}(\gx^3)$ implies an uncancelled RG-scale dependence due to the omission of two-loop thermal masses, and additional resummations are required at the same order.%
\footnote{
These additional resummations generate contributions to the couplings, as well as momentum-dependent field normalisation contributions~\cite{Kajantie:1995dw}. 
}
The root of both problems lies in the Bose enhancement of the low-energy modes, resulting in an enhancement of the effective parameters of these modes \cite{Laine_2016}.

A systematic way to deal with the problems mentioned above is to construct a series of effective field theories describing the thermodynamics at the different relevant energy scales. See ref.~\cite{Gould:2023ovu} for a recent discussion of the possibly relevant scales. For our purposes, it suffices to distinguish the two energy scales, the hard scale and soft scale, see table~\ref{tab:scales}. Let us now discuss the construction of the 3d EFTs following the discussion in chapter \ref{chapter:TFT}.
\begin{table}[h]
\centering
\begin{tabular}{lcc}
\bf{Name} & \bf{Energy scale}  & \bf{Scaling of expansion parameter}   \\
Hard & $\pi T$ & $\frac{g^2}{\pi^2}$   \\
Soft & $gT$ &  $\frac{g}{\pi}$  \\
\end{tabular}
\caption{
Relevant energy scales for the SU(2)cSM model at finite temperature. Parameter $g$ denotes the largest relevant coupling in the theory, in our case the dark gauge coupling $\gx$.
}
\label{tab:scales}
\end{table}

\paragraph{Hard scale}
For the construction of the EFTs, we make use of the partition function given by 
\begin{align}
\label{eq:partition-function}
\mathcal{Z}^{\text{hard}} = \int D\Phi \; 
\exp\left( 
    -\int\limits_0^{1/T} \dd\tau \int \dd^3\mathbf{x} \; \mathcal{L}_E \right).
\end{align}
Here, we remind that $\tau$ is the imaginary time coordinate, and its periodicity is set by the inverse of the temperature, $\beta=\frac1T$, in the units where $k_B=1$. 
Here, the functional integration $\int D\Phi$ is performed over all fields in SU(2)cSM, while $\mathcal{L}_E$ denotes the Euclidean space Lagrangian density. 
In imaginary time formalism, the fields can be written as a sum over momentum modes, with momenta $P = (\omega_n, \mathbf{p})$, and with the Matsubara frequency $\omega_n = 2\pi n T$ for bosons and $\omega_n = (2n+1)\pi T$ for fermions. 
The theory described by eq.~(\ref{eq:partition-function}) contains all momentum modes, but modes with masses larger than $\pi T$ get Boltzmann-suppressed, so we see that the largest relevant energy scale in the HT regime is the hard scale of $\mathcal{O}(\pi T)$.

\paragraph{Soft scale EFT}
Since at high temperatures non-zero Matsubara modes are Boltzmann-suppressed, the only relevant modes for the phase transition dynamics are the zero Matsubara modes.
We are interested in the dark $\SUTwoX$ sector of the model; therefore, we want to obtain an effective description of the dynamics of the gauge boson $X$ and the scalar $\varphi$ zero modes. We can do that by integrating out all the non-zero modes, i.e. those with frequencies $\omega_n \geq \pi T$, which results in an effective field theory for the zero modes. The relevant energy scale will now be the soft scale, see table~\ref{tab:scales}. It is given by the Debye mass $\mD \sim \gx T$. 
The resulting partition function of the 3d soft EFT, which now contains only gauge and scalar fields, is given as:
\begin{align}
    \mathcal{Z}^{\text{soft}} _3 =
    \int \mathcal{D}\Phi_{n=0} ~e^{-\int_\textbf{x} S^{\text{soft}}_3 }, 
\end{align}
where the path integral is only over the zero Matsubara modes, and the action $S^{\text{soft}}_3$ reads:
\begin{align}
\label{eq:EFT-action}
S^{\rmii{soft, dark}}_3 =& \;  \intx  \; 
\Big\{
    \frac{1}{4} F^a_{ij} F^a_{ij} 
    + (D_i \phi)^\dagger (D_i \phi) 
    + \frac{1}{2} (D_i \Xzero^a)^2 
    + V_3(\phi, \Xzero^a) 
\Big\} + f_0.
\end{align}
Note that the $1/T$ factor coming from the integral in the temporal dimension is now absorbed into the 3d Lagrangian, such that the exponent is indeed dimensionless. 
Thus, inside this EFT, the fields are static and three-dimensional; they have no momentum in the $\tau-$direction. The remaining coefficient $f_0$ in the action is a unit operator which is related to pressure in the symmetric phase \cite{Kierkla:2023von, Ekstedt:2024etx}. Let us now discuss the field content of the 3d EFT.

The field $\phi$ is now the 3d scalar field with dimension $[T^{\frac12}]$. Then, the gauge field strength tensor is computed for the spatial gauge modes $\Xspat ^a$. The temporal gauge mode $\Xzero^a$ can be seen as another scalar field that is a triplet of the dark SU(2). Index $i$ corresponds to spatial Lorentz indices, while $a=1,2,3$ denotes the SU(2) index. The covariant derivatives are defined as:
\begin{align}
    D_i \phi &= \partial_i \phi -ig_3 \frac{\tau^a}{2} \Xspat^a \phi, \\
    D_i X_0^a &= \partial_i \Xzero^a +g_3 \varepsilon^{abc} \Xspat^b \Xzero^c .
\end{align}
The tree-level potential of the soft-EFT is a function of $\phi$ and $X_0^a$:
\begin{align}
\label{eq:VEFT-soft-tree}
V_3^{\rmii{soft, tree}}(\phi, X^a_0) &= 
\frac12 \mthree^2 \phi^\dagger \phi + \frac14 \lambda_3 (\phi^\dagger \phi)^2 + \sum_{n=3}^\infty c_{2n} (\phi^\dagger \phi)^n \nonumber  \\
&+ \frac{1}{2} \mDX \Xzero^a \Xzero^a 
+ \frac{1}{4} \kappathree (X^a_0 X^a_0)^2 
+ \frac{1}{4} \hthree \phi^\dagger \phi  \Xzero^a \Xzero^a.
\end{align}
Note that the couplings $\lambda_3, \kappathree, \hthree$ and masses $\mDX, \mthree$ have now dimension $[T]$. 
Higher-order operators in the soft EFT for the doublet $\phi$ with coupling constants $c_n$ correspond to the terms containing the $\zeta$-terms in eq.~\eqref{eq:Jb_X_HT}. We use these marginal operators only as an indicator of a breakdown of the HT expansion: when their effect becomes non-negligible at low temperature, or more importantly, at large field values, the HT expansion starts to break down \cite{Kierkla:2023von}. For a discussion of the higher-order operators in the 3d EFT, see refs.~\cite{Bernardo:2025vkz, Chala:2024xll, Chala:2025xlk}.
The parameters of the 3D theory are obtained by a matching procedure, as we have described in chapter \ref{chapter:TFT}. 
Here, we use {\tt DRalgo}~\cite{Ekstedt:2022bff} for the determination of the parameters of the soft scale EFT for the SU(2)cSM model \cite{Kierkla:2023von}. 
The explicit matching relations between 3d couplings and 4d ones are collected in the appendix \ref{app:DR_matching}. 
The momentum-dependent field normalisation contributions of the hard modes are absorbed into the parameters of the 3d EFT for all fields, rather than including $Z$-factors in the kinetic terms of the soft scale EFT action~\cite{Kajantie:1995dw}. 
For illustration of the matching procedure, see e.g.\ refs.~\cite{Kajantie:1995dw, Schicho:2021gca, Hirvonen:2021zej}, or appendix B.1 in ref.~\cite{Croon:2020cgk}. For more details of the matching procedure in SU(2)cSM see refs \cite{Kierkla:2023von, Kierkla:2025vwp}.
The construction of the 3d EFT is performed in the symmetric phase and relies on the high-temperature expansion for the matching, which assumes that $m/T \ll 1$ for all the fields. 
In models with classical scale invariance, all fields are massless in the symmetric phase, so they are explicitly in the HT regime; however, deep in the broken phase, the HT expansion is broken, as discussed in section~\ref{sec:HT-LT-regimes}.

After these remarks, we can now define a background field of the scalar field\footnote{Note that this background, in general, does not have to be constant.}
\begin{equation}
\phi = \frac{1}{\sqrt 2} \begin{pmatrix} 0 \\ \vthree \end{pmatrix}.
\end{equation}
The masses for the spatial and temporal gauge bosons can then be defined as
\begin{equation} \label{eq:mass-gauge}
\mXthree^2 = \frac 1 4 \gxthree^2 \vthree^2, \qquad \mXtemporal^2 = \mDX^2 + \frac 1 2 \hthree \vthree^2. 
\end{equation}
The tree-level potential for the background field $\vthree$ becomes
\begin{equation}
\label{eq:Veff:softTree}
V_3^{\rmii{soft, tree}}(\vthree) = \frac 1 2 \mthree^2 \vthree^2 + \frac 1 4 \lamthree \vthree^4 ,
\end{equation}
where we have dropped the higher-order operators. As we have written above, these operators are relevant only for field values such that high-temperature expansion breaks down -- they correspond to $\zeta$-terms in the high-T expansion and thus inside the HT region of our interest, these operators can indeed be neglected \cite{Kierkla:2023von}.

\subsection{Nucleation EFT and its validity}
\label{sec:nuclEFT_valid}
Note that since the 4d quartic coupling is negative, its 3d counterpart will also be negative. However, the 3d mass term is not positive due to thermal corrections. 
Thus, by their interplay, the potential given above in eq.~\eqref{eq:Veff:softTree} already contains a barrier. One might be tempted to proceed with the calculation of the nucleation rate using the potential \eqref{eq:Veff:softTree}; however, this would not be fully consistent with the power counting, $\lambda \sim g^3$, relevant for the first-order phase transitions \cite{Kierkla:2023von, Gould:2023ovu}. 
Integrating out gauge modes will result in a term of order $ \mathcal{O}(g^3)$, which is then of the same order as the quartic coupling. Therefore, it should be included in the leading order effective potential and contribute to the bounce solution. 
In fact, in the case of non-conformal models, it is a generic feature of gauge-Higgs type of theories -- the barrier arises from the gauge modes' radiative corrections. The tree-level terms must be affected by corrections. This then implies that to have a thermal first-order phase transition, there must be a mass hierarchy between the nucleating field and other fields, such as gauge fields \cite{Gould:2021ccf}.\footnote{Another possibility of affecting tree-level terms, leading to a first-order transition, would be the presence of a strong coupling \cite{Gould:2021ccf}.}

Integrating out the gauge modes formally results in a new, \textit{nucleation EFT}. Considering gauge corrections up to the one-loop order, i.e. considering only quadratic fluctuations, and keeping terms up to $\order{\gxthree^3}$, the LO effective potential becomes \cite{Kierkla:2023von}:
\begin{align}
\label{eq:Veff-EFT-LO}
\VLO (\vthree) =&\ 
\frac 1 2 \mthree^2 \vthree^2 + \frac 1 4 \lamthree \vthree^4  - \frac{1}{12\pi} 
\Big[ 6 (\mXthree)^{3} + 3 (\mXtemporal)^{3} \Big] 
.
\end{align}
This potential can be connected directly to the 4d effective potential with daisy resummation in eq.~\eqref{eq:V4HT}. The only difference here lies in the form of effective 3d parameters -- after truncating the EFT matching, we can recover fully the form in eq.~\eqref{eq:V4HT}. We show explicitly this truncation in appendix \ref{app:sec:compare3D4D}.
The nucleation EFT is only valid if the integrated gauge modes are heavier than the nucleating scalar field. This hierarchy could be obeyed only for a certain range of background field values, as the gauge spatial modes' mass is proportional to the scalar field background. This means that gauge modes change their energy scale all the way from hard scale to the non-perturbative ultrasoft scale -- they are \textit{scale-shifters} \cite{Gould:2021ccf}. 
Thus, as long as gauge modes are heavier than the scalar for the considered background field values, nucleation EFT remains valid. However, the nucleation EFT inevitably becomes non-local close to the symmetric phase, at some value $\vbreak$ where the scalar field background approaches zero and spatial gauge modes become lighter than the scalar, and eventually massless. 
%

Fortunately, the error introduced from the scale-shifters for $\vthree < \vbreak$ region is subleading and the contribution to the total action from these field values is negligible. 
This was illustrated first in \cite{Gould:2021ccf} for a so-called ``cubic anisotropy model'' and here we will follow their argument, adapting it to the SU(2)$_X$ gauge sector. 
Let us estimate the field value $\vbreak$. To do that, we will consider the LO effective potential for the scalar field background values $\vthree \ll \mDX$, where we have $\hthree \vthree \rightarrow 0$:
\begin{align}
\label{eq:VnuclLO_se_nuclEFT}
    \VLO \simeq 
    &\ \frac 1 2 \mthree^2 \vthree^2 + \frac 1 4 \lamthree \vthree^4  
    - \frac{1}{12\pi} 
    \Big[
        \frac{6}{8} (\gxthree \vthree)^{3} + 
        3\mD^3
    \Big] 
.
\end{align}
To estimate $\vbreak$, it is useful to consider the contributions to the derivative of the effective potential at the top of the barrier, $\vbarrier$. There, all contributions must cancel out, and we have
\begin{align}
    \mthree^2 \vbarrier - \abs{\lamthree} \vbarrier^3 
    -\frac{1}{4\pi} \qty[
        \frac68 \gxthree^3 \vbarrier^2 
    ] = 0.
\end{align}
Then we can express the nucleating mass $\mthree$ as:
\begin{align}
\label{eq:mthree_vbarrier}
     \mthree  
     = 
     \frac{3\gxthree ^3 \vbarrier } {16 \pi \mthree} 
     + \frac{\abs{\lamthree}}{\mthree} \vbarrier^2
\end{align}
We know that at the $\vbreak$ we have
\begin{align}
    \mXthree ^2 (\vbreak) &= 
    \mthree ^2 \\
    \Rightarrow \vbreak &= \frac{2 \mthree}{\gxthree}.
\end{align}
Now, using eq.~\eqref{eq:mthree_vbarrier} we can express $\vbreak$ in terms of $\vbarrier$ as:
\begin{align}
\label{eq:vbreak_estimate}
    \vbreak = \qty(
        \frac{3\gxthree^2  } {8 \pi  \mthree} 
    ) \vbarrier
    +\frac{2\abs{\lamthree} }{\gxthree \mthree} \vbarrier^2
    \simeq 
    \qty(
        \frac{3\gxthree^2  } {8 \pi  \mthree} 
    ) \vbarrier
    .
\end{align}
Now, let us notice that if the nucleating field is perturbative in the false vacuum, then \cite{Gould:2021ccf}:
\begin{align}
    \frac{3\gxthree^2  } {8 \pi  \mthree} \ll 1
\end{align}
which means that as long as perturbation theory works, then $\vbreak \ll \vbarrier$ and we are justified to use nucleation EFT to describe nucleation rate in $\SUTwoX$ sector.
All the terms in the effective actions are proportional to $\vthree$. 
Thus, if we have $\vbreak \ll \vbarrier$, then the breakdown of EFT introduces a negligible error to the action.

It is worth reminding (see chapter \ref{chapter:TFT}) that in non-conformal models, 
there is a partial cancellation between 4d tree-level mass $\mu_{4\rm d}$ and the thermal corrections for the scalar
\begin{align}
   \mthree ^{\rmii{supersoft}} \simeq - |\mu_{4\rm d}|^2 + \# gT^2. 
\end{align}
This cancellation brings the scalar field down to a lower energy scale -- the \textit{supersoft} scale. 
This makes the scalar much lighter than gauge modes outside the symmetric phase. Thus, the nucleation EFT is valid for a larger range of background field values.
In contrast, in the case of conformal models, there is no 4d tree-level mass, and the $m_3$ mass has only a purely thermal contribution $\sim gT^2$, which means that the scalar field belongs to the soft scale. Thus, the breakdown of nucleation EFT occurs earlier in models with CSI.



\section{Nucleation rate in SU(2)cSM}

With the nucleation EFT ready, we are ready to calculate the thermal nucleation rate for the $\SUTwoX$ sector. In this section, we will illustrate the general strategy and show explicitly how to obtain the necessary ingredients using perturbation theory. 

First of all, recall from section \ref{sec:T_bubble_nucl} that thermal bubble nucleation rate $\GammaT$ can be factorised into dynamical and statistical parts (for a review see also e.g.~\cite{Gould:2021ccf, Ekstedt:2022tqk})
\begin{align}
    \GammaT = \Astat \cross \Adyn.
\end{align}
The statistical part captures equilibrium effects, and its computation can be performed within the Matsubara formalism \cite{Laine_2016}.
During the examination of the statistical part, we will introduce an operator for scalar-field, $\phi$, fluctuations, which will also be connected to the dynamical part. 
The dynamical part itself captures out-of-equilibrium dynamics and is associated with the negative mode of the fluctuation operator. 
Thus, we shall focus on the statistical part first and describe the dynamical part later.

To find the statistical part of the thermal nucleation rate, we need to calculate the effective action in the 3d nucleation EFT we have defined (see ref.~\cite{Ekstedt:2022tqk} for a review). In the saddle-point approximation for nucleation rate, this effective action should then be evaluated on a critical bubble. The bubble is a particular background field configuration corresponding to the solution of Euclidean equations of motion, with imposed $O(3)$ symmetry -- the bounce solution \cite{Coleman:1977py, PhysRevD.16.1762}.  Denoting the solution as $\varphib$, we can then express the statistical part as
\begin{align}
    \Astat = A e^{-\Snucl[\varphib]},
\end{align}
where $A$ here contains normalisation factors and scalar fluctuation determinant.
To calculate this expression, we will resort to perturbation theory; we will expand both the effective action as well as the background. Then, the action evaluated on the saddle-point solution up to the next-to-leading order is given as \cite{Ekstedt:2021kyx,Ekstedt:2022ceo,Ekstedt:2022tqk}
\begin{align}
    \Snucl [\varphi_b] = 
    \underbrace{\SLO[\varphibLO]}_{\mathcal{O}(g^3)} + 
    \underbrace{\SNLO[\varphibLO]}_{\mathcal{O}(g^4)} + \cdots,
\end{align}
where the power-counting parameter $g$ corresponds to $\gx$ gauge coupling in our setup. Thus, one is capable of calculating the NLO action using only the LO bounce solution \cite{Ekstedt:2021kyx, Lofgren:2021ogg, Hirvonen:2021zej}.\footnote{
This expansion can be understood as a semi-classical expansion, where the LO bounce solution is a classical background and we evaluate quantum corrections on top of it. Nevertheless, in principle, the bounce field configuration also obtains quantum corrections in the form of ``dumbbell diagrams'' \cite{Gould:2021ccf}, which here are assumed to be small and of higher order than $\mathcal{O}(g^4)$.
} 
Further on, we shall refer to this expansion as \textit{soft expansion}.
Our goal for the next sections is to obtain this effective action up to  NLO order and calculate the statistical part of the nucleation rate. We will also show how to obtain the scalar fluctuation determinant within soft expansion.

\subsection{Nucleation rate at one-loop}
First, let us start with obtaining one-loop corrections within soft EFT.
The best-suited method of computing the 1PI effective action for our purposes is the background field method \cite{Andreassen:2016cvx, Abbott:1981ke, Schwartz:2014sze} we have introduced in chapter \ref{chapter:Seff_vacuum}.
After all, this is exactly what we need to achieve, keeping in mind that our background is eventually going to be an $O(3)$-symmetric solution to the Euclidean equations of motion. 
In the first step, we will compute one-loop corrections, i.e. Gaussian fluctuations on the general background.

We will denote a general inhomogeneous background field as $\phi_b(x)$. We can expand the scalar field  as
\begin{align}
    \phi(x) \longrightarrow \frac{1}{\sqrt{2}} 
    \mqty(
        \phi_1 + i\phi_2 \\
        \phib(x) +  h(x) + i\phi_4
    ),
\end{align}
where $h(x)$ is a fluctuation on top of the general background $\phi_b(x)$ and other fields are Goldstone bosons. Then the effective action can be obtained by integrating out all the fluctuations
\begin{align}
    e^{-\Seff[\phi_b(x)]} \equiv 
    \int \pathD{\phi_i} \pathD{h} \pathD{X^a} \pathD{X^a_0}  
    ~e^{-S^{\text{soft}}_3 [\phib , h]},
\end{align}
where $i=1,2,4$ denotes Goldstone modes.
The action expanded on the background takes the following form (we omit the terms that do not depend on the scalar field here):
\begin{align}
    \SthreeSoft [\phib, h] =& \intr
    \frac12 \qty(D_i \phib )^2
    +\frac12 \qty(  D_i h)^2
    + V_3^{\rmii{soft, tree}}(\phib +h) 
    + \mathcal{L}_{\rmii{GF}}
    +\cdots \\
    =& \intr 
    \frac12 (\partial_i \phib )^2 +
     \frac12 \qty(\gxthree ^2 \frac{\tau^a}{2} \Xspat^a \frac{\tau^b}{2} \Xspat^b \phib ^2)
    +\frac12 \mthree^2  \phib ^2
    + \frac{\lamthree}{4} \phib^4
    + \frac{1}{4} \hthree \phib^2  \Xzero^a \Xzero^a \\
    &+\frac12 \qty(  D_i h)^2
    +\frac12 \mthree^2 h^2
    + \frac{\lambda_3}{4}  h^4
    + \frac{1}{4} \hthree h^2  \Xspat^a \Xspat^a
    + \mathcal{L}_{\rmii{GF}}
    + \mathcal{L}_{\rmii{mix}} +\cdots
\end{align}
The dots denote terms linear in $h$ which correspond to tadpole diagrams and do not contribute to 1PI diagrams. Moreover, such terms would vanish when evaluated at the stationary point of the action. The term $\mathcal{L}_{\rmii{GF}}$ contains the gauge-fixing terms evaluated on the background. 
In general, the term $\mathcal{L}_{\rmii{mix}}$ will contain mixing of the spatial gauge mode and Goldstone modes, for details see appendix~\ref{sec:compDet} and refs.~\cite{Ekstedt:2021kyx, Baacke:1999sc}. In certain cases, such terms can be removed by a gauge choice; however, this depends on the background field under consideration. 
In particular, this would work if the background is constant and homogeneous. However, for a non-homogeneous background, the mixing can be non-trivial.
We will discuss this issue in more detail later. 
In order to capture the one-loop effects, we need to focus on terms that are quadratic in field fluctuations. First, we can define the following fluctuation operators for gauge modes and ghosts
\begin{align}
    \OpXzero[\phib] &=  \partial^2_i +   \mXtemporal^2[\phib],\\
    \OpXG[\phib] &=  \partial^2_i  + \mXthree^2 + W_{\rmii{mix}}[\phib] , \\
    \OpX[\phib] &= \partial^2_i + \mXthree^2,\\
    \Opg[\phib] &=  \partial^2_i  + W_\rmii{g} [\phib].
\end{align}
Here $\OpXzero$ is the operator of temporal gauge fluctuations, $\OpXG$ is the operator for mixing gauge-Goldstone modes, then $\OpX$ is the operator associated with transverse gauge modes, while $\Opg$ denotes the ghost fluctuation operator.
Operators $W_{\rmii{mix}}$ and $W_\rmii{g}$ are the coefficients in terms that are bilinear in relevant fluctuating field, the details can be found in appendix~\ref{sec:compDet} or in refs.~\cite{Andreassen_2015, Ekstedt:2021kyx}.
We can perform the path integral over gauge fields, since they are quadratic and result in Gaussian integrals. The integration will result in fluctuation determinants. For now, we will use a shorthand notation
\begin{align}
\label{eq:detOX}
    \operatorname{det} \OpX 
    \equiv
    \operatorname{det} \OpXG
    \operatorname{det} \OpXT
    \operatorname{det} \Opg .
\end{align}
Then, writing the resulting effective action, we obtain the following expression 
\begin{align}
    e^{-S_{\rm eff}[\phib(x)]} &= 
    \int \pathD{h} \qty[
        (\detOXzero[\phib])^{-\frac32} (\detOX[\phib])^{-\frac32}
        ~e^{-S^{\text{soft}}_3 [\phib + h]} 
    ]. \\
    &= \int \pathD{h} ~e^{
        -S^{\text{soft}}_3 [\phib + h] - \frac32\log (\detOXzero[\phib])
        - \frac32\log (\detOX[\phib])
    }.
\end{align}
The argument of the exponent on the right side is the one-loop action of the nucleation EFT we have introduced in the previous section
\begin{align}
\label{eq:1loop_nuclSeff}
    S_\rmii{nucl} ^\rmii{1-loop} = 
    S^{\text{soft}}_3 [\phib ] 
    + \frac32\log (\detOXzero[\phib])
    + \frac32\log (\detOX[\phib])
\end{align}
Note that in the previous section, we have shown the LO potential, which does not contain the determinants explicitly. The reason is that the loop expansion does not coincide with the soft expansion, i.e. some parts of one-loop terms can be of order $\order{g^4}$. We will illustrate how to explicitly obtain the LO potential from the 1PI nucleation action in the next section. 
We can observe that this is a theory of scalar field only, where the scalar mass is ``dressed'' and contains one-loop effects of heavier gauge modes, i.e. taking the second derivative of the potential will include contributions from the gauge determinants. 

Now, in the analogous way to eq.~\eqref{eq:scalar_det_toy}, we also need to integrate out one-loop, scalar fluctuations, which will result in the scalar fluctuation determinant contributing to the prefactor in the statistical part of the nucleation rate. We define the scalar fluctuation operator as
\begin{align}
    \OpS ^\rmii{1-loop} [\phib]  &\equiv 
    \frac{\delta^2 S_\rmii{nucl} ^\rmii{1-loop}}{\delta \phib^2}
\end{align}
Finally, we can obtain an expression for the statistical part, cf. eq.~\eqref{eq:Astat_Snucl}, containing the one-loop corrections
\begin{align}
\label{eq:1PI_Seff_varphib}
    \Astat ^\rmii{1-loop}
    \equiv 
    \qty|\frac{\detOS[\phi_\rmii{CB}^\rmii{1-loop}]}{\detOS[\phi_\rmii{meta}]}| ^{-\frac12}
    e^{
        -\Delta S_\rmii{nucl} ^\rmii{1-loop} [\phi_\rmii{CB} ^\rmii{1-loop}]
    },
\end{align}
where the critical bubble $\phi_\rmii{CB} ^\rmii{1-loop}$ here is a stationary point of the action $S_\rmii{nucl} ^\rmii{1-loop}$.
A detailed form of the scalar and gauge determinants will be discussed later.\footnote{Note that here, for simplicity, we did not yet remove the zero-modes from the determinant evaluated on the bubble.}
Note that this expression is based on the loop expansion, and has to be yet organised within soft expansion. 
In the next sections, we will derive the effective action at next-to-leading order in soft expansion and show how to include it in the statistical part of the nucleation rate. 
First, we will do that using the derivative expansion, and then discuss its limitations. Then we will illustrate the calculation of fluctuation determinants and show how to include them in the calculation of thermal bubble nucleation rate.

\section{Nucleation rate in derivative expansion}
To obtain the nucleation rate within this nucleation EFT, we want to use the saddle-point approximation for the path integral, as in section~\ref{sec:nucl_rate_model_radiativebarrier}. 
This means we need to find the $O(3)$-symmetric solution to the equation of motion of the Euclidean action, i.e. the critical bubble and then evaluate the effective action, $S_\rmii{nucl}$. This task, however, is now difficult due to the presence of gauge determinants in the effective action. 
Therefore, it is convenient to expand these gauge determinants in the powers of momenta. This approach is called a \textit{derivative expansion}, see e.g. \cite{Kierkla:2025qyz, Garny:2012cg, Ekstedt:2021kyx, Ekstedt:2023sqc, Moss:1985ve, Baacke:1993, Kripfganz_1995} and references therein.\footnote{
Another name one can encounter in the literature is a \textit{gradient expansion}. We will interchangeably talk about derivatives and momenta as they are related by the Fourier transform.
}
This will also allow us to organise the terms within soft expansion.
The gauge determinants expanded in momenta become 
\begin{align}\label{eq:gauge_det_gradexp}
    \log \detOXzero[\phi_b] &\simeq \intx ~
    \Biggl[
    \underbrace{\frac{-1}{12\pi} (3 \mxzero^3 ) }_{\mathcal{O}(p^0)}
    + \underbrace{
        \frac{-1}{64\pi}\frac{\hthree^2 \phib^2}{\mxzero^3} \frac{(\partial_i \phib)^2}{2}
      }_{\mathcal{O}(p^2)} + \mathcal{O}(p^4)
      \Biggr],\\
    \log \detOX [\phi_b] &\simeq \intx ~
    \Biggl[
    \underbrace{\frac{-1}{12\pi} (6 \mXthree(\phi_b)^3 ) }_{\mathcal{O}(p^0)}
    + \underbrace{
        \frac{-11}{16\pi}\frac{\gxthree}{\phib} 
        \frac{(\partial_i \phib)^2}{2}
    }_{\mathcal{O}(p^2)} + \mathcal{O}(p^4)
    \Biggr].
\end{align}
This expansion is valid only if the magnitude of the considered background momenta is smaller than the mass of the mode in the given fluctuation operator, so in the case of gauge modes, the expansion requires 
\begin{align}
\label{eq:gauge_grad_exp_condition}
    \frac{\nabla^2 \phib} {\mXthree ^2} \ll 1 
\end{align}
to be fulfilled.
As the value of the momenta of the fluctuating scalar is of the order of its mass, $\nabla \sim \mthree$, this condition can also be expressed as 
\begin{align}
  \frac{m_3 ^2} { \mXthree^2} \ll 1.   
\end{align}
Notice that then, this condition coincides with the validity of the nucleation EFT, cf. discussion in the section \ref{sec:nuclEFT_valid}. 

For the scalar determinant, it is not possible to use derivative expansion as the expansion parameter in this case is always large, 
\begin{align}
    \frac{\nabla}{\mthree} \simeq \frac{\mthree}{\mthree} =1.
\end{align}
Thus, one should, in principle, evaluate the prefactor containing ratio of scalar determinants explicitly \cite{Ekstedt:2021kyx, Ekstedt:2023sqc}. Instead, in the literature, their contribution is commonly estimated based on dimensional analysis, see e.g.~\cite{LINDE198137, LINDE1983421}. We will comment on that soon.

\subsection{Leading order nucleation rate and bounce solution in soft expansion}
\label{sec:LOnuclrate}
From the perspective of power counting of the soft expansion, we can see that the zero-momentum parts of the gauge determinants are of order $\order{g^3}$ and thus correspond to the leading order of the effective action. 
More specifically, since they are zero-momentum parts, they contribute to the effective potential. Thus, the complete LO effective action within nucleation EFT in derivative expansion is given by:
\begin{align} \label{eq:SLO_EFT}
    \SDRLO [\phib] = \intx \frac12 (\partial_i \phib)^2
    +\VLO[\phib]
    ,
\end{align}
where the potential is
\begin{align}
     \VLO[\phib] 
     &\equiv 
    \frac{1}{2}\mthree ^2 \phib^2 + \frac{\lamthree}{4}\phib^4 
    + \VLOvectors [\phib] \\
     &=
     \frac{1}{2}\mthree ^2 \phib^2 + \frac{\lamthree}{4}\phib^4 
    -\frac{1}{12\pi} \qty(
        6 \mXthree^3+ 3 \mXtemporal^3
    ).
\end{align}
Note that we introduced the following notation for the contributions from spatial and temporal gauge modes
\begin{align}
\VLOvectors = 
\VLOspatial
+ \VLOtemporal 
= 
    -\frac{1}{12\pi} \qty(
        6 \mXthree^3
    )
    -\frac{1}{12\pi} \qty(
     3 \mXtemporal^3
    )
\end{align}
We can see that this expression is exactly the LO potential we have written in eq.~\eqref{eq:VnuclLO_se_nuclEFT}. Back then, we had thus implicitly used the derivative expansion. 
Now, the saddle-point solution $\varphibLO$ can be obtained by imposing $O(3)$ symmetry and varying the Euclidean LO action, which leads to the well-known bounce equation \cite{Coleman:1977py, LINDE198137, LINDE1983421}
\begin{align} \label{eq:bounceLO_eq}
    \dv[2]{\varphibLO}{r} + \frac{2}{r} \dv[]{\varphibLO}{r} = \dv[]{\VLO}{(\varphibLO)}.
\end{align}
To obtain the complete LO nucleation rate in the soft expansion, we first need to evaluate the Euclidean action on the $\varphibLO$ solution
\begin{align} \label{eq:SLO_bounce}
    \SDRLO[\varphibLO ] = 
    \intr
    \qty(
        \frac12 (\partial_r \varphibLO)^2 
        + V^{\rmii{LO}}_3 [\varphibLO(r)]
    ),
\end{align}
where we have used the notation $\intr \equiv 4\pi\int_0 ^\infty \dd{r}  ~r^2 $. 
It should be emphasised that in the soft expansion, the LO nucleation rate does not incorporate the full one-loop contribution of the gauge bosons, but only the terms that appear as corrections to the effective potential. 
The non-zero momenta parts of the determinants (of order $\mathcal{O}(p^2)$ or higher in derivative expansion), on the other hand, scale as $\mathcal{O}(g^4)$ and belong to NLO in the soft-expansion, as we will show in the next section. In general, it is important to distinguish between 
loop expansion, soft expansion, and derivative expansion, as one of them can work well while the other breaks down \cite{Ekstedt:2021kyx, Kierkla:2025qyz}. 

Then, we need to include the scalar fluctuation determinant. The scalar fluctuations take place on the critical bubble, which in our case is the LO bounce solution, $\varphibLO$. 
Thus, the associated fluctuation operator should be a second derivative of $\SDRLO$ and not $S_\rmii{nucl} ^\rmii{1-loop}$,
\begin{align}
\label{eq:Oscalar_LO}
    \OpS [\phib]  &\equiv 
    \frac{\delta^2 \SDRLO}{\delta (\varphibLO)^2} = 
    -\partial_i^2 \varphibLO + m^2_\rmii{eff}[\varphibLO], \\
    m^2_\rmii{eff}[\varphibLO] &\equiv (\VLO[\varphibLO])^{\prime\prime}. 
\end{align}
Since the contribution of the scalar determinant is subleading, at LO in soft expansion, we will resort to a simple approximation.
As the ratio of scalar determinants has mass-dimension three, see sec.~\ref{sec:T_bubble_nucl}, and the temperature should be the largest scale, it is the most common approach to simply use the approximation \cite{LINDE198137, LINDE1983421}
\begin{align}
    \frac{
        \operatorname{det}^\prime \mathcal{O}[\varphibLO]
        }{
        \operatorname{det} \mathcal{O}[\varphiF]
    }
    \simeq T^3,
\end{align}
where $\varphiF =0$ denotes the false-vacuum field configuration in our setting. Then $T^3$ becomes a prefactor in the statistical part of nucleation rate\footnote{Gauge determinants are dimensionless, thus they indeed contribute to the exponential part directly.}. We will later discuss the validity of this approximation in SU(2)cSM. Thus, the leading-order statistical part of the nucleation rate becomes
\begin{align}
    \Astat^\rmii{LO} \equiv T^3 e^{- \Delta \SDRLO[\varphibLO ] },
\end{align}
where $\Delta \SDRLO[\varphibLO ] = \SDRLO[\varphibLO ] - \SDRLO[\varphiF ]$.

\subsection{Next-to-leading order nucleation rate}

Now we will show that the nucleation rate at NLO receives contributions from the non-zero momentum part of the gauge determinants, see e.g. \cite{Hirvonen:2021zej, Gould:2021ccf}. 
These contributions can also be understood as parts of the field renormalisation operators; $Z_\rmii{3}$-factors within 3d nucleation EFT. They can be read off from eq.~\eqref{eq:gauge_det_gradexp}, so the overall $Z_\rmii{3}$-factor at one-loop order reads
\begin{align}\label{eq:Z3_factor_total}
    Z_\rmii{nucl} ^\rmii{1-loop} [\phib] &= 
    1 
    +\underbrace{\frac{-11}{16\pi}\frac{\gxthree}{\phib}}_{\equiv \ZX^\rmii{NLO}}
    +\underbrace{\frac{1}{64\pi}\frac{\hthree^2 \phib^2}{\mXtemporal} }_{\equiv \ZXtemporal^\rmii{NLO}}
    +\dots
    %
\end{align}
where dots denote higher orders in momenta, which we will discuss later. These terms are not yet of order higher than $\order{g^3}$. However, in the soft expansion, the effective action is going to contain the following terms
\begin{align}
    \intr \ZX^\rmii{NLO} \frac{(\partial_i \varphibLO)^2}{2} 
    &=  
    \intr \qty(
        \frac{-11}{16\pi}\frac{\gxthree}{\varphibLO}
    )
    \frac{(\partial_i \varphibLO)^2}{2}, \\  
   \intr \ZXtemporal^\rmii{NLO} \frac{(\partial_i \varphibLO)^2}{2} 
    &= 
    \intr \qty(
        \frac{1}{64\pi}\frac{\hthree^2 (\varphibLO)^2}{\mXtemporal^3}
    )
    \frac{(\partial_i \varphibLO)^2}{2}.
\end{align}
These terms scale indeed like $\order{g^4}$ due to the derivatives over bounce, see ref.~\cite{Hirvonen:2021zej} for details. Moreover, we see that the apparent divergent behaviour in eq.~\eqref{eq:Z3_factor_total} is regularised here by the derivatives, so the total contribution is finite. 
%
At NLO, there is also another term in the action, coming from the two-loop corrections. 
Similarly, to the one-loop corrections, we can expand the two-loop corrections in powers of momenta.
The dominant part of the two-loop gauge correction again corresponds to the zero-momentum term, i.e. correction to the effective potential. 
We can thus again use derivative expansion, and the NLO contribution to the effective potential from the two-loop soft (gauge) modes takes the following form \cite{Kierkla:2023von, Kierkla:2025qyz, Ekstedt:2022bff, Gould:2023ovu}:
\begin{align}
\label{eq:VeftNLO}
\begin{split}
V_\rmii{nucl}^{\rmii{NLO}} &=
    \frac{1}{(4\pi)^2}\frac{3}{64} g^2_{\rmii{$X$},3} \Bigl\{
          8 m^2_{\rmii{$X$},3} \bigl(5-21\ln3 \bigr) 
        - 3 g^2_{\rmii{$X$},3} v^2_{3} \bigl(1-14\ln2 \bigr)
      \\ &
      \hphantom{{}\frac{1}{(4\pi)^2}\frac{3}{64} g^2_{\rmii{$X$},3} \Bigl(}
      + 2 \bigl(80 m^2_{\rmii{$X$},3} - 3 g^2_{\rmii{$X$},3} v^2_{3}\bigr)
        \ln\frac{\muthree}{2 m_{\rmii{$X$},3}}
      \Bigr\}
      \\ &
  + \frac{1}{(4\pi)^2} \Bigl\{
    \frac{3}{4} g^2_{\rmii{$X$},3} \bigl(
        6 m^2_{\rmii{$X_0$},3}
      + 4 m_{\rmii{$X$},3} m_{\rmii{$X_0$},3}
    \bigr) 
    + \frac{15}{4}  \kappa_3^{ } m^2_{\rmii{$X_0$},3}
    - \frac{3}{8} h^2_{3} v^2_3 \Bigl( 1 + 2 \ln\frac{\muthree}{2 m_{\rmii{$X_0$},3}}\Bigr)
    \\ &
    \hphantom{{}\frac{1}{(4\pi)^2}\biggl(}
    - \frac{3}{2} g^2_{\rmii{$X$},3} \bigl(m^2_{\rmii{$X$},3} - 4 m^2_{\rmii{$X_0$},3}\bigr)
      \ln\frac{\muthree}{2 m_{\rmii{$X_0$},3} + m_{\rmii{$X$},3} }
  \Bigr\} 
  \\ &
  - \frac{1}{(4\pi)^2} \Bigl\{
      \frac{15}{4} \mDX^2 \kappa_3^{ }
      + \frac{3}{2} g_{\rmii{$X$},3}^2 \mDX^2\Bigl(3 + 4 \ln\frac{\muthree}{2 \mDX}\Bigr)
  \Bigr\}
  ,
\end{split}
\end{align}
where $\muthree$ is the renormalisation scale of the 3d EFT.
It scales at $\mathcal{O}(g^4)$ order; thus, at NLO order in the soft expansion, we need to account for the zero-momentum part of two-loop gauge corrections (see, e.g.~\cite{Kierkla:2023von, Kierkla:2025qyz, Gould:2023ovu, Gould:2021ccf, Ekstedt:2024etx}). One can also notice the analogy with the toy model in chapter~\ref{chapter:TFT}, where we have needed to include two-loop corrections to obtain the RG-scale invariance.

Taking both the correction to the kinetic term and the effective potential, and evaluating them on the LO bounce background, we obtain the complete expression for the NLO correction to the nucleation rate (in derivative expansion):
\begin{align}\label{eq:Seff_NLO_grad}
    \SDRNLO[\varphibLO] = 
    \intr 
    \qty(\ZX^\rmii{NLO}[\varphibLO] + \ZXtemporal^\rmii{NLO}[\varphibLO]) \frac{(\partial_r \varphibLO)^2}{2}
    + V_3^{\rmii{NLO}}[\varphibLO].
\end{align}
Then, the exponent in the total statistical part $A_{\rmii{stat}}$ of nucleation rate at the NLO in soft expansion contains simply a sum of LO and NLO contributions 
\begin{align} \label{eq:Astat_NLOgrad}
    \Astat \simeq
    T^3
    e^{-\Delta \Snucl [\varphib]}
    =
    T^3 
    e^{
        -\Delta \SDRLO[\varphibLO] - \Delta \SDRNLO [\varphibLO]
         }, 
\end{align}
where again we have approximated the ratio of scalar determinants as $T^3$ based on its dimensions.
This expression is both renormalisation-scale invariant and gauge invariant up to $\mathcal{O}(g^5)$ in soft expansion, as we will discuss later, see also \cite{Kierkla:2023von, Kierkla:2025qyz, Ekstedt:2021kyx, Hirvonen:2021zej, Lofgren:2021ogg}. The expression above is one of the main results of this thesis. It provides a theoretically consistent method of computing thermal nucleation rate at NLO for supercooled phase transition in a classically scale-invariant model \cite{Kierkla:2023von}. 
Nevertheless, this method relies heavily on derivative expansion for the gauge mode contributions. And while the soft expansion is indeed under control, the naive use of derivative expansion may introduce non-negligible errors. We will discuss this issue in detail in the following sections.

\subsection{Region of validity of derivative expansion}
\label{sec:validity_derivexpan}
\begin{figure}
    \centering
    \includegraphics[width=0.8\linewidth]{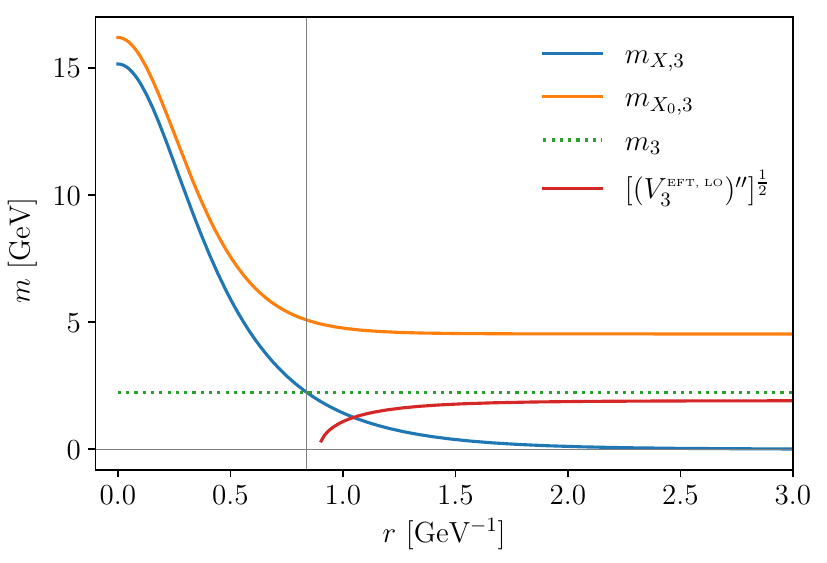}
    \caption{Masses of particles as a function of bubble radius for a benchmark $\mX=10^4$~GeV and $\gx=0.8$. Vertical grey line denotes $\rtail$ where spatial gauge mode mass becomes equal to $\mthree$ mass. 
    }
    \label{fig:scale-shifters}
\end{figure}

Derivative expansion for gauge determinants has an underlying requirement that the masses of gauge modes should be larger than the magnitude of the momenta of the nucleating scalar mode, cf. eq.~\eqref{eq:gauge_grad_exp_condition}. 
This condition, in the context of calculating the nucleation rate (where we would evaluate the determinant on the bounce background), would become
 %
 %
\begin{align}
    \frac{(\partial \varphib)^2} {m_{\rmii{gauge} }^2(\varphib)} \sim
    \frac{\mthree ^2} {m_{\rmii{gauge} }^2(\varphi_b)} \ll 1.
\end{align} 
Here, we have dropped the superscript LO from the bounce solution for notational clarity.
We know that the bounce solution corresponds to the $O(3)$-symmetric field configuration that interpolates between an ``escape point'' configuration located beyond the barrier, and the false vacuum configuration. In other words, it describes a ``bubble'' of some finite radius $R$, where the vev of the scalar field is non-zero inside the bubble (for $r<R$), while outside (for radius $r\gg R$) the field is in the false vacuum state where $\varphiF = 0$. 
Let us now discuss how the field-dependent masses of the relevant modes behave as a function of bubble radius. We plot the masses in figure~\ref{fig:scale-shifters}. 
We can observe that while the temporal gauge mode is still massive in the false vacuum, i.e. $\mxzero(r\rightarrow \infty) = \mDX$, the mass of spatial gauge modes inevitably goes to zero there. This means that derivative expansion \textit{always} breaks down at the large values of $r$ (i.e.\ outside the bubble). 

We can estimate the region of validity of the derivative expansion by estimating the value of the radius, $r_t$, at which the (spatial) gauge mode has the same mass as the nucleating scalar\footnote{Subscript $t$ corresponds to the ``tail'' of the bounce solution. Or the ``trouble'' that one encounters there.}
\begin{align} \label{eq:mxm3_at_rt}
 \mx(\rtail) = \partial \varphib (\rtail) \sim \mthree.
 \end{align}
The expansion is then estimated to be valid only if $r \ll \rtail$. Notice that this situation is analogous to the breakdown of nucleation EFT that we have discussed in section~\ref{sec:nuclEFT_valid}. 
Indeed, if we assume that $(\partial \varphib) \sim \mthree$, then these two breakdowns occur simultaneously. The reason is the scale-shifting nature of spatial gauge modes at the tail of the bounce solution $r\gg\rtail$, and we can identify $\varphib(\rtail) = \abs{\vbreak}$, where $\vbreak$ was estimated in eq.~\eqref{eq:vbreak_estimate}.

\section{Theoretical uncertainties from scale-shifters}
\label{sec:uncertainties_scale-shifters}
Contributions of the ``scale-shifters'' in derivative expansion can be finite (at some order of momenta), just as we showed in the case of spatial gauge modes. However, there would still be an error introduced into the calculation of nucleation rate.
This, in turn, would not only affect the precision of predicted physical parameters of the phase transition, but also the observables such as gravitational wave spectra. Therefore, it is crucial to estimate the errors in nucleation rate and consider ways of improving the calculation. 
It is worth emphasising that in our model, only spatial gauge modes are breaking derivative expansion, thus in this section we will focus solely on their contribution to the effective action.

\subsection{Estimating the tail of bounce solution}

First, we need to estimate the value of the ``tail'', $r_t$, as it then can be used for estimation of the magnitude of the relevant contributions to the effective action. 
We will do it analogously to refs.~\cite{Gould:2021ccf, Kierkla:2025qyz, Hirvonen:2021zej}. Note that for large values of the radial coordinate, the bounce solution can be expressed in an asymptotic form, valid for large radius, $r$, as \cite{Gould:2021ccf, Ekstedt:2023sqc}
\begin{align} \label{eq:varphib_tailform}
    \varphib(r) \longrightarrow \frac{\Ainf}{r}e^{-\mthree r},
\end{align}
where $\Ainf$ is a constant. Using this form and eq.~\eqref{eq:mxm3_at_rt} we get
\begin{align}
     \frac{\Ainf \gxthree }{2 \rtail} e^{-\mthree \rtail} = \mthree,
\end{align}
from which it follows that
\begin{align}
    \mthree \rtail + \log{\mthree \rtail} = \log{\frac{A_\infty \gxthree}{2}}.
\end{align}
Since bounce equation implies that $\mthree \rtail \sim 1$, we arrive at a final estimate \cite{Gould:2021ccf}:
\begin{align}\label{eq:rtail_estimate}
    \rtail \simeq \frac{1}{\mthree} \log{\frac{A_\infty \gxthree}{2}}.
\end{align}
Notice that this expression agrees with an intuition that the ``lighter'' nucleating mass should lead to a larger value of $\rtail$ and thus, a larger region of validity of nucleation EFT and derivative expansion, cf. discussion at the end of section~\ref{sec:nuclEFT_valid}.

\subsection{Scale-shifters at one-loop: effective potential}
\label{sec:scale-shifters1loop_Veff}
Considering first the LO in the soft expansion, we can observe that the scale-shifters can introduce an error to the zero-momentum contribution (effective potential) to the nucleation rate.
Specifically, it appears in the spatial gauge modes contributions evaluated at the tail of the bounce solution. 
Let us denote the total spatial gauge mode contribution to the LO action as: 
\begin{align}
\label{eq:CXdefinition}
    \CX [\varphib] 
    \equiv 
    \intr \VLOspatial[\varphib]
    =
    \intr \frac{-3}{48\pi} \gxthree^3 \varphib^3. 
\end{align}
Considering only radius values at the tail, $r>\rtail$, and using the asymptotic form of the bounce solution from eq.~\eqref{eq:varphib_tailform} we get 
\begin{align} \label{eq:CX_tail}
    \CX [\varphib]\eval_{r>\rtail}  = 
    -\frac{A_{\infty} ^3 \gxthree^3}{4} 
    \int_{\rtail} ^\infty \dd{r}
    \frac{e^{-3\mthree r}}{r} =
    \frac{A_{\infty} ^3 \gxthree^3}{4}  E_1(3\mthree r_t),
\end{align}
where $E_1(x)$ is the exponential integral of the first kind. The $E_1(x)$ function can be expanded for $x\gg 1$ as
\begin{align}
    E_1(x) \xrightarrow{x\gg1} \frac{e^{-x}}{x} \qty(1-\frac{1}{x} + \mathcal{O}\qty(\frac{1}{x^2}) ).
\end{align}
By implementing that form in the previous result in eq.~\eqref{eq:CX_tail}, we arrive at
\begin{align}
    \CX [\varphib]\eval_{r>\rtail}  
    &\simeq 
        \frac{A_{\infty} ^3 \gxthree^3}{4}
        \frac{e^{-3\mthree \rtail}}{3\mthree \rtail} 
        \qty(1 - \frac{1}{3\mthree \rtail} )
        \xrightarrow[]{3\mthree \rtail \gg 1} 0.
\end{align}
Fortunately, this means that the breakdown of derivative expansion introduces only a small error to the total effective action. The contribution at the tail is finite and suppressed exponentially by a factor $e^{-\mthree \rtail}$. 
Thus, this contribution is negligible in comparison to the rest of the terms in the nucleation action at the LO.  
This result also highlights the validity of the LO bounce solution obtained from eq.~\eqref{eq:bounceLO_eq}. 
After all, in principle, it would be possible to extremise the action containing the determinants and omit the use of derivative expansion. Let us denote such a solution as $\varphib^{\rmii{1-loop}}$. 
Since the derivative expansion is under control at LO, it means that the bounce solution from eq.~\eqref{eq:bounceLO_eq} should be very close to the one-loop bounce, $\varphibLO \simeq \varphib^{\rmii{1-loop}}$. 
Moreover, all the differences should be visible only for a large bubble radius, $r>\rtail$. But there, the background field value simultaneously goes to $\varphiF = 0$, making this difference less important than the contribution from the bulk of the bubble. 
Thus, we conclude that it is justified and correct to use the LO action in the derivative expansion to calculate the bubble nucleation rate.

\subsection{Scale-shifters at one-loop: higher orders in derivative expansion}
\label{sec:derivexp_higherorders}

Let us now examine the non-zero momenta part of the spatial gauge determinant. These contributions would be associated with operators emerging at the higher orders in the derivative expansion. 
The determinant evaluated on the bubble and expanded up to fourth order in gradients, reads~\cite{Fraser:1984zb, Andreassen:2016cvx, Kierkla:2025qyz}:
\begin{align}
\label{eq:S3:higher}
\detX  =
 \intr
 \Big[
      \VLOspatial (\varphib)
     &+ \frac{1}{2} Z_{2,3} (\partial_i \varphib)^2
     + \frac{1}{2} Z_{4,3} (\partial^2 \varphib)^2
 \nonumber \\ 
   &+ \frac{1}{2} Y_{3,3} (\partial_i \varphib)^2 (\partial^2 \varphib)
   + \frac{1}{8} Y_{4,3} (\partial_i \varphib)^2(\partial_j \varphib)^2
   + \mathcal{O}(\partial^6)
 \Big].
\end{align}
Here, for non-zero momentum parts we have skipped the ``$X$'' subscript, and instead we introduce another one to count the power in derivatives, i.e. $Z_{2,3} \equiv \ZX$.
As argued in ref.~\cite{Kierkla:2025qyz}, due to the use of spatial rotation and translation symmetry and integration by parts,
the operator basis in eq.~\eqref{eq:S3:higher}
is unique, and depends only on the five independent functions,
$\VLOspatial $,
$Z_{2,3}$,
$Z_{4,3}$,
$Y_{3,3}$, and
$Y_{4,3}$.
The specific spatial gauge- and temporal gauge-modes contributions to the corresponding operators are given explicitly in the appendix of ref.~\cite{Kierkla:2025qyz}.

The first non-zero momentum contribution is the $Z_{2,3}$-operator. Considering one-loop fluctuations only, the operator is given in the form of eq.~\eqref{eq:Z3_factor_total}:  
\begin{align}
    Z_{2,3}(\varphib) &= \ZX =
    1 - \frac{11}{16\pi}\frac{g_{\rmii{$X$},3}}{\varphib},
\end{align}
and we remind that this contribution belongs to  NLO in the soft expansion \cite{Kierkla:2023von, Kierkla:2025qyz}. At next order in derivatives, i.e.\ $\mathcal{O}(\partial^4)$, there are three more operators which read
\begin{align}
\label{eq:higher-Z}
  Z_{4,3}(\phib) &=
    - \frac{23}{80\pi}\frac{1}{g_{\rmii{$X$},3} v_{3}^3}
  \,,
  \\
  Y_{3,3}(\phib) &=
    + \frac{359}{80\pi}\frac{1}{g_{\rmii{$X$},3} v_{3}^4}
  \,,
  \\
  Y_{4,3}(\phib) &=
    - \frac{127}{80\pi}\frac{1}{g_{\rmii{$X$},3} v_{3}^5}
  \,.
\end{align}
Notice that here, $\phib$, denotes the general background, i.e. these expressions can be used more generally. A second remark is that the contributions to the action containing these terms would be above NLO in soft expansion. 
Now, we can again use the asymptotic behaviour of the bounce solution given in eq.~\eqref{eq:varphib_tailform} and quantify these contributions to the nucleation action
\begin{align}
  \label{eq:Z2:int}
  \int_{r\geq\rtail} 
  \frac{(\partial_r \varphib)^2}{\varphib}
  &\approx
    - 4\pi \Ainf \mthree^2 \int_{r\geq\rtail}
    \!{\rm d}r\, e^{-\mthree r}
      \Bigl[r + \dots\Bigr]
  \,,\\
  \int_{r\geq\rtail}
  \frac{(\partial_r ^2\varphib +\frac{2}{r}\varphib)^2}{\varphib^{3}} 
  &\approx
    - \frac{4\pi}{\Ainf} \mthree^4 \int_{r\geq\rtail}
        \!{\rm d}r\, e^{\mthree r}
      \Bigl[r^3 + \dots\Bigr]
  \,,\\
  \int_{r\geq\rtail}
    \frac{(\partial_r \varphib)^2  (\partial_r ^2\varphib +\frac{2}{r}\varphib)}{\varphib^{4}}
  &\approx
    - \frac{4\pi}{\Ainf} \mthree^4 \int_{r\geq\rtail}
        \!{\rm d}r\, e^{\mthree r}
      \Bigl[r^3 + \dots\Bigr]
  \,,\\
  \int_{r\geq\rtail}
    \frac{(\partial_r \phib)^4}{\varphib^{5}} 
  &\approx
    - \frac{4\pi}{\Ainf} \mthree^4 \int_{r\geq\rtail}
        \!{\rm d}r\, e^{\mthree r}
      \Bigl[r^3 + \dots\Bigr]
  \,,
\end{align}
where we have expressed the relevant derivative operators in spherical coordinates.
We can see that all operators, except the first,
diverge at the tail of the bounce solution for $r\geq \rtail$. 
In other words, all operators of $\mathcal{O}(\partial^4)$-order contribute to the effective action in equal magnitude, which suggests that these contributions can be resummed by abandoning the derivative expansion. 

In terms of soft expansion, derivative expansion becomes thus unreliable above NLO, which is the last order where the non-zero momentum contribution is finite. 
Then, the rest of the contributions containing higher derivatives belong to (soft) N$^3$LO order, but at the same time derivative expansion diverges for the one-loop determinant evaluated on the bubble. 
Thus, using the derivative expansion for the one-loop spatial gauge determinant is possible only up to the NLO order in soft expansion. 

\begin{figure}
    \centering
    \includegraphics[width=0.5\linewidth]{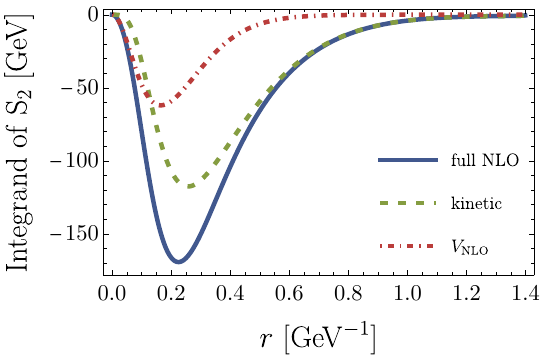}
    \caption{Contributions to the NLO correction to the effective action as a function of radial coordinate along the bubble radius for benchmark $g_X = 0.8$, $M_X = 10$\, TeV, and $T= 14.18$\, GeV. Here, $S_2$ corresponds to $S^{\rmii{NLO}}_\rmii{3}$ in our notation. Green, dashed line denotes $\ZX$-factor contribution, while the red dot-dashed line denotes $V^\rmii{NLO} _\rmii{3}$ contribution, and the blue solid line is their sum. 
    }
    \label{fig:SNLO_gradexp_contributions}
\end{figure}

The remaining question is whether the NLO spatial-gauge contribution introduces a non-negligible error to the nucleation rate. 
This issue is more pronounced if we focus only on the NLO corrections, as the contribution from the $\ZX$-factor (one-loop correction) is dominating over the NLO correction to the effective potential (two-loop correction). 
Both of these contributions to $\SDRNLO$ are shown for an example benchmark point in figure~\ref{fig:SNLO_gradexp_contributions}. 
One can see that the integrand of $\SDRNLO$ at the tail of the bounce solution is dominated solely by the $\ZX$-factor. 

%

\subsection{Scale-shifters at two-loops: effective potential}
\label{sec:der-exp-2loop}

Next, we will estimate the contribution from the two-loop effective potential in the effective action, in the region of breakdown of the derivative expansion.
Here, we will illustrate this behaviour specifically for the soft NLO contribution from spatial gauge modes. 
The temporal gauge modes are always heavier than the nucleating scalar along the bounce solution. Therefore, the derivative expansion always works for them, and
we have also confirmed this numerically for the one-loop effects. As two-loop contributions are suppressed in comparison to LO, we expect the derivative expansion for temporal gauge modes to work there as well.

Now, let us consider the part of
$V_\rmii{nucl}^{\rmii{NLO}} $ that contains only the contributions from the gauge spatial modes:
\begin{align}
	C^{\rmii{NLO}}_{\rmii{$X$},3} [\vthree] &\equiv
	\int_{\vec{x}}    
    \frac{1}{(4\pi)^2}\frac{3}{64} \gxthree^4 \vthree^2
	\Bigl(
			\underbrace{
				11 +42\ln{2} -14\ln{3} 
			}_{\equiv B}
			+34\ln{\frac{\muthree}{ \gxthree \vthree }}
	\Bigr)
  \\ &=
  \underbrace{
				\frac{B}{(4\pi)^2}\frac{3}{64}     \int_{\vec{x}}     \gxthree^4 \vthree^2
		}_{\equiv I_1}
	  +\underbrace{
	  	         \frac{34}{(4\pi)^2} \frac{3}{64} 	\int_{\vec{x}}    \gxthree^4 \vthree^2
	  	         \ln{\frac{\muthree}{ \gxthree \vthree }}
	  	}_{\equiv I_2}
  \,.
\end{align}
To estimate the size of this contribution at the tail of the bounce solution, we can again use the asymptotic form of $\varphib$ in eq.~\eqref{eq:varphib_tailform}.
Then, the integrals $I_1$ and $I_2$ become:
\begin{align}
  I_1[\varphib] &=
	  \frac{B}{(4\pi)}\frac{3}{64} \int_{r\geq R}\!{\rm d}r\, 
	 \gxthree^4 c^2  e^{-2 \mthree r} 
	 = \left[
	 	 \frac{B c^2}{(4\pi)}\frac{3}{64} \gxthree^4 \frac{e^{-2 \mthree r}}{2 \mthree}
	  \right]
	  \xrightarrow{r\rightarrow \infty} 0
    \,, \\[2mm]
	I_2[\varphib] &=
	 \frac{34}{(4\pi)} \frac{3 c^2}{64}  \int_{r\geq R}\!{\rm d}r\, 
	  \gxthree^4   e^{-2 \mthree r} 
	   \ln{ \frac{\muthree}{ \gxthree  \frac{c}{r} e^{-\mthree r}   }}
     \\
	  & =  \frac{34}{(4\pi)} \frac{3 c^2}{64} \gxthree^4
	   \biggl[
	   		\frac{e^{-2 \mthree r} }{4m_3}
	   		-\frac{ \mbox{Ei}(-2 \mthree r)  }{2m_3}
	   		+\frac{ e^{-2 \mthree r}  \ln{ \frac{\muthree}{ \gxthree  \frac{c}{r} e^{-\mthree r}   }}  }{2\mthree }
	  \biggr]
	  \xrightarrow{r\rightarrow \infty} 0
    \,.
\end{align}
Thus, at large values of $r$, where the derivative expansion breaks down, the zero-momentum contribution simultaneously goes to zero. This agrees with the behaviour of $V_\rmii{nucl}^{\rmii{NLO}} $ in figure~\ref{fig:SNLO_gradexp_contributions}. Hence, the error is under control, and it is justified to use the two-loop corrections to the effective potential in the calculation of the nucleation rate.

\subsection{Scale-shifters at two-loops: higher orders in derivative expansion}

In case of one-loop spatial gauge corrections, the derivative expansion was finite up to the first non-zero momentum contribution, i.e. the correction to the $\ZX$-factor. This is not the case for two loops. As it was shown in section~\ref{sec:derivexp_higherorders} (see also ref.~\cite{Kierkla:2025qyz}), the two-loop contribution to the 3d EFT $\ZX$-factor becomes divergent on the tail, as it contains terms such as: 
\begin{align}
  \label{eq:Z2:2loop:int}
  \int_{\vec{x}} \frac{(\partial_i \varphib)^2}{\varphib^{2}}
  &\approx
    - 4\pi \mthree^2 \int_{r\geq \rtail}\!{\rm d}r\,
      \Bigl[r^2 + \dots\Bigr]
  \,.
\end{align}
Therefore, the derivative expansion breaks down even earlier at higher-loop order. Indeed, upon inspection of three-loop gauge contributions to the effective potential \cite{Ekstedt:2024etx}, one can notice terms of the form:
\begin{align}
    V^\rmii{3-loop} \supset \# \log\frac{\mthree}{\mx(\phi)},
\end{align}
which are explicitly divergent on the tail of the LO bounce.
Note that at a given loop order, the finite-momentum contributions are suppressed by the soft expansion parameter in comparison to the zero-momentum contribution. In other words, in soft expansion at a given loop order, the contribution from the effective potential is larger than from the derivative operators. This is a general statement, which is also valid for different models. Therefore, up to the NLO order in soft expansion, one can use two-loop corrections to the effective potential, as they are finite, and the error they introduce is under control, as we have shown. The non-zero-momentum contribution from the one-loop correction, on the other hand, may introduce a non-negligible error, affecting the accuracy of predictions at NLO. 

\begin{figure}[t]
    \centering
    \includegraphics[width=0.7\linewidth]{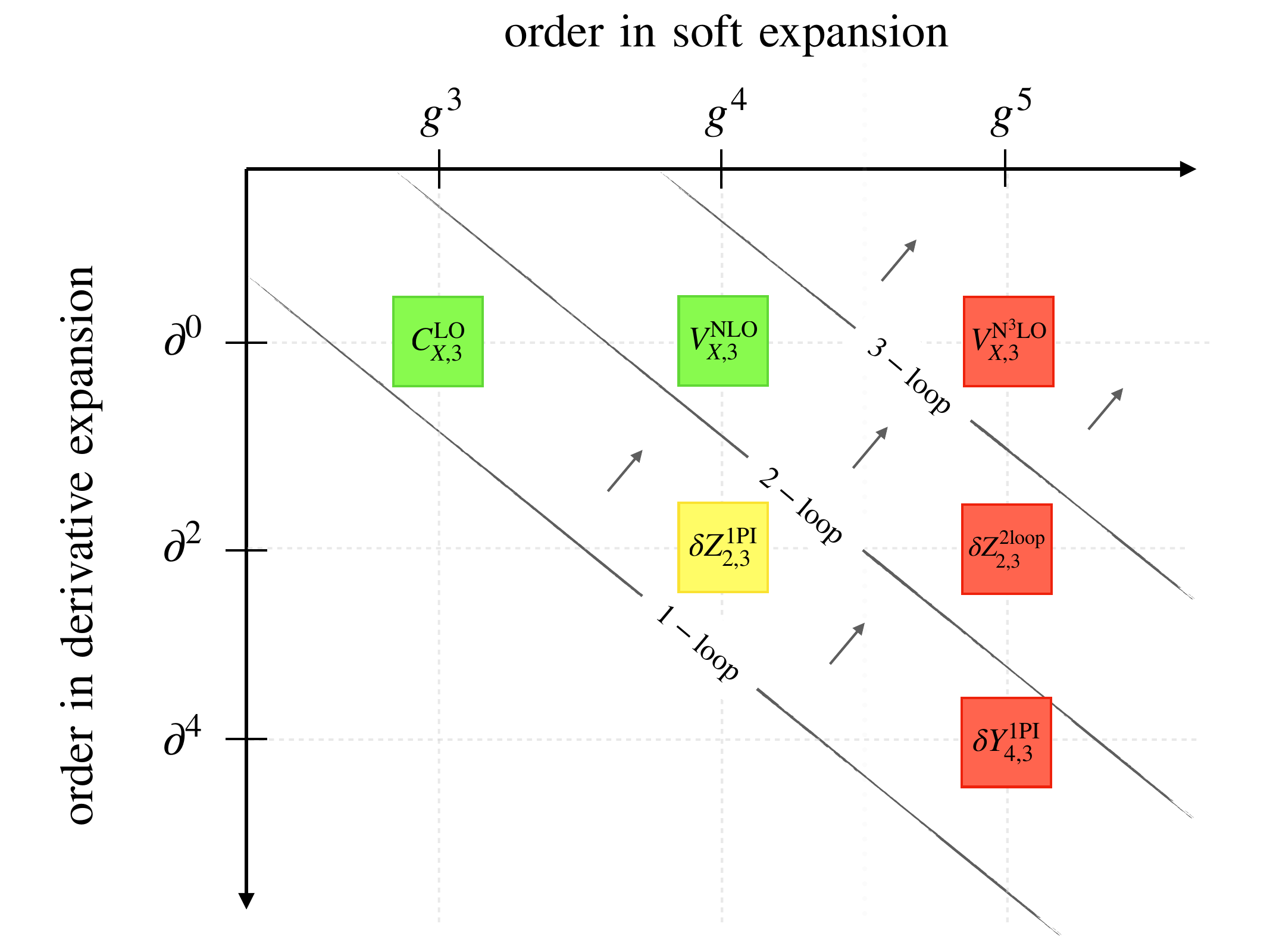}
    \caption{
        Spatial gauge contributions in the derivative expansion up to $\order{\partial^6}$ and up to three-loop fluctuations, i.e., up to $\order{g^5}$ in soft expansion. Colours denote the behaviour of a specific contribution evaluated on the LO bounce solution.  Green colour denotes finite contributions with negligible errors to the nucleation rate. Yellow denotes finite contributions with significant errors. Red denotes divergent contributions.  
    }
    \label{fig:derivexp_triangle}
\end{figure}

The summary of this section is depicted schematically in figure~\ref{fig:derivexp_triangle}. 
Derivative expansion is completely unreliable beyond NLO, and one should either try to modify its usage, e.g. by evaluating the relevant contributions on a truncated bubble radius ($r<\rtail$) or abandon it completely. In the next section, we will demonstrate how to obtain the NLO nucleation rate by abandoning derivative expansion for one-loop effects and computing the fluctuation determinants directly instead.
The treatment of corrections above NLO order in the soft expansion is beyond the scope of this work, but it was introduced for certain classes of models in the literature \cite{Ekstedt:2022tqk, Bezuglov_2019}.   


\section{NLO nucleation rate with functional determinants}
\label{sec:NLO_nuclrate_dets}
In the previous section, we employed derivative expansion to illustrate the calculation of the statistical part of the nucleation rate at NLO. 
We have also highlighted the limitations and drawbacks of this approach, which are caused by the scale-shifting gauge modes. 
Thus, we argued that despite its simplicity, this method is inherently flawed and introduces a non-negligible error to the effective action. 
This exact issue, as well as a proposed remedy, was suggested in~\cite{Gould:2021ccf}. The general idea for curing the calculation is to use derivative expansion where it is under control, and compute full fluctuation determinants for problematic modes. This ``recipe'' can be applied to our setting and is summarised in the following steps
\begin{enumerate}
    \item Compute the bounce in the leading-order approximation with the gauge modes integrated out, i.e., using eq.~\eqref{eq:bounceLO_eq}. 
    \item Evaluate one-loop fluctuation determinants for gauge modes on the LO bounce.
    \item Evaluate the leading-order action on the LO bounce as in eq.~\eqref{eq:SLO_bounce} and subtract the gauge contributions to avoid double-counting.
    \item Employ derivative expansion to evaluate gauge two-loop corrections in the eq.~\eqref{eq:VeftNLO} on the LO bounce.
\end{enumerate}
In this formulation, the gauge bosons contribute to the barrier through their large mass inside the bulk of the bounce and to the fluctuation determinants due to their light fluctuations at the tail of the bounce. 
This procedure allows for capturing the varying behaviour of the spatial gauge fields. At the same time, by obtaining their full one-loop contributions by calculating functional determinants, we do not employ the derivative expansion and thus avoid the errors it introduces. 
Now we will demonstrate this approach in more detail and discuss its theoretical consistency.

\subsection{Statistical part of nucleation rate at NLO}

We will now define the nucleation rate at NLO, which incorporates the treatment of scale-shifters. First, let us recall the one-loop action from eq.~\eqref{eq:1loop_nuclSeff}. This form contains complete one-loop gauge contributions, without relying on the derivative expansion. Our goal is to incorporate it in the soft expansion. 

To begin, let us remind that in previous sections, we have shown that at (soft) LO, the error from scale-shifters is under control. Thus, we can use the LO action $\SDRLO$, as well as its bounce solution, $\varphibLO$.
Then, we have also shown that the two-loop correction to the effective potential can also be trusted, see section~\ref{sec:der-exp-2loop}. Therefore, the nucleation action up to NLO (in the soft expansion), containing complete one-loop effects, can be written as
\begin{align}
    \Snucl[\varphib] &\simeq 
    \SthreeSoft [\varphibLO ] 
    + \frac32\log (\detOXzero[\varphibLO])
    + \frac32\log (\detOX[\varphibLO])
    + \intr \VDRNLO[\varphibLO].
\end{align}
Term $\intr \VDRNLO[\varphibLO]$ contains the NLO soft corrections to the potential from the two-loop gauge effects. Note that this definition of the effective action misses the $\ZX$-factors, 
as now these NLO corrections to the action are included in the functional determinants, without resorting to the derivative expansion. 
Equivalently, we can also use the LO action if we subtract the gauge contributions to avoid double-counting
\begin{align}
    \Snucl[\varphib] \simeq& 
    \SDRLO - \Cvec\\
    &+ \frac32\log (\detOXzero[\varphibLO])
    + \frac32\log (\detOX[\varphibLO])
    + \intr \VDRNLO[\varphibLO] 
    ,
\end{align}
where $\Cvec \equiv \CX + \CXtemporal$ denotes the zero-momentum, one-loop contribution from spatial and temporal gauge modes, cf.~eq.~\eqref{eq:CXdefinition}.\footnote{
Formally, we have not yet defined $\CXtemporal$, but following the logic of eq.~\eqref{eq:CXdefinition}, it is of course $\CXtemporal\equiv \intr \VLOtemporal[\varphibLO]$.
}
These expressions allow for completely removing the large errors from scale-shifting gauge modes. To use it in the calculation of the nucleation rate, we need to consider normalisation to the false vacuum. Following \cite{Kierkla:2025qyz}, we will use the shorthand notation for the ratio of gauge determinants
\begin{align}
\label{eq:detX0}
	\detXzero &\equiv
    \biggl(
        \frac{
            \detOXzero[\varphibLO]
        }{
           \detOXzero[\varphiF]
        }
    \biggr)^{-\frac32}, \\
\label{eq:detX}
    \detX &\equiv 
    \qty(
    \frac{\det \OpX[\varphibLO] }{\det \OpX[\varphiF] }
    )^{-\frac32} 
  .
\end{align}
The $\detOX$ determinant here was also a shorthand notation defined in eq.~\eqref{eq:detOX}. It contains a product of determinants of mixing gauge-Goldstone modes, transverse gauge modes, and ghosts. When evaluated on the bounce solution, the product is  
\begin{align}
    \detOX[\varphibLO] =
    \operatorname{det} \OpXG[\varphibLO]
    \times
    \operatorname{det} \OpXT[\varphibLO]
    \times
    \operatorname{det} \Opg [\varphibLO].
\end{align}
Then, employing our notation for the ratios of determinants, we write explicitly the total contribution from the vector degrees of freedom, Goldstone modes (of the new scalar field $\phi$) and ghosts,
\begin{equation}\label{eq:vecdet}
\detV \equiv
    \detXzero
    \mathrm{det}_{\rm{g}}
    \operatorname{det}_\rmii{$X_T$}
    \operatorname{det}_\rmii{$XG$}
  .
\end{equation} 
where the second term $\mathrm{det}_{\rmi{g}}$ denotes the contribution from the ghosts, and the third $\mathrm{det}_{\rmii{$X_T$}}$ from the transverse polarisation of the spatial gauge mode, neither of which mixes with the other components.
The last term $\mathrm{det}_{\rmii{$XG$}}$ denotes the contribution from the Goldstones and spatial gauge fields, whose contributions mix (see section~\ref{sec:compDet}).
For a non-constant background field, this mixing cannot be removed by a gauge choice~\cite{Ai:2020sru}, and we have checked that the contribution from the mixed determinant is significant \cite{Kierkla:2025qyz} (compared to the approximation used in~\cite{Ekstedt:2023sqc}, where the off-diagonal terms are dropped). This is one of the important results of this thesis. 
We compute its contribution using the Gel'fand--Yaglom method, following~\cite{Ekstedt:2021kyx}, and we summarise the computation in great detail in appendix~\ref{sec:compDet}. The contributions from the temporal gauge modes are computed with {\tt BubbleDet}~\cite{Ekstedt:2023sqc}. 
The contributions from ghosts and the transverse polarisation of the spatial gauge mode are also computed using our Gel'fand--Yaglom implementation; however, using certain gauge choices, one could compute them using {\tt BubbleDet} as well.

Since now we have obtained the NLO nucleation action, we can also write the scalar fluctuation determinant, cf.~eq.~\eqref{eq:Oscalar_LO}. 
The scalar fluctuations contribution is thus defined as
\begin{align}\label{eq:detS_softEFT}
  \detS &\equiv
	\mathcal{I}_\phi 
     \qty|
        \frac{\detOSp[\varphibLO]}{\detOS[\varphiF]} 
     |^{-\frac12}
  \,,&
  \mathcal I_\phi &= \left(\frac{ \SDRLO[\varphibLO] }{2\pi}\right)^{3/2}
  \,,&
  \mathcal O_\phi &= - \partial^2 + (\VLO)^{\prime\prime}
  \,,
\end{align}
here, $\varphiF$ denotes the false vacuum (corresponding to $\vthree = 0$).
Prime on the determinant denotes removal of zero-modes. Indeed, as discussed in chapter \ref{chapter:TFT}, the scalar fluctuation operator contains zero modes that would make the resulting determinant singular. Therefore, one needs to remove them first by going to collective coordinates (see, e.g., \cite{Gould:2023ovu}), which results in the Jacobian factor, here denoted as $\mathcal I_\phi$. 
We remind that in 3d spatial dimensions, there are three zero-modes, thus the ratio of determinants in $\detS$ is dimensionful, it has units of $[T^3]$ while other determinant ratios are dimensionless. 
For practical purposes, the scalar determinant can be computed using publicly available code \texttt{BubbleDet} \cite{Gould:2023ovu}.

Finally, we can compute the statistical part of the nucleation rate at NLO in soft expansion with fluctuation determinants. Including the nucleation action and the prefactor, we obtain \cite{Kierkla:2025qyz, Gould:2021ccf}
\begin{align}
\label{eq:Astat_NLOdet}
\Astat &= 
\detS
~
e^{
    -\Delta \SthreeSoft [\varphibLO]
    -\log\detV
    -\int_{\vec{r}} \Delta \VDRNLO[\varphibLO]
} 
,
\end{align} 
where $\Delta \VDRNLO[\varphibLO] = \VDRNLO[\varphibLO] - \VDRNLO[\varphiF]$. 



\subsection{Dynamical part of nucleation rate}
To obtain the complete thermal nucleation rate, we also need to include the dynamical factor $\Adyn$. Its role is to capture out-of-equilibrium effects, such as dissipative effects (see chapter \ref{chapter:TFT}). 
Assuming Langevin dynamics, the dynamical factor is related to the rate of growth of the critical bubble by a formula
\begin{align}
    \Adyn = \frac{1}{2\pi} \sqrt{\abs{\lambda_-} + \frac{\eta^2}{4} -\frac{\eta}{2}},
\end{align}
where $\lambda_-$ is the negative eigenvalue of the scalar fluctuation determinant we have shown in eq.~\eqref{eq:detS_softEFT}. The associated negative eigenmode of the scalar fluctuation then corresponds to isotropic growth or shrinking of the bubble \cite{Ekstedt:2023sqc, Ekstedt:2022tqk}. Thus, one can obtain $\lambda_-$ easily while calculating the scalar determinant.

The other parameter, $\eta$, is the so-called Langevin damping coefficient. Calculation of this parameter requires solving real-time dynamics of the system, and thus, it is unavailable with the EFT construction we have used so far. 
Therefore, for this work, we will work in a ``no-damping'' approximation \cite{PhysRevLett.46.388, RevModPhys.62.251} and set $\eta=0$, so the dynamical part of the nucleation rate becomes:
\begin{align} \label{eq:Adyn_eta0} 
    \Adyn \simeq \frac{\sqrt{\abs{\lambda_-}}}{2\pi}. 
\end{align}
This approximation is also implemented in the code \texttt{BubbleDet}.

\subsection{Final formula}
The nucleation rate is a product of statistical and dynamical parts. Taking the results from the last two subsections, we arrive at the final formula for thermal nucleation rate \cite{Kierkla:2025qyz}
\begin{align}
\label{eq:GammaT_detNLO}
    \GammaT = 
    \frac{\sqrt{\abs{\lambda_-}}}{2\pi}
    \times
    \detS
    ~
    e^{
        -\Delta \SthreeSoft [\varphibLO]
        -\log\detV
        -\int_{\vec{r}} \Delta \VDRNLO[\varphibLO]
}. 
\end{align}
This is also the most advanced approach used in this thesis, and it presents a current state-of-the-art computation of nucleation rate at NLO in soft expansion. It captures all relevant contributions to a given order, and it minimises the errors from the breakdown of the derivative expansion. The approach that we have shown here was derived for a supercooled transition in SU(2)cSM; however, note that it can be used for any kind of cosmological phase transition. Indeed, from the point of view of 3d EFT, there is no sharp distinction between a parent theory with or without classical scale-invariance. 
Our study emphasises that even for the worst-case scenario, i.e. of scale-shifters in a theory with a massless scalar in 4d, the error can be under control if we abandon a naive use of derivative expansion. 
Thus, the approach represented in eq.~\eqref{eq:GammaT_detNLO} is the most important theoretical result in this thesis. Now, let us further investigate its theoretical consistency. 

\subsection{Theoretical consistency}
In the next chapter, our overall goal will be to use the 3d EFT framework to compute physical, thermodynamical quantities, like the nucleation or percolation temperature. Therefore, the main object of our theoretical interest -- the nucleation rate -- must be both gauge-invariant and renormalisation scale invariant. Here we will discuss these two topics, following \cite{Kierkla:2025qyz}.

\subsubsection*{Gauge-independence}
Gauge-invariance of the nucleation rate, computed by including the contributions from the fluctuation determinants, was demonstrated for the ``SU(2)+Higgs'' model in ref.~\cite{Baacke:1999sc}. 
This is basically the same model as the dark sector of SU(2)cSM, which we have considered here, but there is a difference in a sign of scalar self-coupling. 
The authors of ref~\cite{Baacke:1999sc} illustrate that the gauge fluctuation determinant splits into contributions from physical modes, and from other modes that eventually get cancelled by the Faddeev-Popov ghosts. This cancellation happens when the determinant is evaluated on the solution to the equations of motion, i.e., bounce solution, thus finally leaving no dependence on the gauge fixing parameter $\xi$. 
Other demonstrations of the gauge-independence of the nucleation rate can be found in~\cite{Endo:2017gal}, and in derivative expansion in~\cite{Lofgren:2021ogg, Hirvonen:2021zej, Metaxas_1996}.
The latter references concretely demonstrate that the factor
$e^{-\int_{\vec{x}} V_3^{\rmii{NLO}}[\varphib]}$ is gauge invariant,
in accord with Nielsen-Fukuda-Kugo identities~\cite{Nielsen:1975,Fukuda:1975}.
Following~\cite{Ekstedt:2021kyx}, in our evaluation of the vector determinant, we exploit the gauge-independence to choose the gauge with $\xi=1$ in which the Goldstone-vector mixing becomes more manageable;
see appendix~\ref{sec:compDet}.

\subsubsection*{Renormalisation scale invariance}
The running of the 3d EFT starts at two-loop order, i.e.~the 3d renormalisation scale appears explicitly for the first time in the $\VDRNLO$ term (see chapter \ref{chapter:TFT} for more detailed discussion in a toy model). 
Therefore, working with the action containing only the one-loop effects would result in uncancelled RG-scale dependence, as it is still present implicitly in the LO action of EFT, in the matching relation of the mass $\mthree$ parameter.
Thus, we should include the logarithms at two loops for consistent and reliable results ~\cite{Gould:2021}. In the case of SU(2)cSM, these logarithms are included in the term
$ \VDRNLO$ 
in eq.~\eqref{eq:Astat_NLOdet}.
These two-loop corrections were previously shown in eq.~\eqref{eq:VeftNLO} within the derivative expansion.
The inclusion of the full two-loop corrections without resorting to the derivative expansion is a difficult task due to the inhomogeneous scalar background.%
\footnote{%
  Such computation involves two-loop diagrams with numerically determined
  propagators using Green's functions akin
  to~\cite{Bezuglov_2019,Ekstedt:2022tqk,Dashko:2024spj}.
} 
However, by utilising the super-renormalisability of the 3d EFT, we can still extract the zero-momentum part of two-loop corrections and include the two-loop logarithm containing the renormalisation scale.
To illustrate this, let us now consider a soft expansion of the effective action in the following form
\begin{align}
  -\ln \frac{\Astat}{T^3} &=
  \Seff = 
        \mathcal{C}_0
      + \varepsilon \; \mathcal{C}_1
      + \delta \mathcal{C}_{\frac{3}{2}}
      + \mathcal{O}(\varepsilon^2)
      \,,
\end{align}
where the soft expansion parameter is 
$\varepsilon \sim \gx/\pi$, and the specific terms are:
\begin{itemize}
    \item $\mathcal{C}_0 \equiv S_{\rm 3}^{\rmii{LO}}[\varphib]$
    is the leading order action,
    \item 
    $ 
        \mathcal{C}_1 \equiv  
        \int_{\mathbf{r}} \mathcal{X}[\varphib] 
         + \qty( \log \det_{\rmii{$V$}} -\Cvec[\varphib]  )
    $ 
    contains one-loop effects above LO in soft expansion, and also includes the unknown complete two-loop contribution $\mathcal{X}[\varphib]$
    (which would be calculated without utilising the derivative expansion),
    \item and $\mathcal{C}_{\frac{3}{2}} \equiv - \log \det_{\rmii{$S$}}$ captures the scalar determinant.
\end{itemize}
The total two-loop contribution $\mathcal{X}[\varphib]$ is suppressed compared to LO by the soft expansion parameter, and within the derivative expansion, the leading term is
$\mathcal{X}[\varphib] \approx \VDRNLO[\varphib]$. Similarly, the non-zero momentum part of the one-loop contribution is suppressed within the derivative expansion:
$$
\log \detV - \Cvec[\varphib] \approx \frac{1}{2} Z^{\rmii{NLO}}_3(\varphib)(\partial_i \varphi)^2.
$$
However, note that the above term is suppressed by $\varepsilon \sim \gx/\pi$ in general, without the derivative expansion. After all, these are separate expansions. 
The breakdown of derivative expansion means that we do not control contributions at the tail of the bounce solution, as they seem divergent. At the same time, there is no reason why soft expansion should behave differently in the bulk of the bubble and at the tail. 
Correctness of soft-expansion justifies our treatment, where we have included $\Cvec[\varphi]$ in the LO action, where it contributes to the bounce, and then when we subtract the zero-momentum gauge contribution, the resulting leftover part of the determinant, i.e. $\log \detV- \Cvec[\varphib]$, is suppressed compared to LO. 
For the scalar fluctuations, we have used an independent expansion parameter $\delta$, but since we know how the scalar contributions scale compared to the soft expansion, i.e.\ $\delta \sim \varepsilon^{\frac{3}{2}}$, we can bundle together these two expansions~\cite{Gould:2023ovu}.

The 3d renormalisation scale enters the soft-expansion expressions in dimensional regularisation when we are renormalising the UV divergence of the two-loop sunset diagram through mass renormalisation.
As a consequence, the 3d mass parameter runs with the renormalisation scale
$\muthree$, i.e.\ $\mthree^2 = \mthree^2(\muthree)$ with scaling
\begin{align}
    \muthree \dv{\mthree^2}{\muthree} = 
    \varepsilon \beta_{\mthree^2} 
    + \mathcal{O}(\varepsilon^2),
\end{align}
where the $\beta$-function is given by%
\begin{align}
  \beta_{\mthree^2} = \frac{1}{(4\pi)^2}\left(
    \frac{39}{16} \gxthree^4
  - 6 \gxthree^2 \hthree
  +  \frac{3}{2} \hthree^2 \right)
  \,.
\end{align}
Explicit dependence on $\ln \muthree$ appears at NLO in $\mathcal{C}_1$-term through
two-loop contribution $\mathcal{X}[\varphi]$, as we have stressed before.
The contributions
$\mathcal{C}_0$ and
$\mathcal{C}_{\frac{3}{2}}$ do not depend explicitly on $\muthree$,
but only implicitly through the mass, as well as the bounce, which depends implicitly on the mass. 
Physical quantities are independent of the renormalisation scale, i.e.
\begin{align}
\label{eq:RG-invariance}
 \dv{}{\log \muthree} \Bigl( -\log \frac{\Astat}{T^3} \Bigr) = 0
  \,.
\end{align}
Since this holds order by order in $\varepsilon$,
we get an identity at $\mathcal{O}(\varepsilon)$
\begin{align}
\muthree \frac{\partial}{\partial \muthree} \mathcal{C}_1
= 
\muthree \frac{\partial}{\partial \muthree} \mathcal{X}[\varphib]
=
  -\beta_{\mthree^2} \Big(
      \frac{\partial}{\partial \mthree^2} \mathcal{C}_0
    + \frac{\partial \varphib}{\partial \mthree^2} \underbrace{\frac{\partial}{\partial \varphib} \mathcal{C}_0}_{=0} \Big) 
=  -\beta_{\mthree^2}  \frac{\varphib^2}{2} 
  \,,
\end{align}
where the derivative of $\mathcal{C}_0$ vanishes since the bounce extremises the LO action (and we have left integration over space implicit).
Hence, we conclude that the two-loop result has the following formal structure
\begin{align}
\label{eq:ActionRGOnly} 
\mathcal{X}[\varphib]
&\simeq
      \mathcal{A}
      -\beta_{\mthree^2} \Big(
          \frac{1}{2} \varphib^2 \Big)
        \ln \frac{\muthree}{M_i(\varphib)}
\,,
\end{align}
where $\mathcal{A}$ is a yet undetermined contribution, which does not involve $\ln \muthree$,
and $M_i$ formally correspond to gauge field masses in soft EFT.
The point of this discussion is the realisation that in our perturbative setup,
the UV structure,
i.e.\ the logarithms of the renormalisation scale are not affected by whether or not we apply the derivative expansion, but are completely dictated by the running of the mass within the EFT.

Thereby,
all logarithmic terms of $\mathcal{X}[\varphib]$ have to align with those in $\VDRNLO[\varphib]$, and we have
\begin{align}
\label{eq:VeftNLO-do-derivative-expansion}
\begin{split}
\mathcal{X}[\varphib] &= 
\mathcal{A}
    + \frac{1}{(4\pi)^2}\frac{3}{64} \gxthree^2 
    \Bigl(
          2 \bigl(
                80 \mXthree^2 - 3 \gxthree^2 \vthree^2
            \bigr) 
            \ln\frac{\muthree}{2 \mXthree}
      \Bigr)
      \\ &
  + \frac{1}{(4\pi)^2} 
  \biggl\{
        -\frac{3}{8} \hthree^2 \vthree^2 
        \Bigl(
            2 \log \frac{\muthree}{2 \mXtemporal}
        \Bigr)
        - \frac{3}{2} \gxthree^2 
        (
            \mXthree^2 - 4 \mXtemporal^2
        ) \log \frac{\muthree}{2 \mXtemporal + \mXthree }
  \biggr\} 
  \\ &
  - \frac{1}{(4\pi)^2} 
  \biggl\{
      \frac{3}{2} \gxthree^2 \mDX^2 
      \Bigl(4 \log \frac{\muthree}{2 \mDX} \Bigr)
  \biggr\}
  \,.
\end{split}
\end{align}
This is an expression for the two-loop soft contribution without the derivative expansion. However, since $\mathcal{A}$ is left undetermined, in our approach, cf.~eq.~\eqref{eq:GammaT_detNLO}, we simply resort to a replacement
$\mathcal{X}[\varphib] \rightarrow \VDRNLO[\varphib]$,
i.e.\
we employ the derivative expansion within the two-loop result, as we have shown previously that this contribution is under control on the tail of the bounce solution.

\section{Numerical results for nucleation rate in SU(2)cSM}
\label{sec:nucl_rate_recipes}
Now, we will discuss various approaches to the computation of the bubble nucleation rate that we have described so far. Then, we will show the resulting differences between them, directly for the effective action. 
In the next chapter, after discussing the overall picture of the cosmological transition in SU(2)cSM, we will also discuss the impact on thermodynamical parameters of the transition, and predictions for gravitational wave spectra and observational prospects in future experiments.

Let us now summarise all the approaches we have described so far, and consider their variants, leading to different approximations of the thermal bubble nucleation rate. 
\begin{enumerate}[label=Approx.~\arabic{enumi}:]
\item
  \label{it:daisy}
  [daisy]
  \\
     Effective action is computed using the daisy resummation. The scalar determinant and dynamical prefactor are approximated on dimensional grounds by $T^4$,
    \begin{equation}
    	\GammaT^{[\rmii{daisy}]} \equiv 
        T^4 \cross e^{-\Delta S_{\rm daisy} [\varphib^\rmii{daisy}]} 
      \,,
    \end{equation}
    where the action $S_{\rm daisy}$ is 4-dimensional, and is given as:
    \begin{align}
        S_{\rm daisy}[\varphib^\rmii{daisy}] 
        \equiv 
        \frac{1}{T} \intr 
        \frac12(\partial \varphib^\rmii{daisy})^2 
        + V_4(\varphib^\rmii{daisy}, T),
    \end{align}
    where $V_4$ is the 4d effective potential with daisy-resummation term; as defined in eq.~\eqref{eq:V4_T}. Note that here $\varphib^\rmii{daisy}$ is the $O(3)$-symmetric stationary point of the $S_{\rm daisy}$ action. The normalisation to the metastable phase is given as $\Delta S_{\rm daisy} [\varphib^\rmii{daisy}] =
    S_{\rm daisy} [\varphib^\rmii{daisy}] - S_{\rm daisy} [\varphiF]$. 
    This approach is the most commonly used in the literature; however, it is not robust as it suffers from many theoretical uncertainties, see chapter \ref{chapter:TFT} or ref.~\cite{Croon:2020cgk} for a review.

\item
  \label{it:LO}
  [LO]
  \\[1mm]
  Effective action is the LO of the nucleation EFT (see eq.~\eqref{eq:Veff-EFT-LO}), then the scalar functional determinant and dynamical prefactor are
  approximated on dimensional grounds by $T^4$.
  In the construction of the soft EFT, the matching to the 4d theory is performed at
  two-loop level for the masses, and one-loop level for the couplings.
  Here, the gauge one-loop functional determinants are computed by using the
  derivative expansion and keeping the terms up to and including $\mathcal{O}(\gx^3)$.
  This approach reduces to the [daisy] if the matching relations in the construction of the EFT are
  truncated to the LO, see appendix \ref{app:sec:compare3D4D} for details. 
  Nucleation rate is given by
	\begin{equation}
	\GammaT^{[\rmii{LO}]} = 
    T^4  
    \cross 
    e^{-\Delta\SDRLO[\varphibLO ]}
    ,
\end{equation}
with $\SDRLO[\varphibLO ]$ defined in eq.~\eqref{eq:SLO_bounce}, see also section \ref{sec:LOnuclrate}.

\item
  \label{it:NLO-grad}
  [NLO~$\nabla$]
  \\[1mm]
  Effective action is computed within the nucleation EFT, with soft NLO corrections, in the derivative expansion, see eq.~\eqref{eq:Astat_NLOgrad}.
  The scalar functional determinant and dynamical prefactor are
  approximated on dimensional grounds by $T^4$.
  The effective potential contains one- and two-loop gauge field contributions
  (up to and including $\mathcal O(\gx^4)$). The kinetic term in the action receives
  a correction from a field-dependent NLO part of the $Z_\rmii{3}$-factor defined in eq.~\eqref{eq:Z3_factor_total}. This is the approach used in~\cite{Kierkla:2023von}.
  Nucleation rate is given by
  \begin{equation}
	\GammaT^{[\rmii{NLO $\nabla$}]} =
    T^4
       \times
        e^{-\Delta\SDRLO[\varphibLO] -\Delta\SDRNLO[\varphibLO]}
      ,
  \end{equation}
  with $\SDRNLO[\varphibLO]$ defined in eq.~\eqref{eq:Seff_NLO_grad}.
  This approach relies fully on the derivative expansion and suffers from errors introduced by the scale-shifting gauge modes, see section \ref{sec:uncertainties_scale-shifters} for details. Nevertheless, it is more theoretically consistent than the previous two approaches, as it is both RG-scale invariant and gauge-invariant, and includes all relevant terms up to $\mathcal O(\gx^4)$ in soft expansion.

\item
  \label{it:NLO-det}
  [NLO det]
  \\[1mm]
  The effective action is computed within the nucleation EFT, but the one-loop contributions are obtained by evaluating functional determinants,
  supplemented with soft NLO corrections to the effective potential computed in the gradient expansion.  
  We have defined the statistical part of this approach in eq.~\eqref{eq:Astat_NLOdet}.
  Note that here the momentum-dependent NLO correction is accounted for by the functional determinants, without resorting to the derivative expansion.
  The scalar fluctuation determinant is included, as well as
  the dynamical prefactor evaluated in no-damping approximation using {\tt BubbleDet}, see eq.~\eqref{eq:Adyn_eta0} 
  This is the most accurate approach considered in this work, and
  that is available in the literature so far \cite{Kierkla:2025qyz}. 
  \begin{equation}
    \label{eq: Gamma [NLO det]}
    \GammaT^{[\rmii{NLO det}]} =
    \frac{\sqrt{\abs{\lambda_-}}}{2\pi}
    \times
    \detS
    ~
    e^{
        -\Delta \SthreeSoft [\varphibLO]
        -\log\detV
        -\int_{\vec{r}} \Delta \VDRNLO[\varphibLO]
    }. 
      \,,
  \end{equation}
  with $\detS$, $\detV$ defined in
  eqs.~\eqref{eq:detS_softEFT}--\eqref{eq:detX0} and
  $\VDRNLO[\varphib]$ given in eq.~\eqref{eq:VeftNLO}. This approach is both RG-scale invariant and gauge-invariant, and includes all relevant terms up to $\mathcal O(\gx^4)$ in soft expansion.
  
\begin{enumerate}
\item
  \label{it:NLO-det-T4}
  [NLO det $T^4$]
  \\[1mm]
  In this approach, instead of including the full scalar determinant prefactor and the dynamical prefactor, the prefactor is approximated as $T^4$. Doing so will allow us to study the importance of the scalar prefactor by comparing it with the full [NLO det] method. The thermal nucleation rate is given by
  \begin{equation}
    \GammaT^{[\rmii{NLO det\ }T^4]} = 
    T^4 \times
    \detS \times
    e^{
        -\Delta \SthreeSoft [\varphibLO]
        -\log\detV
        -\int_{\vec{r}} \Delta \VDRNLO[\varphibLO]
    }. 
  \end{equation}
\end{enumerate}

\end{enumerate}
By exhausting these various approaches, we will illustrate which contributions to the thermal nucleation rate are dominant and most important to include to obtain robust predictions for the thermodynamical parameters of a phase transition.

\subsection{\texorpdfstring{Comparing resummation methods\footnote{%
A perhaps more explicit title would be ``Comparing approaches for computing nucleation rate''. However, authors of this dissertation wanted to refer to (classic in certain scientific communities) article by J. Löfgren titled \textit{``Stop comparing resummation methods''}~\cite{Lofgren:2023sep}.%
}}{Comparing resummation methods}}

\begin{figure}[h]
    \centering
    \includegraphics[width=0.7\linewidth]{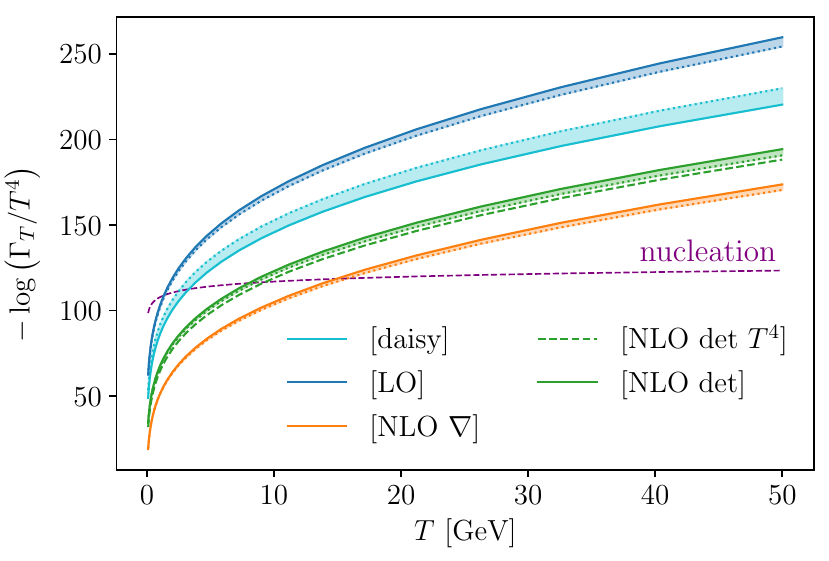}
    \caption{
        Different approaches for computing  $\GammaT/T^4$ for a benchmark point with
        $\gx = 0.8$ and $\mX = 10^4$~GeV.
    		Bands illustrate the sensitivity of different approaches to the choice of 4d RG scale at
        $\mufour = \pi T$ (solid) and
        $\mufour = 7 T$ (dotted). 
    		The \hyperref[it:NLO-det-T4]{[NLO~det~$T^4$]} curve is evaluated at $\mufour = 7T$.
    		The 3d scale is set to $\muthree=T$. 
    }
    \label{fig:action-sample}
\end{figure}

In figure~\ref{fig:action-sample}, we show the logarithm of the nucleation rate (normalised by $T^4$) as a function of temperature. Note that this quantity corresponds directly to the effective action in the nucleation EFT. We compute the action using the methods listed before, for a fixed benchmark point with
$\gx = 0.8$ and
$\mX = 10^4$~GeV, see section~\ref{sec:su2csm_thermal_scale_running} for a discussion of free parameters in SU(2)cSM.
%
%
This allows us to compare the effect of corrections to the action and the scalar determinant in the exponential prefactor.
Moreover, one can easily evaluate the impact of various approximations on the nucleation temperature by looking at the intersection of the $-\log( \GammaT/T^4)$ lines with the approximate nucleation-criterion line (dotted purple) for which
 $\Gamma/H^4=1$.%
\footnote{%
  This is an approximate nucleation criterion more consistent with the scenario of the phase transition proceeding during a phase of thermal inflation
  than the commonly adopted criterion $S_3/T\approx 140$. 
}
In figure\ref{fig:action-sample} we present the approximations to nucleation rate with the following colour coding:
\begin{itemize}
    \item \ref{it:daisy} \hyperref[it:daisy]{[daisy]}: turquoise,
    \item \ref{it:LO} \hyperref[it:LO]{[LO]}: blue,
    \item \ref{it:NLO-grad} \hyperref[it:NLO-grad]{[NLO~$\nabla$]}: orange,
    \item \ref{it:NLO-det} \hyperref[it:NLO-det]{[NLO~det]}: green,
    \item \ref{it:NLO-det-T4} \hyperref[it:NLO-det-T4]{[NLO~det~$T^4$]}: dashed green.
\end{itemize}
The solid line corresponds to the choice of 4D RG-scale
$\mufour = \pi T$, while dotted to $\mufour = 7 T$ within a given approach. 
Let us now comment on the various aspects of these results 
\paragraph{Convergence of soft-expansion}
First, one should note the substantial difference between the results obtained at different orders in the soft expansion:
\hyperref[it:LO]{[LO]} (blue) and
\hyperref[it:NLO-det]{[NLO~det]} (green). This demonstrates that it is mandatory to include higher-order corrections to the nucleation rate to obtain precise results for thermodynamic parameters of the phase transition. 
The large magnitude of NLO corrections in comparison to LO is a consequence of the slow convergence of perturbation theory in Thermal Field Theory. 
However, as the full NNLO calculation is beyond the scope of this work, and the lattice results for models with classical scale invariance are unavailable\footnote{To our knowledge, there is no study of numerical simulation of nucleation rate in 3d in a model with negative quartic coupling. However, recently there were simulations of nucleation in 1+1 dimensions with negative quartic coupling \cite{Pirvu:2024nbe, Hirvonen:2025hqn}.} 
we are unable to confirm the convergence of soft-expansion in SU(2)cSM. Some encouraging results for BSM theories can be found in~\cite{Niemi:2020hto, Gould:2023ovu, Niemi:2024axp, Ekstedt:2024etx},
where, despite large corrections at NLO, the perturbative convergence is very good, and the NLO result is close to the NNLO result. 
It should, however, be stressed that these comparisons were made for equilibrium quantities, and not for the nucleation rate.

\paragraph{Comparison with daisy-resummation}
Next, comparing the green \hyperref[it:NLO-det]{[NLO~det]} and the turquoise \hyperref[it:daisy]{[daisy]} curves, one can observe the significant difference between the simplest approach and the most advanced one.
The nucleation temperature shifts from approximately 3.3~GeV, when computed using the daisy-resummation method, to about 8~GeV when using 3d EFT at NLO with functional determinants. This further emphasises the insufficiency of the daisy approach.
On the other hand, the difference between
\hyperref[it:daisy]{[daisy]} (turquoise) and
\hyperref[it:LO]{[LO]} (blue) is solely caused by a different treatment of the matching relations when constructing the high-temperature 3d EFT. Daisy resummation corresponds to truncated matching relations, as discussed in detail in~\cite{Kierkla:2023von}.  
The \hyperref[it:daisy]{[daisy]} curve is closer to the
\hyperref[it:NLO-det]{[NLO~det]} curve than
\hyperref[it:LO]{[LO]}, which might seem counter-intuitive,
as the matching relations of
\hyperref[it:LO]{[LO]} and
\hyperref[it:NLO-det]{[NLO~det]} are identical. 
It should be stressed that, in principle, there is no correlation between the higher-order corrections in the hard-scale matching relations and the NLO corrections to the effective action in the nucleation EFT. Therefore, it does not have to be the case that they move the resulting curves
``in the same direction'', i.e.~matching at higher orders might increase the nucleation temperature and higher-order corrections in the EFT might decrease it, as seen here.
On the other hand, it is well known that using the EFT approach with higher-order matching relations results in reduced RG-scale dependence in comparison to the daisy-resummation approach, and we indeed observe such behaviour in figure~\ref{fig:action-sample}, where the width of the bands indicates the sensitivity to the RG-scale. The width of \hyperref[it:daisy]{[daisy]} curve is much larger than that of the rest of the EFT curves. 
\paragraph{Errors from scale-shifters }

Now, let us focus on the comparison of the two approaches to the effective action at NLO that we have described in previous sections. First, we have 
the full one-loop effective action, with the fluctuation determinants included
(\hyperref[it:NLO-det]{[NLO~det]}), and then there is
the NLO action computed completely relying on derivative-expansion~\cite{Kierkla:2023von}
(\hyperref[it:NLO-grad]{[NLO~$\nabla$]}).
As using the derivative expansion introduces errors from the scale-shifters, we observe that this method
overestimates the nucleation rate, resulting in even larger differences with respect to \hyperref[it:daisy]{[daisy]} and \hyperref[it:LO]{[LO]}. Moreover, when computing nucleation temperature using \hyperref[it:NLO-grad]{[NLO~$\nabla$]} method, we would underestimate its value by around 5 GeV. We will discuss in more detail the implications of scale-shifters on thermodynamical parameters in the following sections.
Thus, even though the inclusion of the NLO corrections to the action turns out to be crucial for a precise determination of the nucleation rate, when treated within the derivative expansion, it can indeed induce significant uncertainties, and robust predictions require addressing the breakdown of the derivative expansion.

\paragraph{Significance of scalar determinant }
It is commonly assumed that the scalar prefactor constitutes a subleading contribution to the nucleation rate,
cf.\ e.g.~\cite{Lofgren:2021ogg}.
In ref.~\cite{Ekstedt:2021kyx, Ekstedt:2023sqc}
it was shown that in certain scenarios,
however, the prefactor can be important.
In figure~\ref{fig:action-sample} we see that for our benchmark point, the scalar prefactor has a sub-dominant effect on the final result, as the solid (\hyperref[it:NLO-det]{[NLO~det]}) and dashed green lines (\hyperref[it:NLO-det-T4]{[NLO~det~$T^4$]}) are very close to each other. This is in agreement with our expectations following from soft expansion power counting, but we will see in the next subsection that the difference in predictions (with and without the scalar determinant) for percolation temperature can be as large as 40\% for other benchmark points.

\chapter{Phenomenology of supercooled phase transition}
\label{chapter:pheno}

A cosmological, first-order phase transition proceeds through a nucleation, growth, and percolation of bubbles of the broken-symmetry phase, in the Universe still in the symmetric phase.
This bubble nucleation process is driven by the field's quantum tunnelling, or thermal ``jumping'' over a barrier in the effective potential.
In this chapter, we will describe the phase transition in the SU(2)cSM model, starting with the general picture, and then describing all the necessary parameters and mechanisms of the generation of the gravitational wave signal. 
For the calculation of the parameters and GW spectra, we perform a numerical scan of the parameter space of the model. Finally, we will discuss the importance of theoretical consistency for phenomenological predictions. We will compare the most advanced EFT approach presented in this work with the ``daisy-resummation'' approach, which by far is the most common in the literature. 

This chapter thus contains the most important results of this thesis for the phenomenology of supercooled phase transitions. 
Most of the results presented here are also presented in \cite{Kierkla:2022odc, Kierkla:2023von, Kierkla:2025qyz, Kierkla:2025vwp} and here are accompanied by some new, unpublished results.

\section{Transition picture in SU(2)cSM}
\begin{figure}
    \centering
    \includegraphics[width=0.7\linewidth]{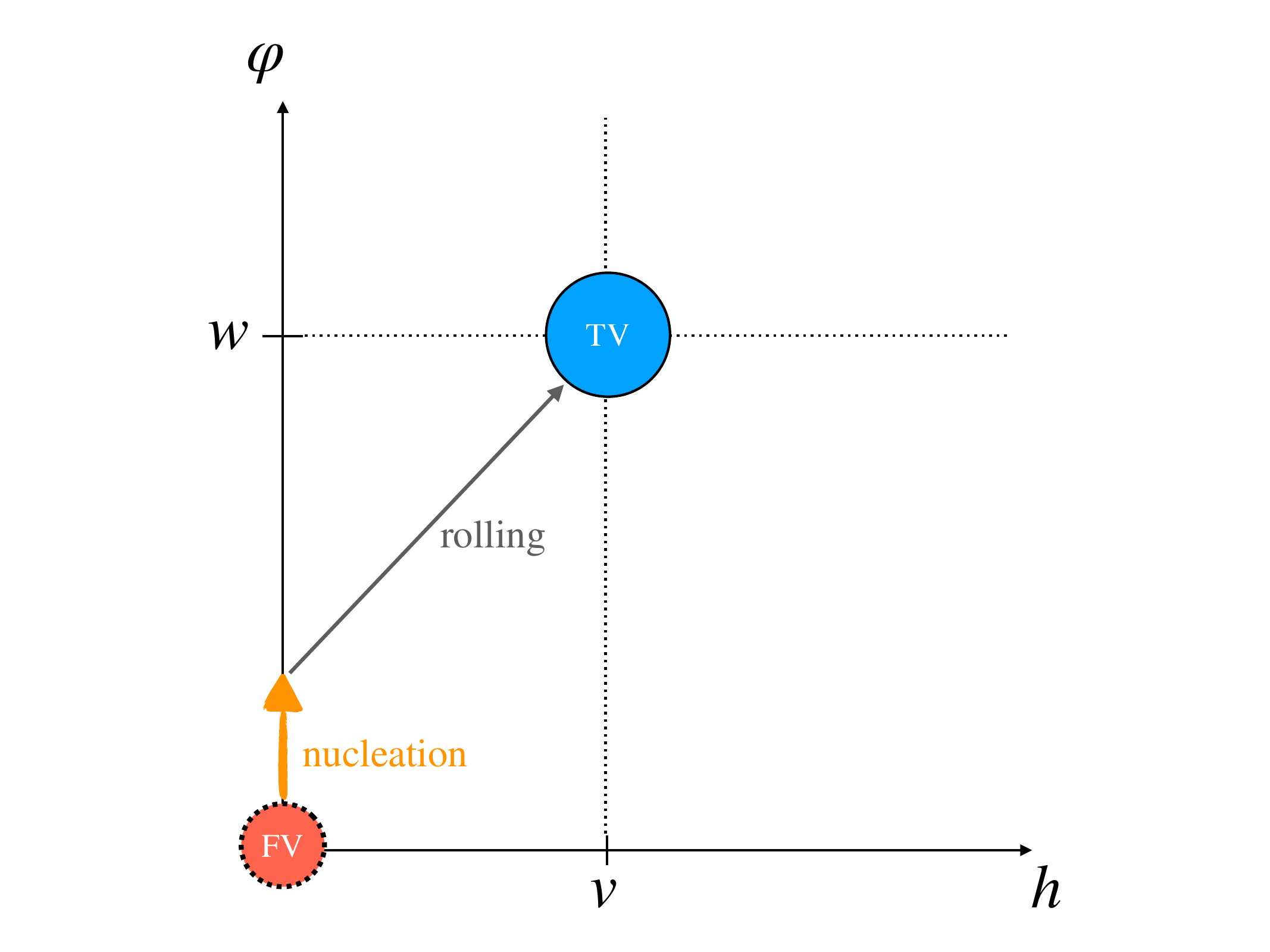}
    \caption{Schematic depiction of the phase transition in SU(2)cSM. Here, the red circle with a dashed contour and label FV corresponds to ``false vacuum'', while the blue circle with a solid contour and label TV corresponds to ``true vacuum''. The orange arrow illustrates the nucleation in the dark sector, while the grey arrow shows the subsequent evolution of the fields inside the growing bubble.}
    \label{fig:transition_su2csm_scheme}
\end{figure}

In principle, in SU(2)cSM, the tunnelling proceeds in the full field space of $h$ and $\f$, i.e.\ we should solve for the critical bubble solution in the two-dimensional field space. 
However, in the case of classically scale-invariant extensions of the Standard Model, the vev hierarchy may be large and the tunnelling happens mostly in the ``dark sector'', i.e., only in the direction of the new scalar field in the field space. 
For the SU(2)cSM model, we have checked \cite{Kierkla:2022odc} with the use of \texttt{CosmoTransitions}~\cite{Wainwright:2011} that for a sample of benchmark points  the Higgs field $h$ value remains zero in the critical bubble solution.\empty
\footnote{
    By zero, we mean that the background value of Higgs in the 2-dimensional bounce solution was numerically close to zero. For the inverted scalars' mass hierarchy, it could be expected that contributions in Higgs directions become larger; however, we do not expect them to affect the nucleation rate drastically.
} 
This means that indeed, the thermal nucleation proceeds only along the $\f$ direction, while the transition in the $h$-direction proceeds via continuous rolling, and happens inside the growing bubble, see figure~\ref{fig:transition_su2csm_scheme} for a schematic depiction. 
Assuming that this holds for the full parameter space, i.e.\ that only the $\f$ direction is relevant for the tunnelling, allows us to obtain better accuracy and simplifies the use of effective theory description for bubble nucleation.\empty
\footnote{ Nevertheless, it is possible to construct the EFT and perform all numerical calculations for the two-field case.}
This was also a common assumption in earlier works, such as~\cite{Hambye:2018qjv, Baldes:2018} and was also checked in ref.~\cite{Prokopec:2018}.
Note that, as was pointed out in ref.~\cite{Prokopec:2018}, this is incompatible with the Gildener--Weinberg approach \cite{Gildener:1976ih}, which only analyses the potential along the direction from the origin of the field space to the true minimum.

Another important remark is that in models with classical scale invariance, the onset of the phase transition can be delayed so much that the QCD phase transition proceeds first, see e.g. ~\cite{Iso:2017uuu, vonHarling:2017, Hambye:2018qjv, Schmitt:2024pby, Sagunski:2023ynd}. Then the quark condensate forms at $T_{\mathrm{QCD}} \approx 0.1\g$, see e.g.~\cite{Braun_2006} and by coupling to the Higgs boson generates an effective mass term in the potential, thus changing the mechanism and dynamics of the phase transition. This kind of behaviour was analysed, e.g.\ in refs.~\cite{Hambye:2018qjv, Baldes:2018,Ellis:2020, Schmitt:2024pby, Sagunski:2023ynd, Kierkla:2025vwp}.
In the present work, we do not consider the QCD-sourced PT and focus on the delayed electroweak phase transition caused by thermal fluctuations. 
%

In what follows, we review the phase transition dynamics in SU(2)cSM, defining the thermal ``milestones'', and obtaining the parameters characterising the phase transition. 
For the computations of parameters, we use self-developed code, optimised for the case of a classically scale-invariant potential. We perform a scan of the parameter space, taking as a starting point the result of the zero-temperature scan discussed in section~\ref{sec:zero-temp}.

\section{Parameters of phase transition in SU(2)cSM}
\label{sec:ptparams_SU2cSM}
In this section, we will discuss the parameters that characterise the phase transition. We will show the parameter space scans in SU(2)cSM for each of them. These parameters are needed to obtain the phenomenological predictions, such as the resulting stochastic gravitational-wave background. 

\subsection{Thermal milestones}

\paragraph{Critical temperature} 
At high temperatures of the hot early Universe, the symmetry is restored, and the effective potential has a single minimum at the origin of the field space. For small field values, it has a parabolic shape, as it is dominated by quadratic thermal corrections. As the Universe cools down, a second minimum is eventually formed. At the critical temperature $\Tc$, the two minima are degenerate, and for lower temperatures, the minimum with broken symmetry becomes the true vacuum. This is the temperature at which the tunnelling from the symmetric phase to the broken phase becomes possible. See also figure~\ref{fig:VPT}.

\begin{figure}[h!t]
\center
\includegraphics[width=.8\textwidth]{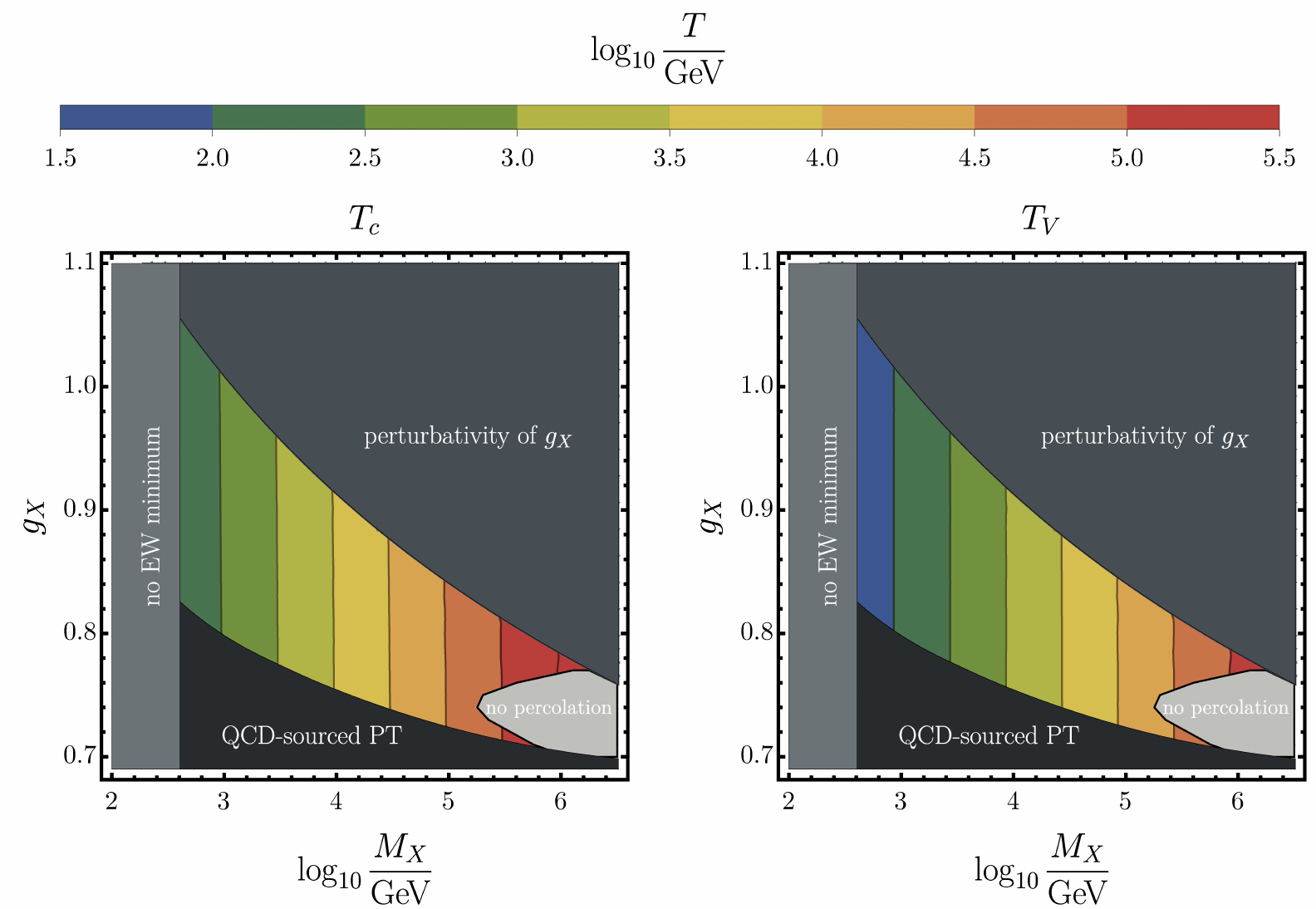}
\caption{The values of the critical temperature $\Tc$ (left panel) and the temperature at which thermal inflation starts $\Tv$ (right panel). Figure taken from \cite{Kierkla:2022odc}.
}
\label{fig:Tc-TV}
\end{figure}

The value of the critical temperature can be obtained by finding a global minimum $w$ of the effective potential $V_4(\varphi, T)$ (as defined in eq.~\eqref{eq:V4_T}), and demanding that the energy difference between two minima, given by\footnote{
In principle, the symmetric minimum $\varphi=0$ is inside the high-temperature regime, thus the free energy there could be substantially affected by higher-order thermal corrections and should be evaluated within the HT EFT. However, since the energy difference is dominated by the effective potential in the true vacuum, omitting these corrections would lead to a negligible error in the value of $\Tc$. Moreover, the value of $\Tc$ does not affect any phenomenological observables of supercooled phase transitions, as the temperature when the transition completes is orders of magnitude below $\Tc$. Later, we will show that explicitly. 
}
\begin{align}
\label{eq:DeltaV_V4}
    \Delta V_4 (T) \equiv \qty|V_4(w, \Tc)  - V_4(0, \Tc)|,
\end{align}
is zero at the critical temperature, i.e., 
\begin{align}
     \Delta V_4 (\Tc) = 0.
\end{align}
The values of the critical temperature obtained numerically for the SU(2)cSM model are shown in the left panel of figure~\ref{fig:Tc-TV}. 
The critical temperature is closer to the electroweak scale $T_c\approx 100$ GeV for a lower mass of the $X$-boson. 
For higher values of $\MX$, the critical temperature becomes significantly higher than in the standard picture of the electroweak phase transition, reaching $10^5-10^6$ GeV for the largest allowed $\MX$. Looking at figure \ref{fig:scalar-couplings}, we can also see that this effect is a consequence of the gradual ``decoupling'' of the dark SU(2) sector from the SM sector as the portal coupling gets smaller and smaller.
Figure~\ref{fig:Tc-TV} contains the same excluded areas as before (see the discussion of figure~\ref{fig:scalar-couplings}), and two new shaded regions. 
The lower left corner (darkest grey) is not analysed because there the PT is sourced by the QCD phase transition, which is beyond the scope of the present work. We will comment on the phenomenology of such transitions below.
The light-grey region around $M_X\approx 10^6\g$ is where the percolation criterion of eq.~\eqref{eq:shrink_criterion} is violated and is discussed in more detail later.

\paragraph{Temperature at which thermal inflation starts}
%
If there is large supercooling, i.e.\ the completion of phase transition is delayed to low temperatures, much below the critical temperature, then eventually energy stored in the false vacuum will trigger a period of thermal inflation. 
It happens because the stored energy, which we will denote $\rho_V$, starts to dominate over the radiation energy density $\rho_R$. This can be explicitly shown by considering the constituents in the Hubble parameter:
\begin{align}
    H^2 = \frac{1}{3 \bar{M}^2_{\text p}} (\rho_R + \rho_V) 
        = \frac{1}{3 \bar{M}^2_{\text p}} \qty(\frac{T^4}{\xi^2_g} +\Delta V),
\end{align}
where $\xi^2_g = 30/(\pi^2 g_*)$ and $g_* = 116.75$ denotes the number of relativistic degrees of freedom (dof) in the plasma,\empty
\footnote{Effective number of dofs $g_*$ at some temperature $T_*$  can be assumed to approximately be the same as the number of entropy degrees of freedom $g_* \simeq g_{*s}$ under the condition that $T_* > 0.1 \rm ~MeV$. Thus, we calculate them as $g_* = 106.75 + 9 + 1 = 116.75$, where 106.75 is the value from the SM, 9 corresponds to $X$-boson dofs and 1 to the new scalar.}
while $\Delta V$ is the difference of the values of the effective potential at both vacua (at the given temperature $T$), and $\bar{M}_{\text p} \simeq 2.4 \cross 10^{18} \mbox{ GeV}$ is the reduced Planck mass.
We see that if the temperature drops to a value such that vacuum energy density gives the dominant contribution to the Hubble parameter, it will result in a vacuum domination period, and thus trigger a period of rapid expansion of the Universe.
The onset of the period of thermal inflation can be estimated as the temperature, which we will denote as $T_V$, at which vacuum and radiation contribute to the energy density equally 
\begin{align}
\label{eq:T-thermal-inflation}
 \Tv \equiv \qty(\xi^2_g \overline{\Delta V} )^\frac{1}{4},
\end{align}
where $\overline{\Delta V} \equiv \Delta V(\Tv)$, i.e., it is given by eq.~\eqref{eq:DeltaV_V4} at $\Tv$. 
Then, by knowing the temperature $T_V$, the Hubble constant can be approximated conveniently:
\begin{align}
H^2_\text{RV} \equiv \frac{1}{3\bar{M}_{\rm{Pl}}^2 \xi^2_g}
\left(
    T^4+T_V^4
\right).
\end{align}
This approximation works very well for SU(2)cSM, and we implement it in our calculations. Moreover, for supercooled transitions, it is also a good approximation to assume that the energy difference is constant below $T_V$ $\Delta V (T<\Tv) \simeq \overline{\Delta V}$. Then, in this limit, the contribution to the Hubble parameter from radiation energy can also be neglected, leading to the simplest approximation of the Hubble parameter \cite{Ellis:2018},
\be
H^2 \simeq H^2_V=\frac{1}{3\bar{M}_{\rm{Pl}}^2}\overline{\Delta V}.
\ee
The values of $\Tv$ obtained for the parameter space of SU(2)cSM are presented in the right panel of figure~\ref{fig:Tc-TV}. Vacuum domination begins at temperatures around an order of magnitude below the critical temperature and lasts until the completion of the phase transition. 

\paragraph{Nucleation temperature} 
Below the critical temperature, nucleation of bubbles of true vacuum becomes possible due to thermal fluctuations or quantum tunnelling. In practice, quantum tunnelling is always subleading in comparison to thermal effects.
Knowing the thermal bubble nucleation rate at a given temperature for SU(2)cSM, we can calculate the \textit{nucleation temperature}, which we will denote as $\Tn$. 
For the calculation of the nucleation rate, we use the most advanced approach described in this work, i.e. \hyperref[it:NLO-det]{[NLO~det]} as defined in eq.~\eqref{eq: Gamma [NLO det]}.
Nucleation temperature is then obtained by estimating the moment when at least one bubble is nucleated per Hubble volume \cite{LINDE198137, LINDE1983421} 
\begin{align} 
\label{eq:Tn_full}
N(T_n) = 
\int^{t}_{t_c} \dd{t} \frac{\Gamma(T)}{H^3(t)} =
\int_{\Tn}^{\Tc} \frac{dT}{T} \frac{\GammaT(T)}{H(T)^4}
=1,
\end{align}
where in the second step we have used the time-temperature adiabatic relation
\begin{align}
    \dv[]{t}{T} = -\frac{1}{T}\frac{1}{H(T)}.
\end{align}
This moment can be interpreted as the onset of the supercooled phase transition.\footnote{Technically, the transition is possible already at the critical temperature. However, the nucleation rate for the temperatures close to $\Tc$ is infinitesimally small in the case of strong supercooling.}
It should be noted that the common criterion for evaluating $\Tn$ as $\Seff(T_\text{n})\approx 140$ is not reliable in the case of strongly supercooled transitions, as it relies on the assumption of radiation domination at the moment of the phase transition (see e.g.~\cite{Baldes:2018}). 
Instead, a better approximation is given by assuming that nucleation happens instantaneously and demands $\GammaT \simeq H^4$. Now, by writing the nucleation rate in a simple form $\Gamma \simeq T^4 e^{-S}$, one arrives at a simpler condition
\begin{align}
    S(\Tn) = 4\log{\qty(\frac{\Tn}{H})},
\end{align}
which can easily be solved numerically. In our numerical analysis, however, we use the complete criterion given in eq.~$\eqref{eq:Tn_full}$.

\paragraph{Percolation temperature} 
After successful nucleation, bubbles begin to grow until they start colliding with each other, leading to percolation. 
Percolation of bubbles can be considered as the moment of completion of the phase transition, see e.g.~\cite{Ellis:2018, Athron:2022mmm}. 
At the same time, this temperature, which we will denote as $T_p$, is also usually associated with the temperature relevant for the production of gravitational waves. We will call this temperature $\Tst$.
In this thesis, we also choose the percolation temperature as the characteristic temperature of the PT $T_* \equiv \Tp$, at which we will evaluate the parameters relevant for the GW generation. Therefore, in this notation, all the quantities with an asterisk in the subscript are evaluated at $\Tp$ temperature.\footnote{ On the other hand, if there is no supercooling, it is a reasonable approximation to set $T_* = \Tn$, see e.g.~\cite{Lewicki:2021pgr}.} 

For the calculation of the percolation temperature, we follow refs.~\cite{Ellis:2018, Ellis:2020, Kierkla:2022odc}.
The probability of finding a point still in the false vacuum at a certain temperature is given by $P(T) = e^{-I(T)}$, where $I(T)$ is the amount of true vacuum volume per unit comoving volume and, in our case, reads as follows \cite{Guth:1981} 
\begin{align} \label{eq:I}
		I(T) = \frac{4\pi}{3} \int_{T}^{T_c} \dd{T^\prime} \frac{1}{T^{\prime 4}}\frac{\GammaT(T^\prime)}{ H(T')}\qty(\int_{T}^{T'} \frac{\dd{\tilde T} }{ H(\tilde T)})^3.
\end{align}
To simplify this expression, we can distinguish between the vacuum and radiation domination periods by expressing the Hubble parameter in the following form \cite{Ellis:2018}
\begin{align} \label{eq:HR_and_HV}
	H(T) \simeq 
\left\{\begin{array}{ll}
	H_{\mathrm{R}}(T)=\frac{T^{2}}{\sqrt{3} \bar{M}_{\mathrm{pl}} \xi_{g}}, &\quad \mathrm{for}\quad T > T_{V}, \\
	H_{\mathrm{V}}=\frac{T_{V}^{2}}{\sqrt{3} \bar{M}_{\mathrm{pl}} \xi_{g}}, &\quad \mathrm{for}\quad T < T_{V}.
\end{array}\right.
\end{align}
Inserting \eqref{eq:HR_and_HV} into \eqref{eq:I}, the integral can be split into two parts, which allows us to write a simplified version of $I(T)$ valid in the region where $T<T_V$ \cite{Ellis:2018}:
\begin{align}
\label{eq:I_RV}
\begin{split}
I_{\mathrm{RV}}(T) \equiv 
\frac{4 \pi}{3 }
\left[
\int_{T_V}^{T_c} \frac{d T^{\prime} \Gamma\left(T^{\prime}\right)}{T^{\prime 6} H_{\mathrm{V}}^4} T_V^2\left(2 T_V-T-\frac{T_V^2}{T^{\prime}}\right)^3
+\int_T^{T_V} \frac{d T^{\prime} \Gamma\left(T^{\prime}\right)}{T^{\prime} H_{\mathrm{V}}^4}\left(1-\frac{T}{T^{\prime}}\right)^3
\right].
\end{split}
\end{align}
When studying strongly supercooled phase transitions, the nucleation rate per Hubble volume is usually infinitesimal above $\Tv$. Therefore, in practice, we can thus drop the integral over $T>\Tv$ in  eq.~\eqref{eq:I_RV} and write a simplified version of $I(T)$ for SU(2)cSM:
\begin{align}
\begin{split}
I_{\rm RV}(T) \simeq
\frac{4 \pi}{3 }
\left[
    \int_T^{T_V} \frac{d T^{\prime}}{T^{\prime}} \frac{\GammaT\left(T^{\prime}\right)}{ H_{\mathrm{RV}}^4}\left(1-\frac{T}{T^{\prime}}\right)^3
\right] .
\end{split}
\end{align}
For numerical purposes, it is also convenient to divide the integration range from $T$ to $\Tn$ and from $\Tn$ to $\Tv$.
Finally, estimation of the temperature $T_p$ is given by a following criterion~\cite{Guth:1981, Athron:2022mmm}
\begin{align} \label{eq:Tp34_criterion}
    I(\Tp) = 0.34,
\end{align}
which can be derived semi-analytically using percolation theory \cite{Vinod:1971}. This criterion can be understood as follows: statistically equal-sized and randomly distributed spherical bubbles percolate if they contain 34\% of the whole three-dimensional Euclidean space volume. Which, in the context of cosmology, means that 34\% of the Universe is already in the new, true vacuum state.\footnote{
It should be emphasised, though, that a concrete percolation criterion is not known for a de Sitter space. Thus, we resort to the aforementioned criterion, assuming it should be an approximation to the correct value during thermal inflation. }
We numerically solve for the value of the percolation temperature using 
\be
\label{eq:percolation-temp}
I_{\rm RV}(T_p) = 0.34.
\ee
%
\begin{figure}[h!t]
\center
\includegraphics[width=.4\textwidth]{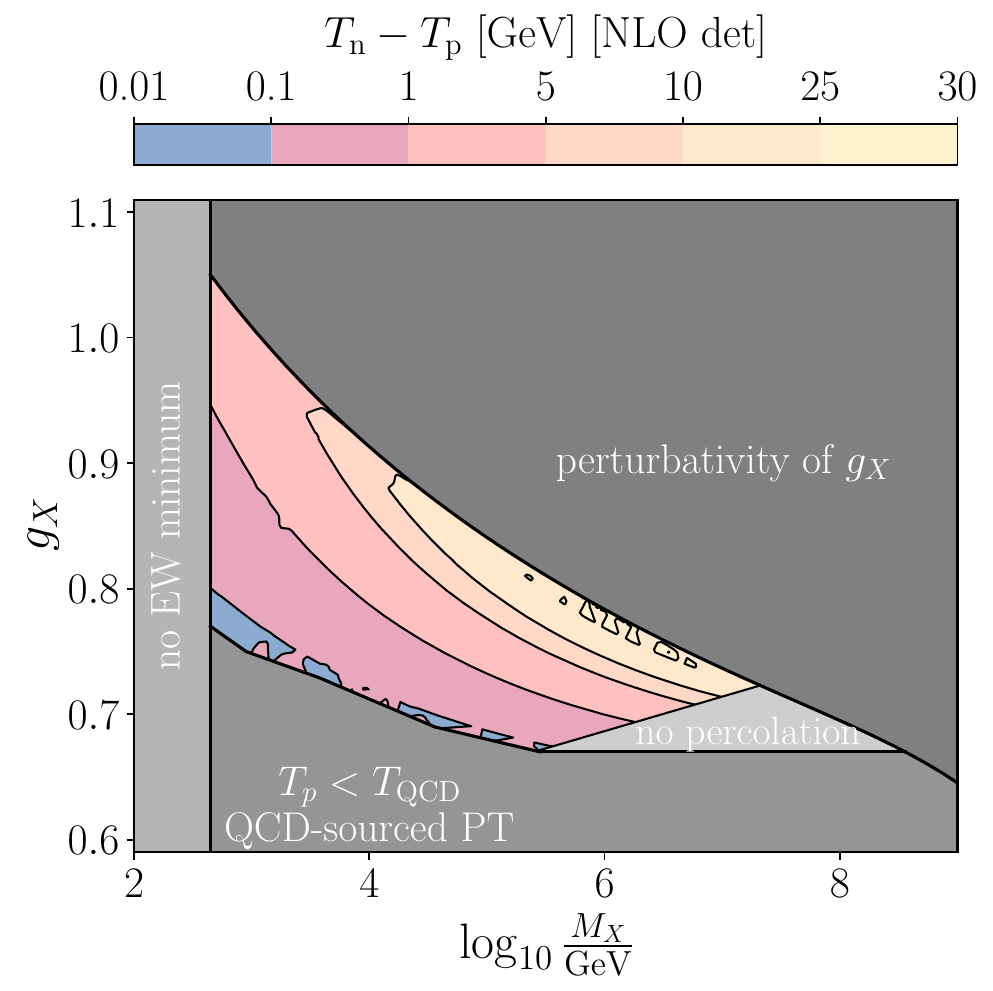}
\includegraphics[width=.4\textwidth]{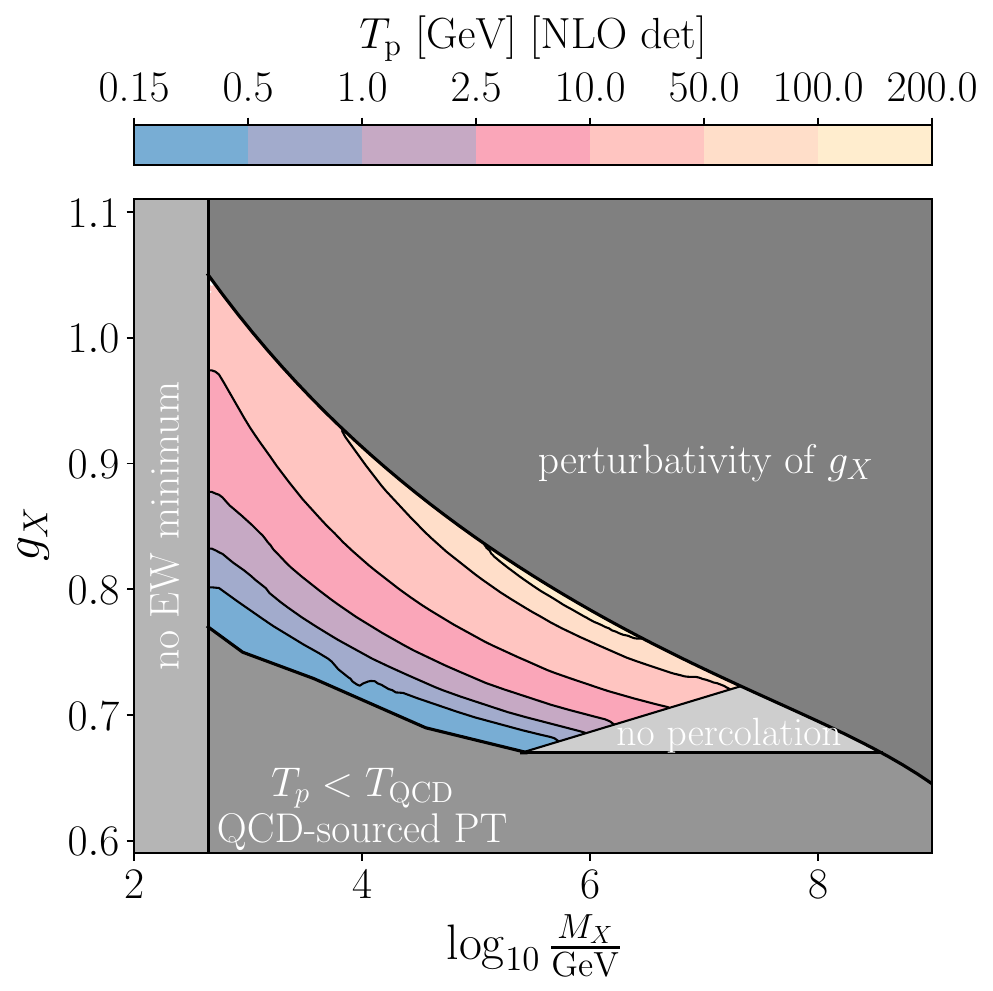}
\caption{The difference between the nucleation temperature $\Tn$ and percolation temperature $\Tp$ (left panel) and the values of percolation temperature $\Tp$ (right panel). 
}
\label{fig:Tn-Tp}
\end{figure}
The results of the scan of the parameter space for $T_p$ can be found in figure~\ref{fig:Tn-Tp}, on the right panel. 
Moreover, we obtain the values of the nucleation temperature $\Tn$  by numerically solving eq.~\eqref{eq:Tn_full} and in the left panel of figure~\ref{fig:Tn-Tp} we show the difference between the values of $\Tn$ and $\Tp$. One can see that these two temperatures are of the same order of magnitude, but they do differ, which is in agreement with our previous arguments. 
One can also appreciate the degree of supercooling present in the studied model by comparing the percolation temperature with the critical temperature $\Tc$ -- the former is orders of magnitude below the latter, see figure~\ref{fig:Tc-TV}.
That is a common feature of classically scale-invariant models, and it results in the production of a strong GW signal due to the large value of $\Delta V$ at percolation, as we will show in the next sections.
Moreover, we find that in the whole available parameter space $\Tv > \Tp$, which means that thermal inflation is always present.

\paragraph{Temperature $\Tshrink$ at which the volume of false vacuum is decreasing}
In a case of strongly supercooled transitions, it is important to ensure successful percolation, as in principle the thermal inflation could prevent the bubbles of true vacuum from filling the entire space. This dictates an additional condition to fulfil to correctly estimate $\Tp$, as one should also ensure that at the estimated percolation temperature the volume of valse vacuum $V_\rmii{FV} \sim a^3 P(T)$ is shrinking \cite{Ellis:2018}
\begin{align} 
\label{eq:shrink_criterion}
    \frac{1}{V_\rmii{FV}} \dv{V_\rmii{FV}}{t_p}\eval_{t_p} = 
    3H(t_p)-\dv{I(t)}{t}\eval_{t_p} = 
    H(T_p)\qty(3+T_p\dv{I(T)}{T}\eval_{T_p}) <0.
\end{align}
SU(2)cSM exhibits a region where this criterion is not satisfied \cite{Kierkla:2022odc, Kierkla:2023von, Kierkla:2025qyz, Kierkla:2025vwp}. It is shown in all figures illustrating the parameter-space scan in light grey (lower-right part of the plots). 
In this region, the phase transition does not complete via bubble percolation at $\Tp$. It may happen later still, as there might exist a temperature $\Tshrink$ at which the volume $V_\rmii{FV}$ starts to shrink \cite{Kierkla:2025vwp, Ellis:2018}  before the QCD phase transition. However, in such a case, there is no criterion to estimate the correct value of $\Tp$, and there is no robust way of setting a value of $\Tst$. 
The inclusion of the percolation criterion changes the picture of the phase transition, as compared to earlier works that studied the SU(2)cSM model in the context of the PT~\cite{Hambye:2018qjv, Marfatia:2020, Baldes:2018} but did not investigate percolation (except for ref.~\cite{Baldes:2018} but there it has been numerically evaluated only for a single point in the parameter space).

Then, it is also possible that the Universe keeps inflating until the onset of the QCD phase transition, and then the PT completes because of the appearance of the quark condensate \cite{vonHarling:2017, Iso:2017uuu, Schmitt:2024pby, Sagunski:2023ynd, Marzo:2018nov}. Then, this region would belong to the same class as the region in the lower part of our parameter space (denoted as QCD-sourced PT). For SU(2)cSM model, values of $\Tshrink$ are always below $\Tqcd$. Therefore, we consider the whole region where the condition in eq.~\eqref{eq:shrink_criterion} is violated to be excluded. For a detailed discussion, see ref.~\cite{Kierkla:2025vwp}.

If the thermal inflation lasts long enough, it may become possible to end the PT via quantum fluctuations, see ref.~\cite{Lewicki:2021xku}. In ref.~\cite{Lewicki:2021xku}, it was shown that before a phase transition is ended by quantum fluctuations, a substantially long period of thermal inflation takes place, and there is a discontinuous change in the number $N$ of $e$-folds of inflation between the region where percolation of bubbles ends the phase transition and the region where quantum fluctuations become significant. 
The first scenario was found to be realised for $N\sim\mathcal{O}(1)$--$\mathcal{O}(10)$, whereas the latter for $N\approx 20$--$50$. We can evaluate the number of $e$-folds until QCD PT by computing $N=\log\frac{\Tv}{\Tqcd}$~\cite{Hambye:2018qjv}. 
We find $N\approx 15$ at the right edge of the $M_X$ axis; therefore, we expect that throughout the presented parameter space, in the case of extended inflation, the phase transition would be eventually completed by QCD effects. 
Extrapolating the results of our scan we expect that the bound from perturbativity of $\gx$ and the region of QCD sourced PT should meet around $\MX \sim 10^8\g$, which would result in $N\approx 18$ so we do not expect the scenario of quantum fluctuations ending the PT to be realised (or possibly it may be realised in a very small part of the parameter space). The detailed study of the boundary between the two scenarios is beyond the scope of this work.

\paragraph{Reheating temperature $\Treh$} The total energy released in the phase transition is associated with the energy difference defined in eq.~\eqref{eq:DeltaV_V4}. To reheat the Universe after the phase of thermal inflation, the energy has to be transferred from the scalar field $\f$ to the relativistic plasma. If reheating is instantaneous, then by energy conservation, almost the whole energy $\Delta V$ is turned into the energy of radiation\footnote{A small portion of this energy is converted into gravitational waves.},
\be
\label{eq:condition-Tr-1}
\Delta V = \rho_R(\Treh)=\rho_R(\Tv),
\ee
where we assume that the number of relativistic degrees of freedom remains constant and use eq.~\eqref{eq:T-thermal-inflation}. This gives (see e.g.\ refs.~\cite{Ellis:2018, Hambye:2018qjv})
\be
\label{eq:Treh=TV}
\Treh=\Tv.
\ee
On the other hand, if at $T_p$ the rate of energy transfer from the $\f$ field to the plasma, $\Gamma_{\f}$, is smaller than the Hubble parameter, $\Gamma_{\f} < H(T_p)$, then the energy will be stored in the scalar field oscillating about the true vacuum and redshift as matter until 
$\Gamma_{\f}$ becomes comparable to the Hubble parameter. In this case, the reheating temperature will read~\cite{Ellis:2019, Hambye:2018qjv}
\be
\Treh=\Tv \sqrt{\frac{\Gamma_{\f}}{H_*}}  .
\label{eq:Treh-general}
\ee
The numerical results for $\Tv$ can be found in figure~\ref{fig:Tc-TV} (right panel). For a more refined treatment of the reheating temperature, see e.g.\ ref.~\cite{Ellis:2020}. 
%
Before the phase transition, the energy is stored in the $\f$ field. Thus, for assessing the efficiency of reheating, we should know the decay rate of the $\f$ field, which quantifies the energy transfer rate from $\f$ to the plasma. For details, see ref.~\cite{Kierkla:2022odc}. Moreover, in the literature~\cite{Hambye:2018qjv, Ellis:2020, Marfatia:2020} one can find various approaches to the computation of $\Gamma_{\f}$, which result in varying predictions for the reheating temperature. In ref.~\cite{Kierkla:2022odc}, these different approximations are discussed and compared. Here, we do not discuss the detailed procedure for computing $\Gamma_{\f}$; we focus on determining the region where reheating is instantaneous, using the values of $\Gamma_\phi$ obtained in \cite{Kierkla:2022odc}.

\begin{figure}
    \centering
    \includegraphics[width=.4\textwidth]{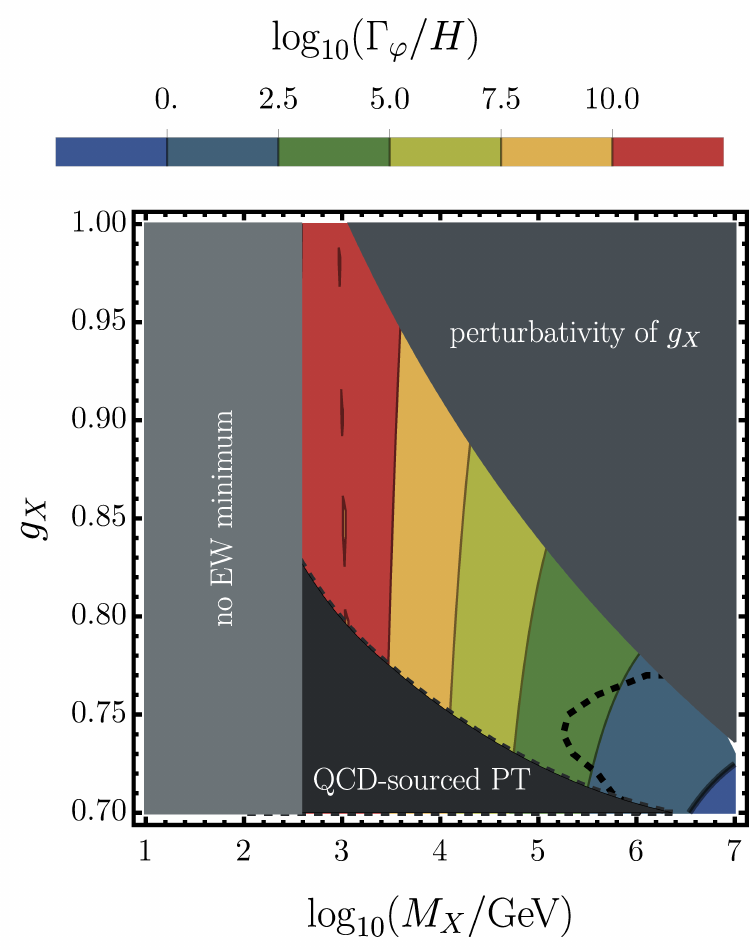}
    \caption{Contour plot of the decimal logarithm of the ratio of the energy transfer rate $\Gamma_{\f}$ to the Hubble parameter $H$. The equality $H=\Gamma_{\f}$ is indicated as a thick black solid line in the lower right corner. The percolation bound is shown as a black dashed line (in other plots, it is shown as a light-grey region). 
    }
    \label{fig:reheating}
\end{figure}

The obtained energy transfer rate $\Gamma_{\f}$ and the Hubble parameter $H$ should be compared to determine whether reheating is instantaneous. This comparison is presented in figure~\ref{fig:reheating}, which shows the logarithm of the ratio of $\Gamma_{\f}$ to $H$ throughout the parameter space. The two parameters, $\Gamma_{\f}$ and $H$, only become equal for large values of the $X$ mass (along the thick black line), in the region where the percolation condition of eq.~\eqref{eq:shrink_criterion} is violated (to the right of the black dashed line). This means that in the region where the phase transition proceeds due to nucleation and percolation of bubbles, which is the focus of the thesis, the reheating is always instantaneous and $\Treh=\Tv$ as stated in eq.~\eqref{eq:Treh=TV}. 


\subsection{Energy budget}

\begin{figure}[h]
\center
\includegraphics[width=.4\textwidth]{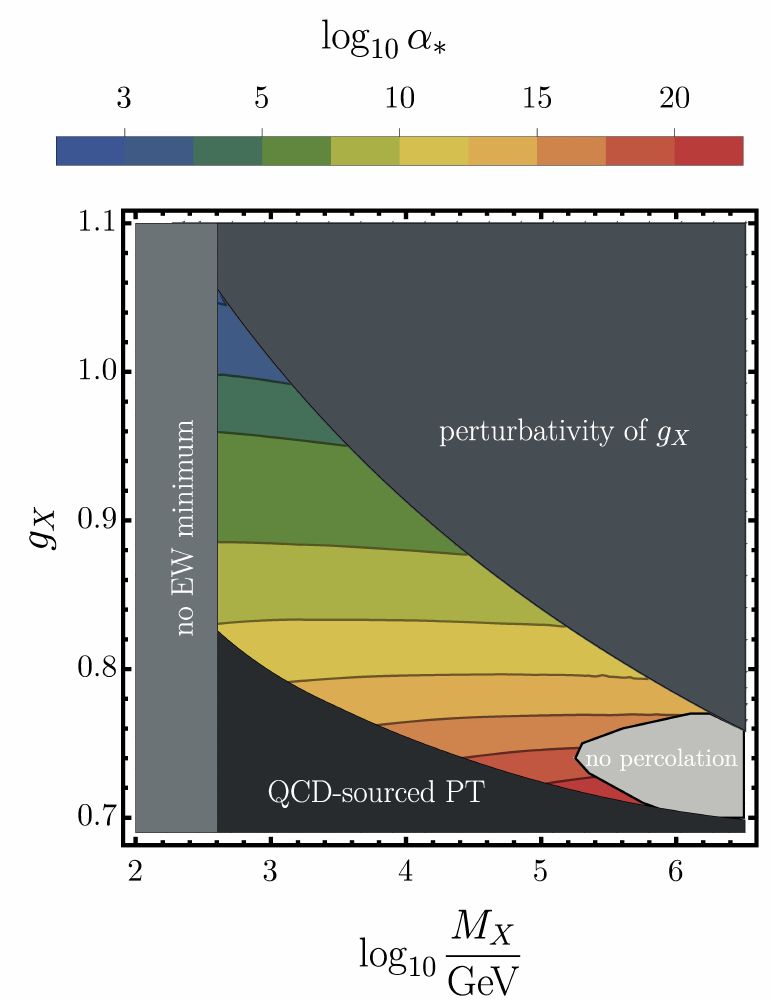}
\caption{The values of transition strength parameter $\alpha_*$. 
}
\label{fig:alpha}
\end{figure}

\paragraph{Transition strength} The latent heat released during the transition consists of the difference in free energy (or effective potential) and entropy variation. However, in the limit of large supercooling, $\Tp\ll \Tc$, the entropy contribution can be neglected~\cite{Espinosa:2010, Marzola:2017} as it is much smaller than the difference in energy. Latent heat is important for the gravitational wave signal, as the relevant quantity, called the transition strength, is the ratio of the latent heat to the energy density of radiation at the time of the transition, $\rho_R(\Tp)$,
\begin{align}
\alpha_* \equiv \frac{\Delta V}{\rho_R(T_p)}.
\label{eq:alpha}
\end{align}
The strength of the PT in the studied model is very large, varying in the range $10^3 - 10^{20}$, see figure~\ref{fig:alpha}. This is due to the large VEV of the new scalar field $\f$ and low percolation temperature.

\paragraph{The bubble-wall speed $v_w$} 
Since the PT in the studied model is extremely strong, as can be appreciated in figure~\ref{fig:alpha}, we can safely assume that the wall velocity is equal to the speed of light $v_w=1$. The reason is that due to the thermal inflation present, the surrounding plasma of particles is diluted, which implies that the friction force exerted on the bubble wall should be small. At the same time, the large amount of energy stored in the false vacuum implies a large driving force. Thus, the bubble should expand fast.
In terms of hydrodynamics, an expanding bubble in a supercooled Universe would always be a detonation \cite{Espinosa:2010, Krajewski:2024zxg, Krajewski:2023clt, Krajewski:2024gma}.

\begin{figure}[h!t]
\center
\includegraphics[width=.45\textwidth]{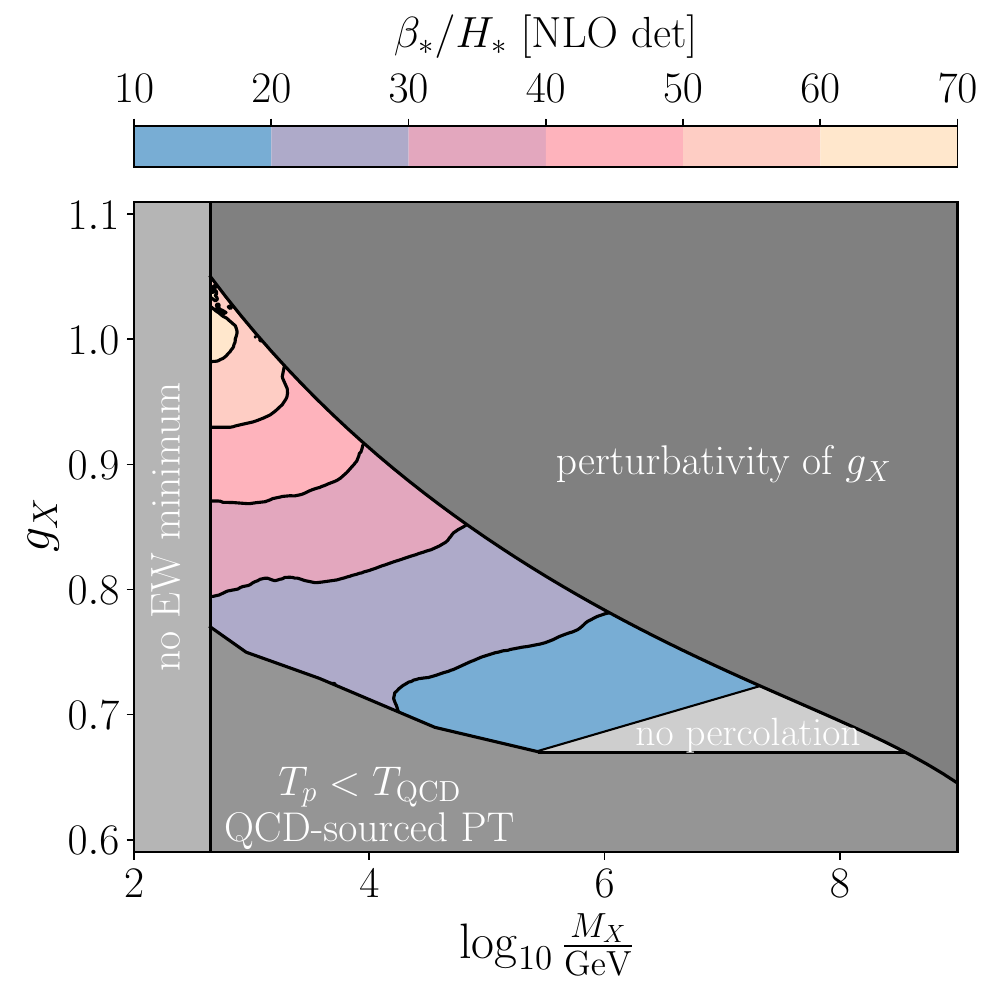}\hspace{20pt}
\includegraphics[width=.45\textwidth]{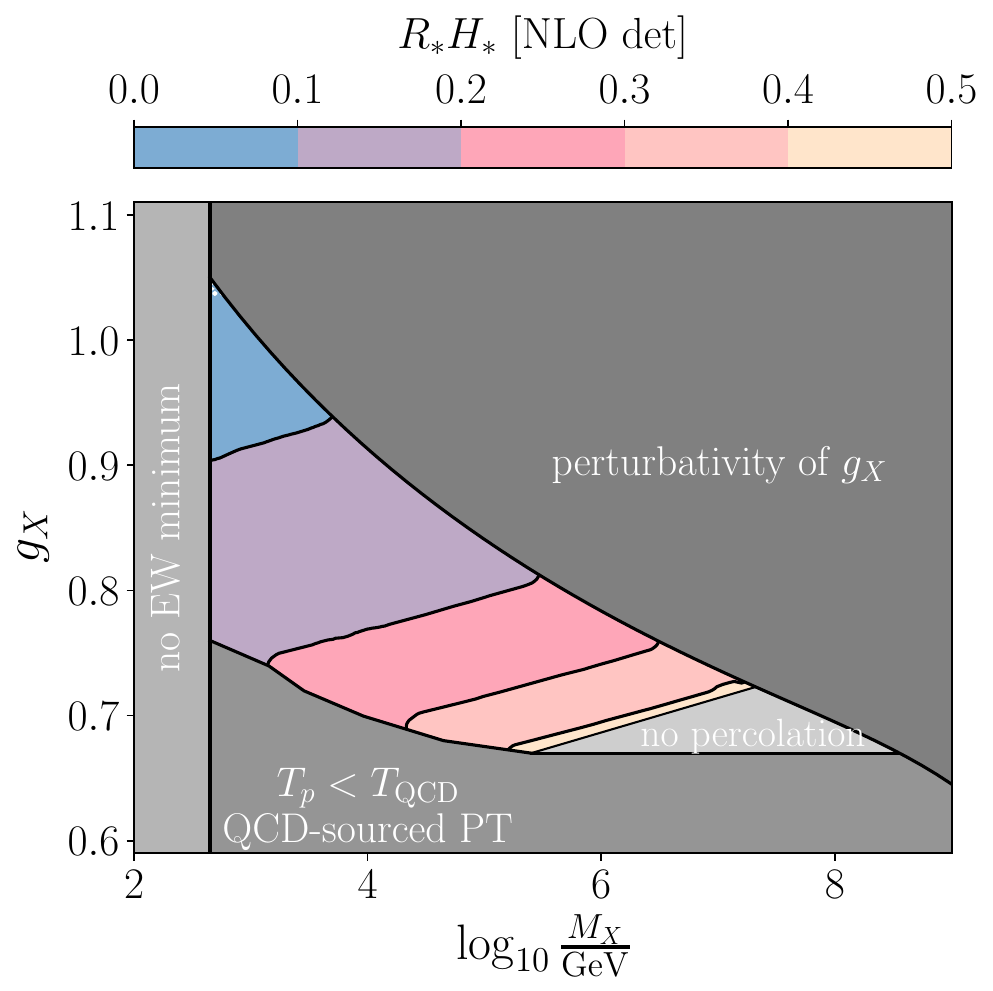}
\caption{The values of inverse time scale $\beta / H_*$ (left panel) and the length scale of the transition $R_*H_*$ (right panel). 
}
\label{fig:beta-RH}
\end{figure}

\paragraph{Inverse time scale of the transition $\beta$} 
Another of the parameters characterising the phase transition, important for the GW computations, is the (inverse) time scale of the transition. It is given simply by a linear fit to the nucleation rate
\be
\Gamma(t) \sim e^{-\beta t},
\ee
which translates to
\be
\frac{\beta}{H_*} = T_p  \qty(\eval{ \frac{\dd S_{\rmii{eff}}}{\dd T}}_{T=T_p}).
\label{eq:beta}
\ee
Since $\beta/H_*$ is computed as the derivative, it can become numerically unstable. In our scan, we paid special attention to smoothing the $\Seff$--curve, but still the results for $\beta/H_*$ computed based on eq.~\eqref{eq:beta} are less reliable than for the other parameters, see figure~\ref{fig:beta-RH} (left panel). Therefore, as an alternative, we will characterise the PT by its length scale, see below.
The plot showing the values of $\beta_*$, however, can help us understand the shape of the region excluded by the percolation criterion of eq.~\eqref{eq:shrink_criterion} as it follows the shape of lines of constant $\beta_*$. 
Parameter $\beta_*$ is given as the derivative of the Euclidean action divided by the temperature. The larger $\beta_*$, the faster $\Seff$ decreases with decreasing temperature, which means that the tunnelling rate increases quickly as the temperature decreases. This works in favour of percolation since a growing tunnelling rate can balance the expansion of the Universe due to thermal inflation. In contrast, if $\beta_*$ is smaller -- as it is near the region excluded by the percolation condition -- the decay rate of the false vacuum does not increase as fast with decreasing temperature. 

\begin{figure}
    \centering
    \includegraphics[width=0.5\linewidth]{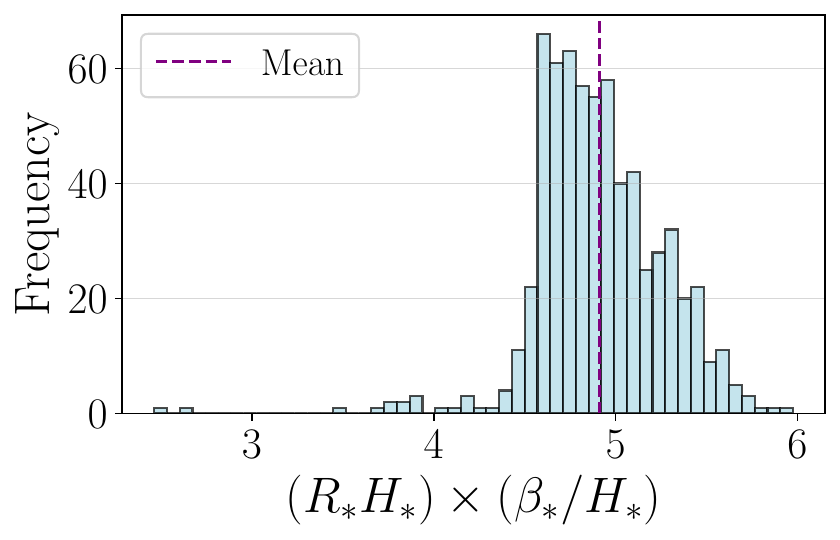}
    \caption{Histogram of the product $(R_* H_*) \times (\beta_* /H_*) $. Mean value is given by 4.91.}
    \label{fig:hist_betaH_RH}
\end{figure}

\paragraph{Length scale of the transition} The length scale of the transition is approximately given by the average bubble radius $R_*$ at the time of percolation. It is approximately given by the cubic root of the inverse of the bubble number density $n_B$ \cite{Turner:1992, Enqvist:1992}:
\be
R_* \equiv n_B^{-1/3} =  \left( T_p^3 \int^{T_c}_{T_p} \frac{\dd{T^\prime}}{T^{\prime 4}} \frac{\Gamma(T^\prime)}{H(T^\prime)} e^{- I(T^\prime)} \right)^{-1/3}.
\ee
For an alternative definition of length--scale relevant for GW production see \cite{Ellis:2018, Athron:2022mmm}.
There is a relation between the average radius $R_*$ and the inverse time scale $\beta$ given by 
\begin{align}
    R_* \simeq \frac{(8\pi)^{\frac{1}{3}} v_w}{\beta_*},
\end{align}
which holds at $\tilde T_p$ such that $I(\tilde T_p)=1$, see ref~\cite{Enqvist:1992}. Using the results presented in figure~\ref{fig:beta-RH}, one can check that the relation does not hold exactly; this is partially due to the percolation condition used, see eq.~\eqref{eq:percolation-temp}. In an article~\cite{Lewicki:2022pdb} it is pointed out that a more adequate relation would be $R_*=5/{\beta_*}$, and this is also in a more reasonable agreement with our numerical results as we find that the product $(R_* H_*) \times (\beta_* /H_*) $ varies between 4.5 and 5.5, and we find the mean value $ \overline{(R_* H_*) \times (\beta_* /H_*) } = 4.91$. See the histogram in figure~\ref{fig:hist_betaH_RH}. 

\paragraph{Lorentz factor of the bubble wall
\label{sec:lorentz_factor}}

The motion of an expanding bubble is directly affected by the surrounding plasma. In a broad picture, each particle coupled to the nucleating field gains mass upon entering the bubble. As a result, the bubbles ``feel'' an effective pressure acting upon them. While bubbles accelerate at the beginning, they eventually reach terminal velocity when the net pressure difference between the effective driving force acting on the wall and the plasma friction is zero.
A crucial question for GW production is whether the bubble wall reaches the terminal velocity before the collision, or it keeps accelerating. Depending on that, the main source of GW is sound waves in the plasma or bubble wall collisions, respectively. To determine the behaviour of the bubble wall before the collision, we have to examine the pressure exerted on it by the surrounding plasma.
The pressure difference across the wall can be expressed as
\be
\Delta P = \Delta V - \PLO - \PLL,
\ee
where $\PLO$ is the leading contribution to pressure accounting for $1\rightarrow1$ transitions~\cite{Bodeker:2009qy} i.e.~particles gaining mass, while $\PLL$ is the contribution associated with $1\rightarrow N$ splittings in the vicinity of the bubble wall~\cite{Bodeker:2017,Hoche:2020ysm,Gouttenoire:2021kjv}. The first contribution quickly reaches a constant value
\be
\PLO= \sum_a k_a c_a \frac{\Delta m_a^2 T_{p}^2}{24}, \quad c_a = \mbox{$1$ $(1/2)$ for bosons (fermions)},
\ee
where the sum runs over particle species, $k_a$ denotes the number of degrees of freedom of a given particle, and $\Delta m_a^2$ is the change in mass across the wall. 
To compute the latter term, $\PLL$, it is necessary to sum contributions from $1\to N$ splitting processes, for arbitrary $N$. That is a difficult task, and the outcome is a subject of debate in the literature. 
According to ref.~\cite{Hoche:2020ysm}, the NLO friction can be expressed as
\be
P_{\mathrm{NLO}}^{(2)}  \sim \gamma^2 \sum_i k_i g_i   T^4_p,\label{eq:friction-Jessica}
\ee
where $g_i$ is the $i$-th boson's gauge coupling. 
On the other hand, ref.~\cite{Gouttenoire:2021kjv} reports the following expression for the NLO friction:
\be
P_{\mathrm{NLO}}^{(1)}  \sim \gamma \sum_i g_i ^2  m_i T^3_p \log{\left( \frac{m_i}{\mu_{\mathrm{ref}}}\right)},\label{eq:friction-Yann}
\ee
where $m_i$ is the mass in the broken phase and $\mu_{\mathrm{ref}}$ is an IR cutoff proportional to $T_p$ (for a detailed discussion see~\cite{Gouttenoire:2021kjv}). 
The crucial difference between the two results lies in the power of the Lorentz $\gamma$ factor. The $\gamma^2$-scaling of eq.~\eqref{eq:friction-Jessica} suggests a faster damping of walls' velocity and thus a constrained possibility of GW production via bubble collisions. 
However, in the case of strongly supercooled transitions, it was shown \cite{Kierkla:2022odc, Lewicki:2022pdb} that either choice does not strongly affect the estimation of the gravitational wave source. Thus, we resort to $\gamma$-scaling, and as suggested in \cite{Gouttenoire:2021kjv, Ai:2025bjw}, we set the IR cut-off to the Debye mass $\mD \sim gT$, see appendix \ref{app:DR_matching}.  

To determine the behaviour of the wall at the moment of the collision, we consider the evolution of the Lorentz factor with the bubble radius~\cite{Ellis:2020, Lewicki:2022pdb}. At the initial stage of expansion, when the $\gamma$ parameter is not large yet, but\footnote{We have checked that the initial radius is indeed much smaller than the final one for the whole parameter space.} $R_0\ll R$, the constant LO term in the friction dominates, leading to the Lorentz factor of the bubble wall increasing linearly \cite{Gouttenoire:2021kjv} 
\begin{align}
    \gamma_{\textrm{run-away}} \simeq R/3R_0,
\end{align} 
where $R_0$ is the initial bubble radius. Its value can be estimated from energy conservation across the critical bubble wall~\cite{Ellis:2019}, and it is given by
\be
R_0 \equiv\left[\frac{3 E_{0, V}}{4 \pi \Delta V(t=0)}\right]^{1/3},
\ee
where $\Delta V(t=0)$ is the energy difference between the centre of the initial bubble and the outside, while $E_{0, V}$ is the potential energy contribution to the energy of the initial bubble. It is a good assumption to identify the initial size with the critical radius when the kinetic and potential energy are equal, which gives $E_{0, V} \simeq \Seff(\varphi=\varphib)/2$. If bubbles collide in this first step of accelerated expansion, it resembles the run-away scenario.

\begin{figure}
\center
\includegraphics[width=.45\textwidth]{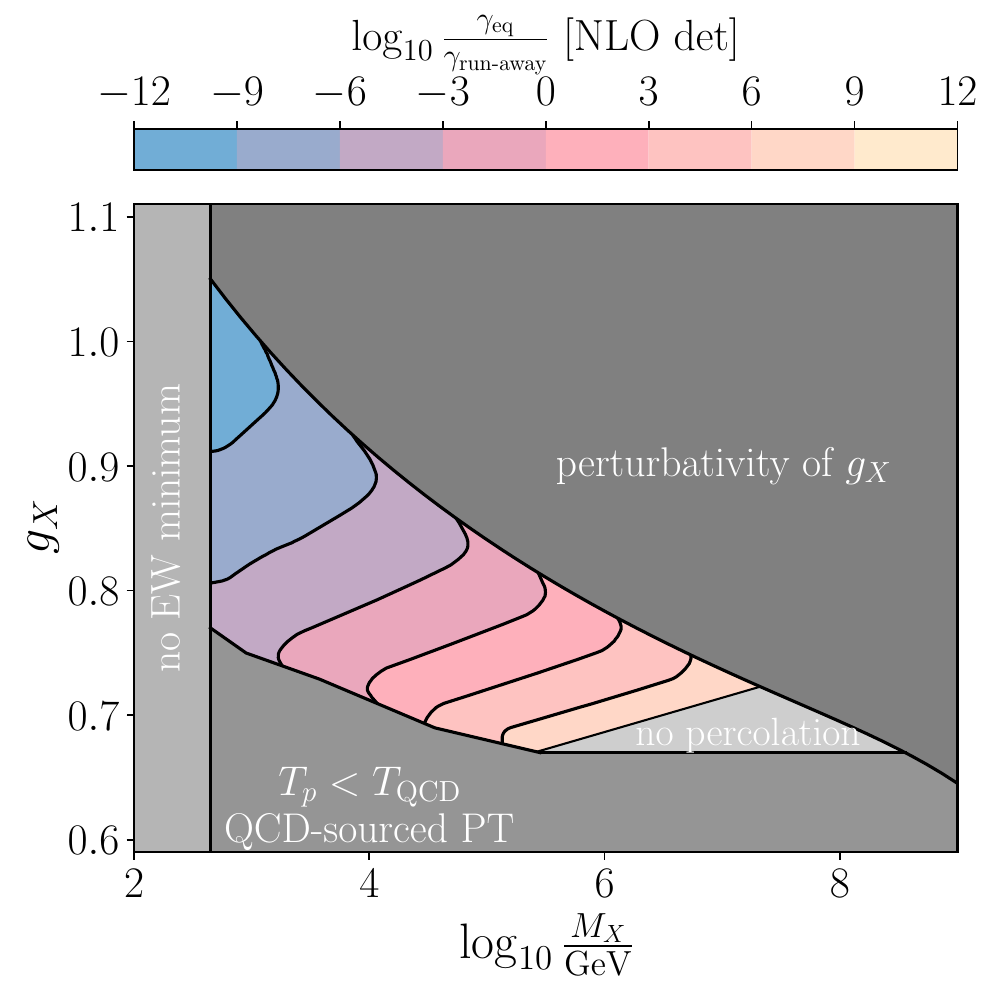}\hspace{20pt}
\includegraphics[width=.45\textwidth]{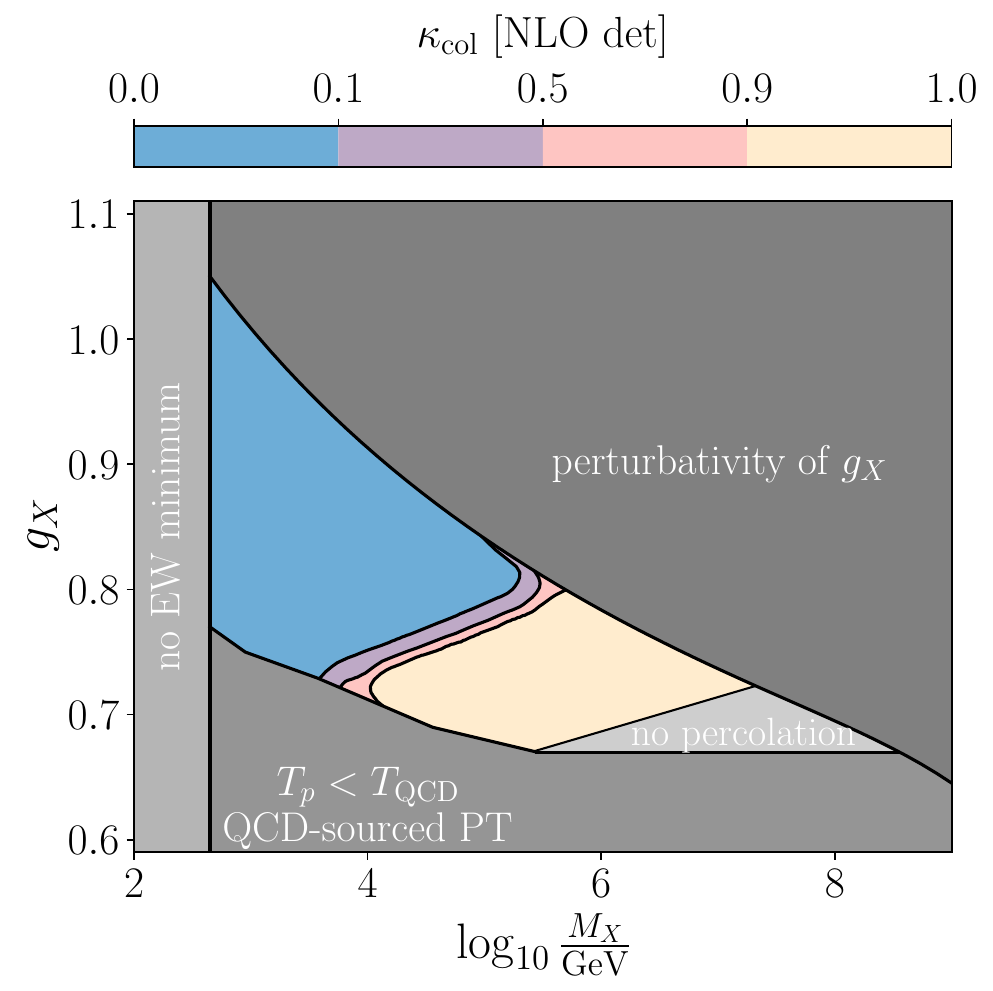}\hspace{20pt}
\caption{Left panel: the ratio of Lorentz factors for terminal velocity and runaway scenarios. The results were obtained by using the friction force of ref.~\cite{Gouttenoire:2021kjv}. Right panel: the values of the efficiency factor $\kappa_{\mathrm{col}}$ of transferring energy of the PT into GW sourced by bubble collisions.}
\label{fig:kappa-sw}
\end{figure}

If the wall expands further, the NLO pressure terms become important, and finally the wall reaches a stationary state where $\Delta P = 0$, which leads to a Lorentz factor of the form
\be
\gamma_{\textrm{eq}} \equiv \left( \frac{\Delta V - P_{\mathrm{LO}}}{P_{\mathrm{NLO}}^{(n)}} \right)^{\frac{1}{n}},
\ee
with $n$ being the power of $\gamma$ in the NLO pressure term. 
Finally, the $\gamma$ factor of the wall at the moment of the collision reads
\be
\gamma_* = \min{(\gamma_{\textrm{eq}}, \gamma_{\textrm{run-away}})}.
\ee
In figure~\ref{fig:kappa-sw}, on the left panel, we show the ratio of $\frac{\gamma_\text{eq}}{\gamma_\textrm{run-away}}$. Physically, when this ratio is smaller than unity, then the bubbles reach terminal velocity; otherwise, they still accelerate upon collisions. We can see that for most of the parameter space, the bubbles indeed reach terminal velocity, while for larger values of $\MX$ the transition becomes more supercooled and a run-away scenario becomes viable. 
%

\paragraph{Energy transfer\label{sec:energy_transfer}}
Knowing the terminal Lorentz factor, it is possible to investigate the transfer of the released energy to the plasma. The efficiency factor $\kappa_{\text{col}}$ at the end of the transition is given as a ratio of the energy stored in the bubble wall to the total released energy. We use the form derived in~\cite{Lewicki:2022pdb}
\be
\kappa_{\textrm{col}} =
\left( 1-\frac{\alpha_{\infty}}{\alpha} \right)
\left( 1-\frac{1}{\gamma_{\text {eq }}^c}\right)
\frac{R_{\text {eq}}}{R_*} \frac{\gamma_*}{\gamma_{\text {eq}}},
\ee
where $\alpha_{\infty} = \PLO/\rho_R$ \cite{Ellis:2020} and $R_{\text {eq}}=3R_0\gamma_{\text{eq}}$ \cite{Lewicki:2022pdb}. 
The energy that goes into sound waves in the plasma is parametrised by the efficiency factor\footnote{Note that the recent results of refs.~\cite{Giese:2020rtr, Giese:2020znk} do not apply to strongly supercooled phase transitions with $\alpha_*>1$.} 
\cite{Espinosa:2010, Hindmarsh:2013, Hindmarsh:2015}
\be
\kappa_{\mathrm{sw}}=\frac{\alpha_{\mathrm{eff}}}{\alpha_*} \frac{\alpha_{\mathrm{eff}}}{0.73+0.083 \sqrt{\alpha_{\mathrm{eff}}}+\alpha_{\mathrm{eff}}}, \quad \text{ with } \quad \alpha_{\mathrm{eff}}=\alpha_*\left(1-\kappa_{\mathrm{col}}\right),
\ee
which for strong supercooled transitions is simply $\kappa_{\mathrm{sw}} \simeq (1-\kappa_{\text{col}})$. 
We show the resulting values of the efficiency factor for bubble collisions in figure~\ref{fig:kappa-sw}, on the right panel.
As expected, the model under consideration allows for the generation of the GW signal both by the sound waves in the plasma and the bubble collisions (recall that $\kappa_{\mathrm{col}}=1-\kappa_{\mathrm{sw}}$). There is also a small region where both sources contribute significantly. However, as argued in ref.~\cite{Lewicki:2022pdb} in the case of very strong transitions, such as in SU(2)cSM, in the region dominated by the sound waves, the resulting plasma shells are thin and move with ultra-relativistic velocity -- they resemble colliding bubble walls. In conclusion, the resulting gravitational wave spectra sourced from sound waves mimic the spectra from bubble collisions in their shape. We will discuss it more in the following sections.

\section{Gravitational Waves phenomenology in SU(2)cSM}
Cosmological first-order phase transition sources a stochastic gravitational wave background. We can distinguish three different sources (or production mechanisms) of gravitational waves: bubble collisions, sound waves in the plasma and turbulent motion in the plasma. Each of the sources can contribute to the full GW spectrum is then a sum of these contributions. For a state-of-the-art review, see ref.~\cite{Caprini:2024hue}. 
%
Determining the shapes of the GW spectra for each source is not an easy task as it requires performing rigorous numerical simulations, see e.g.~\cite{Hindmarsh:2015, Lewicki:2022pdb, Jinno:2022mie}. Therefore, the most common approach for phenomenological purposes consists of modelling spectral shapes $S(f)$ of the GW spectrum as a broken-power-law-type function of frequency with some parameters obtained as numerical fits to the results of simulations. Amplitude and peak frequency of the spectrum are then related to the thermodynamical parameters of the PT: temperature $T_*$, inverse time scale $\beta_*$ (or length scale $R_*$), transition strength $\alpha_*$, wall velocity $v_w$ and efficiency factors. 

In very strong transitions, i.e.~when $\alpha_*\gg 1$, energy stored in the false vacuum dominates over the radiation energy density. Then, if the effective friction acting on the wall is small,  the released energy is going to be transferred mostly to the bubble wall and its kinetic energy. 
Historically, early simulations were thus focusing on the scalar field, and it was assumed that GWs are mainly generated by strong transitions. It was common to use the so-called envelope approximation, where bubble walls were approximated as infinitely thin shells and colliding parts of the bubbles were neglected.
Since then, it has been shown that these colliding parts are crucial for a full understanding of the GW production. The underlying assumption of the envelope approximation is the instantaneous loss of momentum of the shells. In reality, however, it is not instantaneous, and it is not obvious how exactly the shell decays. In this work, we will use the approach of ref.~\cite{Lewicki:2022pdb}, where the momentum decay in the radial direction along the bubble radius is derived as a result of multi-bubble nucleation and collision histories. It is worth noting, however, that current simulations were performed for rather small values of the transition strength parameter. Thus, applying these fits in our case is an extrapolation. 

The transition in SU(2)cSM is indeed very strong, but it should also be noted that even in the case of such strongly supercooled transitions with a vacuum domination period, the energy can still be transferred mostly to the surrounding plasma instead of the bubble walls.
It is the case for the majority of the model's parameter space \cite{Kierkla:2022odc, Ellis:2020}, as we have shown in fig.~\ref{fig:kappa-sw}. However, even in such a scenario when the bubbles acquire terminal velocity at the moment of percolation, they are surrounded by highly relativistic fluid shells. These shells are thin and highly concentrated around the walls, resembling again the ``pure scalar walls''.

In ref.~\cite{Lewicki:2022pdb}, it was shown that indeed the distribution of the energy-momentum tensor in such a scenario looks very similar to the bubble collisions for the gauge theories. Thus, even if the efficiency factor $\kcol$ is small and one would naively use the spectral shape for sound waves, instead, one should still use the bubble collisions templates. In other words, gravitational waves sourced by sound waves in strongly supercooled PTs mimic the spectra of bubble collisions. 

In this section, we will show formulae for GW spectra in SU(2)cSM and calculate the so-called signal-to-noise ratio (SNR) for LISA, which will allow us to assess the observational prospects. For a review of GW from PTs in more general scenarios, see e.g.~\cite{Caprini:2024hue, Caprini_2020, Hindmarsh:2020hop}.

\subsection{GW from bubble collisions in SU(2)cSM} 
\label{sec:collisions}

\paragraph{Spectral shape for bubble collisions}
The spectral shape of GW sourced by colliding bubbles is modelled using a broken power law of the form \cite{Caprini:2024hue}: 
\begin{align}
    S_{\rm col}(f) = \frac{\left(n_1-n_2\right)^{\frac{n_1-n_2}{a_1}}}{\left[-n_2\left(\frac{f}{f_p}\right)^{-\frac{n_1 a_1}{n_1-n_2}}+n_1\left(\frac{f}{f_p}\right)^{-\frac{n_2 a_1}{n_1-n_2}}\right]^{\frac{n_1-n_2}{a_1}}}.
\end{align}
The peak frequency is denoted as $f_p$ and the spectral shape is normalised such that $S(f_p)=1$. The geometric parameters $n_1, n_2, a_1$ are obtained as fits to the results of the simulations. Interpretation of $a_1$ is straightforward, as it describes how ``broad'' the spectrum is around the peak. Then, upon defining $f_b = (-n_2 /n_1)^{1/a_1} f_p$, it can be observed that for frequencies $f\ll f_b$ the spectrum grows as $f^{n_1}$, while for $f\gg f_b$ the spectrum either grows or decreases as $f^{n_2}$ depending on the sign of the $n_2 $ parameter. 
The geometric parameters (in the case of gauge theories) have the following values:
\begin{align} \label{col_params}
    n_1 = 2.4, \quad n_2 = -n_1, \quad a_1 = 1.2, 
\end{align}
notice that these values imply that $f_p = f_b$. 

The amplitude of the GW spectrum $\Omega_{\rm col,p}$ is related to the parameters of the phase transition as follows: 
\begin{align}
\Omega_{\rm col,p} = 
A_\textrm{str}
\qty(\frac{H_*}{ \beta_* })^2 
\qty( \frac{\kappa_{\text{col}} \alpha_* }{1+\alpha_*} )^2, 
\end{align}
where $A_\textrm{str} = 0.05$ is another geometric factor determined by simulations \cite{Lewicki:2022pdb}. Note that in the case of strongly supercooled transitions $\alpha_* \gg 1$ and efficiency factor $\kcol$, can also be of order $1$, which saturates the whole energy budget factor.
The peak frequency is associated with the length/time scale of the transition:
\begin{align} 
    f_{\rm col, *} = \qty(\frac{\Bar{f}}{ \beta_*}) \beta_*,
\end{align}
where the numerical factor $\qty(\Bar{f}/ \beta) = 0.11$ is again obtained from the simulations \cite{Lewicki:2022pdb}.

\paragraph{Redshift of the SGWB signal}
Both amplitude and peak frequency of the GW produced at $T_*$ are affected by the cosmological history of the universe. To obtain predictions for the spectra today (as denoted by the subscript ``$0$''), one has to take the redshift factors into account. Assuming radiation domination immediately after the transition (i.e.~instantaneous reheating, see section~\ref{sec:ptparams_SU2cSM}) one has to apply the following rescaling \cite{Caprini:2024hue}
\begin{align}
    f_0 = \frac{a_*}{a_0}f_* = H_{*,0} \frac{f_*}{H_*} , 
    \quad
    \Omega_0 h^2 = F_{\rm GW,0}h^2 \Omega_*  ,
\end{align}
where $H_{*,0}$ is the value of the Hubble parameter at the time of transition, redshifted to today
\be
H_{*,0} = \frac{a_*}{a_0} H_* 
= 1.65 \cross 10^{-5} \textrm{ Hz} \qty(\frac{T_{\rm reh}}{100 \textrm{ GeV}}) \qty(\frac{g_*}{100})^\frac{1}{6}, 
\ee
while $ F_{\rm GW,0}$ is the redshift factor for the fractional energy density
\begin{align}
     F_{\rm GW,0} h^2= h^2 \qty(\frac{a_r}{a_0})^4 \qty(\frac{H_r}{H_0})^2 =
      1.64 \cross 10^{-5} \qty[\frac{100}{g_*}]^\frac{1}{3} .
\end{align}
The present-day value of the Hubble parameter is given as  $H_0 = 100 h \textrm{ km/s/Mpc}$, where $h = 0.67 $ (value from Planck 2018 results \cite{Planck:2018vyg}) is the dimensionless Hubble parameter.

\paragraph{Complete GW spectrum}

\begin{figure}
    \centering
    \includegraphics[width=0.8\textwidth]{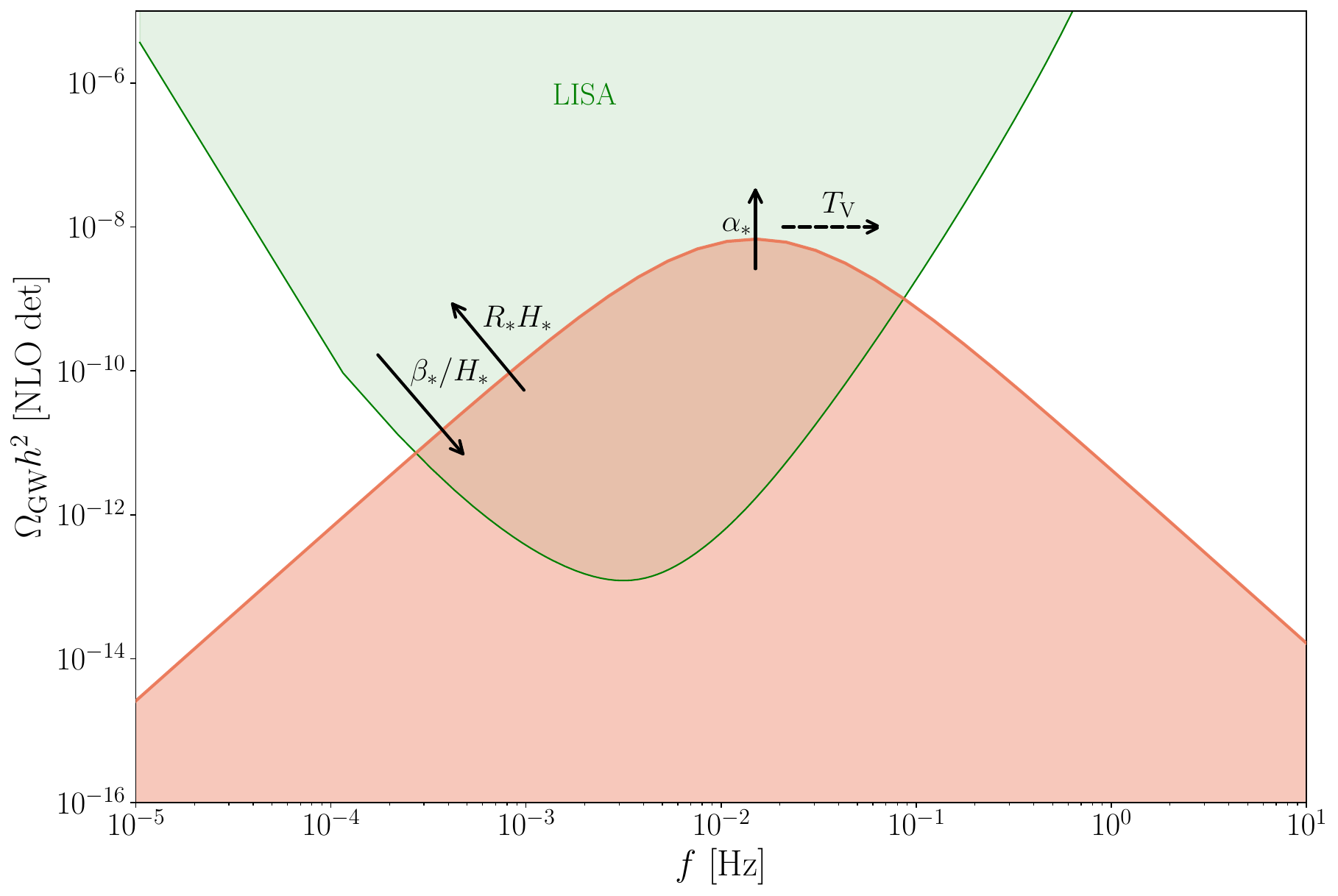}
    \caption{Gravitational wave spectra from the phase transition in SU(2)cSM. 
    Single GW spectrum for  BM2, see table \ref{tab:GW_BMs} for details. Black solid arrows illustrate how a change (arrowheads correspond to an increase) in phase transition parameters would affect the spectrum (direction of the arrow). Dashed arrows illustrate the effect of redshift
    }
    \label{fig:GWspectra_single}
\end{figure}

\begin{figure}
    \centering
    \includegraphics[width=0.8\textwidth]{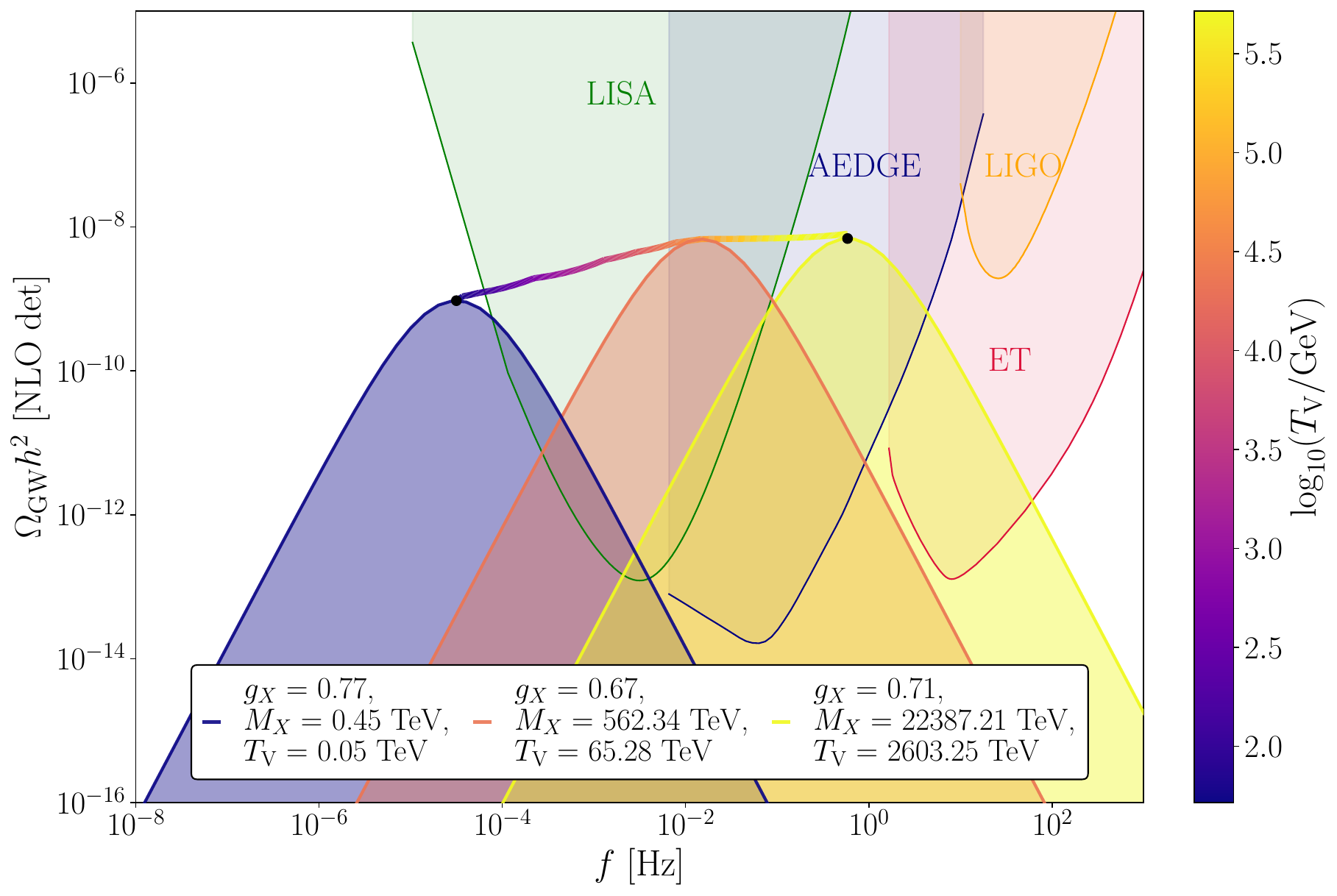}
    \caption{Gravitational wave spectra from the phase transition in SU(2)cSM. 
    Example spectra for three BM points, see table \ref{tab:GW_BMs}. The solid lines connect the peak amplitudes of the largest signals in SU(2)cSM, and the colouring corresponds to a value of $\Tv$, which is the main parameter controlling the redshift of peak frequency. 
    }
    \label{fig:GWspectra_daniel}
\end{figure}

Combining the spectral shape, amplitude, and peak frequency and then applying redshift factors results in the final formulae for the gravitational-wave spectrum sourced by bubble collisions
\begin{align} \label{GWcol_full}
    \Omega_{\rm col,0} h^2 &= 
        1.64 \cdot 10^{-5} \qty[\frac{100}{g_*}]^\frac{1}{3}
        A_\textrm{str}
        \qty(\frac{H_*}{ \beta_* })^2 
        \qty( \frac{\kappa_{\text{col}} \alpha_* }{1+\alpha_*} )^2
        S_{\rm col}(f), \\
    f_{\rm col,0} &= 
    1.67 \cross 10^{-5} \textrm{ Hz} 
    \qty(\frac{T_{\rm reh}}{100 \textrm{ GeV}}) 
    \qty(\frac{g_*}{100})^\frac{1}{6} 
    \qty(\frac{f_*}{H_*}).
\end{align}
In the figure~\ref{fig:GWspectra_single} we show the example GW spectrum for a benchmark point in the parameter space of SU(2)cSM, where we denote the total spectra as $\Omega_{\rm GW} h^2$.
We show the schematic dependence of the spectrum on the phase transition parameters: amplitude grows with $\alpha_*$, peak frequency depends on transition time or length scale; however, due to the redshift, it is mainly specified by the reheating temperature $\Treh$. 
The coloured regions on the plots correspond to integrated sensitivities of chosen detectors.\footnote{Author thank Marek Lewicki for sharing the data points for the sensitivity curves}
In the figure~\ref{fig:GWspectra_daniel}, we show spectra for two more benchmarks. The details of benchmark points used are stored in table~\ref{tab:GW_BMs}. We also illustrate how the spectra with maximal amplitudes move when changing the value of the reheating temperature. This illustrates how the GW signal in SU(2)cSM moves inside and outside the LISA sensitivity range. As the reheating temperature is directly associated with the mass of the dark gauge boson, the position of the peak frequency also gives insight into the energy scale of new physics in the dark sector.
\begin{table}[ht]
    \centering
    \begin{tabular}{ |c|c|c|c|c|c|c| }
    \hline
            & $\gx$ & $\mx$ [GeV] & $\Tv$ [GeV] & $\Tp$ [GeV] & $\RH$ & $\kcol$ \\ 
        \hline
        BM1 & 0.77  & 446.68      & 51.87       & 0.19        & 0.17  & $1.85 \times 10^{-5}$ \\ 
        \hline
        BM2 & 0.67  & 562341.33   & 65281.66    & 0.25        & 0.46 & 1  \\ 
        \hline
        BM3 & 0.71  & 22387211.39 & 2603249.92  & 34.04       & 0.46 & 1 \\
    \hline
    \end{tabular}
    \caption{Details of benchmark points used for gravitational waves spectra plots. These benchmarks belong to a set for which we find the largest amplitude of the GW signal at a given frequency. BM1 is the benchmark with the smallest value of $\Tv$, BM2 has the largest amplitude in the whole parameter space of SU(2)cSM, while BM3 corresponds to the largest redshift, i.e. the largest value of $\Tv$. }
    \label{tab:GW_BMs}
\end{table}

\begin{figure}[ht]
    \centering
    \includegraphics[width=0.45\textwidth]{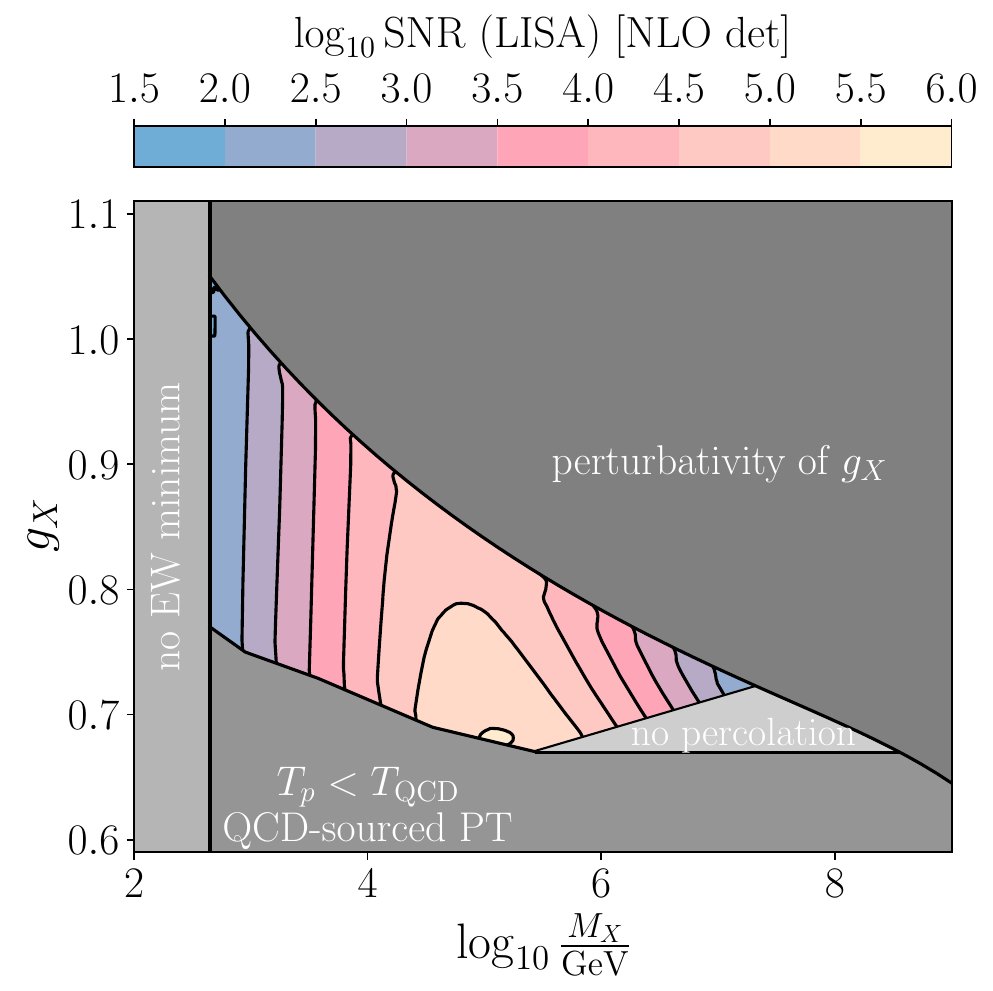}
    \caption{SNR for LISA for SU(2)cSM model calculated with \hrefNLOdet~approach.}
    \label{fig:SNRLISA_noSelfNoise_NLOdet}
\end{figure}

\paragraph{Signal-to-noise ratio for LISA}
To assess the observability of a signal, we compute the signal-to-noise (SNR) ratio for the LISA detector that has the best prospects of observing the predicted signal. We calculate the SNR using the usual formula \cite{Caprini_2020, Robson_2019}:
\begin{align} 
\label{eq:SNR}
\mbox{SNR} = \sqrt{
\mathcal{T} 
\int_{f_{\rm{min}}}^{f_{\rm{max}}}
\dd f \left[ \frac{h^2\Omega_{\rm{GW}}(f)}{h^2\Omega_{\rm{Sens}}(f)} \right]^2
},
\end{align}
where $\mathcal{T}$ is the duration of collecting data and $h^2\Omega_{\rm{Sens}}(f)$ is the sensitivity  of the detector.
The values of the SNR were obtained using the approach described, e.g. in \cite{Caprini:2015}. This approach is based on the assumption that the so-called ``self-noise'' of the GW signal is negligible. For the SU(2)cSM model, the GW signal is generically strong; therefore, the obtained SNR values may be overestimated.\footnote{We thank Kai Schmitz for bringing this to our attention.} 
Nevertheless, including the self-noise factor would not lead to values of SNR <10. 
One can see that by estimating the self-noise by simply including the $h^2\Omega_{\rm{GW}}(f)$ in the denominator of eq.~\eqref{eq:SNR}. Then the smallest value would correspond to the saturated SNR (i.e., such that the fraction in the integrand is one) which will result in SNR$\sim \order{10^3}$.\footnote{We thank Mauro Pieroni for a discussion on this topic.}
%
For calculations we have used data collecting durations as $\mathcal{T}_{\rm{LISA}}$ = 75 \% $\cdot$ 4 years \cite{Caprini_2020}.
We will assume that a signal could be observed if $\mathrm{SNR}>10$, which is the usual criterion. In figure~\ref{fig:SNRLISA_noSelfNoise_NLOdet} we show the resulting SNR for LISA.
First of all, we observe that the predicted signal is strong and always within the reach of LISA, which agrees with previous estimates in the literature \cite{Kierkla:2022odc, Kierkla:2023von}. 
To be more specific, values of SNR are much greater than 10 for the entire available parameter space, which means that a first-order phase transition in the SU(2)cSM model will be visible at LISA.
We observe slightly lower SNR, around 10, at the edges of the parameter space corresponding to (small) large $M_X$. The reason is that the peak frequency of the spectrum is (lower) higher for (lower) higher reheating temperatures. The latter grows with $M_X$, so for the (smallest) largest values of $M_X$ allowed at NLO, the signal moves out of the sensitivity range of LISA. Moreover, high values of SNR would allow for a reconstruction of phase transition parameters such as $\Tp$ or $R_*H_*$ with a very good precision, even reaching the accuracy of 1\% \cite{Gonstal:2025qky}.




\section{Impact of higher-order corrections on the phenomenology}
Figure~\ref{fig:action-sample} served as an illustration of different approaches to
compute thermal nucleation rate $ \GammaT$.
Now, to generalise the discussion on theoretical consistency and to scrutinise the importance of a diligent calculation of nucleation rate, we will discuss phenomenological results of the parameter-space scan in the SU(2)cSM model, following \cite{Kierkla:2025vwp, Kierkla:2023von} for various approaches in computing nucleation rate.
We will show the results for the percolation temperature, $\Tp$, and the average bubble radius at percolation normalised to the Hubble radius, $R_*H_*$, efficiency factor $\kcol$, and SNR for LISA. We will compare a subset of the approaches listed in the previous section. For the calculation of $\Tp$ and $\RH$ we shall follow~\cite{Kierkla:2022odc, Kierkla:2023von} as summarized in previous section \ref{sec:ptparams_SU2cSM}. The relative differences for $\Tp$ presented in the following figures are defined as follows
\begin{align}
	\label{eq:delta}
  \delta^{i}_{j} \Tp &=\frac{\bigl|\Tp^{i}-\Tp^{j}\bigr|}{\Tp^{j}}
  \,, &
  {\rm where}
  \quad
  \begin{aligned}
    i &\in \bigl\{
	[{\rm NLO\ det}], [{\rm NLO}\ \nabla], [{\rm NLO\ det\ }T^4], [\rm daisy]
	\bigr\}
  \,,\\
    j &\in \bigl\{[{\rm NLO \ det} ], [{\rm NLO\ det}\ T^4]\bigr\}
  \,.
  \end{aligned}
\end{align}
The relative differences for
$R_*H_*$, $\delta^i_j R_*H_*$, are defined in analogy.

The overall goal of this thesis is to improve the description of supercooled phase transitions and estimates of the GW signal.
Therefore, first, we will focus on the difference between using the daisy resummation approach, which is widely used in the literature,  versus high-T EFTs.
Then, we will further investigate the EFT approach and scrutinise the validity of the derivative expansion, and its impact on quantities that are directly dependent on the nucleation rate, such as $\Tp$ and $R_* H_*$, by comparing \hrefNLOgrad ~to the \hrefNLOdet ~case.

Note that $\Tp$, while being a physical parameter describing the phase transition and setting the temperature scale for evaluating $R_*H_*$, does not directly influence the GW spectrum.
This is true for models with a very strong phase transition, where the relevant temperature determining the peak frequency is the reheating temperature, see section \ref{sec:ptparams_SU2cSM} and ref.~\cite{Kierkla:2022odc}.
Nevertheless, we find it instructive to study the dependence of $\Tp$ on the different approximations, as this is the most direct measure of the validity of the different approximation schemes.



\subsection{Phenomenological uncertainties in daisy resummation approach}

\begin{figure}
\centering
    \includegraphics[width=.45\textwidth]{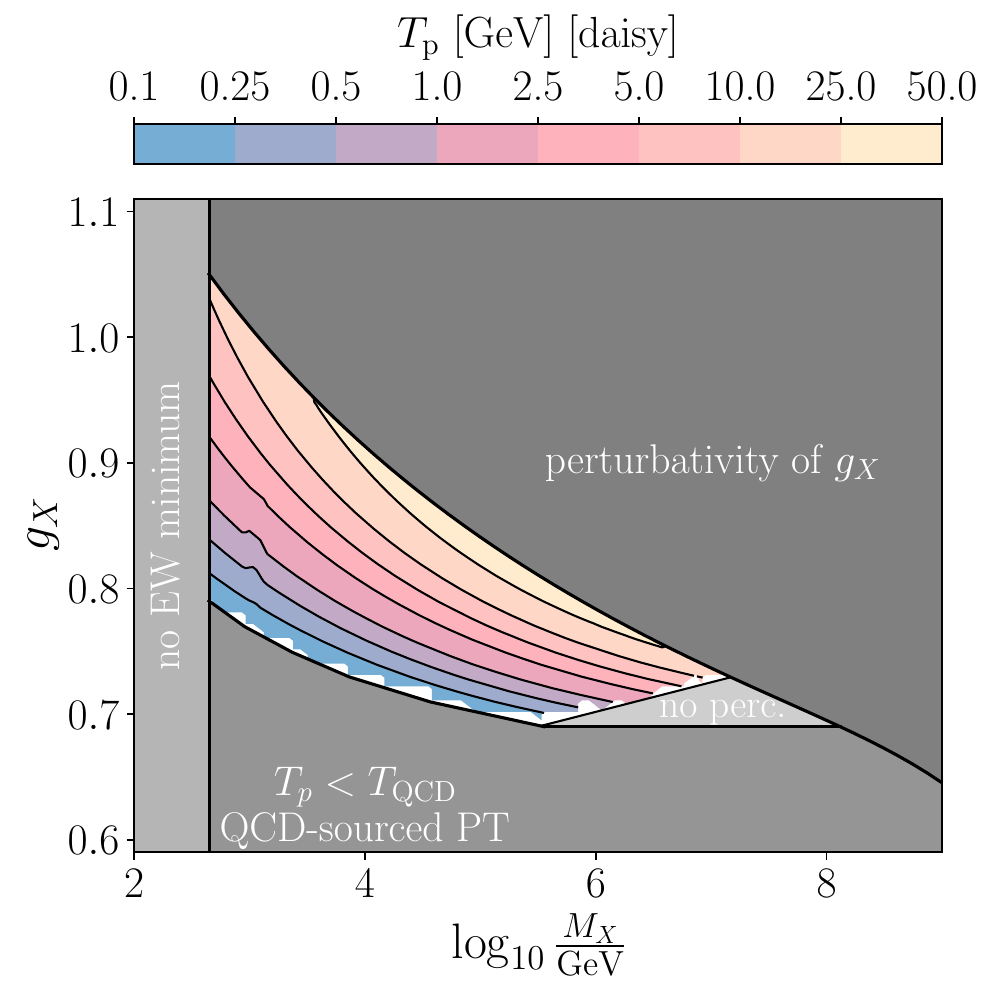}
  	\includegraphics[width=.46\textwidth]{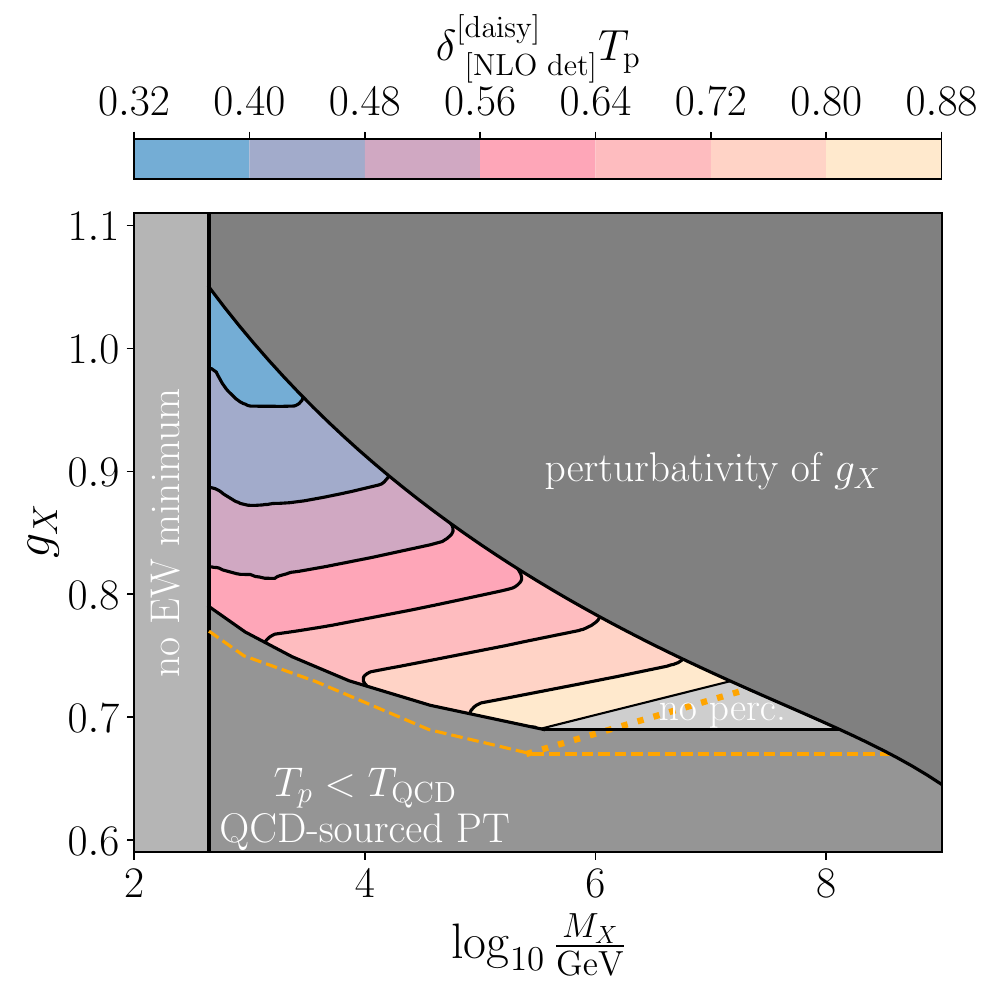}
	\caption{%
       Percolation temperature.
	   Left panel: \hyperref[it:daisy]{[daisy]} results for $\Tp$.
	   Right panel: relative difference between \hyperref[it:NLO-det]{[NLO~det]} and \hyperref[it:daisy]{[daisy]} results for $\Tp$. The orange dashed and dotted lines indicate the excluded regions obtained with \hyperref[it:NLO-det]{[NLO~det]}.
    }
	\label{fig:Tp_daisy_and_Tp_rdiff_with_detNLO}
\end{figure}

\paragraph{Percolation temperature from \hyperref[it:daisy]{[daisy]} }
Below we present the results of the scan of the parameter space obtained using the daisy-resummation approach, which we denote as \hyperref[it:daisy]{[daisy]} (see sec. \ref{sec:nucl_rate_recipes}). 
In figure~\ref{fig:Tp_daisy_and_Tp_rdiff_with_detNLO}, on the left panel, we present the values of percolation temperature, where the 4d renormalisation scale is set to $\mu_4 = \pi T$. As the  \hyperref[it:daisy]{[daisy]} approach misses a sizeable contribution to the effective action in comparison to \hyperref[it:NLO-det]{[NLO~det]}, it leads to smaller values of nucleation rate at a given temperature. A consequence of this is larger supercooling, the parameter space scan shows that the maximal percolation temperature obtained with the \hyperref[it:daisy]{[daisy]} approach is around 50 GeV, while for \hyperref[it:NLO-det]{[NLO~det]} it was up to 200 GeV, cf.~right panel of figure~\ref{fig:Tn-Tp}. This has direct phenomenological consequences, as predictions made with the \hyperref[it:daisy]{[daisy]} approach imply a smaller available parameter space for electroweak transition without QCD condensate present. 

On the right panel, we show the relative difference of $\Tp$ between with two approaches, normalised to \hyperref[it:NLO-det]{[NLO~det]}. 
The difference varies from 30\% in the low-mass, large-coupling region, up to 88\% in the small-coupling, large-mass corner of parameter space.

As was mentioned in \cite{Kierkla:2023von}, this behaviour might seem counter-intuitive. One would expect the largest corrections between the two methods for large couplings, but note that such intuition applies the coupling at the thermal scale. 
It should be stressed here that the coupling and mass displayed in figure~\ref{fig:Tp_daisy_and_Tp_rdiff_with_detNLO} are the ``input'' parameters, defined at the scale $\mu_4 = \MX$. 
To calculate nucleation rate, and then calculate $\Tp$, they need to be RG-evolved to the thermal scale (for details see section~\ref{sec:su2csm_thermal_scale_running}). 
In the large-mass corner, the coupling becomes significantly larger at the thermal scale, which explains why the difference between the LO and NLO approaches is largest in this part of parameter space. 
Let us point out again that the value of $T_p$ does not directly affect the GW spectrum in models with classical scale invariance, so the large corrections we find in the \hyperref[it:NLO-det]{[NLO~det]} description are not reflected in a strong modification of the GW signal, as we will show below. However, it signals that the differences between the descriptions at different orders in perturbation theory are non-negligible.

One can also observe the change in the overall allowed region. Constraints from scans performed with daisy-resummation are indicated by the orange dashed (QCD sourced PT) and dotted (no percolation) lines. The region of non-percolation is shifted, and also the region where the phase transition is expected to be sourced by QCD effects is pushed to lower values of $\gx$ at NLO as expected from the arguments above. Therefore, the predictions for the GW signal in this region could be significantly altered by using the \hyperref[it:daisy]{[daisy]} or \hyperref[it:NLO-det]{[NLO~det]} approach.

\begin{figure}
\centering
    \includegraphics[width=.45\textwidth]{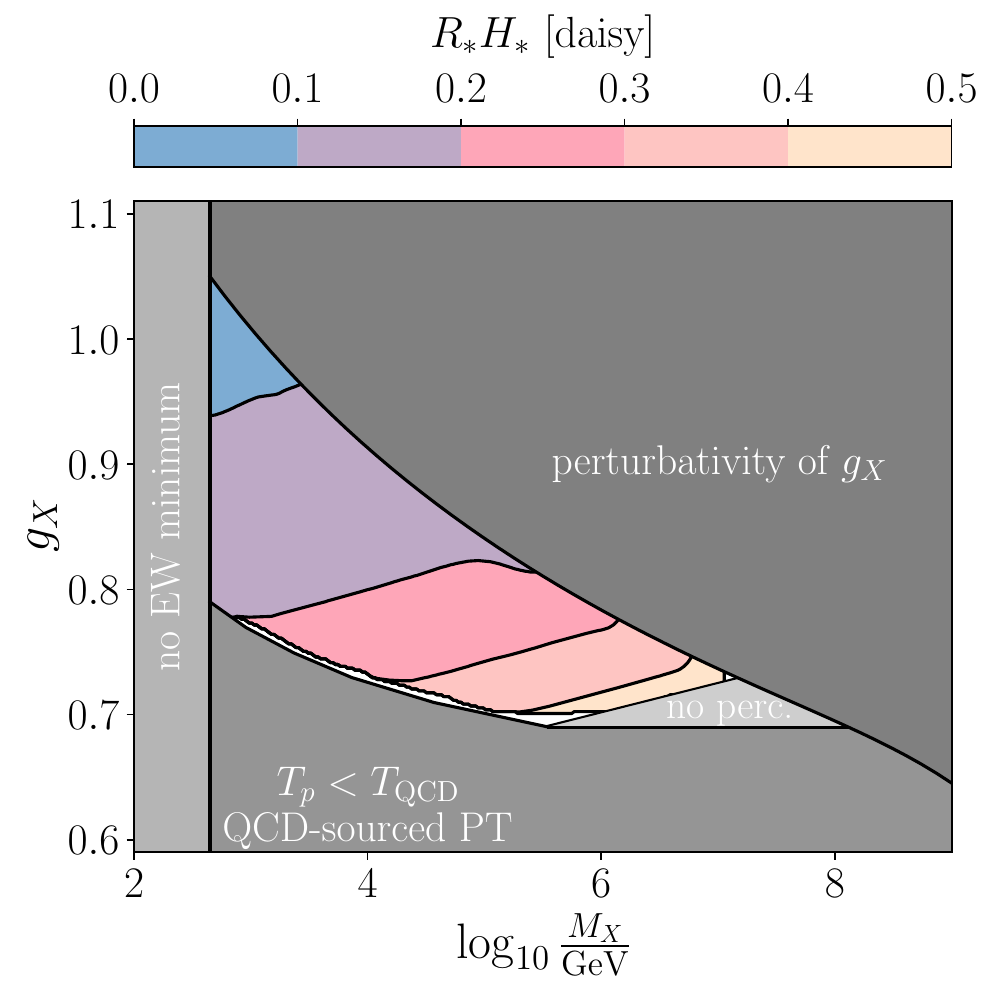}
  	\includegraphics[width=.46\textwidth]{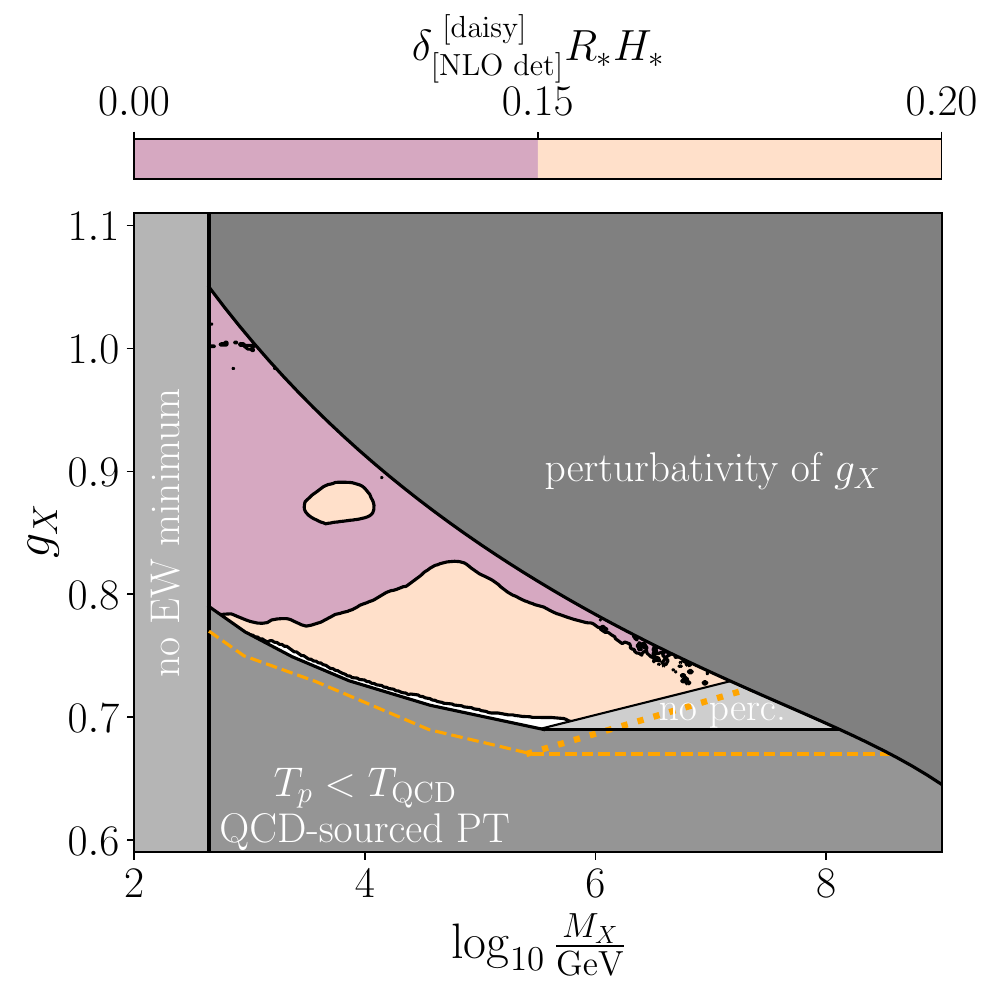}
	\caption{%
       Predictions for the Hubble normalised average bubble radius for different approximations to the NLO nucleation rate.
	   Left panel: \hyperref[it:daisy]{[daisy]} results for $\RH$.
	   Right panel: relative difference between \hyperref[it:NLO-det]{[NLO~det]} and \hyperref[it:daisy]{[daisy]} results for $\RH$. The orange dashed and dotted lines indicate the excluded regions obtained with \hyperref[it:NLO-det]{[NLO~det]}.
       The ``sand island'' in the upper-left part of the plot is probably due to numerical artefacts and instead should belong to the purple region. 
    }
	\label{fig:RH_daisy_and_RH_rdiff_with_detNLO}
\end{figure}

\paragraph{Length scale from \hyperref[it:daisy]{[daisy]} }
In figure~\ref{fig:RH_daisy_and_RH_rdiff_with_detNLO}, on the left panel, we further present the values of average bubble radius normalised to Hubble, obtained in the \hyperref[it:daisy]{[daisy]} approach. The results show larger bubbles through the parameter space in comparison to the results of \hyperref[it:NLO-det]{[NLO~det]}. This behaviour is another consequence of the arguments given above -- missing corrections lead to a smaller value of the effective action, which translates to larger supercooling and slower transitions. 
The right panel of figure~\ref{fig:RH_daisy_and_RH_rdiff_with_detNLO} displays the relative difference of $\RH$. Here we see that the differences are smaller than for $\Tp$, as they vary between $\mathcal O(10 \%)-O(20 \%)$.
%

\paragraph{Source of GW from \hyperref[it:daisy]{[daisy]} }

\begin{figure}
\center
\includegraphics[width=.45\textwidth]{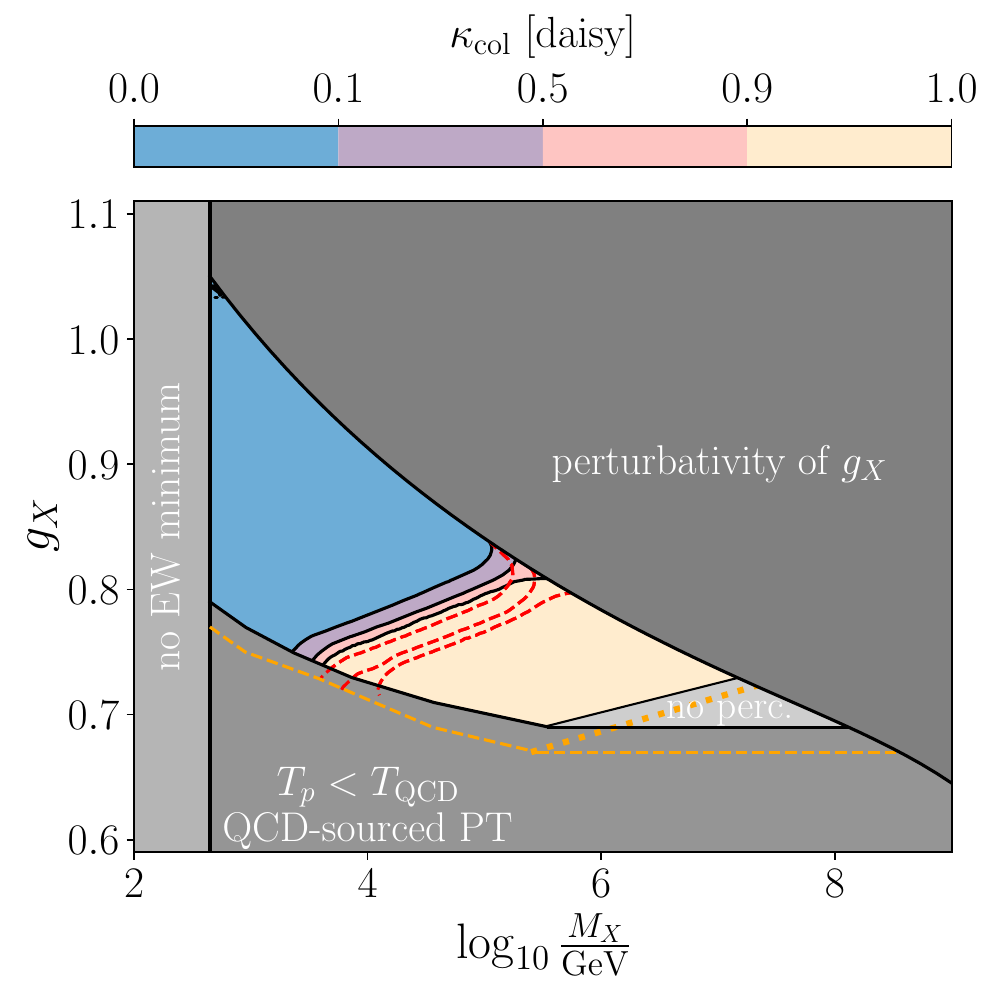}\hspace{20pt}
\caption{
The values of the efficiency factor $\kappa_{\mathrm{col}}$ of transferring energy of the PT into GW sourced by bubble collisions obtained in \hyperref[it:daisy]{[daisy]} approach.  The orange dashed and dotted lines indicate the excluded regions obtained with \hyperref[it:NLO-det]{[NLO~det]}. The red lines indicate contours of $\kappa_\rmii{col} = \{0.1, 0.5, 0.9\}$ (from above to below) obtained with \hyperref[it:NLO-det]{[NLO~det]}. 
}
\label{fig:daisy_kcol}
\end{figure}

Missing higher-order corrections will also affect the predictions of the source of the gravitational waves \cite{Kierkla:2023von}. In figure~\ref{fig:daisy_kcol} we show predictions for the bubble collisions efficiency factor obtained in the \hyperref[it:daisy]{[daisy]} approach. Yet again, we see that a consequence of ``artificially augmented''  supercooling results in a larger part of the parameter space, where $\kcol\sim 1$, which corresponds to the run-away regime. To illustrate this difference explicitly, we show red dashed lines which correspond to contours $\kappa_\rmii{col} = \{0.1, 0.5, 0.9\}$ from figure~\ref{fig:kappa-sw} obtained in \hrefNLOdet approach.  

\begin{figure}
    \centering
    \includegraphics[width=0.45\textwidth]{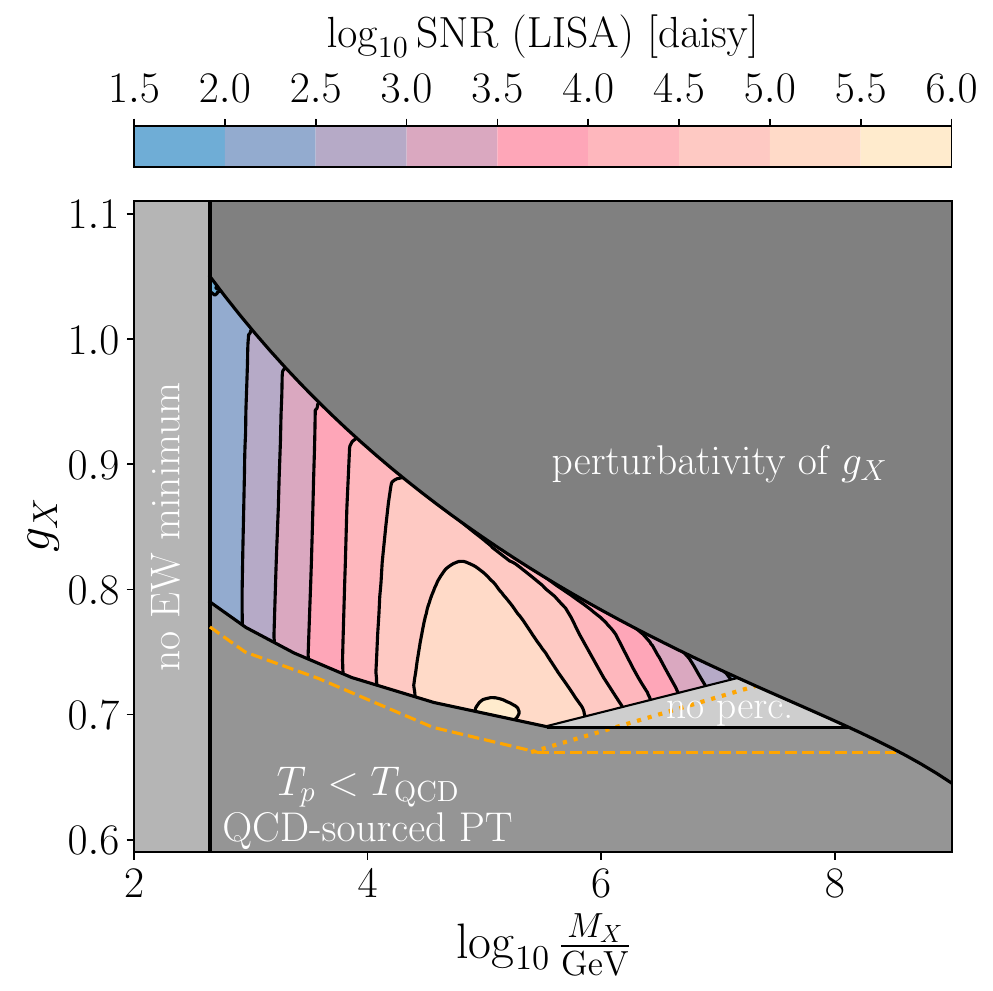}
    \caption{Signal-to-noise ratio for LISA for SU(2)cSM model calculated with [daisy] approach.}
    \label{fig:SNRLISA_noSelfNoise_daisy}
\end{figure}

\paragraph{SNR from \hyperref[it:daisy]{[daisy]} }
Finally, in figure~\ref{fig:SNRLISA_noSelfNoise_daisy} we show the LISA SNR obtained in the \hyperref[it:daisy]{[daisy]} approach. As the peak of the signal is determined mostly by the value of reheating temperature (which in our case corresponds to $\Tv$), higher-order thermal corrections to the nucleation rate do not affect SNR very much. The resulting SNR plot looks almost the same as figure~\ref{fig:SNRLISA_noSelfNoise_NLOdet}, and the main difference between the two comes from the overall ``shape'' of the parameter space.  
However, the theoretical diligence is still important as the potential observed signal would correspond to a specific model, with specific values of parameters. If the SNR for this specific benchmark is large, then we need to be able to reconstruct the parameters, and for that, the theoretical errors must be smaller than the accuracy of reconstruction.

\subsection{Dependence on the 4d renormalisation scale}
\label{sec:RGscaleDep}

The fundamental flaw of the daisy resummation approach is the lack of theoretical consistency. Aside from imaginary parts in the effective potential and gauge dependence, the nucleation rate computed in this way suffers from an uncancelled RG-scale dependence, see references \cite{Lofgren:2023sep, Croon:2020cgk}. This issue can be cured by the inclusion of certain two-loop level diagrams, as we have seen in chapter \ref{chapter:TFT}. 
Resolution of this problem is achieved in the 3d NLO effective potential (with the matching also performed at the two-loop level).
\footnote{All the other problems mentioned are also automatically resolved within the 3d EFT framework when using soft expansion. Consistent use of perturbation theory comes with many rewards.} 
There, the RG-scale dependence cancels up to terms of order higher than the order to which we compute. Moreover, the potential (and the full action) are independent of the 3D scale, up to higher-order corrections. As was shown in the literature~\cite{Croon:2020cgk,Gould:2021},
omission of significant perturbative corrections
(revealed by the scale dependence) is the main source of uncertainty in predicting the GW signals. We are now in a position to check the RG-scale sensitivity of the NLO predictions and to contrast it with the \hyperref[it:daisy]{[daisy]} result. 
Note that with the RG-improvement procedure implemented in this work, when we say that we change the 4d scale, in fact, we mean that we change the thermal cut-off in the running in eq.~\eqref{eq:mu4_final}. 

\begin{figure} 
\centering
\includegraphics[width=.45\textwidth]{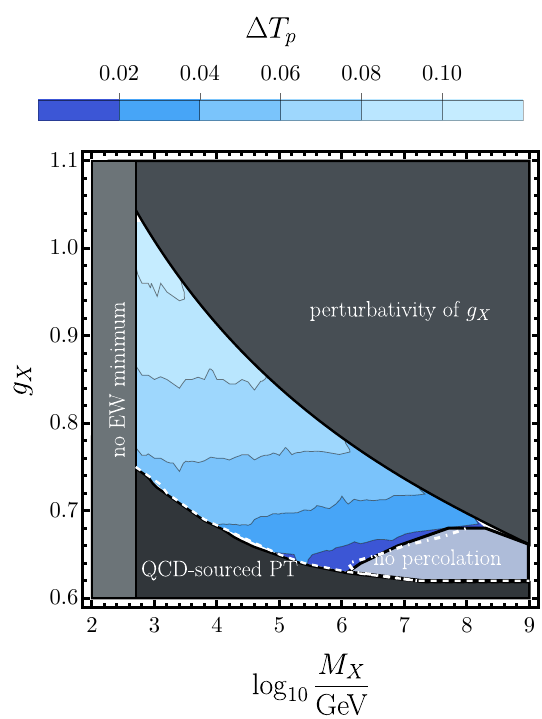}\hspace{2pt}
\includegraphics[width=.45\textwidth]{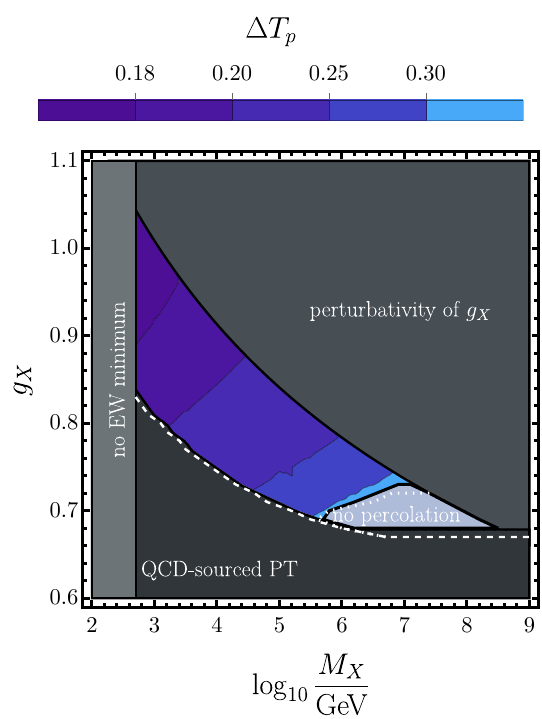}
\caption{
  Left panel: Absolute value of the differences in $\Tp$ obtained from the 
  \hyperref[it:NLO-grad]{[NLO~$\nabla$]} action with $\mu_4=\pi T$ and $\mu_4=2\pi T$, normalised by the result with $\mu_4=\pi T$. 
  Right panel: Absolute value of the differences in $\Tp$ obtained from the \hyperref[it:daisy]{[daisy]} action with $\mu_4=\pi T$ and $\mu_4=2\pi T$, normalised by the result with $\mu_4=\pi T$. 
  }
\label{fig:4d-scale-dep}
\end{figure}

Figure~\ref{fig:4d-scale-dep} (left panel) presents the relative difference in $\Tp$ obtained from the NLO action at two different scales, $\mu_4=\pi T$ and $\mu_4=2\pi T$.\footnote{Note that for this plot we use \hrefNLOgrad approach from \cite{Kierkla:2023von}. While this approach suffers from the breakdown of derivative expansion, its 4d RG scale dependence is the same as in the case of \hrefNLOdet approach, which allows us to use results of \cite{Kierkla:2023von} to illustrate this feature.} 
We observe a mild dependence on the 4D scale, the result for $\Tp$ changing, between the two RG-scales, by at most 10\%. The changes in $R_*H_*$ are much smaller, and they never exceed $2\%$; therefore, we do not show the plot illustrating this difference. We have seen that the predicted SNR for LISA for this model is large, implying that thermal parameters can be reconstructed with very good precision. It therefore needs to be determined if the $2\%$ uncertainty in $R_*H_*$ leads to an observable difference.

For comparison, figure~\ref{fig:4d-scale-dep} (right panel) presents the dependence on the scale of the \hrefdaisy results. We can see that the change in results for $\Tp$ is much larger -- the relative difference is approximately between 15\% and 30\%. This confirms our earlier claims that the inclusion of the NLO corrections cancels the residual scale dependence present at LO. The RG-scale dependence of the bubble radius is again milder and is of the order of 5\% in the whole parameter space, which is again larger than the uncertainty in the NLO result.
In both approaches, daisy and NLO, the overall allowed region is only slightly modified by changing the 4D scale as indicated in figure~\ref{fig:4d-scale-dep} by the white dashed and dotted lines.

\subsection{Accuracy of results in the EFT approach}

After dissecting the shortcomings of the \hrefdaisy~approach, it is time to investigate possible errors in computing the NLO nucleation rate within the EFT framework. As we have shown in the previous chapter, one of the alarming issues comes from the fact that gauge modes are scale-shifters -- the inclusion of gauge loop corrections always leads to the breakdown of the derivative expansion. Here, we will quantify the error from this breakdown by showing how it affects the obtained phase transition parameters. 
We also discuss the error from approximating the prefactor in the tunnelling rate on dimensional grounds as $~T^4$. Further, we show how the inclusion of the Jacobian factor can lead to worse agreement with the full scalar fluctuation determinant.

\paragraph{Impact of using gradient expansion on percolation temperature }

\begin{figure}[!htbp]
\centering
  	\includegraphics[width=.45\textwidth]{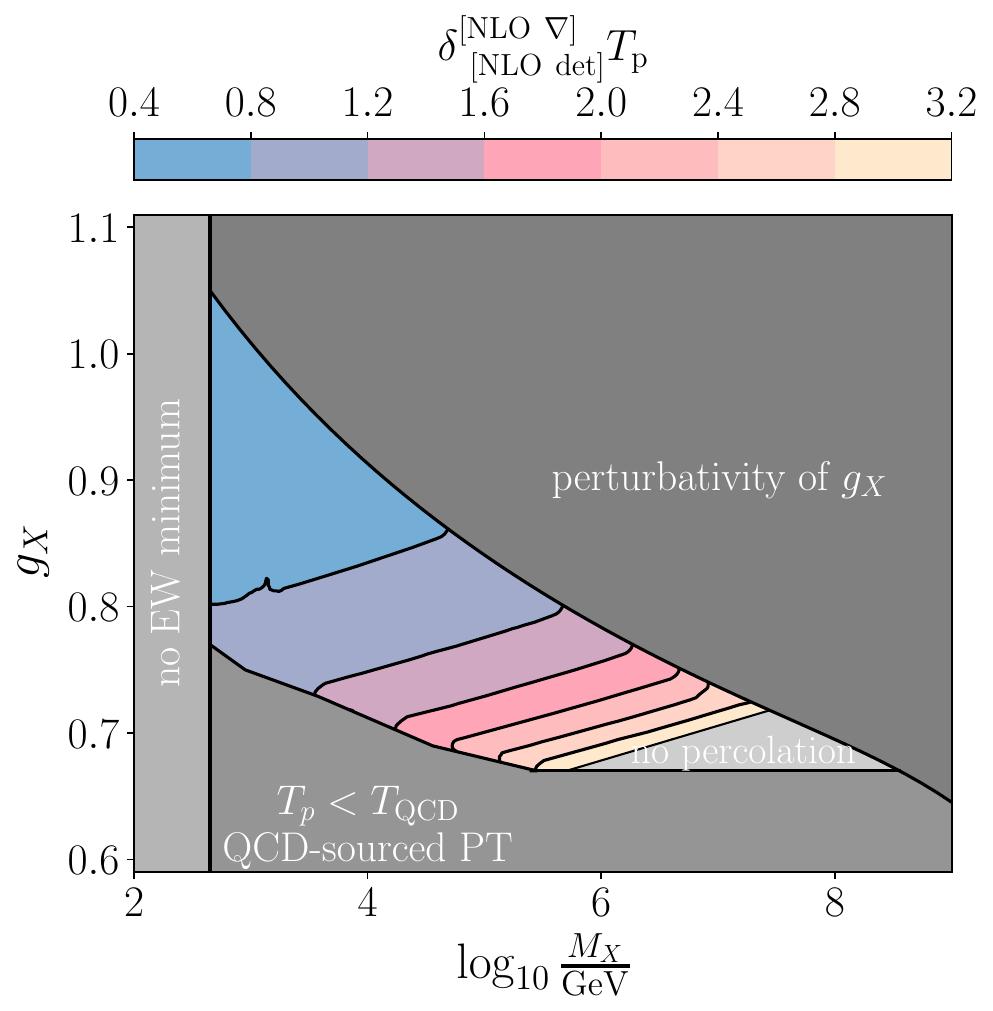}
	\caption{%
	   Relative difference between \hyperref[it:NLO-det]{[NLO~det]} and \hyperref[it:NLO-grad]{[NLO~$\nabla$]} results for $\Tp$. 
    }
	\label{fig:rdiff_Tp_NLOdet_detH_vs_NLOgrad}
\end{figure}

A quantitative comparison between the
\hyperref[it:NLO-det]{[NLO~det]} and
\hyperref[it:NLO-grad]{[NLO~$\nabla$]} approaches is made in
the  figure~\ref{fig:rdiff_Tp_NLOdet_detH_vs_NLOgrad}.
The inclusion of the full fluctuation determinant consistently decreases the percolation temperature compared to the gradient-expanded approach, and the significant difference between \hyperref[it:NLO-det]{[NLO~det]} and \hyperref[it:NLO-grad]{[NLO~$\nabla$]} observed for a single benchmark point in figure~\ref{fig:action-sample} persists throughout the entire parameter space. 
As in~\cite{Kierkla:2023von}, we see that the corrections are the largest in the region of large $\mX$ and small $\gX$, where the phase transition is the strongest
and the slowest (i.e.\ the size of bubbles at collision is the largest, see figure~\ref{fig:beta-RH}).
Throughout the plot, the difference in the percolation temperatures varies from
40\% to more than 300\%, demonstrating the importance of including the fluctuation determinants.

\begin{figure}
	\centering
  	\includegraphics[width=.45\textwidth]{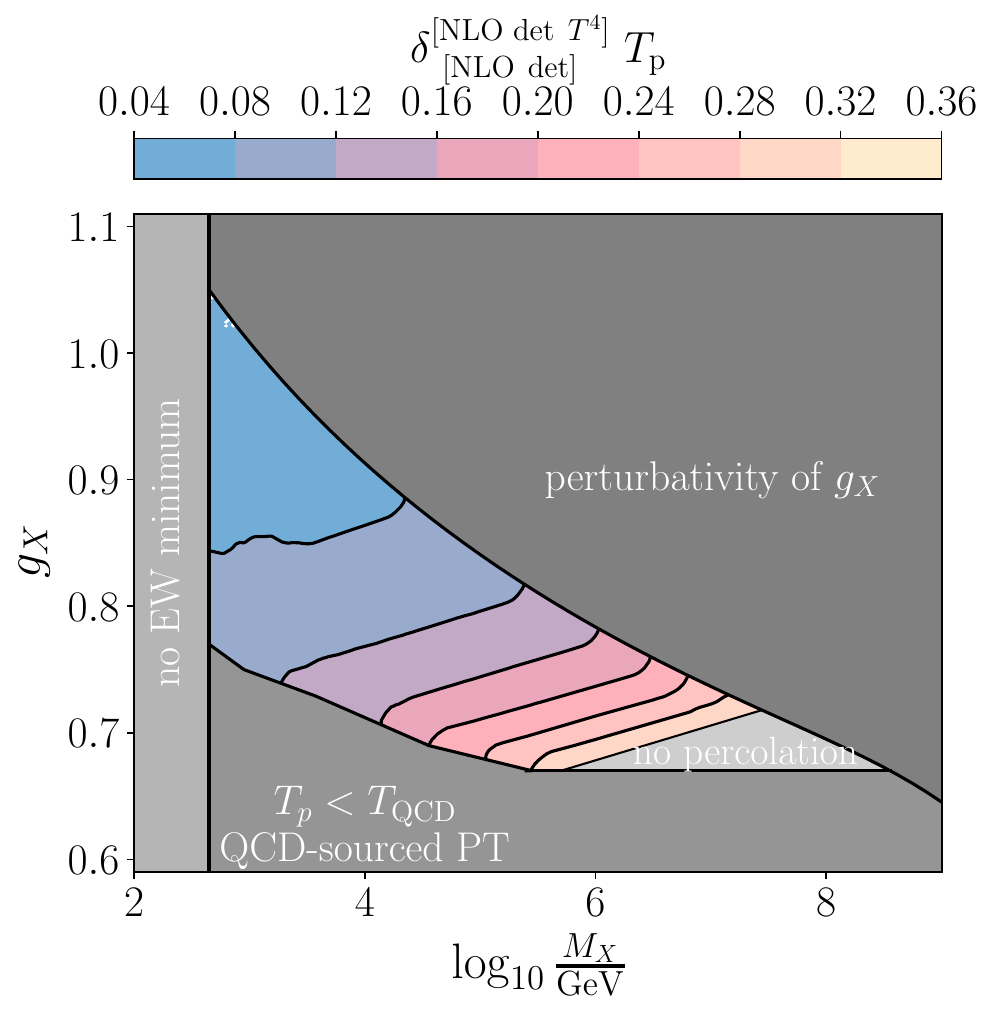}%
	\includegraphics[width=.45\textwidth]{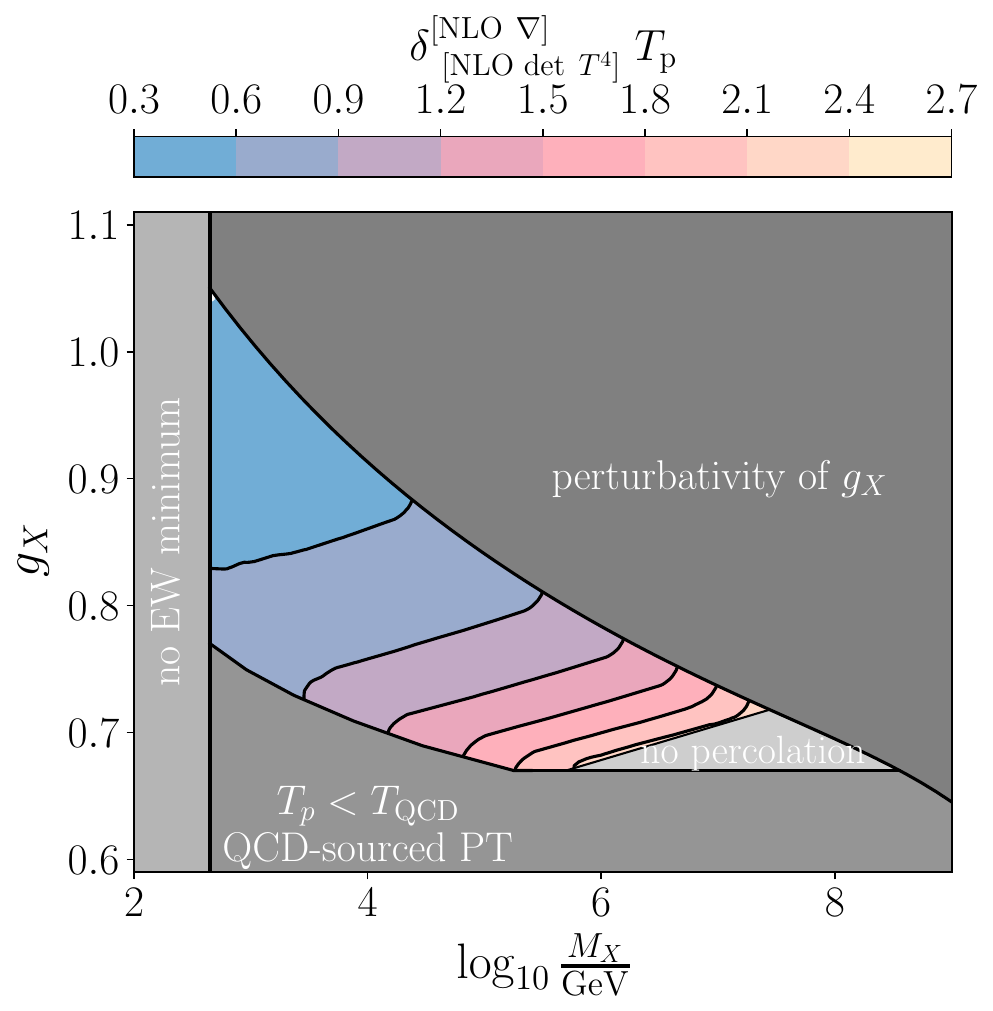}
	\caption{%
		Comparison of different approximations to the nucleation rate and the corresponding percolation temperature.
		Left panel: relative difference between \hyperref[it:NLO-det]{[NLO~det]} and \hyperref[it:NLO-det-T4]{[NLO~det~$T^4$]},
		right panel: relative difference between \hyperref[it:NLO-det-T4]{[NLO~det~$T^4$]} and \hyperref[it:NLO-grad]{[NLO~$\nabla$]}.
	}
	\label{fig:delta-Tp}
\end{figure}

In figure~\ref{fig:delta-Tp}, we investigate the difference between the full
\hyperref[it:NLO-det]{[NLO~det]} and
\hyperref[it:NLO-grad]{[NLO~$\nabla$]} approaches in further detail.
We distinguish two sources of the relative difference observed in the  figure~\ref{fig:rdiff_Tp_NLOdet_detH_vs_NLOgrad}: the approximation of the scalar determinant and contribution to $A_{\rm dyn}$ by $T^4$ as well as the gradient expansion for the gauge modes.
In the left panel of figure~\ref{fig:delta-Tp},
we evaluate the impact of the former approximation.
We see that the scalar pre-factor constitutes
a sub-dominant contribution to the observed differences.
This is in agreement with our expectations, as the one-loop contribution from the scalar is a higher-order effect than the one-loop contribution from the gauge modes.
Nevertheless, the relative differences associated with the scalar pre-factor are non-negligible and range from 4\% to almost 40\%.
The right panel of figure~\ref{fig:delta-Tp} shows the difference between
\hyperref[it:NLO-grad]{[NLO~$\nabla$]} and
\hyperref[it:NLO-det-T4]{[NLO~det~$T^4$]}.
In both approaches, the scalar pre-factor is approximated as $T^4$, and the graph thus allows us to observe directly the effect of the gradient expansion applied to the gauge, ghost and Goldstone modes.
We observe that the relative differences are much larger than in the left panel of the graph, and we can thus conclude that the gradient expansion applied to the gauge, ghost and Goldstone modes is the dominant source of difference observed in the figure~\ref{fig:rdiff_Tp_NLOdet_detH_vs_NLOgrad},
confirming our concerns about the validity of the derivative expansion at the bubble tail.
Indeed, the main cause of the breakdown of the derivative expansion is the spatial gauge modes, which do not have a Debye mass that would protect them from being massless in the symmetric phase.
We have checked explicitly that applying the derivative expansion to the temporal gauge modes only leads to negligible deviations of order 0.01\%.

We emphasise that in the discussed plots, the relative differences between the more and less precise results are largest in the large-mass, small-coupling region.
Part of the reason for this behaviour was explained in~\cite{Kierkla:2023von}:
as the couplings are defined at the scale
$\mu_4 = \mX$, they have to be RG-evolved to the thermal scale.
As a result, the region of large $\MX$ corresponds to relatively large couplings, which explains why the perturbative expansion is less accurate in this region. 
In the region of the largest supercooling (large $\RH$, small $\beta_*/H_*$), an additional effect enters.
Small $\beta_*/H_*$ means that the action is flat, thus a small change in the action results in a large change in the temperature, see figure~\ref{fig:action-sample}. 

\paragraph{Impact of including Jacobian factor on predictions of percolation temperature}

A commonly used approximation for the scalar determinant contribution also contains the $\left( \frac{\Seff}{2\pi} \right)^{3/2}$ factor, which originates from the removal of scalar zero modes.%
 Formally, this factor corresponds to
a Jacobian obtained by going to collective coordinates;
see e.g.~\cite{Ekstedt:2023sqc}.
In our soft EFT, this factor appears in the nucleation rate as
$\mathcal I_\phi$ in eq.~\eqref{eq:detS_softEFT}.
Then, following the usual approach, one would approximate the prefactor containing the scalar determinant and the dynamical contribution as:
	\begin{align}
		A_{\rm dyn}\detS  \simeq T\cdot T^3 
        \left( 
		  \frac{ \SLO[\varphib] }{2\pi} 
		\right)^{3/2}
    \,.
\end{align}
However, we have found that omitting this Jacobian factor leads to a better approximation of the scalar determinant in the SU(2)cSM model. We claim that the reason for this is that when computing the scalar determinant, the Jacobian factor gets cancelled. Therefore, if one does not aim for computing the determinant, then in general there is no reason to include the Jacobian factor. However, as shown in ref.~\cite{Ekstedt:2021kyx} for a certain class of models, with SMEFT-like potentials, the Jacobian factor may lead to a better approximation of the full determinant. Therefore, one should always be cautious when making such approximations. 

\begin{figure}
	\centering
	\includegraphics[width=.45\textwidth]{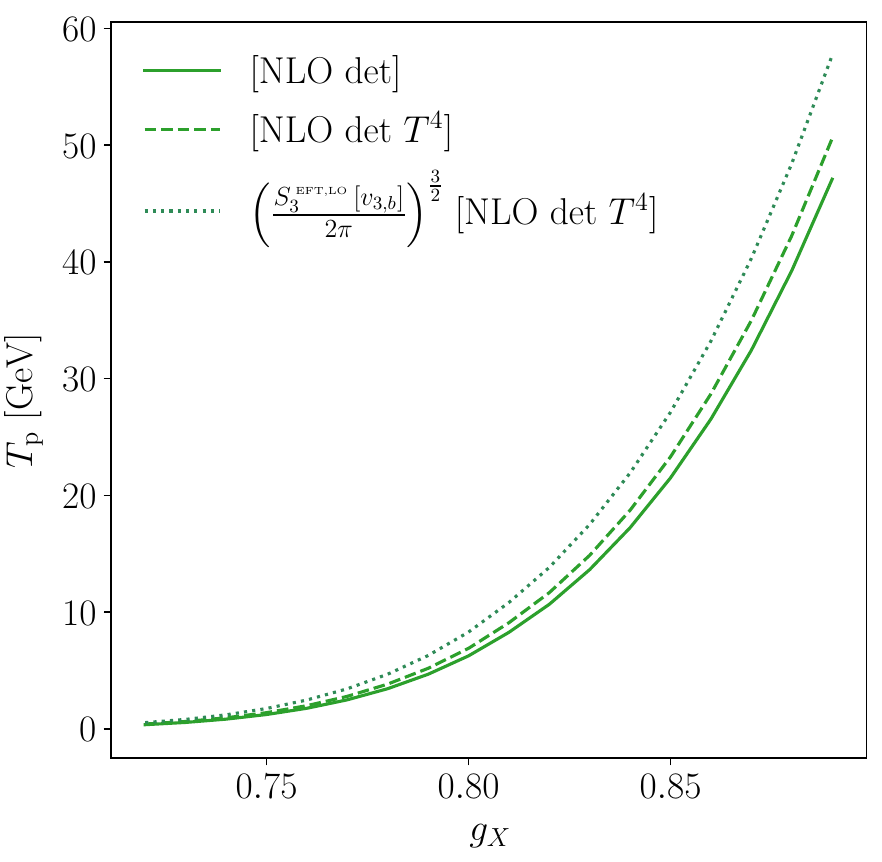}
	\caption{%
    Percolation temperature as a function of $\gX$ for fixed $\MX = 10^{4.05} $~GeV,
    obtained with different approximations to the scalar determinant
    in the prefactor of the nucleation rate.
    Here,
    $\mathcal I_\phi = \bigl(S_{\rm 3}^{\rmii{EFT,LO}}[v_{3,b}]/(2\pi)\bigr)^{3/2}$
    is the Jacobian factor obtained by the removal of scalar zero modes, see also eq.~\eqref{eq:detS_softEFT}. 
    }
	\label{fig:Tp_NLOdet_J}
\end{figure}

In figure~\ref{fig:Tp_NLOdet_J},
we show that predictions for the percolation temperature for a slice of
constant $\mX$ obtained by using solely $T^4$ match
the \hyperref[it:NLO-det]{[NLO~det]} results noticeably better.
The reason for this is that the full scalar contribution is indeed an $\mathcal{O}(1)$ number times $T^3$ for SU(2)cSM, while the term 
$(\SLO[\varphib])/(2\pi))^{3/2}$ alone is not $\mathcal{O}(1)$ thus leading to an overestimation of the nucleation rate.\footnote{In the partial wave decomposition of scalar determinant (see e.g. \cite{Ekstedt:2023sqc}) the total contribution is proportional to the mass of scalar field cubic. Since in the case of classically scale-invariant models $\mthree \sim gT$ we find that using $\detS \sim T^3$ leads to a better approximation of the determinant.} 
We have confirmed that the same conclusion holds throughout the entire parameter space.

\paragraph{Impact of higher-order corrections  on average bubble radius }

\begin{figure}[htpb!]
\centering
    \includegraphics[width=.45\textwidth]{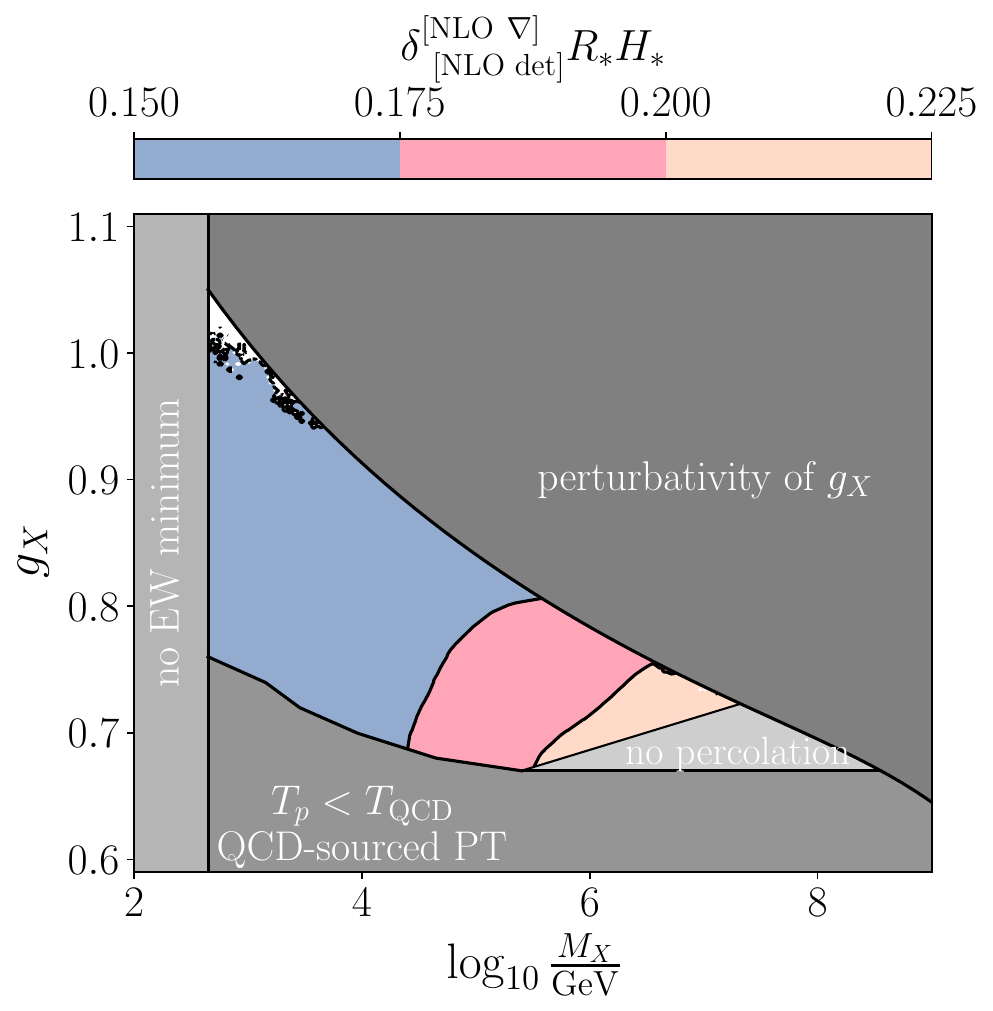}
	\caption{%
        Mean bubble radius at percolation, normalised to Hubble scale, $R_*H_*$. Relative difference between
        \hyperref[it:NLO-det]{[NLO~det]} and
        \hyperref[it:NLO-grad]{[NLO~$\nabla$]}.
        The small white region in the corner of the parameter space is due to numerical instabilities that do not influence the overall picture. 
    }
	\label{fig:rdiff_RH_NLOdet_detH_vs_NLOgrad}
\end{figure}

In the figure~\ref{fig:rdiff_RH_NLOdet_detH_vs_NLOgrad} we demonstrate the relative differences between
\hyperref[it:NLO-det]{[NLO~det]} and
\hyperref[it:NLO-grad]{[NLO~$\nabla$]} for the average bubble radius normalised to Hubble parameter.
As noted in~\cite{Kierkla:2023von}, the relative differences are much smaller than for $\Tp$, and
they range from 15\%  to about 25\%.
We note that again the main source of deviations is the gradient expansion of the gauge, ghost and Goldstone modes, and the approximation of the scalar pre-factor by $T^4$
causes at most a 6\% difference (and less in almost the entire parameter space).
The observed differences between
\hyperref[it:NLO-det]{[NLO~det]} and
\hyperref[it:NLO-grad]{[NLO~$\nabla$]} would not result in large changes in the predictions for GW spectra. However, as indicated by~\cite{Gonstal:2025qky}, the accuracy of reconstructing $\beta_*/H_*$ (when marginalised over other parameters), which is directly related to $R_*H_*$, can be below 10\% throughout the interesting parameter space for supercooled phase transitions.
Then, the theoretical error associated with using the gradient expansion, not to mention the daisy resummation, would exceed the error of reconstruction. 


\chapter{Summary}
\label{chapter:summary}

This dissertation is aimed at studying a supercooled phase transition in a classically scale-invariant extension of the Standard Model. 
The main motivation for this study is the strong prospect of observing a gravitational wave signal from a supercooled first-order transition. In case of a detection of the GW background from such a transition, the parameters of the underlying BSM theory could be reconstructed with very good accuracy.

We have studied a concrete BSM scenario, a model called SU(2)cSM, which extends the conformal Standard Model with an additional dark gauge sector. First, we have discussed the theoretical apparatus for studying quantum fields at finite temperature, and then we have discussed the details of the SU(2)cSM model. In the next step, we have explicitly shown how to perform the calculation of the thermal nucleation rate using various approaches and discussed the theoretical uncertainties associated with the naive use of certain approximations. 
Finally, we conducted a phenomenological study of the supercooled phase transition and provided robust predictions of the resulting gravitational wave signals. We also analyse the impact of theoretical uncertainties on physical observables, such as the temperature of transition completion or the shape of gravitational wave spectra. The main results of the dissertation are collected below.

\section{Main results}

\paragraph{RG-scale dependence in SU(2)cSM}
We have performed a careful analysis of the effective potential in a model with classical scale-invariance. In particular, we have paid attention to a consistent expansion in the power of couplings. Furthermore, we have investigated the issue of RG-scale dependence in such theories. We have shown how to use an RG-improved effective potential to describe theory at relevant scales present, such as 
the electroweak scale for scalar mass generation, the scale of the mass of the new scalar for its decay during reheating, and IR scales associated with the false vacuum. 

\paragraph{Parameter space of SU(2)cSM}
We studied the allowed parameter space for the electroweak phase transition in SU(2)cSM under the following constraints. 

First, we have asserted the correct value of the electroweak minimum (including correct values of Higgs mass and vev). This resulted in the exclusion of the region of parameter space where the mass of the new gauge boson is smaller than $\MX \sim 400$~GeV. 

We then considered the perturbativity of the dark gauge coupling in the symmetric phase. This resulted in another constrained region. This region corresponds to a large part of the parameter space where both the dark gauge coupling and mass of the dark gauge boson are large. We have quantified the bound using the condition $\gX(\mu=\MZ) = 1.15 $. We need to stress that formally this region is not an exclusion; however, the regular methods based on perturbative expansion are not applicable there. 

We have also delineated a region where the electroweak phase transition would occur in the presence of a quark condensate, i.e., below the temperature of the QCD phase transition. This region is not phenomenologically excluded, but a precise description of the phase transition would require additional study of the QCD transition, which is beyond the scope of this work.

Finally, we have required a completion of the phase transition via percolation of bubbles. This resulted in the exclusion of a corner of the parameter space for masses of dark gauge bosons larger than $6$ TeV and small couplings. 

\paragraph{Dimensional reduction for supercooling}

We have shown how to include higher-order thermal corrections in the description of the supercooled phase transition. 
In particular, we have reconciled the high-temperature dimensional reduction with the description of the supercooled phase transition. 
We have demonstrated how to separate computations of phase transition parameters by splitting them into two groups: large-field-related and small-field-related. 
The former correspond to the low-temperature limit of the effective potential, where no resummations are required. In this regime, the analysis can be performed using the standard zero-temperature approach, as outlined, for instance, in ref.~\cite{Kierkla:2022odc}. The latter, in contrast, pertain to the high-temperature regime, where accurate computations demand a high-temperature effective field theory that systematically incorporates the hierarchy between energy scales induced by the thermal background.

In studying the connection between the high- and low-temperature limits, we clarified several aspects relevant to the computation of thermal phase transition parameters within the four-dimensional framework. 
In particular, by examining the interplay between the renormalisation group (RG) improvement of the potential and the scale cancellations between the high-temperature contributions and the zero-temperature part, we identified the thermal scale, $\mu \sim T$, as the preferred renormalisation scale for phase transition computations in the 4D theory (see section~\ref{sec:HT-LT-regimes}). While this may appear as a straightforward observation, it is not a common practice in many analyses found in the literature. In addition, we highlighted the importance of field normalisation in determining phase-transition-related observables.

We further verified that the bounce configuration, corresponding to the tunnelling trajectory of the scalar field, always lies within the domain of validity of the high-temperature regime. However, along the bounce path, the particle masses vary significantly, which complicates the treatment of mass hierarchies. 
The constructed effective field theory is expected to become unreliable in the tail region of the bounce, where the gauge field mass becomes smaller than that of the scalar field. 
In this region, the gauge-field contributions to the action should ideally be included beyond the derivative expansion.

\paragraph{NLO nucleation rate}
We have calculated the next-to-leading order thermal nucleation rate for a supercooled phase transition. First, we established that incorporating higher-order terms in the soft expansion is essential for obtaining quantitatively reliable results. Second, we demonstrated that the derivative (gradient) expansion introduces sizeable uncertainties at NLO, leading to an underestimation of the nucleation rate. We have also illustrated that derivative expansion breaks completely above $\order{\partial^2}$ at the one-loop level, and discussed the breakdown for higher-loop gauge fluctuations. 

\paragraph{Scalar fluctuation determinant}
Moreover, we have included the full scalar contribution to the effective action. 
However, its effect remains subdominant compared to the uncertainties arising from the gradient expansion of the gauge fluctuation determinant. Finally, we verified that, within our model, the often-employed approximation for the exponential prefactor involving the Jacobian $(S_{3}^{\mathrm{LO}} / 2\pi)^{3/2}$ performs worse than the simpler $T^4$ prefactor derived from dimensional analysis--although this conclusion is model-dependent, as illustrated in, e.g., ref.~\cite{Ekstedt:2021kyx}.

\paragraph{Gravitational wave predictions}
We have obtained state-of-the-art predictions for the gravitational wave signal resulting from the supercooled phase transition in SU(2)cSM. 
We have discussed the results of current simulations and used the fits to show example spectra. We then calculated the signal-to-noise ratio for the LISA experiment. 
We have shown that this model generically produces a strong, observable signal across the entire considered parameter space. Thus, it can be tested by future experiments. 

\paragraph{Impact of theoretical uncertainties on GW signal}
Lastly, we have investigated the propagation of errors stemming from theoretical uncertainties for the predictions of gravitational wave spectra. We have explicitly compared our NLO calculations with the commonly used daisy-resummation approach. 
We have shown that relying on this simplified method leads to large uncertainties for phase transition parameters. However, it does not drastically affect the GW predictions. The reason for this is the fact that the signal is mostly dependent on the reheating temperature. 
This parameter is calculated in the low-temperature regime, where higher-order thermal corrections are subleading. Nevertheless, in the prospect of a loud signal, it would be possible to reconstruct the parameters of transition with good accuracy. Thus, theoretical diligence is necessary for obtaining information about the underlying BSM theory. 
We have also checked the influence of using derivative expansion at NLO. We have found that it also leads to sizeable uncertainties for predictions of phase transition parameters, although the resulting error is smaller than the one caused by omitting NLO corrections to the nucleation rate.  

\section{Open questions}
The overall goal of this dissertation is to contribute to the theoretical understanding of cosmological phase transitions in the prospects of future experiments. Nevertheless, there are still open questions and possible avenues for further research on this topic. Below, we discuss some research problems that may extend the study presented in this thesis.

\paragraph{Going beyond NLO}
First, a straightforward question is about extending the calculation of nucleation rate beyond NLO. Such a calculation would require computing corrections to the critical bubble, which can be described by a class of ``dumbbell'' diagrams \cite{Ekstedt:2022tqk, Gould:2021ccf}. Moreover, some studies suggest that in radiative barrier models, the nucleation rate computed in soft expansion diverges quickly from the lattice results as soon as larger values of gauge couplings are considered \cite{Ekstedt:2022ceo}. Thus, including higher-order corrections might require reorganisation of the perturbative expansion. 


\paragraph{Going beyond high-temperature expansion}
Recently, there was a study that obtained a dimensionally reduced theory without relying on high-temperature expansion \cite{Navarrete:2025yxy}. 
Instead, the authors have applied a sophisticated technique called ``loop-tree duality'', used previously in the studies of hot-QCD. Moreover, the authors were able to obtain three-loop corrections within this framework and compared the results to the regular approach based on high-temperature dimensional reduction. 
This new framework may become a paradigm shift in the study of cosmological phase transitions, and it may be even more suitable for supercooled phase transitions in comparison to high-temperature dimensional reduction. 

\paragraph{Comparison to lattice}
Finally, we must acknowledge that our result for the nucleation rate comes with a certain degree of uncertainty. 
The size of NLO corrections in comparison to LO indicates a slow convergence of the soft expansion, or it may suggest that there is no convergence at all. However, encouraging results for other BSM theories can be found in refs.~\cite{Ekstedt:2024etx, Gould:2023ovu, Niemi:2020hto, Niemi:2024axp}. There, despite the slow convergence of perturbative expansion, eventually the perturbative results agree with the lattice.
In the case of classically scale-invariant theories, the answer to this question could be provided by a lattice study of 3d EFT in SU(2)cSM. As we have shown in this thesis, such EFT always contains a negative quartic coupling, which makes the potential unbounded from below. Therefore, there is no available lattice study of such a model. However, while this feature is problematic for simulating equilibrium thermodynamics, it is not a problem for studying nucleation rates alone. 

\paragraph{Numerical tools for computing gauge determinants}
All of the theoretical frameworks used in this thesis could be applied to any other phase transition in a BSM model. We have shown that at NLO, the largest uncertainty comes from the breakdown of the derivative expansion. This issue is solved by numerically calculating the gauge fluctuation determinant. There exists a ready numerical package that allows for the calculation of one-dimensional fluctuation determinants, such as the scalar determinant \cite{Ekstedt:2023sqc}. 
However, in the case of the gauge determinant, such a computation is non-trivial due to the mixing between gauge and Goldstone modes. It would then be a great contribution to the community if there were improved numerical tools that could compute such determinants as well.  

\paragraph{Parameter reconstruction}
In this thesis, we were guided by the principle of theoretical consistency, and we showed how, starting from the field-theoretic framework, we can obtain phenomenological predictions. However, it is also important to study this pipeline in a reverse strategy. Thus, it would be interesting to investigate the impact of theoretical uncertainties on the reconstruction of model parameters from the gravitational wave signal. In particular, one could conduct a similar study to ref.~\cite{Lewicki:2024xan}, but focusing on the case of the supercooled phase transition and conformal BSM models. 

\paragraph{Other phenomenological consequences} 
The theoretical framework described in this thesis can also be applied to obtain predictions for different phenomena. Over the past few years, supercooled phase transitions gained interest as a mechanism for generating primordial black holes, see e.g.~\cite{Lewicki:2023ioy, Lewicki:2024ghw, Lewicki:2024sfw, Gouttenoire:2023naa} and references therein. All of these studies relied on simplified approaches to computing the nucleation rate. The resulting theoretical uncertainties can propagate into the predictions of the abundance of such black holes, see ref.~\cite{Kierkla:2025vwp}.

Another interesting avenue is the possible explanation of GW observed by experiments studying Pulsar Timing Arrays (PTA). It has been shown that the GW signal from the first-order phase transition happening around the QCD scale serves as a prime candidate for explaining the data, see e.g.~\cite{Ellis:2023oxs, Afzal_2023, Madge:2023dxc}. Moreover, numerous studies have considered such a transition in a classically scale-invariant BSM scenario, see e.g.~\cite{Balan:2025uke, Costa:2025csj, Chatrchyan:2025wop, Goncalves:2025uwh}. However, most of these studies, too, relied on simplified methods. Thus, it would be interesting to see if the hypothesis that PTA data explained by cosmological PT still holds in more detailed computations, or if it is merely a mirage.

\appendix
\chapter{Beta functions for SU(2)cSM}
\label{app:betafunc_su2csm}

The $\beta$ functions for the scalar couplings $\lambda_{\field}$, $\lambda_{h\field}$, $\lambda_h$ and the new gauge coupling $g_X$ read~\cite{Khoze:2014xha, Carone:2013, Hambye:2013}
\begin{displaymath}
\begin{aligned}
\beta_{\lambda_h} &= \frac{1}{8\pi^2}\left[12\lambda_h^2+\lambda_{h\field}^2+\frac{\lambda_h}{2}\left(-9g_2^2-3g_1^{ 2}+12Y_t^2\right)+\frac{3g_2^4}{8}+\frac{3(g_2^2+g_1^{ 2})^2}{16}-3Y_t^4\right],\\
\beta_{\lambda_{h\field}} &= \frac{1}{8\pi^2}\left[6\lambda_h\lambda_{h\field}+2\lambda_{h\field}^2+6\lambda_{h\field}\lambda_{\field}+\frac{\lambda_{h\field}}{4}\left(-9g_2^2-3g_1^{ 2}+12Y_t^2-9g^2_{X}\right)\right], \\
\beta_{\lambda_{\field}} &= \frac{1}{8\pi^2}\left[\lambda_{h\field}^2+12\lambda_{\field}^2-\frac{9}{2}\lambda_{\field}g^2_{X}+\frac{9g^4_{X}}{16}\right],\\
\beta_{g_X} &= \frac{1}{16\pi^2}\left[-\frac{43}{6}g^3_{X} - \frac{1}{(4\pi)^2}\frac{259}{6}g^5_{X}\right],
\end{aligned}
\end{displaymath}
where $g_2$ is the SU(2) gauge coupling, $g_1$ the U(1) gauge coupling and $Y_t$ the top Yukawa coupling. The 4d renormalisation factor for the scalar field $\field$ is given by
$$
Z(t)=\exp\left(- \frac{1}{2}\int_0^t \dd x\ \gamma_{\field}(x)\right),
$$
where  $t=\log\frac{\mu}{\mu_0}$ and
$$
\gamma_{\field}(x)=-\frac{9 g_X^2(x)}{32 \pi ^2}.
$$
These $\beta$ functions are also obtained automatically when using {\tt DRalgo}~\cite{Ekstedt:2022bff} for dimensional reduction.

\chapter{Dimensional reduction for SU(2)cSM}
\label{sec:DR-details}

Customarily in the dimensional reduction literature, the scale of the phase transition is referred to as the ultrasoft scale and the corresponding EFT is the one where the soft scale temporal gauge field components are integrated out~\cite{Kajantie:1995dw}, along with other potential soft fields~\cite{Niemi:2018asa}.  As pointed out in ref.~\cite{Gould:2023ovu}, only in the near vicinity of a second-order phase transition, the scalar fields are expected to become ultrasoft. For first-order transitions, the fields driving the transition are expected to live either at the supersoft or the soft scales.
In this thesis, we have indeed organised perturbation theory by treating the nucleating field as soft and constructed the potential and nucleation EFT in the broken phase by integrating out the gauge fields, both spatial and temporal components on equal footing. 

\section{Relation to one-loop thermal effective potential}
\label{app:sec:compare3D4D}

One can confirm straightforwardly that eq.~\eqref{eq:V4HT} in the HT expansion is reproduced exactly by eq.~\eqref{eq:Veff-EFT-LO} when the matching relations (c.f.\ appendix~\ref{app:DR_matching}) are truncated as follows
\begin{align}
\lambda_{3} &= T \Big( \lambda_{\field} + \frac{1}{(4\pi)^2} \frac{3}{8} g^4_X (1 - \frac{3}{2} L_b) \Big), \label{eq:lambda3-LO}\\
g_{X,3}^2 &= g_X^2 T, \label{eq:gX3-LO}\\
h_3 &= \frac{1}{2} \gx^ T, \\ 
\label{eq:mDsq-lo}
m_{D,X}^2 &= \frac{5}{6} \gx^2 T^2, \\
\label{eq:m3sq-lo}
m^2_{3} &= \frac{3}{16} \gx^2   T^2,
\end{align}
and taking into account dimensional scalings $\vthree=\field/\sqrt{T}$ and $\VLO(\field, T) = T V_3$. Note that the Coleman-Weinberg term is captured by the $L_b$ term in $\lambda_3$, where $L_b = 2 \gamma_E - 2\log{[4\pi]} + \log{\left[\frac{\mu_4^2}{T^2} \right]}$ (see eq.~\eqref{eq:LbLf}). We note that while we simply took eq.~\eqref{eq:daisy} for leading daisy resummation from the literature, in eq.~\eqref{eq:Veff-EFT-LO} we actually derived it in the 3D EFT approach and we can clearly see the physics behind it: this term originates from the fact that hard modes screen the soft zero mode, and once this hard scale screening is accounted for by the soft mode mass at one-loop order, the EFT automatically resums this contribution to all orders. In the EFT, a one-loop computation with the resummed propagator is easy and two-loop diagrams are also straightforward, and we exploit this fully in the next section.

\section{Normalisation of the field}
\label{sec:rescaling-field-4D}

To define the theory and the potential, we begin by choosing the values of the mass $\MX$ of the $X$ boson and its coupling $g_X$. They are defined at scale $\mu_4=M_X$. 
As described in section~\ref{sec:su2csm_thermal_scale_running} (see also references~\cite{Kierkla:2022odc, Kierkla:2025vwp}), we can obtain the values of all the couplings at the electroweak scale, set by the mass $M_Z$. We define the theory at that scale, which is reflected in the choice of the reference scale $\mu_0 = \MZ$. 
As a consequence, the field $\field$ is defined at $\mu_4=M_Z$, i.e.\ at this scale it is canonically normalised (for a comprehensive discussion of the scale-dependence of scalar fields, see ref.~\cite{Andreassen_2015}). 
However, if we consider RG-improved effective potential, we would then evolve both the couplings and fields along their RG flows. 
Therefore, at some other scale, $\mu_4\neq\mu_0$, the field is no longer canonically normalised, as it is rescaled by the field renormalisation factor, $\sqrt{Z(t)}$. 
In the HT regime, as we have shown in section~\ref{sec:HT-LT-regimes}, the running is ``frozen'' at $\mu_4 \sim T$ and the normalisation of the field is given by $\sqrt{Z(\log\frac{\kappa T}{M_Z})}$, where $\kappa$ is some constant of order $\order{\pi}$.
Note that the usual bounce equation is derived from an action containing a canonically normalised scalar field. Therefore, this equation is not immediately applicable for the RG-improved potential evolved down to the thermal scale. 
One solution is simply re-deriving the bounce equation, so the $\sqrt{Z}$-factor would appear there as well. However, since we use the effective potential in the HT regime only to describe the tunnelling,  we can also redefine the field. Thus, we introduce a new field that is defined at the scale $\mu=\kappa T$ and is thus canonically normalised
\begin{align}
    \field_{\kappa T}=\sqrt{Z(\log{\frac{\kappa T}{M_Z}})}\field_{M_Z}.
\end{align}
In the main text, we do not differentiate between these fields, and we do not introduce extra subscripts on the field symbol $\field$ for simplicity of notation.%
Overall, the above discussion of $\sqrt{Z}$-factor may not seem so important, as $Z$ stays close to 1, within the perturbative regime of the theory. However, in the IR scales relevant for nucleation, it turns out that it affects the results visibly. Therefore, we take the $\sqrt{Z}$-factor into account in the ``4d computations''. 
This issue has not been appreciated in the literature, see for example ref.~\cite{Kierkla:2022odc}.

Finally, regarding the construction of 3d theory, we remind that it is constructed at the thermal scale $\mu_{4}=\kappa T$ (the values of the couplings at this scale are obtained by running from $\mu=M_Z$), thus the 3d field naturally is defined at this scale and needs no rescaling.

\section{Matching relations}
\label{app:DR_matching}
We perform next-to-leading (NLO) matching when integrating out the hard modes. This corresponds to one-loop matching for the couplings and two-loop matching for the masses. The scalar field couplings are given by
\begin{align}
\lamthree &= T \bigg( \lambda_{\varphi} + \frac{1}{(4\pi)^2} \Big( \frac{3}{8} \gx^4 + L_b ( -\frac{9}{16} \gx^4 - 12 \lambda_{\varphi} + \frac{9}{2} \gx^2\lambda_{\varphi} -  \lambda^2_{h\varphi} ) \Big) \bigg), \\
\gxthree^2 &= \gx^2T \Big(1 + \frac{1}{(4\pi)^2}\gx^2( \frac{2}{3} + \frac{43}{6}L_b) \Big),\label{eq:lambda3} \\
\hthree &= \frac{1}{2} \gx^2  T  \bigg(1 + \frac{1}{(4\pi)^2} \Big( \gx^2(\frac{17}{2} + \frac{43}{6}L_b) + 12 \lambda_{\varphi} \Big) \bigg), \\
\kappathree &= \frac{1}{(4\pi)^2} \frac{17}{3} \gx^4 T,
\end{align}
and the masses by
\begin{align}
\mDX^2  &= \frac{5}{6} \gx^2 T^2 \bigg( 1 + \frac{1}{(4\pi)^2} T^2 \Big( \gx^2 (\frac{69}{20} + \frac{43}{6}L_b) + \frac{2}{5}(3\lambda_\varphi  +  \lambda_{h\varphi}) \Big) \bigg),\label{eq:mDX} \\
\mthree^2 &= T^2 \bigg\{ \frac{3}{16} \gx^2+ \frac{1}{6}(3 \lambda_\varphi + \lambda_{h\varphi} ) \nonumber \\
& + \frac{1}{(4\pi)^2} \bigg( \frac{167}{96} g^4_{X} + \frac{1}{12}(3g^2_{2} + g^2_{1} + 3 L_f y^2_t) \lambda_{h\varphi} + \frac{3}{4} g^2_{X} \lambda_{\varphi} \nonumber \\
& + L_b \Big( -\frac{47}{32} g^4_{X} - \frac{9}{4} g^2_{X} \lambda_{\varphi} - \frac{1}{8} (9 g^2_{2} + 3g^2_{1} + 3 g^2_{X} + 6 y^2 + 8  \lambda_{h} + 8  \lambda_{\varphi} ) \lambda_{h\varphi} + \frac{1}{6} \lambda^2_{h\varphi}  \Big) \nonumber \\
& + \gamma \Big( \frac{81}{32} g^4_{X} + \frac{1}{2} (3g^2_{2} + g^2_{1}) \lambda_{h\varphi} - \lambda^2_{h\varphi} + \frac{9}{2} g^2_{X} \lambda_{\varphi} - 6 \lambda^2_{\varphi} \Big) \nonumber \\ 
& + \ln(A) \Big( -\frac{243}{2} g^4_{X} - 6(3g^2_{2} + g^2_{1})\lambda_{h\varphi} + 12 \lambda^2_{h\varphi} - 54 g^2_{X} \lambda_{2} + 72 \lambda^2_{\varphi} \Big)  \bigg) \bigg\} \nonumber \\
& + \frac{1}{(4\pi)^2} \ln\Big( \frac{\mu_3}{\mu_4} \Big) \bigg( -\frac{36}{16}g^4_{X,3}  + \frac{3}{2} \lambda^2_{VL8} - 6 g^2_{X,3} \lambda_{VL8}  \bigg),\label{eq:3DMass}
\end{align}
where
\begin{align}
\lambda_{h\varphi,3} & = \lambda_{h\varphi}T\bigg(1 + \frac{1}{(4\pi)^2} \Big( -3 y_t^2 L_f + L_b(\frac{3}{4}g_1^2 + \frac{9}{4} g_2^2 + \frac{9}{4}g_X^2 - 2(3 \lambda_h + \lambda_{h\varphi} + 3 \lambda_{\varphi}) ) \Big)\bigg), \\
g_{1,3} &= \sqrt{g_1^2 T - \frac{g_1^4T(L_b + 40 L_f)}{96 \pi^2}}, \\
g_{2,3} &= \sqrt{g_2^2 T + \frac{g_2^4 T(4 + 43 L_b - 24 L_f)}{96 \pi^2}}.
\end{align}
$L_b$ and $L_f$ are defined as
\begin{equation}\label{eq:LbLf}
L_b = 2 \gamma_E - 2\log{[4\pi]} + \log{\left[\frac{\mu_4^2}{T^2} \right]}, \qquad L_f = L_b + 4\log{[2]},
\end{equation}
with $\gamma_E$ the Euler-Mascheroni's constant and $A$ the Glaisher's constant.

The effective masses depend on the effective couplings between the scalar and the several temporal gauge modes, given by
\begin{align}
\lambda_{VL5} &= \frac{g_1^2 T \lambda_{h\varphi}}{8\pi^2}, \\
\lambda_{VL6} &= \frac{g_2^2 T \lambda_{h\varphi}}{8\pi^2}, \\
\lambda_{VL8} &= \frac{\gx^2 T (\gx^2(51 + 43 L_b) + 96 \pi^2 + 72 \lambda_\varphi)}{1928\pi^2},
\end{align}
where $\lambda_{VL5},\lambda_{VL6}$ denote couplings between $\varphi$ and the SM gauge fields, and $\lambda_{VL8}$ the coupling between $\varphi$ and the dark gauge field.

\chapter{Fluctuation determinants}
\section{One-loop fluctuation determinant via {\tt BubbleDet} }
\label{sec:app-NLO-rate}

We can compute the parts of the rate in eq.~\eqref{eq:Astat_NLOdet}
that involve no kinetic mixing between different fields
using {\tt BubbleDet}~\cite{Ekstedt:2023sqc}.
First, we obtain the bounce solution $\varphib ^{\rmii {LO}}$ and the corresponding action from the leading order potential $\VLO$ of the nucleation EFT.
We may use the bounce solver of {\tt CosmoTransitions}~\cite{Wainwright:2011} for finding the critical bubble.
Then the \path{BubbleConfig.fromCosmoTransitions} method can automatically obtain the relevant information about the bubble background. Then,
{\tt BubbleDet} computes the statistical part of the nucleation rate
using the Gel'fand-Yaglom method.
It returns 
\begin{equation}
	{\tt findDeterminant()} = \sum_i
	\left [
		\frac{{\rm dof}(d,s_i,n_i)}{2} \ln{\left| \frac{\det'(-\partial^2 + W_i(r))}{\det (-\partial^2 + W_i(\infty))}\right|} - \ln \mathcal I_i \mathcal V_i
	\right]
	 - \ln A_{\rm dyn}
   \,,
\end{equation}
where the sum runs over the contributing fields,
$d$ denotes the dimensions,
$s_i$ is the spin of the field and
$n_i$ is its internal degrees of freedom.
The remaining notation has been defined in section~\ref{sec:NLO_nuclrate_dets}, and note that here and in the following, we use $W$ to denote the non-derivative part of the operators $\mathcal O$.
Note that the contributions from all fields are dimensionless, except for the contribution from
the scalar undergoing the phase transition, which has mass dimension~3. 
It is also important to pass {\tt Thermal = true} to {\tt BubbleDet.BubbleDeterminant} in case of nucleating scalar to get the dynamical prefactor. 
The output of {\tt findDeterminant()} can then be included in the nucleation rate in the following way:
\begin{align}
\label{eq:GammaNLO}
	\Gamma_T = T^4 e^{-S_0 - \sum_i \left [\frac{{\rm dof}(d,s_i,n_i)}{2} \ln{\left| \frac{\det'(-\nabla^2 + W_i(r))}{\det (-\nabla^2 + W_i(\infty))}\right|} - \ln \mathcal I_i \mathcal V_i \right] - \ln A_{\rm dyn} - 4\ln T}
  ,
\end{align}
where $S_0$ corresponds here to the LO action. {\tt BubbleDet} also has the option to apply the derivative expansion.
This approximates the result of {\tt findDeterminant()} as
\begin{equation}
	S_1  = \int d^dx\left(V_{(1)}(\varphib) - V_{(1)}(\varphiF) \right) + \int d^dx \left
    (\frac 1 2 Z_{(1)}(\varphib) \nabla_\mu \varphib \nabla_\mu \varphib 
    \right)
  \,,
\end{equation}
where $S_1$ is the total one-loop correction. The first part is the LO contribution, and the second part is the NLO contribution. If the gradient expansion works well, this should reproduce the same result as the full fluctuation determinant. This could be illustrated for the fluctuation determinant of the temporal gauge mode, where the derivative expansion works reliably because the temporal gauge mode is always heavier than the nucleating scalar.

\section{Gel'fand-Yaglom method for fluctuation determinants}
\label{sec:GY}

Let us now review how the fluctuation determinant appearing in the nucleation rate can be computed. 
First, we will discuss the case where the determinant is a product of single-field fluctuation determinants.
This case was implemented in {\tt BubbleDet}. 
Then we discuss the case of multiple mixing fields, cf.\ section~\ref{sec:NLO_nuclrate_dets}.
To address mixing, we have implemented the approach discussed in~\cite{Ekstedt:2021kyx}, which we will review here.

\subsection{Fluctuation determinant of a single field}

First, we review the computation of ratios of fluctuation determinants of the form
\begin{align}
  &
	\frac{
			\mathcal {\rm det} \mathcal O_a(\varphib)
		}{
			 {\rm det} \mathcal O_a (\varphiF)
		}
    \,, &
    \mathcal O_a(\f) &= -\partial^2 + W(\varphi)
    \,.
\end{align}
Since the following discussion is not unique to thermal phase transitions,
we will denote the background field as $\f$ rather than $\phib$,
and we will use $\varphiF$ and $\varphib$ to denote the false vacuum and bounce solution respectively.
In the following, we drop the subscript $a$ to avoid notational clutter.
Note that for the current discussion, we are assuming that the fluctuation operator does not have zero modes.
This condition does not hold for the scalar field and Goldstone bosons, and we refer the reader to
e.g.~\cite{Baacke:1993, McKane:1995vp, Dunne:2005rt, Baacke:2008zx, Falco:2017ceh, Ekstedt:2021kyx, Ekstedt:2023sqc,Bhattacharya:2024chz}
for a description of how to deal with these zero modes.

We can use the spherical symmetry of the bounce solution to expand the eigenfunctions in the basis of spherical harmonics~\cite{Baacke:1993, Baacke:2008zx, Ekstedt:2023sqc}, and write the fluctuation determinant as
\begin{align}
  \label{eq:detPartialWave}
	\frac{
			\mathcal
      {\rm det} \mathcal O(\varphib(r))}{
			{\rm det} \mathcal O(\varphiF(r))
		}
  &=
	 \prod_{l=0}^\infty \left( \frac{
      {\rm det} \mathcal O^{l}(\varphib(r))}{
      {\rm det} \mathcal O^{l}(\varphiF(r))}\right)^{{\rm deg}(l)}
  \,,&
  {\rm deg}(l) &= \frac{(d+2l-2)\Gamma(d+l-2)}{\Gamma(d-l)\Gamma(l+1)}
  \,,
\end{align}
where 
\begin{align}
  \mathcal O^{l}(\f (r) ) &= -\partial_l^2 + W(\f (r))
  \,,&
  \partial_l^2 &= \partial_r^2 + \frac{d-1}{r}\partial_r - \frac{l(l+d-2)}{r^2}
  \,.
\end{align}
We are interested in the interval $r\in [0, \infty]$.
Then, the eigenfunctions $\psi_l$ of the fluctuation operators should satisfy the following boundary conditions
\begin{align}
  \label{eq:BCs}
  \psi_l(r)\rvert_{r=0}&=0
  \,, &
  \lim_{r \rightarrow \infty} \psi_l(r) &= 0
  \,.
\end{align} 

Rather than finding the infinite set of eigenvalues to evaluate the determinant for each value of $l$, we will make use of the Gel'fand-Yaglom theorem, which greatly simplifies the computation.
We quickly review it here.%
\footnote{%
  In fact, here we use the theorem in a form that is suitable when
  $W$ corresponds to a matrix, which was first introduced by
  Forman~\cite{Forman1987},
  while the Gel'fand-Yaglom theorem corresponds to the one-dimensional case.
}

We are interested in the ratio of determinants
\begin{equation}
	\frac{{\rm det}[-L_2 + W^{(1)}(r)]}{{\rm \det}[-L_2 + W^{(2)}(r)]}
  \,,
\end{equation}
where $L_2$ denotes a second order differential operator (as $\partial_l ^2 $ in our case).
For simplicity, we write the $W^{(i)}$ as a function of $r$,
rather than functions of solutions to the equation of motion.
The eigenfunctions $\psi^{(i)}_\lambda$ (the label $i$ denotes whether the eigenvalue corresponds to the operator with $W^{(1)}$ or $W^{(2)}$)
defined by 
\begin{align}
	\bigl(-L_2 + W^{(i)}(r)\bigr) \psi^{(i)}_\lambda = \lambda \psi^{(i)}_\lambda
  \,,
  \label{eq:eigenPsi}
\end{align}
satisfy some boundary conditions at $r = r_0,r_1$, which can be written in the following form
\begin{equation}
	M \begin{pmatrix} \psi^{(i)}_\lambda(r_0) \\ \dot \psi^{(i)}_\lambda(r_0) \end{pmatrix} + N\begin{pmatrix} \psi^{(i)}_\lambda(r_1) \\ \dot \psi^{(i)}_\lambda(r_1) \end{pmatrix} = 0
  \,,
\end{equation}
where $M,N$ are $2\times 2 $ matrices, i.e.\
two rows correspond to
two boundary conditions and two columns to the pair of
$(\psi^{(i)}(r), \partial \psi^{(i)}(r)) \equiv \Psi^{(i)}$.
In principle, $M$ and $N$ should be given a label $i$, but since they are identical for $i=1,2$, for our case of interest, we drop this here. 

Now, let $y^{(i)}_{n} (r)$ denote fundamental solutions of the homogeneous equation:
\begin{equation}
	\bigl(-L_2 + W^{(i)}(r)\bigr) y^{(i)}_n(r) = 0
  \,,
\end{equation}
where $n = 1,2$ runs over the number of fundamental solutions.
These fundamental solutions are characterised by their initial conditions:
\begin{align}
  y^{(i)}_1(0) &=1\,,& \dot y^{(i)}_1(0) &= 0\,,\\
	y^{(i)}_2(0) &=0\,,& \dot y^{(i)}_2(0) &= 1\,.
\end{align}
The number of fundamental solutions is 2 times the dimension of $W^{(i)}(r)$.
We can now define the matrix $Y$ containing all fundamental solutions and
their derivatives, and also use it to express the initial conditions:
\begin{align}
  Y^{(i)}(r) &=
	\begin{pmatrix}
		y^{(i)}_1(r) && y^{(i)}_2(r) \\
		\dot y^{(i)}_1(r) && \dot y^{(i)}_2(r)
	\end{pmatrix}
  \,,&
  Y^{(i)}(0)&=1.
\end{align}
Gel'fand-Yaglom theorem now states that the ratio of determinants of fluctuation operators can be obtained as
\begin{align}
	\frac{{\rm det}[-L_2 + W^{(1)}(r)]}{{\rm \det}[-L_2 + W^{(2)}(r)]} 
	= \frac{
			{\rm det}[M + N Y^{(1)}(r_1)]
		}{
			{\rm det}[M + NY^{(2)}(r_1)]
		}
  \,.
\end{align}
Let us now apply this result to the determinant for $\mathcal O^{l}$ operators with the boundary conditions as in eq.~\eqref{eq:BCs}. First, we see that the matrices $M$ and $N$ have the form
\begin{align}
    M &= \begin{pmatrix} 1 && 0 \\ 0 && 0 \end{pmatrix}
    \,, &
    N &= \begin{pmatrix} 0 && 0 \\ 1 && 0 \end{pmatrix}
    \,,
\end{align}
for $r_0=0$ and $r_1=\infty$.
Then we can obtain:
\begin{align}
	M+NY^{(i)}(\infty) =
  \begin{pmatrix} 1 && 0 \\ y^{(i)}_1(\infty) && y^{(i)}_2(\infty) \end{pmatrix}
  \,,
\end{align}
which leads to the simple expression 
\begin{equation}
	\det [ M+NY^{(i)}(\infty)] = y^{(i)}_2(\infty)
  \,.
\end{equation}
This gives the desired ratio of determinants simply by
\begin{align}
	\frac{{\rm det}[-\partial_l ^2 + W^{(1)}(r)]}{{\rm \det}[-\partial_l ^2 + W^{(2)}(r)]}
	= \frac{ y_2 ^{(1)}(\infty) }{ y_2 ^{(2)}(\infty) }
  \,.
\end{align}
This can directly be applied to the ratio of
determinants appearing in eq.~\eqref{eq:detPartialWave}.
The fundamental solution $y_2 ^{i}$ of the $i$-th operator we are interested in is defined as:
\begin{align}
  \bigl[-\partial_l ^2 + W^{(i)}(r)\bigr] y^{(i)}_2 (r) &= 0
  \,, &
  \mbox{with boundary conditions}
  \quad
  y_2 (0) &= 0
  \, \mbox{ and }
  \,
  \dot{y}_2(0) =1
  \,.
\end{align}
Thus, in a nutshell, Gel'fand--Yaglom theorem simplifies the problem of calculating a ratio of two determinants to solving two differential equations and taking infinity limits of the solutions.
In our application, $W^{(1)}$ corresponds to $W(\f_b)$ and $W^{(2)}$ to $W(\f_{\rmii{F}})$.
 Since $W^{(2)}$ is constant, one can solve for $y_2 ^{(2)}$ analytically, which further simplifies the numerical implementation. 
 
In {\tt Bubbledet}, the differential equations for the low-$l$ modes are solved numerically using Gel'fand--Yaglom theorem, and a WKB approximation is used to determine the solutions for large $l$.%
\footnote{%
  Technically {\tt Bubbledet} also uses the WKB approximation for low-$l$
  values to speed up the convergence of computations
  (see section 4.2 of~\cite{Ekstedt:2023sqc}).
}

\subsection{Generalisation for matrices -- Forman theorem}

Before delving into the more complex case of the mixing determinant, let us first briefly discuss the generalisation of
the theorem to higher dimensions ---
typically, cases where $\mathcal{O}$ is a matrix acting on multiple fields.
This generalisation was developed by Forman~\cite{Forman1987} and is known as
\textit{Forman theorem}, see e.g.~\cite{McKane:1995vp, Falco:2017ceh}.

Let us promote $\mathcal{O}$ to a $k \times k $ matrix (in the case relevant to us, $k = 3$).
We can also rewrite the boundary conditions of
 $\Psi^{(i)} = (
    \psi^{(i)}_1, \cdots ,
    \psi^{(i)}_k,
    \partial \psi^{(i)}_1 ,\cdots ,
    \partial \psi^{(i)}_k)$
 $=(\psi^{(i)}, \partial \psi^{(i)})_b$
 at points $r_0, r_1$ in 
terms of matrices $M, N$ of dimensions $2k \times 2k$
\begin{align}\label{eq:bc_M_N}
	M^{ab} \Psi^{(i)}_b(r_0) + N^{ab} \Psi^{(i)}_b(r_1)=0
  \,,
\end{align}
where $a,b = 1,\dots,2k$.

The $2k \times 2k$ matrix $Y^{(i)}$ is now given by
\begin{equation}
	Y^{(i)}(x) = \begin{pmatrix} \vec y_1^{(i)}(r) && \vec y_2^{(i)} (r) && \cdots && \vec y_{2r}^{(i)} (r)  \\
	\dot{\vec y}_1^{(i)}(r) && \dot{\vec y}_2^{(i)} (r) && \cdots && \dot{\vec y}_{2r}^{(i)} (r)
	 \end{pmatrix}
   \,,
\end{equation}
where the $\vec y_{1,2, \cdots k}^{(i)}(x)$ are $k$-component solutions to
$\mathcal O\, \vec{y}_{1,2, \cdots k}^{(i)} = 0$.
The initial conditions  are $Y^{(i)}(0)=1$. 
Then the ratio of determinants can be computed as:
\begin{align}
	\frac{{\rm set}[-L_2 + W^{(1)}(r)]}{{\rm set}[-L_2 + W^{(2)}(r)]} =
	\frac{
		\det \left[ M + N Y^{(1)}(\infty) \right]
	}{
		\det \left[ M + N Y^{(2)}(\infty) \right]
	}
  \,.
\end{align}

\subsection{Application to the Goldstone-vector determinant}
\label{sec:compDet}

The goal of this section is to illustrate the computation of the vector determinant $\det_{\rmii{$V$}}$, which (in $R_\xi$ gauge) is a product of the following terms,
cf.\ eq.~\eqref{eq:vecdet}:
\begin{align*}
  \rmdet_{\rmii{$X_0$}}\,,
  &&& \text{the determinant of the temporal gauge mode}\,,\nn
		\rmdet_{\rmii{$XG$}}\,,
  &&& \text{the determinant of mixing gauge and Goldstone modes}\,,\nn
		\rmdet_{\rmii{$X_T$}}\,,
  &&& \text{the determinant of transverse gauge modes}\,,\nn
		\rmdet_{\rmi{g}}\,,
  &&& \text{the determinant of the ghost modes}\,.
\end{align*}
The temporal gauge mode
does not kinetically mix with the other fields, and the determinant can be calculated using Gel'fand-Yaglom theorem, or using {\tt{BubbleDet}} by treating the $X_0$ as a massive scalar without zero modes. As we shall show in a moment, the transverse gauge modes do not mix with Goldstones and the determinant of
$\det_\rmii{$X_T$} $ reduces to a one-dimensional problem as well.
Let us now discuss how to obtain the remaining part, the vector-Goldstone determinant.

We shall follow the approach of~\cite{Ekstedt:2021kyx}.
We work explicitly in $d=3$.
First, we decompose the gauge and
Goldstone fields into partial waves in the following way:
\begin{align} \label{eq:gauge_goldstone_basis}
  X_\mu(x) &=\sum_{l=0}^{\infty}\left[ \aS(r) n_\mu+\aL(r) \frac{r}{\sqrt{l(l+1)}} 
		\partial_\mu+\aT(r) \epsilon_{\mu \nu \alpha} x_v \partial_\alpha\right] Y_{l m}(\phi, \theta)
  \,,
  \\[2mm]
  \Phi_{\rmii{$G$}}(x) &=\aG(r) Y_{l m}(\phi, \theta)
  \,,\qquad 
  n_\mu=\frac{x_\mu}{r}
    \,.
\end{align}
The transverse gauge fluctuation $\aT$ is independent of the others,
i.e.\ the corresponding fluctuation operator becomes
\begin{equation}
		\mathcal{O}^l _{\rmii{$X_T$}} (\varphib(r))  
	= \left(-\partial^2_l +  \WX(\varphib(r)) \right)
  \,,
\end{equation}
and thus can be computed using one-dimensional 
Gel'fand-Yaglom theorem. 
The ghost determinant, which follows from the ghost operator
\begin{align}
	\mathcal{O}^l_{\rmi{g}}  (\varphib(r))
	= \Bigl(-\partial^2_l + \frac{1}{\xi} \WX( \varphib(r)) \Bigr)
  \,,
\end{align}
can be made identical to the transverse mode operator by a gauge choice $\xi=1$.

The remaining part of the determinant can be written in
the $(\aS, \aL, \aG )$ basis.
In the 
$R_\xi$-gauge, the determinant decomposed in partial waves is 
\begin{equation}
	\operatorname{det}_{\rmii{$XG$}} = \left[ 
		\prod_{l=0}^{\infty}
    \frac{
      \det\mathcal{O}^l _{\rmii{$XG$}} ( \varphib(r) ) }{
      \det\mathcal{O}^l _{\rmii{$XG$}} ( \varphiF(r) ) }
	\right]^{-\frac12}
  \,.
\end{equation}
The fluctuation operator can be written in a matrix form
\begin{align}
  \mathcal{O}^l _{\rmii{$XG$}} =&
	\begin{bmatrix}
		-\partial^2_l +\frac{2}{r^2}+\WX &
		-2 \frac{L}{r^2} &
		( \WX ^\prime)^{\frac{1}{2}} -(\WX)^{\frac{1}{2}} \partial \\
		-2 \frac{L}{r^2} & -\partial^2_l+ \WX & -(\WX)^{\frac{1}{2}} \frac{L}{r} \\
		(\WX)^{\frac{1}{2}} \frac{2}{r}+2 ( \WX ^\prime)^{\frac{1}{2}} +(\WX)^{\frac{1}{2}} \partial_r & 
		-(\WX)^{\frac{1}{2}} \frac{L}{r^2} & -\partial^2_l+\WG
	\end{bmatrix}
  \nn[2mm]
	 &+\Bigl(1-\frac{1}{\xi}\Bigr)
	\begin{bmatrix}
		\partial_r^2+\frac{2}{r} \partial_r-\frac{2}{r^2} & -\frac{L}{r}\left(\partial_r-\frac{1}{r}\right) & 0 \\
		\frac{L}{r}\left(\frac{2}{r}+\partial_r\right) & -\frac{L}{r^2} & 0 \\
		0 & 0 & 0
	\end{bmatrix}
  \nn[2mm]
	 &+\frac{1}{\xi}
	\begin{bmatrix}
		0 & 0 & (\WX ^\prime)^{\frac{1}{2}}+(\WX)^{\frac{1}{2}} \partial_r \\
		0 & 0 & (\WX)^{\frac{1}{2}} \frac{L}{r} \\
		-(\WX)^{\frac{1}{2}} \partial_r-(\WX)^{\frac{1}{2}} \frac{2}{r} & (\WX)^{\frac{1}{2}} \frac{L}{r} & 
		\WX 
	\end{bmatrix}
  \,,
\end{align}
where we omitted the superscript on $W$,
but whenever relevant, the $W$-s will be distinguished by the argument
$\varphib$ or $\varphiF = 0$.
Additionally, we used
\begin{align}
  \WX &= 4 m_{\rmii{$X$},3}^2
  \,,&
  \WG &= \frac{1}{\varphib(r)} 
  \bigl( 
    \VLO [\varphib(r)]
  \bigr)'
  \,,&
  L^2 &= l(l+1)
  \,.
\end{align}
As noted in~\cite{Ekstedt:2021kyx}, the matrix greatly simplifies for the choice $\xi = 1$,
\begin{align}
	\mathcal{O}^l _{\rmii{$XG$}} = 
	\begin{bmatrix}
		-\partial_l^2+\frac{2}{r^2}+\WX & -2 \frac{L}{r^2} & 2 ( \WX ^\prime)^{\frac{1}{2}} \\
		-2 \frac{L}{r^2} & -\partial_l^2+\WX & 0 \\
		2  ( \WX ^\prime)^{\frac{1}{2}}& 0 & -\partial_l^2+\WG+\WX
	\end{bmatrix}.
\end{align}
Now, it is convenient to change the basis as
$\mathcal M^l \rightarrow \mathcal R^{-1} \mathcal M^l \mathcal R$  using the matrix:
\begin{align}
	\mathcal{R}=\frac{1}{\sqrt{2 l+1}}\left(\begin{array}{ccc}
		\sqrt{l} & -\sqrt{l+1} & 0 \\
		\sqrt{l+1} & \sqrt{l} & 0 \\
		0 & 0 & 1
	\end{array}\right).
\end{align}
The operator for vector-Goldstone fluctuations becomes:
\begin{align}
	\mathcal{O}^l _{\rmii{$XG$}} =
	\begin{bmatrix}
    -\nabla_l^{-}+\WX & 0 & \hphantom{+} 2  \sqrt{\frac{l}{2 l+1}} \WX\prime  \\
		0 & -\nabla_l^{+}+ \WX & - 2  \sqrt{\frac{l+1}{2 l+1}} \WX^\prime  \\
		2  \sqrt{\frac{l}{2 l+1}} \WX^\prime  & -2  \sqrt{\frac{l+1}{2 l+1}} \WX^\prime  & 
		-\partial_l^2+ \WG + \WX
	\end{bmatrix},
\end{align}
with 
\begin{equation}
	\nabla_l^{\pm} = \partial_r^2 + \frac{2}{r} \partial_r - \frac{(l \pm 1)(l + 1 \pm 1)}{r^2}
  \,.
\end{equation}
We will now discuss how to find the false-vacuum and bounce solutions for $l\neq 0$.
The partial waves $l=0$ correspond to zero modes.
A description of their treatment can be found in~\cite{Ekstedt:2021kyx}.

\subsubsection{Calculating the determinant of $\mathcal{O}^l _{\rmii{$XG$}}$ for the case $l\neq 0$}

As we have mentioned before, one can find the analytical expression for operators with constant $W$s. This is indeed the case for the determinant evaluated on the false vacuum solutions.
Now we shall derive these false-vacuum solutions and then show how to use them to obtain the final ratio of determinants.

\paragraph{Obtaining the false-vacuum solutions}

Consider the false-vacuum fluctuation operator, where $\varphiF = 0$:
\begin{align}
  \mathcal{O}^l _{\rmii{$XG$}} (\varphiF) =
	\begin{bmatrix}
		-\nabla_l^{-} & 0 & 0 \\
		0 & -\nabla_l^{+}  & 0 \\
		0 & 0 & -\partial_l^2+ \WG(0)
	\end{bmatrix}.
\end{align}
We see that all the terms proportional to $\WX$ are now zero, since gauge bosons do not have a mass in the false vacuum.
Now we want to find the analytical solutions in our basis for the false-vacuum case, i.e.\
$a_{\alpha;\rmii{F}} \equiv  (\aSfv, \aLfv, \aGfv)$,
and use them to calculate the desired ratio of determinants.
The solutions will be distinct for every $l$, but to avoid clutter, we drop the $l$-label.

We see that all of the operators have a similar form, and indeed, in practice, we need to solve only the equation for $\aGfv$
\begin{align}
	\left[
		-\partial_l ^2 + \WG(0)
	\right]
	\aGfv (r) = 0
  \,,
\end{align}
where $\WG(0) = \mF ^2$, with $\mF$ the mass of Goldstone bosons in the false vacuum.
Note,
that $\aGfv$ corresponds to the homogeneous solution that was denoted by $\psi_0$
before (cf.\ eq.~\eqref{eq:eigenPsi}).
Then the other solutions can be obtained by setting $\mF = 0$ and
$l\rightarrow l\pm1$. Let us thus focus on the $\aGfv(r)$
case.
We need to solve the  equation:
\begin{align}
	-\left[
    \partial_r^2
  + \frac{2}{r} \partial_r
  - \frac{l(l+1)}{r^2}
- \mF ^2 \right]  \aGfv(r) = 0
  \,,
\end{align}
with boundary conditions such that $a_{\rmii{$G$}}(r) \rightarrow r^l$ as $r \rightarrow 0$.
Note that other solutions also should obey that boundary condition.
We can now use a trick and write the
$\aGfv(r) = r^l \tilde{a}_{\rmii{$G$}}(r) $.
Then the equation becomes:
\begin{align}
	  r^{l-1} \left[
	  r \partial_r^2 + 2 (l+1)\partial_r - r \mF ^2
  \right]\tilde{a}_{\rmii{$G$}}(r) = 0
  \,,
\end{align}
and its solution is
\begin{align}
  \tilde{a}_{\rmii{$G$}}(r) &= 
	  C_1 r^{\frac{1}{2} (-2 l-1)} J_{\frac{1}{2} (2 l+1)}(-i \mF r)
	+ C_2 r^{\frac{1}{2} (-2 l-1)} Y_{\frac{1}{2} (2 l+1)}(-i \mF r)
  \,,\\[2mm] 
	\Rightarrow \aGfv(r) &= \frac{
      C_1 J_{l+\frac{1}{2}}(-i \mF r)
    + C_2 Y_{l+\frac{1}{2}}(-i \mF r)}{\sqrt{r}}
  \,,
\end{align}
where
$J_\beta$,
$Y_\beta$ denote Bessel functions of the first and second kind, respectively.

Let us now determine the values of both $C_1$ and $C_2$ from the boundary conditions.
First of all, the second term in $\aGfv$ blows up at zero, so we have to set $C_2 =0$.
Then, to determine $C_1$, first we use the relation
between
the Bessel function $J_\beta$ and
the modified Bessel function $I_\beta$:
\begin{align}
	J_\beta (i z) = i^\beta I_\beta (z).
\end{align}
Setting $z=-\mF r$ and
$\beta = l +\frac{1}{2}$ (note that $\beta$ is non-integer) and using
$I_\beta (-x) = e^{i\pi \beta} I_\beta (x) $,
we further get:
\begin{align}
	  J_{l +\frac{1}{2}} (-i \mF r) &=
    i^{l +\frac{1}{2}} I_{l +\frac{1}{2}}(-\mF r)
	= i^{l +\frac{1}{2}} e^{i\pi (l +\frac{1}{2}) } I_{l +\frac{1}{2}}( \mF r)
	= i^{l+\frac{3}{2}} (-1)^l I_{l +\frac{1}{2}}( \mF r)
  \,.
\end{align}
Since the solution for $\aGfv(r)$ then takes the form
\begin{eqnarray}
  \aGfv(r) &=&
    \frac{C_1 i^{l+\frac{3}{2}} (-1)^l I_{l+\frac{1}{2}}(\mF r)}{\sqrt{r}}
  \nn &\stackrel{r\to0}{=}&
    \frac{C_1 (-1)^l i^{l+\frac{3}{2}} 2^{-l-\frac{1}{2}} (m r)^{l+\frac{1}{2}}}{\sqrt{r}
	\Gamma \left(l+\frac{3}{2}\right)} + \mathcal{O}(r^2)
  \stackrel{!}{=} r^l
   \,,
\end{eqnarray}
we could find $C_1$
by first expanding the solution in a series around $r=0$
and second using the boundary condition $\aGfv (r\rightarrow0)=r^l$,
to obtain
\begin{align}
	C_1 = (-1)^{\frac{5}{4}-l} i^{-l} (2r)^{l+\frac{1}{2}}  \Gamma
	\Bigl(l+\frac{3}{2}\Bigr) (m r)^{-l-\frac{1}{2}}
  \,.
\end{align}
Finally, by collecting all pieces together, the full solution becomes:
\begin{align}
	\aGfv(r) = \frac{2^{l+\frac{1}{2}} \mF ^{-l-\frac{1}{2}} \Gamma \left(l+\frac{3}{2}\right)
	I_{l+\frac{1}{2}}(\mF r)}{\sqrt{r}}
  \,.
\end{align}

As mentioned at the beginning, the remaining solutions can be easily obtained by expanding around $\mF=0$ and setting
$l\rightarrow l\pm1$:
\begin{align}
  \aSfv(r) &= r^{l-1}
  \,,&
  \aLfv(r) &= r^{l+1}
  \,.
\end{align}

\paragraph{Obtaining the solutions for fluctuations around the bounce}

Our goal now is to find fundamental solutions
$y$, which we can then use to compute the determinant.
First, the set of homogeneous equations for the fluctuation operator of our interest is
\begin{align}
	\mathcal{O}^l _{\rmii{$XG$}} (\varphib) a^l _{\alpha} =
	\begin{bmatrix}
		  - \nabla_l^{-}+\WX & 0 & \hphantom{+} 2 \sqrt{\frac{l}{2 l+1}} \WX ^\prime
    \\
		0 & -\nabla_l^{+}+ \WX & -2 \sqrt{\frac{l+1}{2 l+1}} \WX^\prime
    \\
    2  \sqrt{\frac{l}{2 l+1}} \WX^\prime  &
    - 2\sqrt{\frac{l+1}{2 l+1}} \WX^\prime &
		- \partial^2_l + \WG + \WX
	\end{bmatrix}
	\begin{bmatrix}
		\aS ^l \\ \aL^l \\ \aG^l
	\end{bmatrix}
	= 0
  \,.
\end{align}
Note,
that again the solutions $a^l_\alpha$ correspond to the homogeneous solutions that we denoted by $\psi_0$ above (cf.\ eq.~\eqref{eq:eigenPsi}).
As we want to calculate the ratio of determinants associated with the bounce and false vacuum solutions, it will be useful to consider the solutions normalised by the false vacuum ones
\begin{align}
	T^l _\alpha (r) \equiv
	\frac{
		 a^l _{\alpha; b} (r)
	}{
    a^l _{\alpha;\rmii{F}} (r)
	}
  \,.
\end{align}
As before, for $a_{\alpha; b} ^l$, we impose boundary conditions such that
$a_{\alpha;b} ^l (r) \rightarrow r^l$ as $r \rightarrow 0$.
Thus, the $T_\alpha ^l$ obey the following boundary conditions:
\begin{align}
  \partial_r T^l_\alpha(0) &= 0
  \,,&
  T^l_\alpha(\infty) &= 0
  \,.
\end{align}

As in the previous one-dimensional Gel'fand-Yaglom example,
we pick $r_0=0$ and $r_1 = \infty$ and express the boundary conditions in terms of
$M$ and $N$ matrices, which in our $d=3$ case are $6\times 6$
dimensional:
\begin{align}
	\underbrace{
		\begin{bmatrix}
			0 & 0 & 0 & 0 & 0 & 0 \\
			0 & 0 & 0 & 0 & 0 & 0 \\
			0 & 0 & 0 & 0 & 0 & 0 \\
			0 & 0 & 0 & 1 & 0 & 0 \\
			0 & 0 & 0 & 0 & 1 & 0 \\
			0 & 0 & 0 & 0 & 0 & 1 \\
		\end{bmatrix}
	}_{M}
	\begin{bmatrix}
		\TS^l (0) \\ \TL^l(0) \\ \TG^l(0) \\ \partial \TS^l(0) \\ \partial \TL^l(0) \\ \partial \TG^l(0)
	\end{bmatrix}
	+ \underbrace{
		\begin{bmatrix}
			1 & 0 & 0 & 0 & 0 & 0 \\
			0 & 1 & 0 & 0 & 0 & 0 \\
			0 & 0 & 1 & 0 & 0 & 0 \\
			0 & 0 & 0 & 0 & 0 & 0 \\
			0 & 0 & 0 & 0 & 0 & 0 \\
			0 & 0 & 0 & 0 & 0 & 0 \\
		\end{bmatrix}
	}_{N}
	\begin{bmatrix}
		\TS^l(\infty) \\
    \TL^l(\infty) \\
    \TG^l(\infty) \\
    \partial \TS^l(\infty) \\
    \partial \TL^l(\infty) \\ 
		\partial \TG^l(\infty)
	\end{bmatrix} 
	=0
  \,.
\end{align}
The last necessary ingredient is the matrix $Y$ containing fundamental solutions.
For a $3\times3$ fluctuation
operator there exist $6$ fundamental solutions $y^i = (\yS, \yL, \yG)^i$ which are defined as:
\begin{align}
	Y(r) = 
	\begin{bmatrix}
		\yS^1 & \yS^2 &\dots & \yS^6 \\ 
		\yL^1 & \yL^2 & \dots & \yL^6 \\
		\yG^1 & \yG^2 & \dots & \yG^6 \\
		\partial \yS^1 & \partial \yS^2 & \dots & \partial \yS^6 \\ 
		\partial \yL^1 & \partial \yL^2 & \dots & \partial \yL^6 \\
		\partial \yG^1 & \partial \yG^2 & \dots & \partial \yG^6 
	\end{bmatrix}
	\xrightarrow{r\rightarrow0}  1
  \,.
\end{align}
Now, according to Forman's theorem, we need to compute the following combination of fundamental solutions:
\begin{align}
	\label{eq:detVG_final_detY}
	\det[ M+NY(\infty)] &= 
	\begin{vmatrix}
		\yS^1 & \yS^2 & \yS^3 & 0 & 0 & 0 \\
		\yL^1& \yL^2 & \yL^3 & 0 & 0 & 0 \\
		\yG^1 & \yG^2 & \yG^3 & 0 & 0 & 0 \\
		0 & 0 & 0 & 1 & 0 & 0 \\
		0 & 0 & 0 & 0 & 1 & 0 \\
		0 & 0 & 0 & 0 & 0 & 1 \\
	\end{vmatrix} _{r\rightarrow \infty}
  \\[2mm]
	&= \Bigl[
      \yS^1 \left( \yL^2 \yG^3 - \yL^3 \yG^2 \right)
	  - \yS^2 \left( \yL^1 \yG^3 - \yL^3 \yG^1 \right)
	  + \yS^3 \left( \yL^1 \yG^2 - \yL^2 \yG^1 \right)
  \Bigr]_{r\rightarrow \infty}
  \,.
  \nonumber
\end{align}
Thus, we see that for the $l$-th operator, we need to
obtain three sets of solutions $(\yS, \yL, \yG)$, and to find the ratio of fluctuation determinants, simply evaluate
the expression~\eqref{eq:detVG_final_detY}.

\paragraph{Equations for fundamental solutions}
We want to find the fundamental solutions associated with equations for $T^l$ that we have introduced before. Using the false-vacuum solutions $a^l_{\alpha;\rmii{F}}$, we then write the equations as follows
\begin{align}
    \bigl[-\nabla_l^{-}+\WX \bigr] \TS^l
  + \biggl[2  \sqrt{\frac{l}{2 l+1}} \WX ^\prime \biggr]
  \biggl(\frac{\aGfv^l }{\aSfv^l}\biggr) \TG^l &= 0
    \,, \\[2mm]
    \bigl[-\nabla_l^{+}+ \WX \bigr] \TL^l
	+ \biggl[-2  \sqrt{\frac{l+1}{2 l+1}} \WX^\prime \biggr]
  \biggl(\frac{ \aGfv^l}{\aLfv^l}\biggr)  \TG^l &= 0
    \,, \\[2mm]
    \bigl[-\partial_l^2+ \WG + \WX \bigr]
    \biggl(\frac{\aGfv^l }{r^l} \biggr) \TG^l
	+ \biggl[2 \sqrt{\frac{l}{2 l+1}} \WX^\prime \biggr]
    \biggl(\frac{\aSfv ^l}{r^l} \biggr)
		\TS^l
    &
    \nn
  - \biggl[2 \sqrt{\frac{l+1}{2 l+1}} \WX^\prime \biggr]
    \biggl(\frac{\aLfv^l}{r^l} \biggr)
    \TL^l &= 0
    \,.
\end{align}
These equations can be solved numerically, and by choosing the initial conditions, we can obtain the three sets of
fundamental solutions that we need:
\begin{align}
  y^1(r) &=
	\begin{bmatrix}
		\yS^1(r) \\ \yL^1(r) \\ \yG^1(r)
	\end{bmatrix}=
	\begin{bmatrix}
		\TS^l(r)\\ \TL^l(r)\\ \TG^l(r)
	\end{bmatrix}^1
  \,, &
  \mbox{with }
	\begin{bmatrix}
		\TS^l(0)\\ \TL^l(0)\\ \TG^l(0)
	\end{bmatrix}^1
  &=
	\begin{bmatrix}
		1\\ 0\\ 0
	\end{bmatrix}
  &
  \mbox{ and }
	\begin{bmatrix}
		\partial \TS^l(0)\\ \partial\TL^l(0)\\ \partial\TG^l(0)
	\end{bmatrix}^1
  &=
	\begin{bmatrix}
		0\\ 0\\ 0
	\end{bmatrix}
  \,,\\[2mm]
  y^2(r) &=
	\begin{bmatrix}
		\yS^2(r) \\ \yL^2(r) \\ \yG^2(r)
	\end{bmatrix}=
	\begin{bmatrix}
		\TS^l(r)\\ \TL^l(r)\\ \TG^l(r)
	\end{bmatrix}^2
  \,, &
  \mbox{with }
	\begin{bmatrix}
		\TS^l(0)\\ \TL^l(0)\\ \TG^l(0)
	\end{bmatrix}^2
  &=
	\begin{bmatrix}
		0\\ 1\\ 0
	\end{bmatrix}
  &
	\mbox{and }
	\begin{bmatrix}
		\partial \TS^l(0)\\ \partial\TL^l(0)\\ \partial\TG^l(0)
	\end{bmatrix}^1
  &=
	\begin{bmatrix}
		0\\ 0\\ 0
	\end{bmatrix}
  \,,\\[2mm]
  y^3(r) &= \begin{bmatrix}
		\yS^3(r) \\ \yL^3(r) \\ \yG^3(r)
	\end{bmatrix}=
	\begin{bmatrix}
		\TS^l(r)\\ \TL^l(r)\\ \TG^l(r)
	\end{bmatrix}^2
  \,, &
  \mbox{with }
	\begin{bmatrix}
		\TS^l(0)\\ \TL^l(0)\\ \TG^l(0)
	\end{bmatrix}^3
  &=
	\begin{bmatrix}
		0\\ 0\\ 1
	\end{bmatrix}
  &
	\mbox{and }
	\begin{bmatrix}
		\partial \TS^l(0)\\ \partial\TL^l(0)\\ \partial\TG^l(0)
	\end{bmatrix}^1
  &=
	\begin{bmatrix}
		0\\ 0\\ 0
	\end{bmatrix}
  \,.
\end{align}
After finding these solutions numerically, all we need to do is to evaluate their
$r\rightarrow\infty$
limits and use the expression from eq.~\eqref{eq:detVG_final_detY}.
Thus, the final result for the $l$-th ratio of vector-Goldstone mixing modes fluctuation determinants is:
\begin{align}
  \frac{ \det\mathcal{O}^l _{\rmii{$XG$}} (v_{3,b}) }{ \det\mathcal{O}^l _{\rmii{$XG$}} (v_{3,\rmii{F}}) } &=
	\frac{
		\rm{Det} \mathcal{M}^l
	}{
    \rm{Det} \mathcal{M}^l _\rmii{FV} 
	}
  \nn &
	=
	 [\yS^1 \left( \yL^2 \yG^3 - \yL^3 \yG^2 \right)
	- \yS^2 \left( \yL^1 \yG^3 - \yL^3 \yG^1 \right)
	+ \yS^3 \left( \yL^1 \yG^2 - \yL^2 \yG^1 \right)]_{r\rightarrow \infty}
  \,.
  \nonumber
\end{align}

\subsection{One-loop vector contribution to the effective action}

Knowing how to obtain the value of the gauge-Goldstone determinant, we now want to explicitly write how it can be used to calculate the nucleation rate.
The functional determinant we have obtained is a one-loop (spatial) gauge contribution to the effective action. It is a natural choice to include it in the exponential of the nucleation rate, so the rate in eq.~\eqref{eq: Gamma [NLO det]} would contain the following term:
\begin{equation}
	\GammaT^{[\rmii{NLO det}]} \sim
	e^{
		-\SDRLO[\varphib]
    -  \ln \detX
	}
  \,.
\end{equation}
Then, taking together all the results from previous sections, we arrive at the following expression for the vector-Goldstone determinant:
\begin{align}
  \ln \detX =
	\underbrace{
		\ln\left( -2 \phi _\infty \varphi_b (r=0)  \right)
	}_{l=0} +
	\sum^{\infty } _{l=1} 
	(2 l +1)
	\biggl(
		\ln
			\frac{ \det\mathcal{O}^l _{\rmii{$XG$}} (\varphib) }{ \det\mathcal{O}^l _{\rmii{$XG$}} (\varphi_{\rmii{F}}) }
		- \ln
			\frac{ \det\mathcal{O}^l _{\rmii{$X_T$}} (\varphib) }{ \det\mathcal{O}^l _{\rmii{$X_T$}}(\varphi_{\rmii{F}}) }
	\biggr)
\,  ,
\end{align}
where we have used the fact that the ghost contribution is minus 2 times the contribution from the transverse gauge mode. The zero mode is contained in the $l=0$ term in the partial wave decomposition, and the first term in the expression  above can be evaluated by using
{\tt BubbleDet} and the function
{\tt findLogPhiInfinity()};
for details see~\cite{Ekstedt:2023sqc}. 

In practice, the sum over $l$ is evaluated up to some finite $L_{\rm{max}}$ and then one uses a WKB approximation for
$l > L_{\rm{max}}$~\cite{Ekstedt:2023sqc, Ekstedt:2021kyx}.%
\footnote{See~\cite{Baratella:2025dum} for recent progress in regularising the functional determinants in $D=4$.}
Here, we write the leading order terms explicitly:
\begin{align}
	\ln \rmdet_{\rmii{$X$}} \xrightarrow{l \gg 1}
  -3 \int {\rm d}r~r ( \Delta \WX(r))
  - \int {\rm d}r~r ( \Delta \WG(r) -\Delta \WX(r) )
	+ \mathcal{O}(l^{-3}) \, ,
\end{align}
where $\Delta W_a(r) \equiv W_a(r) - W_a(\infty)$.

\backmatter

\addcontentsline{toc}{chapter}{Bibliography}
\printbibliography

\end{document}